\documentclass[12pt]{book}
\usepackage[english]{babel}
\usepackage[utf8]{inputenc}
\usepackage[OT1]{fontenc}
\usepackage{natbib}
\usepackage{pdfpages}
\usepackage{amsfonts, amsmath, amsthm, amssymb}
\usepackage{graphicx}
\usepackage[font=small,labelfont=bf]{caption,subcaption}
\usepackage{listings}
\usepackage[margin=3cm]{geometry}
\usepackage{xcolor}
\usepackage{physics}
\usepackage{tabularx}
\usepackage{hyperref}
\usepackage{fancyhdr}
\usepackage{float}
\makeatletter
\def\cleardoublepage{\clearpage\if@twoside \ifodd\c@page\else
  \hbox{}\thispagestyle{empty}\newpage\if@twocolumn\hbox{}\newpage\fi\fi\fi}
\makeatother

\title{Effects of energetic particles produced by magnetic reconnection on discs of young stars}
\date{\today}
\author{Brunn Valentin}
\begin{document}
%\maketitle 
\includepdf[pages=-]{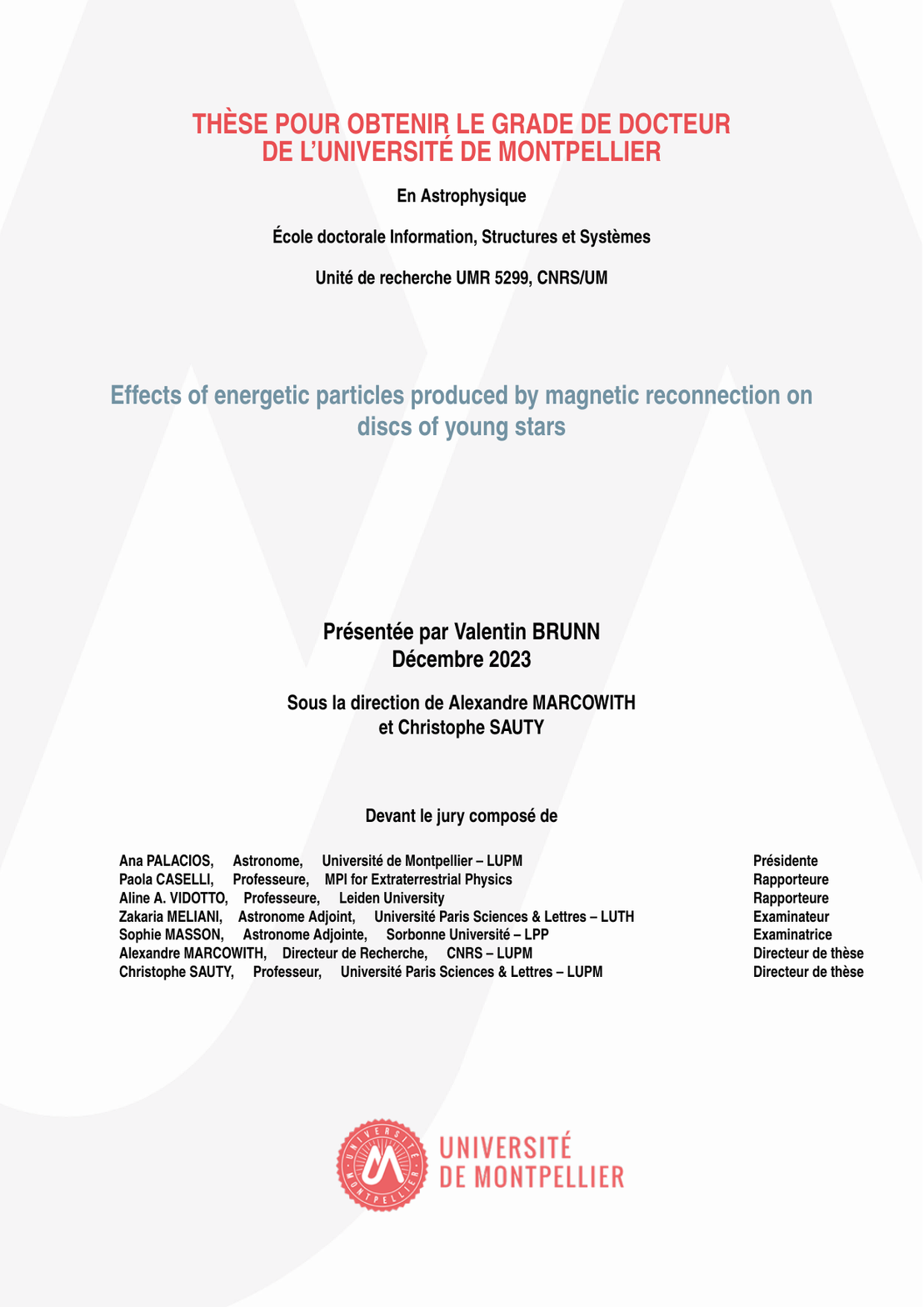} 
\newpage

\chapter*{Abstract}
{\small
T Tauri stars, young analogues of Sun-like stars, are surrounded by protoplanetary discs of dust and gas. These stars and their discs are crucial for understanding stellar evolution and the formation of planets in low-mass systems. These stars exhibit significant variability, notably emitting intense X-ray flares due to magnetic reconnection events. During these events, magnetic energy is converted into kinetic energy of particles. Some of these particles then heat the plasma of the underlying chromosphere to several tens of millions of degrees, emitting the observed X-rays. A portion of particles are thought to escape the chromosphere to interact with the surrounding circumstellar environment. The question is what impact do particles produced by magnetic reconnection events have on the discs of young stars.

The complex characteristics of protoplanetary discs around T Tauri stars require an interdisciplinary strategy to enhance our understanding of these objects. This thesis contribute to establish a framework combining observational methodologies, chemical and dynamic models of protoplanetary discs, and the mechanics of acceleration and transport of energetic particles. This synergy aims to demonstrate their impact on the dynamics and chemistry of the disc and possibly on its associated jets.

We first introduce the modelling of T Tauri stars and their discs, taking into account current observational constraints, such as the mass distribution of the disc and their thermal structure. Then, we focus on the role of ionisation in disc dynamics, including its sources. Not only from standard sources like stellar radiation and galactic cosmic rays but also from non-thermal ionisation due to magnetic reconnection events. We then examine energetic particles accelerated during magnetic reconnection events in T Tauri flares as an alternative source of ionisation, requiring the construction of a theoretical model based on solar flares due to the amplified magnetic and luminous properties of young stars. Next, we analyse the transport of energetic particles in the accretion disc, introducing two transport models based on the particle column density. Then, we present a study using the thermochemical ProDiMO code, revealing that particles from magnetic reconnection events could significantly contribute to the ionisation of the inner disc. Finally, we present a complementary study taking into account temporal factors, showing that considering these particles could increase the ionisation rate as well as viscosity, accretion rate, volumetric heating rate, and chemical complexity of inner protoplanetary discs as well as the launching mechanism of winds and jets.

This thesis demonstrates that magnetic reconnection events are fundamental to understanding the chemistry and dynamics of inner protoplanetary discs.}
\chapter*{Résumé}
{\small
Les étoiles T Tauri, jeunes analogues d'étoiles similaires au Soleil, sont entourées de disques protoplanétaires de poussière et de gaz. Ces étoiles et leurs disques sont essentiels pour comprendre l'évolution stellaire et la formation des planètes dans les systèmes de faible masse. Ces étoiles présentent une forte variabilité, émettant notamment d'intenses éruptions de rayons X dues à des événements de reconnexion magnétique. Lors de ces événements, l'énergie magnétique est transformée en énergie cinétique des particules. Certaines de ces particules chauffent ensuite le plasma de la chromosphère sous-jacente à plusieurs dizaines de millions de degrés, émettant les rayons X observés. D'autres particules sont censées s'échapper de la chromosphère pour interagir avec l'environnement circumstellaire environnant. La question est de savoir quel est l'impact des particules produites par des événements de reconnexion magnétique sur les disques des jeunes étoiles.

Les caractéristiques complexes des disques protoplanétaires autour des étoiles T Tauri nécessitent une stratégie interdisciplinaire pour améliorer notre compréhension de ces objets. Cette thèse a contribué à construire un cadre combinant des méthodologies observationnelles, des modèles chimiques et dynamiques de disques protoplanétaires, et la mécanique de l'accélération et du transport des particules énergétiques. Cette synergie vise à démontrer leur impact sur la dynamique et la chimie du disque et éventuellement sur ses jets associés.

Nous introduisons d'abord la modélisation des étoiles T Tauri et de leurs disques, en tenant compte des contraintes observationnelles actuelles, comme la distribution de masse du disque et leur structure thermique. Ensuite, nous examinons en profondeur le rôle de l'ionisation dans la dynamique du disque, y compris son origine. Non seulement par des sources standard comme le rayonnement stellaire et les rayons cosmiques galactiques mais aussi par l'ionisation non thermique due aux événements de reconnexion magnétique. Nous examinons ensuite les événements de reconnexion magnétique dans les éruptions T Tauri comme une source d'ionisation alternative, nécessitant des considérations théoriques distinctes des flares solaire en raison de leurs propriétés magnétiques et lumineuses décuplés. Puis, nous analysons le transport des particules énergétiques dans le disque d'accrétion, introduisant deux modèles de transport basés sur la densité de colonne des particules. Ensuite, nous présentons une étude utilisant le code ProDiMO, révélant que les particules issues d'événements de reconnexion magnétique pourraient contribuer significativement à l'ionisation du disque interne. Enfin, nous présentons une étude complémentaire prenant en compte des facteurs temporels, montrant que la prise en compte de ces particules pourrait augmenter le taux d'ionisation ainsi que la viscosité, le taux d'accrétion, le taux de chauffage volumétrique et la complexité chimique des disques protoplanétaires internes.

Cette thèse montre que les événements de reconnexion magnétique sont fondamentaux pour comprendre la chimie et la dynamique des disques protoplanétaires internes.}

\chapter*{Remerciements}
Je tiens d’abord à exprimer ma profonde reconnaissance à mes directeurs, Alexandre et Christophe, pour les innombrables échanges enrichissants, pour m'avoir éclairé et soutenu tout au long de cette aventure qu'a été ma thèse. Leur engagement envers mon travail et leur soutien constant ont été une source d'inspiration inestimable. Démarrer cette thèse en plein confinement a présenté des défis uniques, mais ils ont été des piliers de force et de réconfort dans ces moments d'incertitude. Je leur suis particulièrement reconnaissant pour la liberté de recherche qu'ils m'ont accordée, me permettant de suivre les sentiers que je choisissais, tout en sachant qu'ils étaient là, prêts à m'offrir un soutien infaillible face à chaque nouvelle interrogation qui m'animait. Leur capacité à insuffler confiance et motivation dans les périodes difficiles a été un moteur essentiel à ma persévérance. Leur exigence académique a été un cadeau précieux, me poussant à dépasser mes limites et à atteindre une qualité de travail que je n'aurais jamais cru possible. Grâce à leur insistance sur la rigueur méthodologique, la précision scientifique et la clarté de l'expression, ce manuscrit est devenu une source de fierté immense pour moi. Je leur suis tellement reconnaissant pour tout cela. Merci pour tout, infiniment.

Je tiens également à exprimer ma profonde gratitude envers les enseignants du département de physique de Montpellier, qui ont jalonné et enrichi mon parcours universitaire dès ma première année de licence jusqu'à l'achèvement de mon master. Leur inspiration a été un phare dans mon voyage académique, guidant chacun de mes pas avec sagesse et passion. Devenus collègues au LUPM, Bertrand, Eric J., et notre cher disparu Eric N., ont été des mentors, leur influence touche à l'essence même de ma passion pour l’astrophysique.

Je tiens à remercier Julien M. et Vincent, pour nos échanges scientifiques tout au long de ma thèse mais aussi pour leur humanisme profond. Ils m’ont inspiré par leur engagement social résolu et leur détermination à promouvoir les valeurs de l'université et de la science, œuvrant sans relâche pour un savoir universellement accessible. À leurs côtés, je me dresse fièrement pour défendre ces principes fondamentaux face aux décisions qui menacent de saper les services publics et l’université. Nous partageons une vision où la science et l'astrophysique, embrassent pleinement leurs rôles politique, vers un avenir où le savoir éclaire et libère les esprits.

Je tiens à adresser mes remerciements chaleureux à Chadi, mon "frère de thèse", ainsi qu'à Nathan, Jay, Karim, et Claire. Vous avez été les collègues parfaits, non seulement pour les moments de réflexion mais aussi pour votre soutien et ces précieuses pauses d'évasion au sein du laboratoire. Votre présence a illuminé cette aventure, transformant les défis en moments de partage et d'apprentissage. 

Je suis rempli d'une joie immense et d'une gratitude débordante en pensant à tous mes amis qui ont rendu mon aventure doctorale absolument merveilleuse ! Grâce à eux, ces trois années de thèse ont été un véritable festival de bonheur. Les moments partagés sous le même toit à la colocation de l'Avenue de Lodève sont gravés dans ma mémoire : notre équipe incroyable, Rémi, Margaux, Johann, Anaïs, Pierre, Simon, Marius, Caroline, Aziliz, Justine, nos voisins Tanguy, Lisa, Laure, Élise, Lucas, Camille, toutes celles et ceux qui sont venus et repartis et particulièrement Oscar, vous avez tous contribué à cette fabuleuse épopée !

Et Anaëlle, quelle force tu as été pour moi ! Une véritable source d'inspiration, tu as illuminé les derniers moments de ma thèse. A mes côtés, ta vitalité m'a insufflé l'énergie pour aborder ce travail avec sérénité et bien-être. Merci du fond du cœur.

Je souhaite exprimer ma profonde gratitude envers mes parents, qui ont été un pilier constant de soutien tout au long de mon parcours scolaire et académique, culminant avec la réalisation de ce manuscrit. Ils ont éveillé en moi la joie de la curiosité et la découverte et la valeur du travail accompli dans le bien-être. Leur exemple de résilience et de persévérance est une source d'inspiration constante et leur bienveillance et leur intérêt pour mes projets m'ont inspiré la confiance nécessaire pour me lancer dans cette aventure avec courage. Pour tout cela et pour l'amour qu'ils m'ont donné, je leur suis éternellement reconnaissant.

Je souhaite terminer en remerciant ma sœur. Merci d'avoir été à mes côtés tout au long de la rédaction de ce manuscrit, de m'avoir soutenu quotidiennement. Je n'aurais pas pu imaginer un meilleur endroit pour rédiger ma thèse que dans notre colocation entre frère et sœur. Ton écoute, ta patience et ta gentillesse ont été d'un soutien inestimable, me guidant à travers cette aventure académique. Tu as été la force, l'amour et la sagesse qui m'ont accompagné. Aujourd'hui, alors que j'écris les derniers mots de ma thèse, tu commences à écrire la tienne. Je serai là pour toi, prêt à te soutenir dans tes décisions, comme tu l'as fait pour moi. Ma reconnaissance pour tout ce que tu as fait pour moi est sans limites.

Merci à toutes et à tous, cette thèse vous est dédiée.

\newpage
\tableofcontents
\newpage
\chapter*{Introduction}
T Tauri stars, considered as the young counterparts of Sun-like stars, are subject to long lasting questions in Astrophysics. These celestial bodies are young stellar objects surrounded by a protoplanetary discs of dust and gas. T Tauri stars and their surrounding discs hold the clues to the processes that govern both stellar evolution and planet formation in low mass stellar systems. As these stars are extremely active, it is the interplay between accretion, ejection, magnetism, and chemical interactions that is the focal point for our study. This accretion-ejection process is not a simple, smooth inflow and outflow of material but is highly variable, often modulated by magnetic fields and complex kinematics. Accretion is inextricably linked to ejection phenomena, jets and outflows, that appear to be its natural consequences or even more, a requirement for accretion to occur. These ejection processes not only serve as a pressure-release mechanism but also carry away angular momentum, thereby further helping accretion.

T Tauri stars also serve as a unique window into the early stages of planetary system development. These young stellar objects offer the opportunity to study the complex processes that lead to planet formation. Accretion and ejection mechanisms in these systems not only influence the central star growth but also significantly affect the distribution of material in the surrounding disc. This, in turn, sets the stage for the genesis of planetary bodies. The accretion processes in T Tauri systems are fundamental for regulating the availability of material that could eventually coalesce into planets. Understanding these processes offers insight into how much mass is fed into the inner regions of the protoplanetary disc, which is crucial for the formation and migration of potential planetary cores. These mechanisms are not merely gravitational in nature but are often controlled by complex magnetic interactions that channel material onto the star. The ejection phenomena are equally important for planetary formation. These outflows by removing angular momentum from the inner disc, facilitate further accretion, but also potentially redistribute material and set conditions for planetary spacing and stability. Planet formation, even if it is of great interest in the community of T Tauri stars specialists, will however not be treated here. We will however discuss at the very end of this thesis some perspective application of our models to the field of planet formation.

In comparison to the parent molecular clouds, the protoplanetary disc gas is much denser, screening the ionising radiation, leading to a lower ionisation degree, which is the relative abundance of electrons to neutral molecules. But even with this reduced ionisation degree, it is still expected to be high enough to allow partial coupling with the magnetic fields in discs. This coupling could instigate MHD instabilities, contribute to disc winds launching, and enable the necessary angular momentum transfer for mass accretion \citep{2013ApJ...767...30B,suzuki2014magnetohydrodynamic}. The ionisation degree is the key parameter determining the ionisation state of a disc, influencing both the physical and chemical dynamics of protoplanetary discs, which in turn shape the formation of planetary systems \citep{1988PThPS..96..151U}. The main carriers of positive charge differ between various disc regions. Consequently, we can use atomic and molecular ions as observational tracers of the protoplanetary disc structure. The better understanding of these structures requires detailed numerical models of the chemistry of the disc. Astrochemical models, necessary to interpret observed molecular abundances, and non-ideal MHD codes, which simulate the disc dynamics, both use ionisation rates as a core parameter. This rate measures the number of ionisation per unit of time and is determined by the interaction between ionisation sources (UV, X-rays, CR) and molecules within the disc. Understanding the spatial distribution of the ionisation rates in T Tauri discs is crucial, as it directly impacts the disc chemistry, thermal balance, and dynamics. The ionisation rates vary significantly with the radial distance from the central star, as well as the vertical height above the disc midplane, due to the varying contributions of the different ionisation sources. As we will see, the outer disc ionisation is mainly dominated by external sources, such as galactic cosmic rays and interstellar radiation, while in the inner regions of the disc (within a few au), the ionisation rates are primarily determined by the internal sources, such as continuous stellar X or UV radiation. 

T Tauri stars are known to exhibit significant photometric and spectroscopic variability. They are particularly renowned for emitting intense X-ray flares. An X-ray flare refers to a sudden and strong increase in X-ray emissions from a star. In T Tauri stars these flares can be more than three orders of magnitude brighter than those observed in main sequence stars. The mechanism driving these flares is thought to be analogous to solar flares, but far more intense. Both in the Sun and in T Tauri stars, magnetic reconnection events are thought to trigger these flares. %This heightened X-ray activity is due to a much more active corona. 
During magnetic reconnection events, the stellar corona magnetic energy is partly released as kinetic energy of supra-thermal particles. Some of these particles subsequently heat the plasma of the underlying chromosphere to tens of millions of degrees, emitting the observed X-rays. The other part of the particles is expected to escape from the chromosphere and to interact with the surrounding circumstellar environment.\\

The issue that arises is to estimate the impact of particles produced by magnetic reconnection events on the discs of young stars.\\

To answer to this question, we first discuss the models and current observational constraints we have on T Tauri systems. In Chapter \ref{C:TTAURI}, although we address constraints on the central star, we mainly focus on the models and constraints of the surrounding disc. We will see in this chapter that ionisation is a crucial property guiding disc dynamics and chemistry. Chapter \ref{C:ionisation} delves into the sources and effects of ionisation on disc dynamics and chemistry, in this chapter, we propose {\it a so far unconsidered source of disc ionisation, namely the flares caused by magnetic reconnection events}. Chapter \ref{C:Reconnection} explores particle acceleration by magnetic reconnection in T Tauri systems with the objective of studying the effects that these particles can have on the disc. Chapter \ref{C:Propagation} presents the processes of particle transport and interaction within protostellar/planetary discs. Chapter \ref{C:PublicationI} offers a stationary parametric study of ionisation rates produced by flares in the inner disc region. Chapter \ref{C:PublicationII} extends this stationary model by considering temporal effects, allowing us to estimate average ionisation rates produced by flares and, at the same time, by using the thermo-chemical code  {\tt ProDiMO}, to understand in a better way the effect of these rates on the disc chemistry and dynamics. The results of Chapter \ref{C:PublicationI} have been published in 2023, and those of Chapter \ref{C:PublicationII} will be published shortly. We conclude with Chapter \ref{C:Conclusion} by discussing the future perspectives offered by this study on the impact of magnetic reconnection events on the discs and jets of young stars.

\chapter{T Tauri Stars}\label{C:TTAURI}
\section{Introduction}

A T Tauri star is a type of young, pre-main-sequence star that is in the process of gravitational contraction before reaching the main sequence phase of its evolution. These stars are characterised by the presence of a surrounding protoplanetary disc. T Tauri stars exhibit strong magnetic activity and complex magnetic structure that channel disc material onto the star. Understanding these processes offers insight into how much mass is fed into the inner regions of the protoplanetary disc, which is crucial for the formation and migration of potential planetary cores. The ejection phenomena are equally important for planetary formation as they, remove angular momentum from the inner disc, facilitate further accretion and also potentially, redistribute disc material. %Thus, the accretion and ejection processes in T Tauri stars are directly linked to the prospects and conditions for planet formation. 
These mechanisms play a crucial role in shaping the protoplanetary disc physical and chemical environment. This environment, rich in complex molecules, is the very nursery where new planets and the building blocks of life are forming.

The goal of this chapter is to lay a solid foundation for the modeling of T Tauri stars and their protoplanetary discs based on observational constrains.

In this chapter, %\textcolor{magenta}{CS I would start here} 
our exploration is twofold. Firstly, we focus in Sec. \ref{sec:ConstrainingCentralStar}, on constraining the physical properties of the central T Tauri star itself. We aim at setting the context of our current understanding of these objects. The introductory historical overview will transition into a detailed examination of the physical characteristics of T Tauri stars, including their mass, radius, luminosity, spectral features and magnetic fields. Understanding the nature of these stars is crucial for any further discourse on their surrounding environment. Secondly, in Sec. \ref{sec:ProtoplanetaryDiscs}, we dive into the complex world of protoplanetary discs. We discuss how astrophysical observations have helped us to constrain their chemical and thermal structures. Such constraints are fundamental in shaping our theories and models. Furthermore, we delve into existing computational and analytical models that simulate the structure and evolution of these discs.

\section{Constraining the physical properties of the central star} \label{sec:ConstrainingCentralStar}
\subsection{Early Observations and classification of Young Stellar Objects}

\subsubsection{Historical context and early observations of T Tauri stars}

The prototype 
%%CS Prototype or First observed?
of what we call T Tauri stars, was first identified as a variable star in the Taurus star-forming region by \citet{1852MNRAS..13...33H}. T Tauri stars were first identified as a distinct class of young stellar objects by \citet{1945ApJ...102..168J}. These objects were observed to have irregular variations in brightness, strong spectral emission lines, and an excess of infrared emission. Since Joy's initial discovery, extensive research has been conducted to better understand these mysterious objects.

The first complete catalogue of T Tauri stars was compiled by \citet{1962AdA&A...1...47H}, who expanded Joy's initial observations. Subsequent studies by \citet{haro1969flare} and \citet{cohen1979observational} contributed to the classification of T Tauri stars in various star-forming regions. The advent of more advanced observing techniques and instruments in the second half of the 20th century facilitated the discovery of more T Tauri stars, providing information on their properties and environment \citep{1998apsf.book.....H}.

In particular, the development of infrared astronomy in the 1980s led to the discovery of many new young stars buried in molecular clouds, which were previously obscured by dust \citep{1989ApJ...340..823W}. The launch of the Infrared Astronomical Satellite (IRAS) in 1983 made a huge step in our understanding of these objects by providing high-resolution infrared data, allowing researchers to study their circumstellar environments and to better understand their early evolutionary stages \citep{beichman1988infrared}.

The Hubble Space Telescope (HST), launched in 1990, has also played a crucial role in the study of T Tauri stars by providing high-resolution images and spectroscopy in a wide range of wavelengths \citep{krist2008multi,ardila2002observations}. This has greatly improved our knowledge of their physical properties, as well as the processes that govern their formation and evolution.

Ground-based observatories, such as the Very Large Telescope (VLT) and the Keck Observatory, have also contributed to the advancement of research on T Tauri stars. High-resolution spectroscopic studies have allowed detailed analysis of their spectral characteristics, enabling the study of their atmospheres, accretion processes and circumstellar environments \citep{2000ApJ...535L..47M,folha2001near,calvet2004mass}. In addition, these ground-based facilities monitored the variability of T Tauri stars over time, leading to a better understanding of their dynamical nature and the mechanisms behind the observed fluctuations \citep{alencar2002variability,2008A&A...479..827G}.

The study of T Tauri stars has been enhanced by the advent of large-scale surveys, such as the Two Micron All-Sky Survey (2MASS; \citealt{skrutskie2006two}), the Sloan Digital Sky Survey (SDSS; \citealt{york2000sloan}), and the Wide-field Infrared Survey Explorer (WISE; \citealt{wright2010wide}). These surveys have identified T Tauri stars over large portions of the sky and have facilitated the study of their demography and spatial distribution within star-forming regions (e.g., \citealt{2018AJ....156..271L,rebull2011new}).

More recently, the launch of the Gaia satellite in 2013 provided unprecedented astrometric and photometric data for a large number of T Tauri stars, allowing precise distance measurements and a finer understanding of their physical properties and evolutionary state (e.g. \citealt{gagne2018banyan,2019A&A...632A..16F}). The combination of Gaia data with other multi-wavelength surveys has driven the discovery and characterisation of previously unknown T Tauri stars and enabled more comprehensive studies of their formation and evolution within their host star-forming regions (e.g. \citealt{2018AJ....156..271L,2019AJ....158...54E}).

Over the last decade, the field has seen significant improvements in terms of observational capabilities, thanks in particular to the commissioning of the Atacama Large Millimetre/Submillimetre Array (ALMA) and the arrival of a new generation of high-contrast optical and infrared instruments. These facilities have enabled both statistical studies of discs at moderate resolution $ \sim 0.1" - 0.4",$ equivalent to tens of au in nearby star forming regions. They also allowed imaging studies of discs at high resolution, up to  $\sim 20$ mas, equivalent to 1 au in the nearest observable discs at radio light wavelengths. The results of these recent observations and the upcoming data of the James Webb Space Telescope (JWST), combined with advances in physical and chemical modelling of the discs, will provide a much more detailed and complex picture of the planet- and star-forming environment.

\subsubsection{Overview of the star formation process}

\paragraph{The hierarchical structure of cloud complexes:}
\begin{figure}[h!]
    \centering
    %\captionsetup{width=0.9\linewidth}
    \includegraphics[width=1\linewidth]{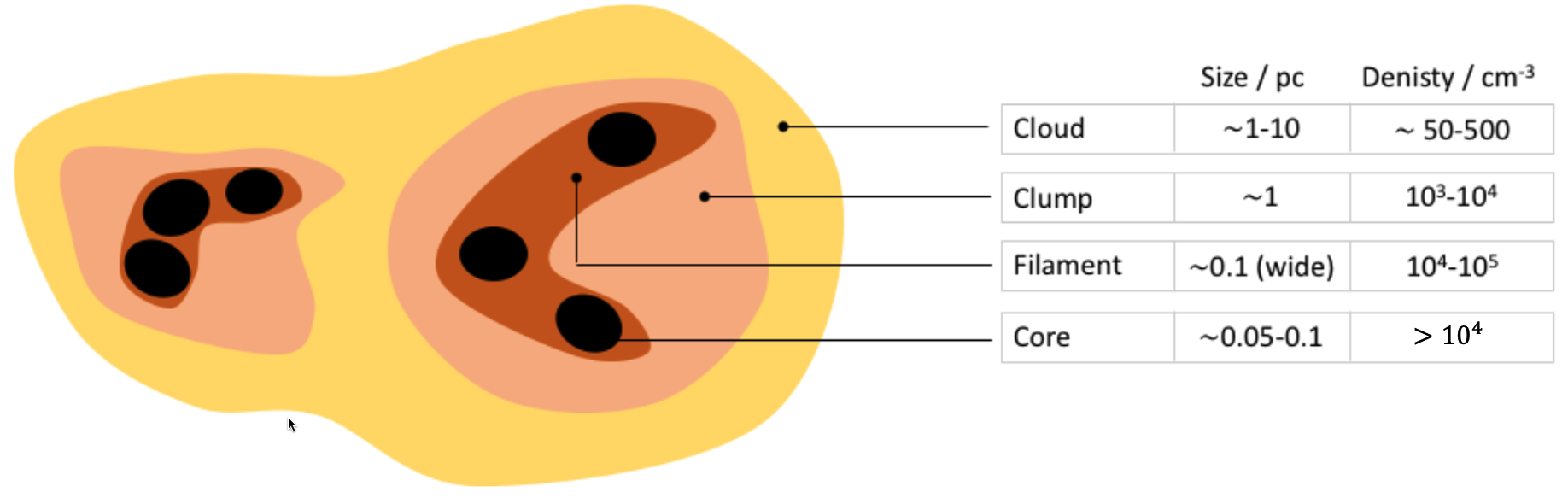}
    \caption{The hierarchical structure of molecular clouds span from diffuse areas to dense cores where stars can form. Temperatures range from these scales, fluctuating from tens of Kelvin down to just a few Kelvin. Ionisation levels also differ significantly, from nearly complete ionisation to almost no ionisation. Figure adapted from \citet{owen2023secret}.}
    \label{fig:HierachicalStructureStarFormation}
\end{figure}
The hierarchical structure of molecular cloud complexes can be divided into four main components, molecular clouds, clumps, filaments and cores, see Fig. \ref{fig:HierachicalStructureStarFormation}. Each component plays a crucial role in the star formation process.

\begin{itemize}
    \item \textbf{Molecular clouds} — Molecular clouds, or giant molecular clouds (GMCs), are vast regions of gas and dust with high densities and low temperatures. These clouds mainly consist of molecular hydrogen (H$_2$), with traces of other molecules, such as CO and dust particles \citep{2017ApJ...845..116W}. Their size can vary from 5 up to 200 parsecs in diameter and contain masses between $10^4$ and $10^6$ solar masses (M$_\odot$) \citep{2000prpl.conf...97W}. 

    \item \textbf{Clumps} — Within molecular clouds, it is possible to observe denser regions called clumps or clusters. Clumps typically extend over a few parsecs and have typical mass in the range $10-10^3$ M$_\odot$  \citep{2007ARA&A..45..339B}. These structures can be gravitationally bound or transient, depending on the balance between self-gravity and internal pressure. Observations of nearby cloud complexes indicate that embedded clusters account for a significant (70–90\%) fraction of all stars formed in GMCs \citep{doi:10.1146/annurev.astro.41.011802.094844}.

    \item \textbf{Filaments} — Filaments are elongated, dense structures within molecular clouds. They are considered as the backbone of cloud complexes, connecting and feeding the dense cores where star formation occurs \citep{2014prpl.conf...27A}. Filaments can extend several parsecs in length and are typically 0.1-0.3 parsecs wide \citep{2011A&A...529L...6A}. The critical line mass (mass per unit length) for gravitational instability in an isothermal filament is given by \citep{nagai1998origin},
    \begin{equation}
    M_{\rm{line}}^{\rm{crit}} = 2 \frac{c_s^2}{G} \simeq 2.6 \times 10^{17} \frac{T}{100~\rm{K}}~\rm{g/cm},
    \end{equation}
    where $c_s$ is the local sound speed and $G$ is the gravitational constant. When the actual line mass of a filament exceeds this critical value, the filament becomes gravitationally unstable, leading to the formation of dense cores and ultimately stars \citep{2000MNRAS.311..105F}. 
    \item \textbf{Cores} — Cores are the densest regions of molecular cloud complexes and represent the smallest scale of the hierarchical structures. The nuclei are typically between 0.01 and 0.1 parsecs in size and range in mass from less than 1 M$_\odot$ to several solar masses \citep{2007prpl.conf...17D}.

    The Jeans mass $M_J$ and Jeans length $\lambda_J$ are critical parameters for determining the stability of the core \citep{2008ApJ...684..395H},

    \begin{equation}
    M_J = \frac{4 \pi \rho}{3} \left(\frac{\pi c_s^2}{G \rho}\right)^{3/2} = 1 \left(\frac{T}{10~\rm K}\right) ^{3/2}\left(\frac{\mu}{2.33}\right)^{-1/2}\left(\frac{n}{10^4 ~ \rm cm^{-3}}\right)^{-1/2}~ \rm M_\odot \ ,
    \end{equation}

    where 
    %$c_s$ is the sound speed, $G$ is the gravitational constant and 
    $\rho$ is the gas mass density, $n$ the gas particle density, $\mu$ the mean molecular weight and T the temperature. In the absence of turbulence, when the actual mass of a core exceeds the Jeans mass, it becomes gravitationaly unstable and may collapse to form a protostar \citep{2005MNRAS.359..211L}.
\end{itemize}

The hierarchical structure of cloud complexes, which spans from molecular clouds to cores, can be characterised by the balance between gravitational forces and internal pressures within these structures. This balance determines their stability and their capacity to form stars. However, the actual dynamics of these objects are much more intricate, and are better described by the interplay among thermal, gravitational, turbulent, and magnetic processes \citep{diego2015probing}. Thus, the hierarchical structure of cloud complexes is inherently magneto-hydrodynamic (MHD). Understanding these complex relationships and the physical properties of each component is essential for a comprehensive understanding of star formation.

\paragraph{Evolution from protostar to main sequence star:}

\begin{itemize}
    \item \textbf{Mass accretion and removing of angular momentum} — During the protostellar phase, the protostar increases its mass by accreting matter from its surrounding envelope. This accretion process determines the final mass, angular momentum and other properties of the resulting main sequence star. As the protostar accretes matter and increases in mass, its internal structure and properties evolve. The protostar contracts and heats up, increasing its core temperature and pressure \citep{2013EAS....62....3H}. 

    The accretion process during the protostellar phase is often associated with the formation of bipolar flows and jets, which remove the excess angular momentum from the system and regulate the accretion process (e.g. \citealt{2014prpl.conf..451F}). The formation of jets and outflows is not yet fully understood, but is thought to be closely related to MHD processes and the interaction between the protostar, the accreting disc and the incoming material (e.g. \citealt{2006A&A...453..785F,2013A&A...550A..99Z}). 

    \item \textbf{Dissipation of the protostellar envelope}  — As the protostar continues to accrete material from its surrounding envelope, the mass of the envelope decreases and the optical depth of the envelope also decreases, allowing radiation from the central protostar to escape more easily (e.g. \citealt{1987ApJ...312..788A,1999ApJ...525..330Y}). Eventually, the envelope is completely dissipated by a combination of accretion, radiative feedback and outflows, revealing the central protostar and the surrounding disc (e.g. \citealt{2000prpl.conf..377C}).

    \item \textbf{Deuterium burning} — As the protostar contracts and heats up during its evolution, the central temperature increases, reaching values sufficient for the start of deuterium burning. Deuterium burning occurs at a temperature of about $10^6$ K and is an important step in the evolution of main sequence stars (e.g. \citealt{1994ApJS...90..467D,1999ApJ...525..772P}). The energy produced by deuterium burning slows down the contraction of the protostar and stabilises it temporarily, allowing the young star to embark on the path of main-sequence evolution.
 
    \item \textbf{Disc dispersal mechanism} — The dispersion of circumstellar discs around T Tauri stars marks the end of the star and planet formation process. Disc dispersal mechanisms include viscous accretion, photoevaporation, and magnetic outflows (e.g., \citealt{2001MNRAS.328..485C,2006MNRAS.369..229A,2013ApJ...767...30B,2016ApJ...821...80B}). The time scale of dispersion is typically a few million years  \citep{2001ApJ...553L.153H,2014ApJ...793L..34P,2023MNRAS.519.3958I}).
\end{itemize}
\begin{figure}[h!]
    \centering
    \includegraphics[width=0.8\linewidth]{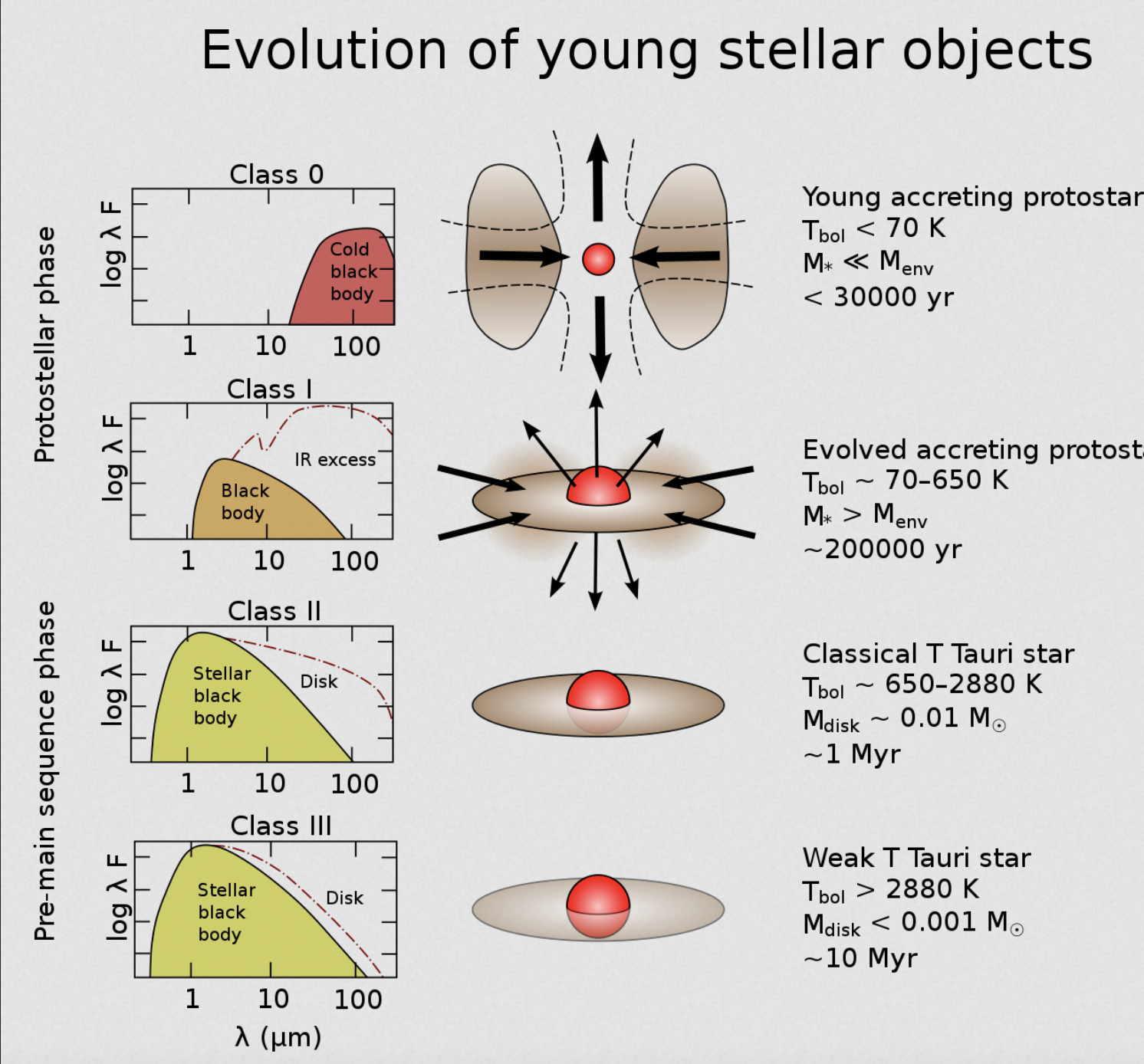}
    \caption{Empirical sequence for the formation and circumstellar evolution of a single star, ranging from a Class 0 protostellar phase to a Class III PMS. This sequence is based on the shape of the spectral energy distribution, represented on the left. The schematic shape of the objects is showed in the center, with arrows indicating strong accretion and ejection processes. The typical bolometric temperature and mass of the circumstellar material (i.e., envelope + disc) are indicated on the right. This figure is extracted from Vallastro, via Wikimedia Commons\protect\footnotemark.}
    \label{fig:TTclass}
\end{figure}
\footnotetext{\url{https://commons.wikimedia.org/wiki/File:Evolution_of_young_stellar_objects.svg}}
\subsubsection{Young stellar object classification:} 
Young Stellar Objects (YSOs) often exhibit more emission in the infrared than anticipated from a pre-main-sequence star (PMS) photosphere. This infrared excess is attributed to the presence of dust near the star. Its intensity is the basis of an empirical classification scheme for YSOs \citep{1984ApJ...287..610L}. In order to define the slope of the spectral energy distribution (SED) between near-IR and mid-IR wavelengths we use the following equation,
\begin{equation}
    \alpha_{IR} \equiv \frac{d \log (\nu F_\nu)}{d \log (\nu)} \equiv \frac{d \log (\lambda F_\lambda)}{d \log (\lambda)}
\end{equation}
The reference wavelength band for the determination of $\alpha_{IR}$ in the near- and mid-IR vary among studies but are typically around $2 \mu m$ and $25 \mu m$. Based on the derivation of $\alpha_{IR}$, four or five classes of YSOs are recognised. Figure \ref{fig:TTclass} shows this classification based on the infrared Spectral Energy Distribution (SED).

\begin{itemize}
\begin{figure}[h!]
    \centering
    \includegraphics[width=0.5\linewidth]{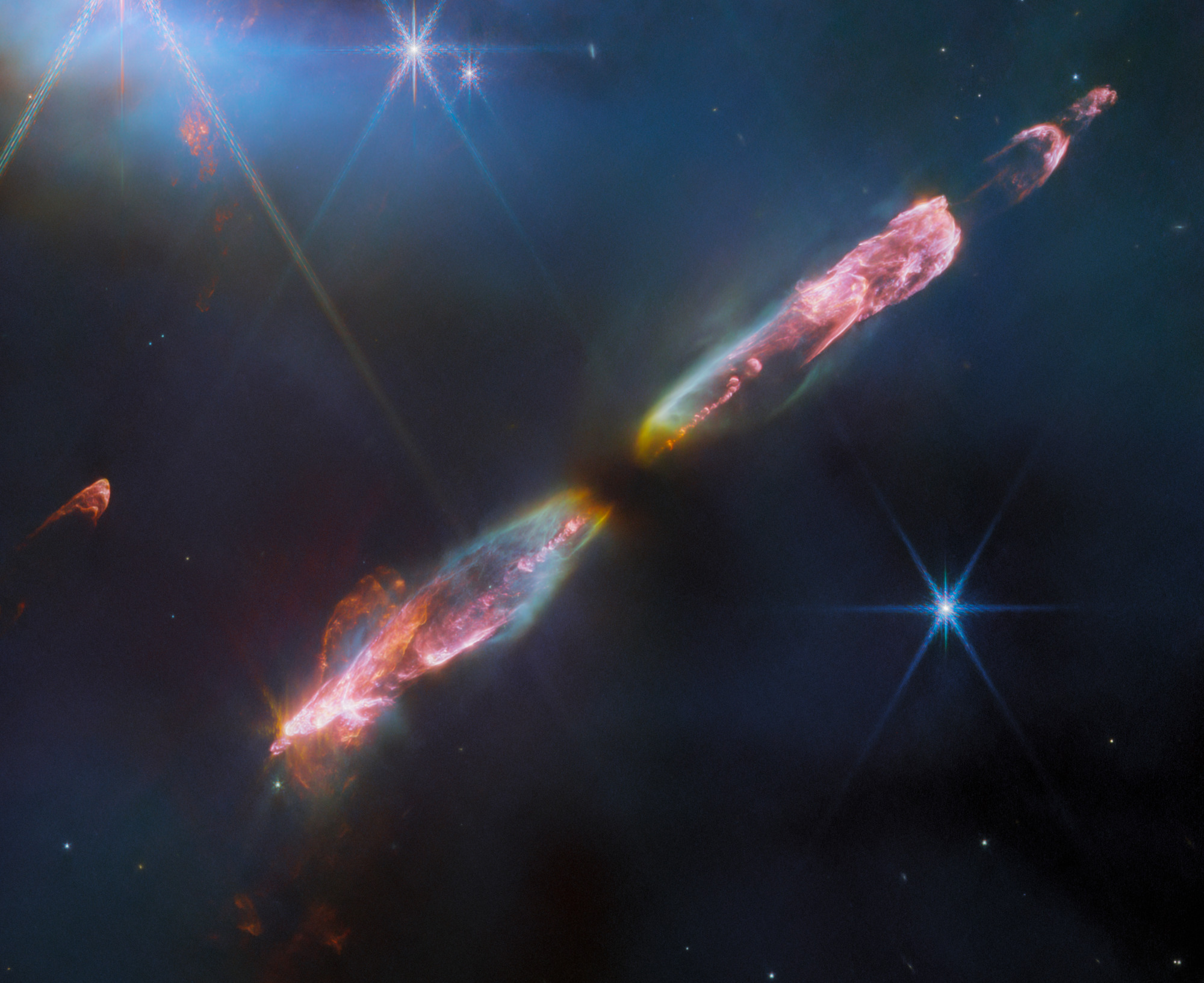}
    \caption{The James Webb Space Telescope offers a high-resolution, near-infrared glimpse of the embedded Class 0, Herbig-Haro 211. Unveiling details of a young stellar outflow. The image displays a sequence of bow shocks towards the southeast (bottom-left) and northwest (top-right), along with the bipolar jet fuelling them. Molecular hydrogen, carbon monoxide, and silicon monoxide, radiate infrared light. This light, captured by Webb, traces the architecture of these outflows. Credits: ESA/Webb, NASA, CSA, Tom Ray\protect\footnotemark}
    \label{fig:ResolvedClass0}
\end{figure}
\footnotetext{\url{https://webbtelescope.org/contents/media/images/2023/141/01H9NWH9JEBFPKVD3M1RRTGGQJ}}
%CS Normalement il faut mettre la date à laquelle tu as consulté la page web.
\item Class 0: These YSOs are deeply embedded in their native molecular cloud and are not observable at visible wavelengths. They have strong submillimetre emission and are thought to represent the youngest and most deeply embedded stage of YSO evolution. There is a possibility that Class 0 YSO do not have a star yet e.g. see \citet{2001A&A...365..165C}. In particular Class 0 have an infalling envelop, which hides a protostellar core that may not yet radiate as a star, but display strong collimated jets, see Fig. \ref{fig:ResolvedClass0}.

\begin{figure}[h!]
    \centering
    \includegraphics[width=0.5\linewidth]{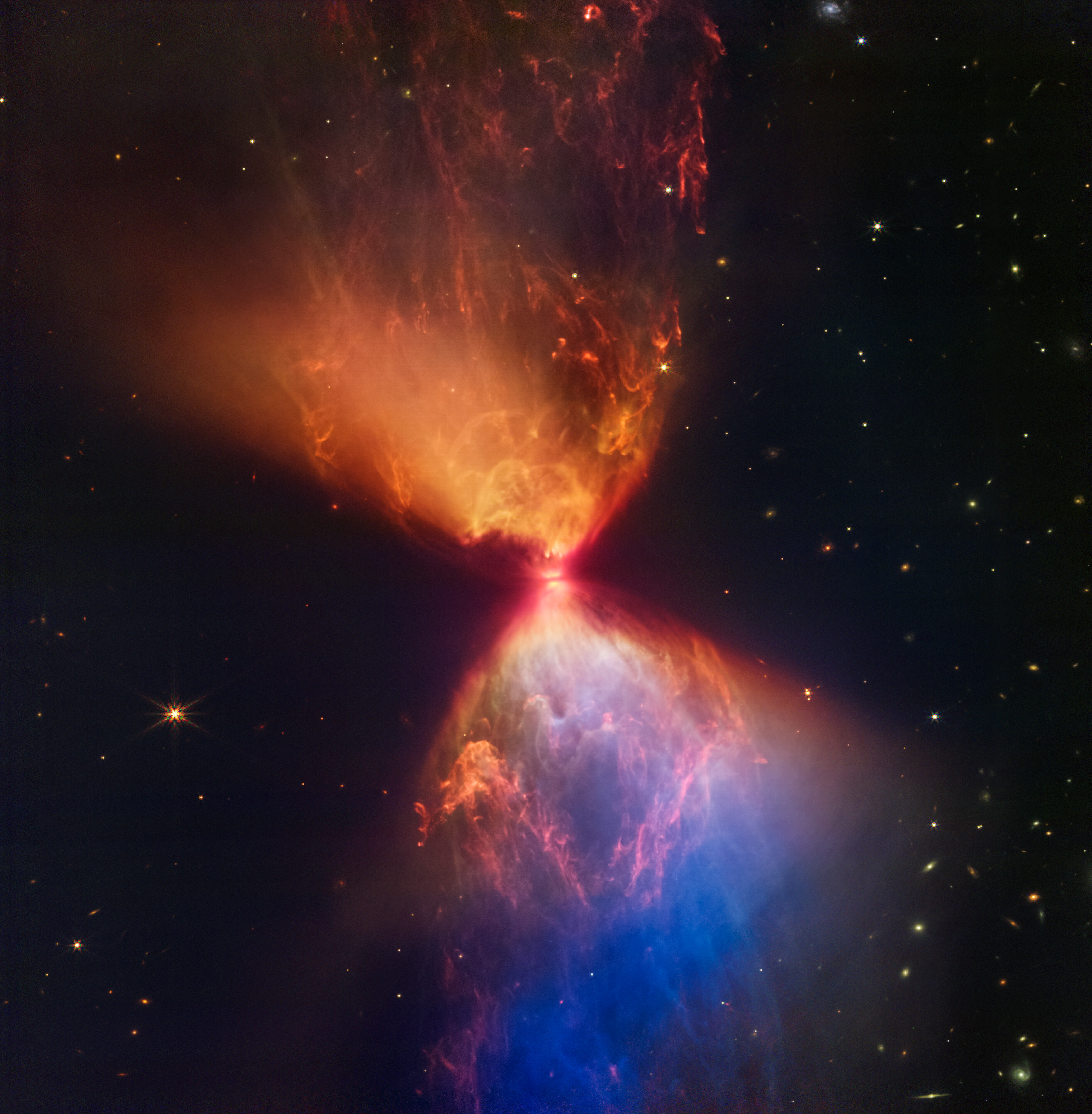}
    \caption{The James Webb Space Telescope Near-Infrared Camera (NIRCam), pictures the protostar inside the cloud L1527, shedding light on the genesis of a Class I star. Located in the Taurus star-forming region, these clouds are visible in infrared light.
    The protostar remains tucked away within the "neck" of this hourglass formation. A protoplanetary disc, viewed edge-on, appears as a dark streak across the neck's centre. Light emitted from the protostar seeps above and beneath this disc, lighting up cavities in the enveloping gas and dust. Credits: ESA/Webb, NASA, CSA\protect\footnotemark}
    \label{fig:ResolvedClassI}
\end{figure}
\footnotetext{\url{https://webbtelescope.org/contents/news-releases/2022/news-2022-055?page=1&Tag=Nebulas}}
\item Class I: These objects exhibit a rising SED in the mid-IR, corresponding to $\alpha_{IR}>0$. This rise indicates the presence of a large amount of circumstellar material, which is typically in the form of an infalling envelope surrounding the protostar, see Fig. \ref{fig:ResolvedClassI}.

\item Class II: With $-1.6 < \alpha_{IR} < 0$, Class II YSOs display a weaker infrared excess. These objects are commonly associated with classical T Tauri stars (CTTS) and are thought to be in a stage where a circumstellar disc is still present and connected to the magnetosphere, see Fig. \ref{fig:ResolvedClassII}. The infalling envelope has been mostly dissipated. Classical T Tauri stars are characterised by strong hydrogen emission lines, often accompanied by other important emission features, such as Ca II, He I and Fe II (e.g.\citealt{1962AdA&A...1...47H,1989ARA&A..27..351B}). These spectral features are indicative of an active accretion process, in which material from the circumstellar disc is transported to the surface of the star, releasing radiative energy. CTTS generally exhibit irregular photometric variability, due to fluctuations in the accretion rate, also due to the presence of hot spots and cold surface features (e.g. \citealt{1994AJ....108.1906H,2002ApJ...571..378A}). %Observations of CTTS provide information about the early stages of star formation and the physical mechanisms responsible for the transfer of mass and angular momentum between the star and its surrounding disc. 
In addition, CTTS are often associated with jets and outflows, which are thought to be related to accretion processes and magnetic fields, see Sec. \ref{sec:Physical properties of the central star}.

\begin{figure}[h!]
    \centering
    \includegraphics[width=0.5\linewidth]{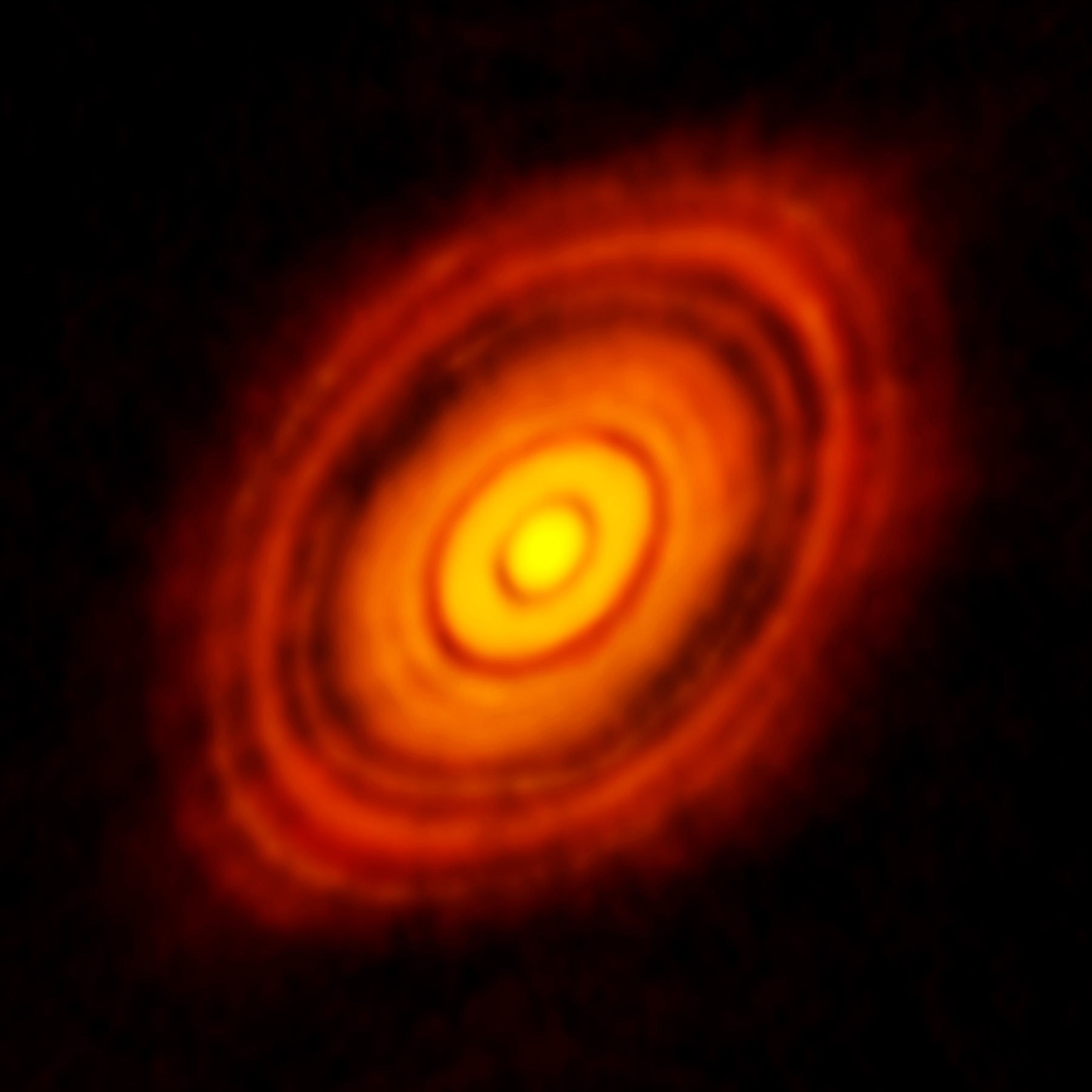}
    \caption{The figure displays the protoplanetary disc encircling the Class I/II star, HL Tauri in high resolution by ALMA. These observations by ALMA unveil previously unseen, complex structures in the disc, highlighting potential spots where planets might form within the system dark gaps. Credits:ALMA (ESO/NAOJ/NRAO)\protect\footnotemark}
    \label{fig:ResolvedClassII}
\end{figure}

\item Class III: These YSOs have a weak infrared excess, with $\alpha_{IR} < -1.6$. They are associated with weak-lined T Tauri stars (WTTS), and their SED is dominated by the photospheric emission from the central star. The weak infrared excess suggests that the gas of the circumstellar disc has mostly dissipated or is optically thin, leaving a disc mostly composed by dust, planetesimals and planets. WTTS have weak or absent hydrogen emission lines, indicating that they are not actively accreting matter from a circumstellar disc. WTTS typically have a lower infrared emission excess, consistent with the presence of a less massive or more evolved disc than CTTS \citep{1997AJ....114..288M}. Although WTTS are not as strongly influenced by accretion processes as CTTS, they still exhibit magnetic activity, such as chromospheric and coronal emission, fuelled by their convective interior and rapid rotation (e.g., \citealt{2000ApJ...532.1097M,2007A&A...468..529G}).
\footnotetext{\url{https://www.eso.org/public/images/eso1436a/}}
%\item Class Flat: Some studies include a separate class of YSOs with flat SEDs, where $-0.3 < \alpha_{IR} < 0.3$. These objects are thought to be in a transitional phase between Class I and Class II, with both an infalling envelope and a circumstellar disc present.
%\textcolor{magenta}{CS then put it in between I and II. In any case, transitions are probably continuous. There is also the Hayashi tracks for T Tauri and the Mass accretion/loss rate versus time - the references should be in my lecture.} {\bf AM: I would keep it here because this is not a standard classification but Hayashi's paper should be a bit commented.}
\end{itemize}

This classification scheme provides a useful framework for understanding the evolution of YSOs, as it is based on the observable properties of their SEDs. As an YSO evolves, the infrared excess and the slope of the SED change, reflecting the gradual dissipation of the circumstellar material and the transition from an envelope-dominated to a disc-dominated system (e.g. \citealt{2007A&A...473L..21B,2010ApJ...710..597S}. However, it is important to note that the SED may be influenced by various factors, such as the inclination angle, the dust properties, and the presence of outflows or jets, which render the interpretation of the YSO classification more complex \citep{2019A&A...622A.149G}.

Understanding the underlying processes that govern the transition between CTTS and WTTS and the factors that may influence their observed properties, such as the initial mass and angular momentum of the protostellar core, the efficiency of angular momentum transport in the disc, and the role of magnetic fields in regulating accretion (e.g. \citealt{2014ApJ...786...97H,2016ARA&A..54..135H}) is a very active field of research. Multi-wavelength observations and detailed modelling of T Tauri stars will continue to play a crucial role in understanding the complex interaction between stars and their circumstellar environments during the early stages of stellar evolution. All along this thesis we will focus on the T Tauri stage of evolution, namely class I and II YSOs.

%CS I thought T TAuri stars were only class II and III so that you were mostly on class II

\subsection{Physical properties of the central star}\label{sec:Physical properties of the central star}
The determination of the basic observables, stellar temperature ($T_{\text{{eff}}}$) and luminosity ($L_*$), is essential to derive the physical properties such as stellar mass ($M_*$). In young stars with discs, accretion processes produce an excess emission filling absorption lines, called veiling. The contribution of veiling due to accretion is non-negligible, and must be accounted for when determining the photospheric parameters. In turn, this allows to measure the accretion luminosity ($L_{\text{{acc}}}$) and infer the mass accretion rate ($\dot{M}_{\text{{acc}}}$). Here we discuss the methods currently used to measure the stellar and accretion properties for populations of young stars with discs.

\subsubsection{Spectral types, stellar and accretion luminosity}

The luminosity of T Tauri stars is usually between less than one to ten times the Sun luminosity, but its determination is not straightforward. The determination of stellar properties for Pre-Main-Sequence (PMS) stars was first carried out using optical spectroscopy (e.g., \citealt{cohen1979observational,kenyon1995pre,1997AJ....113.1733H}). However because the total luminosity of the system includes the contribution of accretion, ie $L=L_{phot}+L_{acc}$, it was soon realised that for these young, extincted, and accreting stars it is important to simultaneously describe the expected underlying photospheric emission, $L_{phot}$ and the continuum excess due to accretion, $L_{acc}$, as well as a correct determination of the extinction. This requires the use of broad wavelength coverage and absolute flux-calibrated spectra. The spectral range from $\lambda \sim 4000 - 7000\, \text{\AA}$ allows spectral types (SpT) to be accurately determined \citep{2014ApJ...786...97H,fang2021improved}. But extending the coverage to $\lambda < 4000\, \text{\AA}$, allows to include the Balmer jump and continuum region. In addition, with the HST, the near-ultraviolet (near-UV) region, considerably improves the determination of the extinction and the contribution of the excess emission due to accretion (e.g., \citealt{herczeg2008uv,manara2013accurate}).

The best stellar templates for the photospheric properties of young stars are non-accreting PMS stars \citep{gullbring1998disc}. Indeed, non-accreting stars have similar gravity and chromospheric emission as accreting PMS stars \citep{2014ApJ...786...97H}. But the chromospheric activity in these targets is much higher than in main-sequence stars. Analyses by \citet{manara2017extensive} have shown how this chromospheric emission scales with stellar temperature in PMS stars. This allows one to discriminate between emission lines dominated by chromospheric or accretion-related emission. 

The effective surface temperatures of T Tauri stars are generally between 3000 K and 6000 K, with lower temperatures corresponding to lower mass stars (e.g. \citealt{2007ApJS..173..104L,2008ApJ...681..594H,2013A&A...551A.107M}). However, the conversion from a spectral type to a value of $T_{\text{eff}}$ has been a subject of discussion in recent years. Simultaneous measurements of $T_{\text{eff}}$ from comparison with synthetic spectra and SpT from empirical templates \citep{manara2021penellope} highlight the limits of previously used relations (e.g., \citealt{luhman2003new}). New relations have been empirically calibrated by \citet{2014ApJ...786...97H}; these should be used to convert SpT into $T_{\text{eff}}$ for PMS stars.

Along with the improvements in modelling, there have been significant advances in spectroscopic capability. One particular advancement is the X-Shooter instrument, on the VLT. It can simultaneously cover a wide wavelength range of \( \lambda \sim 0.3 - 2.5 \) $\mu$m at a medium resolution (R \( \sim10,000 - 20,000 \)) \citep{vernet2011x}. Thanks to its sensitivity and its location in the Southern Hemisphere, the instrument is being used to survey several star-forming regions. The wide wavelength coverage of X-Shooter with absolute flux-calibration is essential, it enables the determination of the accretion luminosity from the UV-excess in the Balmer continuum region \citep{manara2013accurate}. It can also be used to determine the brightness of various permitted transmission lines. These include the high-n Balmer lines series in the near-UV, the Bracket lines series in the near-infrared (near-IR), as well as emission lines of helium and calcium \citep{alcala2014x}.

\citet{alcala2014x} demonstrated that the line luminosities are more reliable tracers of \(L_{\text{acc}}\) than the measurement of the width of the H$_\alpha$ line previously used. The new line-to-accretion luminosity relations \citep{alcala2017x} can be applied to spectroscopic data sets not only in the Balmer continuum region but for a number of other emission lines.

The comparison of accretion luminosity determinations from lines at different wavelengths also allows an independent determination of the extinction \citep{pinilla2021bright}. However, it is important to recognise that only proper inclusion of the impact of extinction and veiling due to accretion at all stages of the analysis can overcome the degeneracy between these parameters. The UV-excess is crucial for determining the excess due to accretion (e.g., \citealt{2013A&A...558A.114M,2014ApJ...786...97H}). This means that methods based on assumptions about stellar temperatures, extinction, or veiling might lead to larger degenerate uncertainties in the derived parameters.

In addition, both temperature and luminosity estimates are influenced by the presence of stellar spots. When spots are present, different values for these parameters are obtained. The values can vary depending on whether high-resolution blue spectra are used, or medium- to low-resolution spectra at the reddest optical wavelength and in the near-IR \citep{gully2017placing}. In addition to this, stellar variability also affects the measured luminosity \( L_* \).

\subsubsection{Determination of mass and age}
\begin{figure}[h!]
    \centering
    \includegraphics[width=0.7\linewidth]{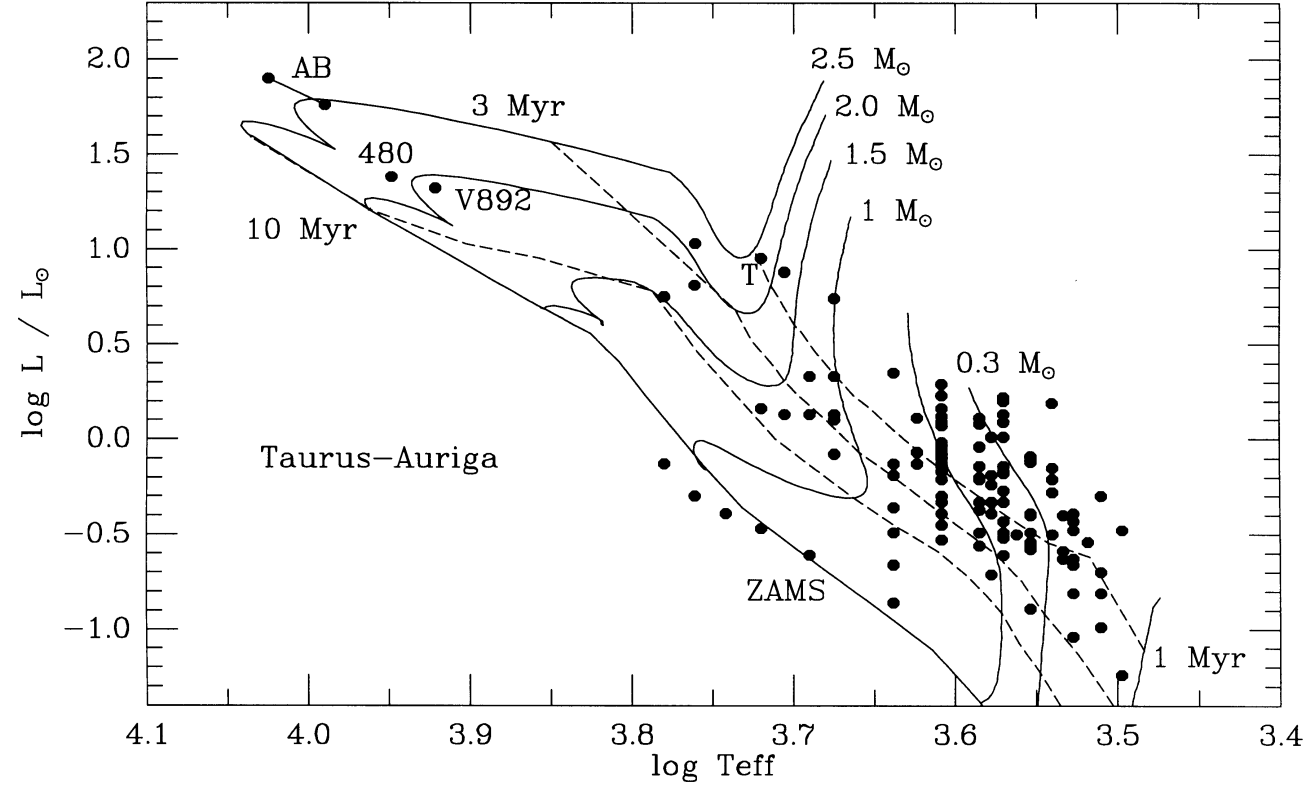}
    \caption{H-R diagram of objects in the Taurus molecular cloud complex. Superimposed in solid lines, the stellar evolution tracks for the pre-main sequence. Stars with lower masses (i.e low $T_{\rm eff}$) follow almost vertical evolutionary tracks until they reach the main sequence; they are the Hayashi tracks. The dashed lines, labeled in years, are isochrones at different ages.} 
    \label{fig:Hayashitracks}
\end{figure}

The classical method to determine the stellar mass $M_*$ and stellar age involves comparing the position of Pre-Main-Sequence (PMS) stars in the Hertzsprung-Russel diagram (H-R) with PMS evolutionary model tracks. The position in the H-R diagram of a T Tauri star, defined by its effective temperature and luminosity, provides information on its mass and evolutionary state, allowing comparisons with theoretical models of PMS evolution, see Figure \ref{fig:Hayashitracks}. The first evolutionary model for low mass stars has been developped by \citet{1961PASJ...13..450H}, hence the name "Hayashi tracks" of the evolutionary path followed by T Tauri stars on HR diagrams.

A significant issue that these models seek to address is the large spread in luminosity ($L_*$) at a given temperature. This spread is observed in nearby clusters, even with advanced analysis methods \citep{herczeg2015empirical,alcala2017x}. The basic assumption to determine a cluster age is that all stars in it have the same age. The luminosity spread might indicate a real age spread or might be due to missing physical mechanisms. Recent Gaia-based analyses support the idea of an age spread in specific regions, especially between on-cloud and off-cloud populations \citep{esplin2022census}. 

\citet{baraffe2015new} updated the models to include new assumptions on atmospheric conventions and metallicity. Some models have begun to include accretion effects, both prior to and during PMS evolution. \citet{feiden2016magnetic} created new PMS evolutionary models including the effect of magnetic fields on PMS star evolution, showing promising alignment with data for high-luminosity and low-mass stars.

The effect of stellar spots also affects the position of a PMS star on the H-R diagram. This has been shown by \citet{somers2020spots} who reported significant changes in the inferred stellar age, and in some instances, the value of $M_*$ due to the modelling of spots.

In recent times, ALMA has enabled the use of dynamical stellar mass estimates to test models. This is achieved through spectrally-resolved observations of CO emission from discs, see Sect. \ref{sec:discmass}.

The results from these works are diverse, some studies indicate better agreement with dynamical mass estimates when using magnetic models (evolutionary models including magnetic fields) in the range of $M_*\approx 0.4 - 1 M_\odot$ \citep{simon2019masses} or $M_*\approx 0.6 - 1.3 M_\odot$ \citep{braun2021dynamical}. In contrast, non-magnetic models align better with lower-mass stars \citep{braun2021dynamical}.

However, there are limitations in these comparisons. Recent works highlight the uncertainty and discrepancies that can arise from comparing dynamical masses measured from different molecules \citep{premnath2020dynamical}. 

\subsubsection{Mass accretion rates}
\begin{figure}[h!]
    \centering
    \includegraphics[width=1.\linewidth]{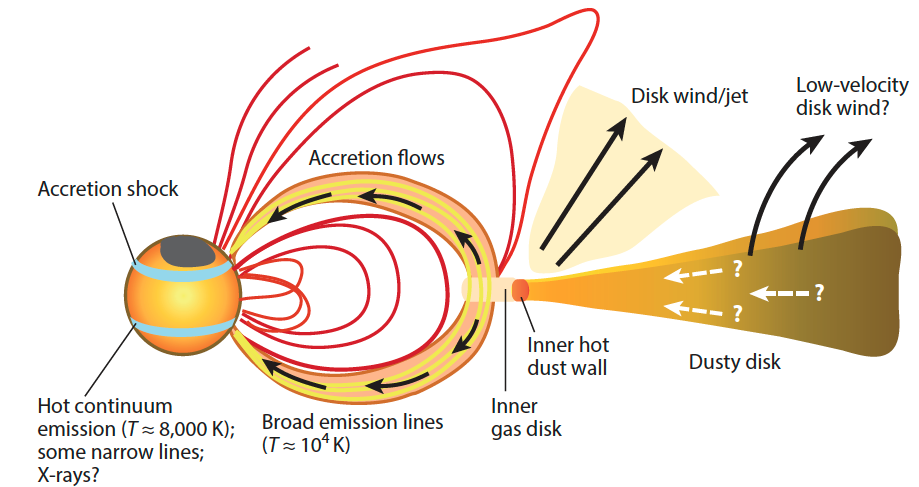}
    \caption{A schematic view of a young star accretion process from a surrounding disc through the stellar magnetosphere presents a complex scene.  %Meanwhile, other magnetic field lines, unconnected to the disc, can generate coronal X-ray emission. %\textcolor{magenta}{CS the rest of the comments may be should be shifted to the main text? This would shorten the caption? BTW DO YOU HAVE in this plot some scaling in au, that could help you } 
     The figure is reproduce from \citet{hartmann2016accretion}}
    \label{fig:MagnetosphericAccretion}
\end{figure}

Figure \ref{fig:MagnetosphericAccretion} illustrates a model of accretion onto young low-mass stars, typically within the age range of 1 to 10 million years and with masses around or below one solar mass \citep{hartmann2016accretion}. In this model, the material from the circumstellar disc, consisting of dust and gas, is driven inward to an approximate radius of 0.1 au. The exact mechanisms governing the transport of mass and angular momentum remains unknown; as discussed in Sect. \ref{sec:ModelingProtoplanetaryDiscs} it is likely a mix of viscous and wind driven accretion. Within approximately 0.1 au, the temperature of the disc exceeds roughly 1,000 K due to the central star radiation, leading to the sublimation of dust. At this boundary, referred to as the dust sublimation radius, the inner edge re-emits the absorbed energy, accounting for a significant portion of the detected near-IR excesses. Farther from the star, within about 1 au, the inner disc is responsible for a bipolar flow or jet, energised by the accretion process. 

The star strong magnetic field not only produces substantial starspots but also truncates the accretion disc at a distance of a few stellar radii. Guided by the magnetic field lines, matter is channelled onto the star in accretion columns or funnel flows. The gas in these accretion columns heats up to around or above 8,000 K, though the exact heating mechanism remains unidentified but is presumably of magnetic origin \citep{hartmann2016accretion}. This leads to the observed broad H$_\alpha$ emission lines.

Within this framework, magnetic field lines connected to the disc guide material to the star at velocities approaching free fall. As the infalling gas reaches near free-fall, it speeds to roughly 300 km$/$s, producing a shock at the stellar photosphere and heating the gas to temperatures of the order of $10^6$ K. Most of the ensuing X-ray emission is absorbed and re-emitted at lower temperatures, resulting in strong ultraviolet-optical continuum excesses and some relatively narrow emission lines. A good description of the excess emission due to accretion is equally important for describing the observed spectra of accreting young stars. The complex structure of the accretion shock region, see Fig. \ref{fig:MagnetosphericAccretion}, has been discussed in \citet{hartmann2016accretion}. \citet{robinson2019multiepoch} recently revised these shock models, including a treatment of the postshock and preshock regions with CLOUDY \citep{ferland20172017}. This leads to higher emissivity of the postshock region, helping the match of the measured veiling at optical wavelengths.

The accretion rate in YSO can be determined using various diagnostics. Direct methods of accretion analysis involve the observation of gas that becomes heated and shocked as it slows down and is shocked at the stellar photosphere. The primary diagnostics for this process are the ultraviolet continuum and multiple emission lines across the ultraviolet and optical spectra, measurable in low-extinction scenarios with a favourable system geometry. In more obscured or embedded sources, proxy lines in the red optical and infrared spectrum are employed. In cases where it is impossible to detect the star directly, we have to rely on the reprocessed emission that appears at infrared or millimetre wavelengths. Protoplanetary discs are observed to supply their central stars with material at rates typically within the range of $10^{-10}$ and $10^{-7}$ solar masses per year \citep{2014A&A...570A..82V,hartmann2016accretion,2016A&A...591L...3M}.\\

Mass accretion rates depend on the type of accretion process, each type of accretion can be identified by observational markers. We now present the main accretion processes along with their detection methods.

\paragraph{Magnetospheric accretion:}\label{sec:magnetosphericaccretion}
In CTTS, the gas accretion onto the star is guided by the stellar magnetic fields from the disc. The accretion shock itself is predominantly or completely hidden beneath the photosphere, where it results in X-ray heating of the neighboring photosphere. This heating effect, along with energy from the accretion funnel flow, leads to hydrogen recombination, H continuum, and line emission, which are most effectively observed at optical and ultraviolet wavelengths. The inner disc may develop warps that are frequently linked to the magnetospheric flow, potentially obstructing stellar light or radiation emanating from the accretion flow.

To measure accretion rates, one can either model the continuum excess emission, which demands a fine determination of the stellar parameters, accretion geometry, and temperature structure, or resort to proxy emission line diagnostics. In the latter method, extinction-corrected line fluxes are calibrated to the accretion shock models, as outlined in reviews by \citealt{2016ARA&A..54..135H} and \citealt{2023ASPC..534..355F}. Monitoring changes in accretion can be achieved by observing variations in either the continuum brightness or line fluxes.

By using the stellar parameters $M_{\star}$ and $R_{\star}$, with the latter often deduced from $T_{\text{eff}}$ and $L_{\star}$, it is possible to transform the assessed $L_{\text{acc}}$, derived either from UV-excess or line emission, into $\dot{M}_{\text{acc}}$ \citep{2016ARA&A..54..135H}. 

In this conversion process, the predominant sources of uncertainty arise from the stellar parameters, especially the ratio $M_{\star}/R_{\star}$, and potential fluctuations in the accretion rate. In recent years, the typical variability of accretion has become a focal point of research. A consensus among various studies indicates that accretion variability in disc-bearing pre-main-sequence stars generally reaches a peak at roughly a factor of 3 over timescales spanning from days to weeks \citep{2016ARA&A..54..135H,manara2021penellope}. There are indications, however, that continuous variability may be more pronounced in some objects. With the precise distances now provided by Gaia, the combined uncertainties in defining stellar and accretion properties lead to a total fractional uncertainty in individual $\dot{M}_{\text{acc}}$ measurements at any specific time of about a factor of 3 \citep{2014A&A...561A...2A, 2017A&A...600A..20A}.

\paragraph{The passively heated dust disc:} Within the disc, the dust temperature exhibits a decrease, ranging from the dust sublimation temperature of approximately $1400$ K at the innermost radii to roughly $10 - 20$ K in the outer disc. In the context of low-accretion discs, the primary mode of heating is passive, where the re-emission of stellar light at the disc surface overshadows the energy locally released through the diffusion of material across the disc. Consequently, the disc temperature is at its peak at the surface. The warmest dust is found at the innermost regions, emitting in the near and mid-infrared. Conversely, the outer disc maintains a lower temperature and is responsible for the prevailing long-wavelength emission \citep{chiang1997spectral}.  For passively heated discs, in regions where radiation raises the temperature to values \( \gtrsim 1000 \) K, the thermal energy is sufficient to ionise elements with low ionisation potentials, such as alkali. This high ionisation allows matter and the magnetic field to be sufficiently coupled to trigger MHD instabilities, such as the magneto-rotational instability (MRI). We will discuss the initiation of this instability in greater detail in Sec. \ref{sect:role of mri}.

\paragraph{The viscously heated gas disc:} For discs that undergo fast accretion, the heating stemming from viscous processes may elevate the disc temperatures to levels that exceed that of the star, encompassing areas significantly larger than the star itself ($L_{acc}>L_{phot}$). Consequently, the luminosity of the system becomes primarily governed by the accretion. When the accretion rate attains a sufficiently high level, the accretion flow in the disc innermost region has the potential to overpower the magnetic pressure, thereby crushing the magnetosphere and disrupting the magnetospheric flow \citep{hartmann1998accretion}. These viscously heated discs are most frequently observed at optical and infrared wavelengths, where shorter wavelengths are indicative of the hotter disc material in proximity to the star. Accretion rates are deduced through the application of models that operate on the straightforward assumption of a correlation between luminosity and accretion rate. The transition from a common low-state accretion disc to a high-state viscously heated disc manifests as substantial increases in brightness, predominantly noticeable at optical and infrared wavelengths. Such alterations in brightness can be directly translated into changes in the accretion rate, provided that potential corresponding changes in extinction are duly considered \citep{2023ASPC..534..355F}.
%\textcolor{red}{CS it is not clear to me if you are describing CTTS or FUORs?} 

%\subsubsection{Mass accretion rates}

%\subsubsection{Mass, Radius and luminosity}

%\textcolor{magenta}{CS Should you describe here the X-ray luminosity or after introducing the magnetic field?}

\subsubsection{Magnetic fields and accretion}

\begin{figure}[h!]
    \centering
    \includegraphics[width=1.\linewidth]{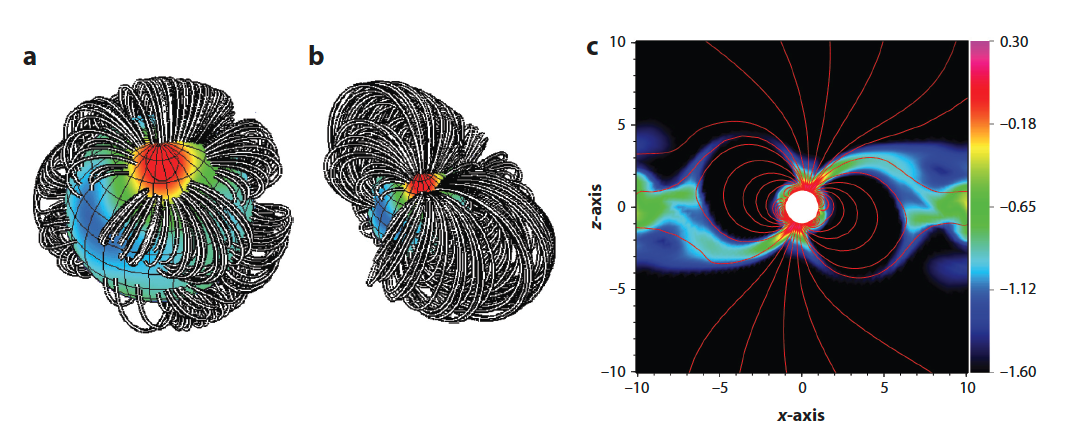}
    \caption{The magnetic field structure of V2129 Oph is characterised by two distinct aspects: (a) intricate high-order fields, and (b) a large-scale, dipole-like magnetic field \citep{donati2007magnetic, gregory2008non}. Additionally, (c) an illustration of the density in the accretion flow of V2129 Oph is provided, taken from \citet{alencar2012accretion}. This is based on simulations by \citet{romanova2011global}, illustrating how matter would travel along the magnetic field lines. The figure is reproduced from \citet{hartmann2016accretion}.}
    \label{fig:MagneticStructureofTTauri}
\end{figure}

In the inner disc, the simplistic notion of uniform accretion columns presented above corresponds to the infall of matter along a dipolar magnetic field (some magnetic field topologies are depicted in Fig. \ref{fig:MagneticStructureofTTauri}). Nonetheless, the reality of accretion flows is far more complex. These flows are neither uniform in density nor in temperature, and can be viewed as a conglomeration of individual accretion columns at a preliminary approximation. Furthermore, T Tauri stars of solar mass typically exhibit surface-averaged magnetic field strengths ranging from 1 to 2 kG \citep{johns2007magnetic, 2009ARA&A..47..333D}, a magnitude adequate to induce magnetospheric truncation radii near corotation if the field were exclusively dipolar. However, a significant portion of the field is arranged in quadrupolar and higher-order configurations, encompassing small-scale fields spread across the entire stellar surface \citep{donati2008large, chen2013spectropolarimetry}, which implies a substantial weakening of the dipolar component. 
These results may be taken cautiously as the authors observed mostly "isolated" stars, ie bright and not too embedded. This could not apply to all CTTS although it is usually assumed by the community. 

Zeeman–Doppler imaging techniques, involving the analysis of rotational modulation in the polarisation of photospheric lines and in narrow emission line components, have been employed to reconstruct the global structure of the magnetic field and the spatial distribution of accretion spots. These methodologies have unveiled magnetic fields characterised by strong multipolar components, and distributions of spots that are uneven across latitude and longitude, predominantly favouring the magnetic poles \citep{donati2008large}.

Some of these field lines might become twisted due to differential rotation between the disc and the star, causing them to bulge outwards or possibly even eject matter. These processes leading to very strong flaring events observed in X-rays \citep{Feigelson99,getman2008a,getman2021a} is the central question of this thesis. We further discuss T Tauri flares in Chapter \ref{C:Reconnection}.

Figure \ref{fig:MagneticStructureofTTauri} illustrates the reconstructed magnetic field of V2129 Oph, distinctly showing both the intricate 1.2-kG octupolar surface field and the 0.35-kG dipolar large-scale field. This large-scale field deviates by roughly 20° from the alignment of the disc, giving rise to a funnel flow that converges near the star pole \citep{donati2007magnetic, gregory2008non}. In these representations of the magnetic field, quadrupolar fields are considered negligible. Their existence would guide the accretion flow towards the star equatorial regions, conflicting with the observed polar accretion spots. The Zeeman–Doppler imaging studies reveal weaker dipole fields, resulting in reduced truncation radii compared to what would be seen if the total magnetic flux were taken into account \citep{bessolaz2008accretion, johnstone2014classical}. However, the strengths of the dipole fields in dark spots may be underestimated, depending on considerations of the surface filling factor \citep{chen2013spectropolarimetry}. In any case, a truncation radius near or inside the corotation radius corresponds to the inner disc structure described in \citet{hartmann2016accretion}, including the inner radius determined from the CO emission \citep{najita2003gas, salyk2011co}. Such truncation radii also align with the locations of extinction effects, which are induced by disc warps associated with accretion along inclined dipoles, and exhibit periods corresponding approximately to the corotation period \citep{bouvier2007magnetospheric, mcginnis2015csi}.

MHD simulations, carried out in both two and three dimensions for rotating stars with tilted dipolar fields, demonstrate that matter is funnelled toward the star in two distinct paths at high tilt angles (the angle between the rotation axis and the dipolar magnetic field axis), and in multiple channels at low tilt angles. These simulations show that factors like the distribution of hot spots, the covering area of these spots, as well as their temperature and density distribution, all vary depending on the tilt angle and the mass accretion rate, see \citet{2014EPJWC..6405001R} and references therein.

Models that take into account more complex magnetic field structures (for instance, the example provided in Fig. \ref{fig:MagneticStructureofTTauri}) illustrate that gas initially follows the dipole field lines. In certain cases, this results in well-structured funnel flowing in a stable accretion regime \citep{bouvier2007magnetospheric,kurosawa2013spectral}. As the flow approaches the stellar surface, strong octupolar fields modify the material trajectory. The point where the flow meets the star surface, called the flow footpoint, is situated at high latitudes when both dipolar and octupolar fields are predominant. If quadrupolar or higher-order fields are more pronounced, the footpoint is located at mid-latitudes \citep{romanova2011global,adams2011magnetically,johnstone2014classical}.

\section{Protoplanetary discs}\label{sec:ProtoplanetaryDiscs}
A circumstellar disc is a general term for ring-shaped accumulation of matter composed of gas, dust, planetesimals, asteroids, on collision fragments in orbit around a star. Such discs come in different types depending the disc age, with the primary ones being, protoplanetary discs, transition discs and debris discs.

A protoplanetary disc is a specific type of circumstellar disc. And all our study focuses on this type of discs. Protoplanetary discs are composed of gas and dust and are typically a few hundred to a thousand au in diameter, and their masses are typically a few percent of the mass of the central star. The dust in these discs consists of micro-to-mm sized grains made up of silicates and carbonaceous material, and the gas is mostly molecular hydrogen, with helium and trace amounts of heavier elements \citep{2023ARA&A..61..287O}.

The structure of these discs is such that the temperature and density decrease with increasing distance from the star. Near the star, the disc is hot and dense, but moving outwards, the temperature drops and the density decreases. This temperature gradient leads to a "snow line," a point beyond which water can freeze onto grains (see Fig. \ref{fig:DustGasStructure}), affecting the radial material composition of the disc \citep{2023ASPC..534..501M}.
The study of these discs has been greatly aided by the recent observations from telescopes such as ALMA, which can observe the light emitted by the dust grains and gas in these discs. Observations of the gas reveal the dynamics within the disc, including potential planet formation processes \citep{2021ApJS..257....1O}. 

In addition, disc dispersal mechanisms, which dictate the lifetime of these discs, are crucial for understanding the timescales for planet formation. Mechanisms such as photoevaporation by high-energy photons from the central star, winds and accretion onto the central star can all lead to disc dispersal \citep{2023ASPC..534..465L}. 

\subsection{Constraining protoplanetary disc structure}

Millimetre interferometry stands out as an exceptional method to measure the main characteristics of discs, specifically their masses, sizes, and large-scale spatial features \citep{williams2011protoplanetary}. To that aim, ALMA offers a unique blend of sensitivity and resolution. %This combination has led to comprehensive surveys of these essential properties across the majority of significant nearby star-forming regions.

ALMA surveys have provided notable enhancements in detection rates and the number of explored regions. Most of these investigations have been conducted in ALMA 6/7 Band, which ranges from approximately $890~ \mu$m to $1.3$ mm. 

The millimetre continuum and several millimetre emission lines are employed to gauge the overall disc masses \citep{pessah2017formation}. However, this approach does have its challenges. The methods are indirect since most of the disc mass, found in unobservable cold H$_2$ gas, cannot be directly measured. Furthermore, the conversion of the observable emissions into disc mass involves substantial assumptions. These include factors like dust opacity, gas-to-dust ratio, chemical compositions, temperature, and optical depth, which unfortunately remain poorly constrained for the majority of discs \citep{2023ASPC..534..501M}. 

\begin{figure}[h!]
    \centering
    \includegraphics[width=\linewidth]{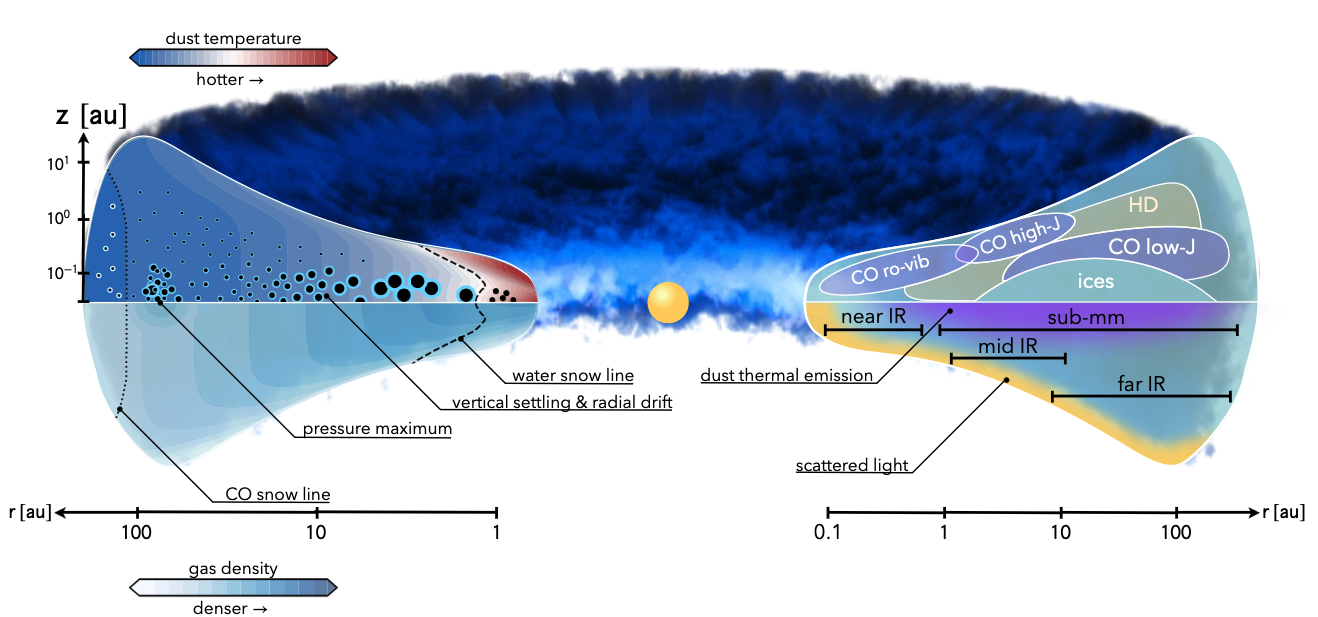}
    \caption{The figure shows the temperature of dust and the density of gas in protoplanetary discs. Black circles of varying sizes indicate the spread and relative size of dust particles. Within the water snow line, represented by a dashed curve, only uncoated grains are found. Beyond this line, grains coated with H$_2$O are outlined in blue, while CO-coated grains, found beyond the CO snow-line indicated by a dotted curve, are outlined in white. On the right side, the top panel simplifies the emission regions of key simple molecules. The bottom panel emphasises the main regions for dust thermal and scattered light emissions, coloured in purple and yellow respectively. Additionally, the emission wavelength range is provided. The axes represent the logarithmic distance from the central star, in both radial and vertical directions. This figure is reproduced from \citet{2023ASPC..534..501M}. }
    \label{fig:DustGasStructure}
\end{figure}

Figure \ref{fig:DustGasStructure} and the following sections highlight that in discs the materials, their physical states, and the necessary observational methods to track them differ significantly depending on their location. While advances in data quality and the increase in sample size have been largely beneficial for our understanding, they may have raised more questions than they have resolved.

\subsubsection{Disc Mass and composition}\label{sec:discmass}

\paragraph{Dust Mass:}
The total amount of observable solid material, referred to as dust, along with the sizes of these dust particles, are critical for understanding planet formation. The total amount of dust indicates how much material is available for forming terrestrial planets or giant planet cores. The sizes of the dust particles are key in determining their behavior in the gas, which in turn affects how they accumulate and form the building blocks of planets.

However, accurately measuring both the dust surface density and particle sizes is a complex task. Both quantities are often intertwined with the disc physical structure and the particles composition, affecting their optical properties. Therefore, interpreting observations of the disc continuum flux requires modeling and substantial assumptions to deduce the desired physical values.

At the most fundamental level, this involves converting the amount of material and its optical properties into the emitted continuum intensity, and vice versa. The relationship between these factors is not straightforward, making the assessment of the crucial details of dust a significant challenge in the study of planet formation.\\

The intensity of a plane-parallel layer with uniform temperature and opacity is given by the equation:

\begin{equation}
    I_\nu = B_\nu (T_{\text{dust}}) (1 - e^{-\tau_\nu})
\label{eq:outgoingintensity}
\end{equation}

Here, \( B_\nu \) is the Planck spectrum at the dust temperature, \( T_{\text{dust}} \), and \( \tau_\nu \) is the optical depth.

In the case of optically thick dust emission (\( \tau_\nu \gg 1 \)), the intensity becomes,

\begin{equation}
    I_\nu = B_\nu (T_{\text{dust}})
    \label{eq:intensityoptthick}
\end{equation}

This allows the dust temperature to be determined from the observed intensity \( I_\nu \).

On the other hand, if the emission is optically thin (\( \tau_\nu \ll 1 \)), the outgoing intensity is given by,

\begin{equation}
     I_\nu = B_\nu (T) \tau_\nu = B_\nu (T) \Sigma_d \kappa_{\text{abs}}.
     \label{eq:intensityoptthin}
\end{equation}
We define $\Sigma$ as the surface density,
\begin{equation}
\Sigma = 2 \int_0^ {H(R)} \rho(R,Z) dZ,
\end{equation}
where $H$ is the disc height at radius $R$ and $\rho$ is the mass density. The dust (resp. gas) surface density $\Sigma_d$ (resp. $\Sigma_g$) is computed from the dust (resp. gas) mass density $\rho_d$ (resp. $\rho_g$).
The frequency dependence of Eq. \eqref{eq:intensityoptthin} comes from two factors:
\begin{itemize}
    \item The Planck spectrum, which in the (sub-)millimetre wavelength is approximately in the Rayleigh-Jeans limit and is proportional to \( \nu^2 \).
    \item  The opacity \( \kappa_{\text{abs}} \), a mass absorption coefficient in units of cm$^2/$g is a pure material property often described as a power-law function of the frequency \( \kappa_{\text{abs}} \propto \nu^\beta \) (though \( \beta \) may be wavelength-dependent).
\end{itemize}

These equations offer a theoretical framework for understanding how dust emission behaves across different optical depths, and they provide insights into how the temperature, opacity, and other properties of the dust might be determined from observed intensities.

From Eq. \eqref{eq:intensityoptthin}, we can derive that measuring the dust surface density, temperature profile, or opacity within the disc would be possible if the other two parameters were known. Unfortunately, this is rarely the case with astrophysical sources. However, disentangling the components on the right-hand side of Eq. \eqref{eq:intensityoptthin} is theoretically feasible. The process would involve, first using 
%CS I checked there is a difference between use and utilise. Though utilise is more formal and could be used here, as you are not using the model above its limits, to use is safer.
a given or parameterised model of the opacity, then applying a corresponding model for the temperature and also having sufficient wavelength coverage to constrain all parameters within these models (e.g., \citealt{carrasco2019radial}).

The following discussion emphasises the methods and challenges of measuring the dust mass and its associated complexities.

We introduce an approximate method to estimate the disc mass if mean values, such as the temperature of the disc \( \bar{T_d} \) and the frequency averaged dust opacity \( \bar{\kappa} \), are known. And we further assume that the emission is optically thin. The expression for this flux-to-mass conversion is given by,

\begin{equation}
    F_\nu = \frac{{B_\nu (\bar{T_d}) \bar{\kappa}}}{{d^2}} M_{\text{{dust}}},
\end{equation}
where \( d \) is the distance to the source, and \( M_{\text{{dust}}} = \int 2\pi r \Sigma_d dr \) is the total dust mass of the disc. The integrals run over the entire disc.

This method was first proposed by \citet{1983QJRAS..24..267H} and has since been widely used and discussed. While it offers a relatively simple way to approximate mass, this assumption may fall short in accuracy if the discs being compared differ significantly in average temperature, size, or grain properties.

It is important to recognise that this approach is highly simplified and subject to various uncertainties. The assumption of known average values for temperature and dust opacity, and the fact that the emission must be optically thin, limits the application of this formula. Moreover, if the discs being compared have considerable differences in temperature or grain properties, the approximation can lead to incorrect conclusions about their relative masses. Even though this method has been commonly employed, it must be used with caution and an understanding of its limitations, see \citet{2023ASPC..534..501M} for a discussion.

%\textcolor{red}{Un peu surpris qu'il n'y ait pas d'autre méthode pour déterminer la masse de poussières dans un disque. Peut être Vincent aurait un avis là dessus ?}

\paragraph{Gas Mass:}
Determining the gas mass of protoplanetary discs, which is generally assumed to make up the majority of the disc mass, is a key unsolved issue in the study of star and planet formation. The mass of the disc is critical in defining its physics and evolution, ranging from its creation to the eventual formation of planets. Even though it is necessary to define the gas masses within these discs, the measurement of them is far to be straightforward, a point further elaborated in \citet{2017ASSL..445....1B} .

Molecular hydrogen (H$_2$) is the primary gaseous element found in these discs, but its emission is weak due to specific molecular physics. The considerable energy gap in its fundamental ground state transition leads to very faint H$_2$ emission in cold areas. This is particularly true in the outer regions of protoplanetary discs, where the typical gas temperatures are $20-30$ K. Only at much higher temperatures, above 100 K and near the central star, does H$_2$ emission become noticeable. However, this emission does not give much information about the overall mass of the disc \citep{thi2001h2,bary2008quiescent}. As the direct detection of cold molecular hydrogen is extremely challenging, observers must use indirect methods to trace the gas mass.
\begin{itemize}
    \item \textbf{HD Emission} The molecule most similar to H$_{2}$ is hydrogen deuteride (HD), though it is less abundant. HD chemistry resembles that of H$_{2}$ in that it does not freeze out onto grains. In contrast, other molecules, like CO and less volatile substances, are unable to remain in the gas phase at low temperatures and adhere to icy grains through a process known as freeze-out \citep{2023ARA&A..61..287O}. 

    HD shares another characteristic with H$_{2}$, it can shield itself from photodissociating UV photons, though it does so with less efficiency \citep{wolcott2011suppression}. The ratio of HD to H$_{2}$ is roughly \(3 \times 10^{-5}\) \citep{prodanovic2010deuterium}.

    Unlike H$_{2}$, HD has a slight dipole moment that enables dipole transitions \((\Delta J = 1)\). To excite HD fundamental rotational level (J = 1 - 0), 128 K of energy is needed, making its expected emission much larger than that of H$_{2}$ in the temperature range between 20 and 100 K. While this line does not directly track the majority of cold H$_{2}$ gas, its emission can be used to gauge the overall gas mass in the disc, relying on physical-chemical models of the disc structure. This method has been elaborated upon in studies like \citet{bergin2013old,kama2016volatile}.
    
    The fundamental rotational transition of HD is observed at \(112 \, \mu\text{m}\), and was surveyed with the PACS instrument on board the Herschel Space Observatory. This transition has been a focus for a number of nearby and luminous protoplanetary discs, yielding only three notable detection: TW Hya \citep{bergin2013old}, DM Tau, and GM Aur \citep{mcclure2016mass}. The detection of HD in TW Hya was particularly precise, revealing a surprisingly substantial disc mass of \(0.06 \, M_\odot\) for an old disc of approximately \(10\) Myr. Nonetheless, difficulties arise when translating HD mass into the overall disc mass.

    Initially, it is worth noting that the emitting region of the HD \(112 \, \mu\text{m}\) line is located above the midplane, where the gas temperature is above \(30\) K. This condition necessitates a detailed understanding of the disc vertical structure to estimate the disc mass \citep{trapman2017far}. A second challenge is that currently, to our knowledge, there exists no instrument, neither in use nor planned, capable of covering the HD \(J = 1 - 0\) transition, necessary for an extensive unbiased sample of discs \citep{kamp2021formation}.

    \item \textbf{CO Emission} Carbon monoxide (CO) and its less abundant isotopologues are frequently utilised as tracers of gas properties, structure, and dynamics within discs. Being the second most abundant molecule after H$_2$, CO serves as the principal gas-phase container of interstellar carbon. Moreover, due to its chemical stability, well-understood and comparably uncomplicated interstellar chemistry, CO can be effortlessly integrated into physical-chemical models with varying degrees of complexity \citep{woitke2011unusual,williams2014parametric}. These characteristics render CO an attractive gas mass tracer, enabling its distribution to be more directly correlated to that of molecular hydrogen, all while minimally depending on assumptions regarding disc chemistry \citep{kamp2017consistent}.

    \citet{aikawa2002warm} offered a straightforward depiction of CO distribution within discs, portraying CO as being prevalent within a warm molecular layer defined by two boundaries. The lower boundary is characterised by CO freeze-out within the cold disc midplane, and the upper boundary is marked by photo-dissociation due to UV photons originating from either the central star or an external radiation source. The former process diminishes the quantity of gas-phase CO, leading it to freeze onto dust particles at \(20\,K\), specific to pure CO ice, possessing a laboratory-measured binding energy of \(855\,K\) \citep{bisschop2006desorption}. This value, however, can range between \(17\,K\) and \(30\,K\), depending on the assumed density and binding energy in composite ices \citep{cleeves2014ancient}. Photo-dissociation of CO is regulated by line processes, initiated by the particular absorption of photons into predissociative excited states, making it susceptible to self-shielding \citep{1988ApJ...334..771V}.

    Above all, CO serves as a highly convenient mass tracer, owing to its substantial abundance and detectability at both millimetre and submillimetre wavelengths. The predominant isotopologue of carbon monoxide \(^{12}\text{CO}\), is so abundant in discs that its emission is primarily optically thick, yielding insights mainly into the disc temperature and dynamics. Although subject to significant uncertainties due to the high optical depth of the lines, \(^{12}\text{CO}\) emission has been utilised in prior investigations to approximate the mass of cold gas by employing rudimentary equations \citep{thi2001h2}. To assess column densities more accurately, it is essential to make use of more optically thin markers, like the low-level rotational emission of \(^{13}\text{CO}\) and \(^{18}\text{CO}\). These less abundant isotopologues of \(^{12}\text{CO}\), are susceptible to collective shielding \citep{visser2009photodissociation}, which selectively influences their prevalence. Processes selective to isotopes, such as isotope-selective photodissociation, have been validated as significant when interpreting \(^{13}\text{CO}\) and \(^{18}\text{CO}\) emissions for disc mass measurement. This leads to the possibility to underestimate mass by as much as two orders of magnitude if these processes are overlooked. These mechanisms were incorporated into physical-chemical models by \citet{miotello2014protoplanetary}, while \citet{miotello2016determining} furnished both simulated CO fluxes and analytical guidelines.

    A key obstacle in determining gas masses through CO isotopologue lines arises from the ambiguities associated with the C/H ratio within discs. Analyses of the HD fundamental line in TW Hya revealed that CO-derived gas masses could be anywhere from one to two orders of magnitude less than HD-inferred disc masses, even when considering isotope-selective procedures and CO freeze-out \citep{2013Natur.493..644B, cleeves2015constraining,trapman2017far}. This apparent inconsistency may be attributed to the sequestering of carbon and oxygen-rich volatiles as ice within more massive bodies, consequently resulting in diminished observed CO fluxes. Intriguingly, this disparity between CO and HD is absent in the warmer discs surrounding Herbig stars \citep{kama2020mass}, reinforcing the conjecture that volatile entrapment occurs more efficiently when freeze-out is prevalent.
    
    \item \textbf{Direct Measurement} Ideally, the goal is to directly determine the mass of a disc without depending on indirect chemical tracers. This can be explored by studying the gravitational mass of the disc through dynamical investigations. Initially suggested by \citet{rosenfeld2013spatially}, this approach states that the self-gravity of the disc is associated with its orbital velocity. This velocity can be identified through observations of gas that are both spatially and spectrally resolved. It will deviate from purely Keplerian rotation if the gas tracer emission layer is raised from the midplane and if there is a substantial negative pressure gradient (leading to sub-Keplerian orbital velocity), or when the gravitational potential is significantly affected by the disc mass i.e. not only affected by the central star (resulting in super-Keplerian velocity). Because the observations need to be spatially resolved, and because the disc mass enclosed within the observation radius should be significant compared to the stellar mass, this method is only feasible at a large distance from the star.

    \citet{veronesi2021dynamical} applied this concept by modelling the CO rotation curve of the massive Elias 2-27 disc. They used a theoretical rotation curve that took into account both the disc self-gravity and the star contribution to the gravitational potential. They found a more accurate alignment with the observed data, allowing them to make the first dynamical estimation of the disc mass. They concluded that the disc mass is 17\% of the star mass, suggesting susceptibility to gravitational instabilities.

    This technique is promising, as it allows the determination of gravitational mass without depending on the emission from a particular tracer. However, by its nature, such measurements can only be performed on discs that are massive in comparison to their host stars, and it necessitates high spectral and spatial resolution, which, in turn, demands extended integration times.
\end{itemize}

\citet{kama2020mass} have lately reported constraints on the mass upper limits of Herbig discs, constraining nearly all the discs to a gas mass \(M_{\text{gas}} \leq 0.1M_\odot \), thereby excluding the possibility of global gravitational instability. Within this study, one particular disc has been pinpointed with an especially stringent limitation on the disc mass, set at \(M_{\text{gas}} \leq 0.067 ~ M_\odot\), which translates into a gas-to-dust ratio of \(\leq 100\).

To summarise, determining the total mass of a disc is both fundamental for disc modelling and very challenging. Each method outlined previously has its limitations and is based on various assumptions. It is important to be aware of these assumption when utilising disc mass measurements.

\subsubsection{Disc Radial Structure}
The way in which disc mass is distributed radially is central in determining the locations and initial masses of forming planets within a disc. As the disc evolves over its lifetime, significant changes are expected in this distribution, though the precise details of this evolution remain an unresolved question. Various processes contribute to this evolution, including mass accretion onto the central star \citep{2016ARA&A..54..135H}, dissipation of part of the disc content through mechanisms like high-energy radiation-driven winds or magnetic torques \citep{2016SSRv..205..125G} and external photoevaporation \citep{clarke2007response}.

The dominance of a particular process will considerably alter the disc radial structure, significantly affecting factors such as the surface density distribution, denoted as $\Sigma(R)$. A reliable methods to measure $\Sigma(R)$ would be extremely valuable to model protoplanetary discs, however, there is currently no consensus on the most effective tracers and observational techniques to employ.

\paragraph{Constraining the radial surface density of the disc gas:}
Estimating the gas surface density in discs is challenging and shares certain limitations with gas mass evaluations. Three main factors prevent the determination of a consistent 'global' gas column density estimation: the dust continuum optical depth, the optical depth of the gas tracer, and the excitation conditions of the gas.

In the outer disc, more than a few tens of au, the environment is typically optically thin at radio wavelengths. However, emissions from molecular lines, such as CO submillimetre lines, are often optically thick in at least part of the disc. A common approach to address this challenge is to select a rare isotopologue line that may be optically thin.

Recent high-resolution observations, as precise as approximately 20 au, carried out as part of the MAPS survey \citep{law2021molecules}, have shed more light on this subject. The modelling of combined gas and dust by \citet{zhang2021molecules} has been utilised to determine gas surface density profiles. For regions beyond approximately 150 au, many of these profiles follow power laws and exhibit a tapered outer edge, which is consistent with the pattern expected from viscous disc evolution.

An illustrative example of this kind of study is the work on the CO column density in the inner disc ($5-21$ au) around the star TW Hya. This was achieved by analysing C$^{18}$O and $^{13}$C$^{18}$O lines, as demonstrated by \citet{zhang2017mass}. While C$^{18}$O was observed to be optically thick, tracing temperature, $^{13}$C$^{18}$O is optically thin tracing the surface density. By leveraging the resolved $^{13}$C$^{18}$O rotational emission, along with the empirically derived temperatures, and utilising HD J = 1 — 0 to normalise the total gas surface density, the authors were able to calculate a gas surface density expression of 

\begin{equation}
    \Sigma_{\text{gas}} = 13^{+8}_{-5} \times \left(\frac{r}{20.5 \text{au}}\, \right)^{-0.87^{+0.38}_{-0.26}}\, \text{g cm}^{-2} 
\end{equation}
within 20 au of the star. This study emphasises the possibility to pin down the distribution of gas with CO, despite the challenges in firmly constraining its absolute normalisation factor.

A supplementary method to estimate the gas surface density in the inner region of a disc (within approximately 10 au) involves the use of higher excitation CO ro-vibrational lines that detect gas temperatures ranging from several hundred up to 1000 K. These lines are frequently identified in discs surrounding T Tauri stars \citep{najita2003gas,salyk2011co}. However, to generate accurate surface density profiles, it is again essential to detect rare CO isotopologue line profiles, such as $^{13}$CO, C$^{18}$O and  $^{13}$C$^{18}$O. So far, this approach is predominantly constrained to the more luminous and massive counterparts, namely the Herbig discs (e.g., \citealt{carmona2014constraining,carmona2017gas}). 

\begin{figure}[h!]
    \centering
    \includegraphics[width=0.7\linewidth]{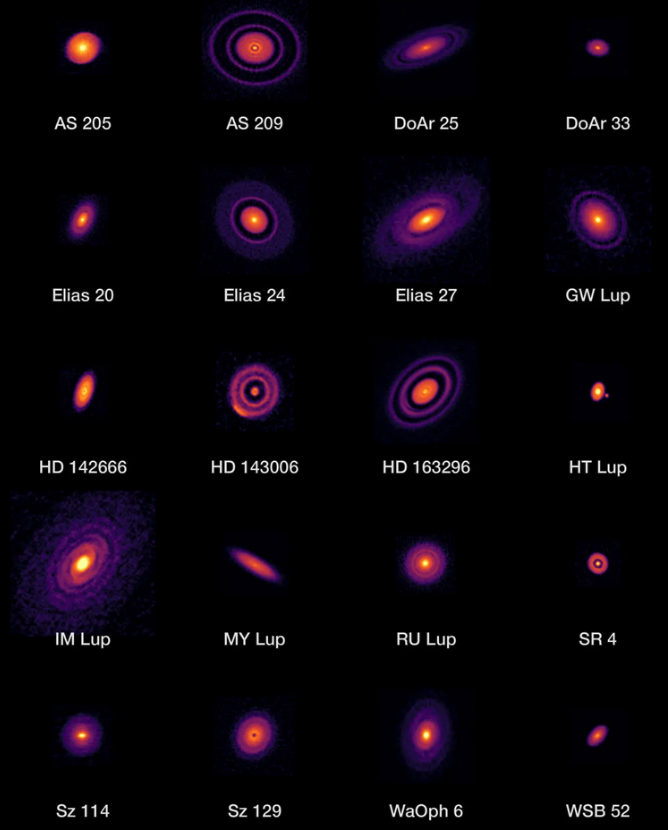}
    \caption{ALMA's high-resolution images of nearby protoplanetary discs, which are results of the Disk Substructures at High Angular Resolution Project (DSHARP). Credit: ALMA (ESO/NAOJ/NRAO)\protect\footnotemark}
    \label{fig:VariousPPD}
\end{figure}

In summary, the current understanding of radial dust surface density in discs remains elusive. The high-resolution observations hint at the potential presence of substantial substructures that could obscure true dust mass in optically thick areas. Such substructures are diverse, from shallow rings to well-demarcated narrow bands, see Fig. \ref{fig:VariousPPD}. However, some discs exhibit no discernible substructures, further complicating our comprehension of their role in dust emissions.\footnotetext{\url{https://www.almaobservatory.org/en/press-releases/alma-campaign-provides-unprecedented-views-of-the-birth-of-planets/}} On the other hand, it emerges that many observed dust gaps are not accompanied by corresponding gaps in gas; this revelation influences the comprehension of the mechanisms creating these gaps. For instance, if dust rings are due to gas pressure traps and the drift of dust grains, there should be an observable relationship between the luminosity of the dust continuous emission and the gas surface density or pressure, as posited by studies like those by \citet{2018ApJ...869L..46D}. In the disc inner regions, specifically within 50 astronomical units and in alignment with the Minimum Mass Solar Nebula, the observed data does not illustrate a straightforward power law for the gas surface density profile. Instead, notable sub-structures are evident, as highlighted in \citet{2021ApJS..257....5Z}.

\paragraph{Constrains on the radial thermal structure:}
Gas temperature within a disc can be determined through direct methods such as analysing gas emission line fluxes, radial emission profiles or through indirect (chemical) techniques like examining molecular emission maps that trace specific snow lines. In the latter, the gas emission specifically pinpoints part of the radial and vertical regions of the disc. Thus, using a single tracer only allows the reconstruction of the gas temperature profile in a specific portion of the disc, where it primarily emits from. To construct a full understanding, multiple tracers are required, ideally with overlapping coverage.

Discussing the direct gas line methods, they can be grouped by their radial origin.
\begin{itemize}
    \item \textbf{Inner disc (less than 3 au) } — Initial estimates of the typical inner disc gas temperatures were obtained from emissions of CO, OH, and water in the near-IR. These results were in agreement with the theoretical understanding from thermo-chemical disc models. Further validation came from the analysis of Spitzer spectra, which reinforced that typical gas temperatures in the disc surface up to a few au range from 500 to 900 K. The gas temperature exhibits a range with different mid-IR molecules. For instance, water fits best with temperatures around 500 K, whereas CO and OH display a wide range, from 500 to 1700 K. CO$_2$ and HCN lie in between, with typical temperatures of about 700 to 800 K.

   Since Local Thermodynamic Equilibrium (LTE) slab models include a number of free parameters (like gas temperature T, column density N, emitting area A), these results can be somewhat degenerate without spatially resolved spectral line profiles. Thermodynamically, different molecules are expected to come from varying heights in the disc \citep{woitke2018modelling}, arranged by height from high to low, OH, CO, H$_2$O, CO$_2$, HCN. Therefore, discovering different temperatures for these molecules coincides with detailed disc models. 

   Generally, mid-IR observations only probe the column of gas above the local dust continuum. Moreover, molecules like water might be significantly affected by thermal dust emission within the inner disc \citep{kamp2013uncertainties}, so the measured water temperature might more accurately reflect the standard dust temperature at a few au.

   The determination of gas temperature profiles within a few au has been refined with spectral line profiles, particularly by utilising CO ro-vibrational lines. Studies have produced temperature estimates ranging from 700 to 1700 K for typical emitting regions of less than or equal to 0.2 au \citep{salyk2011co,bast2011single,brown2013vlt}.

    The advent of new VLT instruments such as GRAVITY and MATISSE, and the forthcoming ELT/METIS instrument, will allow the study of spatially resolved gas line emission in inner T Tauri discs.

    \item \textbf{Middle disc region (3-10 au)} —  The estimation of gas temperatures is achieved through diverse wavelength ranges. Low excitation water lines in the mid-IR probe this area. The higher J rotational lines of CO, investigated in the far-IR are also relevant \citep{2023ASPC..534..501M}.

    An innovative indirect method was employed by \citet{zhang2013evidence} to determine a temperature reference in the disc surrounding TW Hya from the water snow line location. By utilising an emission line “drop-out” method across mid-IR to far-IR water lines, they determined the surface snow line location to be at 4.2 au.

    \item \textbf{Outer disc region (10 - a few hundred au)} —  The far-IR wavelength range is abundant in gas cooling lines of common species like OH, H$_2$O, and CO. \citet{kamp2013uncertainties} discovered that these far-IR lines from the disc surrounding TW Hya are highly sensitive to temperature, including [O I] and water. The brightness of the [O I] 63 $\mu$m line, and consequently the disc surface temperature between 30 to 100 au, is explained by the UV excess and X-ray emissions from T Tauri stars, combined with disc flaring (the reflection of the radiation of the star on the disc surface). The presence of any shadowing by an inner disc, such as puffed-up geometry or misalignment, can influence the temperature in the outer disc surface and the subsequent line emission \citep{woitke2019consistent}.

    In more recent studies, a combination of 2D thermo-chemical disc modelling and ALMA observations of CN, HCN, and HNC has been applied to explore the relationships between disc irradiation, such as UV excess and geometry, and gas temperature \citep{cazzoletti2018cn,long2021exploring}. For the face-on disc TW Hya, multiple research efforts have attempted to separate the disc temperature and surface density profiles through the examination of CO isotopologue lines. A steep temperature profile inside 30 au followed by a flat plateau, around 20 K, up to approximately 65 au was determined by \citet{schwarz2016radial}, suggesting the detection of the CO iceline surface.

    \citet{zhang2017mass} employed the spatial distribution of the even rarer isotopologue $^{13}$C$^{18}$O to estimate the midplane CO iceline location at $20.5 \pm 1.3$ au, revealing a power law index of approximately,
    \begin{equation}
        T(R)\propto R^{-0.5}
    \end{equation}
    for the midplane gas temperature inside that CO iceline. Additionally, peak brightness maps of a single optically thick CO line can be employed to ascertain radial disc temperature profiles when dealing with high spatial resolution data, as demonstrated by \citet{weaver2018empirical}.

    These observational and theoretical analyses highlight the complexities in isolating the radial temperature profile from the vertical temperature gradient. It becomes apparent that both aspects must be simultaneously constrained, as separating them through observation proves to be a challenging task.
\end{itemize}
In summary, there has only been a small amount of studies on the quantitative interpretation of temperatures obtained from gas observations. This shortfall is attributed partly to the absence of reliable gas data and the requirement for consistent multi-molecule/line datasets. However, with the recent influx of data from the MAPS ALMA Large programme \citep{2021ApJS..257....1O}, this void is beginning to be addressed with forthcoming observations.

\subsubsection{Disc Vertical Structure}
\paragraph{Constrains on the dust and gas vertical structure:}

\begin{figure}[h!]
    \centering
    \includegraphics[width=0.6\linewidth]{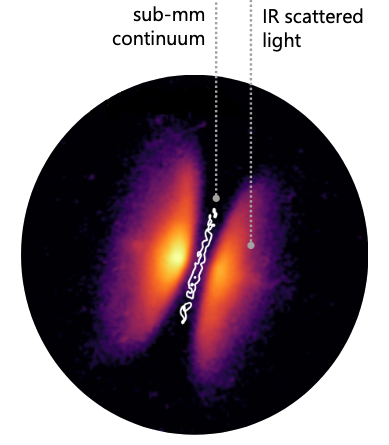}
    \caption{\citet{villenave2020observations} observations of the edge-on disc Tau 042021 reveal sub-millimetre continuum emissions from millimetre-sized grains at the centre, flanked by scattered light from smaller dust particles at the top. The continuum emissions at millimetre wavelengths indicate a flat, settled midplane. In contrast, the scattered light emissions, represented by the colour map, display a flared and vertically extended structure, whereas the millimetre emissions, outlined in white, depict a remarkably flat midplane. The figure is adapted from \citet{2023ASPC..534..501M}.}
    \label{fig:DustSettling}
\end{figure}

The vertical distribution of dust within a disc is influenced by a variety of factors. Gas and any related turbulent movements push dust upward, while gravitational forces pull the grains toward the midplane.

The vertical arrangement of dust grains is closely connected to the characteristics of the surrounding gas. Essentially, the vertical structure of the gas is governed by the equilibrium between the vertical pressure gradient and the gravitational pull toward the midplane. This relationship defines the gas scale height $H$ at radius $R$ assuming a standard Shakura-Sunyaev Keplerian geometrically thin disc, $H/R \ll 1$, where, 

\begin{equation}
    H = \frac{{c_s}}{{\sqrt{{G M_{\ast}/r^3}}}} = \frac{{c_s}}{{\Omega_K}},
    \label{eq:scaleheight}
\end{equation}
where \( c_s \) is the sound speed, $\Omega_K$ the Keplerian pulsation, \( G \) is the gravitational constant again, \( M_{\ast} \) is the mass of the central star, and \( r \) is the distance from the central star. The vertical distribution of the dust depends greatly on how tightly the grains are bound to the gas. If the dust is well-coupled to the gas, it will mimic the gas profile. If it is decoupled, the dust will tend to settle toward the midplane \citep{dubrulle1995dust}, see Fig. \ref{fig:DustSettling}.

This coupling is dependent on the size of the dust particles, which can be characterised by their Stokes number. This number is roughly the ratio of the time required for the dust to adjust to the gas velocity to the orbital time scale. The difference in vertical distribution depending on the grain size can be seen in both infrared and millimetre-wave observations. Observations in the optically thick infrared indicate that micron-sized grains of dust have been stirred up, while mostly optically thick millimetre-wave observations trace settled millimetre-sized dust grains near the mid-plane. This distinction can be observed visually by comparing the images captured by ALMA, SPHERE, and HST (e.g. \citealt{avenhaus2018discs,andrews2018disc,villenave2020observations}), see Fig. \ref{fig:DustSettling}.

High-resolution images of inclined axisymmetric rings have enabled geometrical estimations of the rings vertical extent. For instance, the HL Tau system is consistent with a geometrically thin disc, $H/R \ll 1$, where $H$ is the disc height at radius R \citep{2016ApJ...816...25P}. However, this is not universally observed, as seen in HD 163296, which exhibits both puffed-up and settled rings \citep{doi2021estimate}, where in this case Eq. \eqref{eq:scaleheight} is not valid.

This raises the problem of understanding why the micron-sized grains are puffed up. Analyses of molecular line emissions have revealed that disc turbulence, can be measured from the $\alpha$ parameter. The turbulent velocity can be expressed as $v_{turb}\approx \sqrt{\alpha}  c_s$ \citep{2011ApJ...727...85H} or $v_{turb}\approx \alpha c_s$ \citep{2013ApJ...775...73S}. From grain observations, the values of $\alpha$ are found to be notably weak in the outer disc. For instance, in TW Hya \( \alpha < 0.007 \) and HL Tau \( \alpha \approx 3 \times 10^{-4} \) (see \citealt{teague2016measuring,flaherty2018turbulence,2016ApJ...816...25P}). The theoretical determination and effects of $\alpha$ is a central question of this thesis. Its influence on accretion mechanisms will be explored in Section \ref{sec:ModelingProtoplanetaryDiscs}, while a methodology for estimating \( \alpha \) will be introduced in Sect. \ref{sec:Derivationofaccretionrate}. Additionally, the impact of energetic particles generated by stellar flares on the value of \( \alpha \) will be examined in Chapter \ref{C:PublicationII}.

In the inner disc (within 10 au), turbulence can be measured using near-IR molecular lines, but this method has only been successful for one Herbig disc, revealing a higher $\alpha$ value in the inner disc compared to typical ALMA measurements for the outer disc. This may indicate that MRI is more efficient inside than outside.

In this context, the role of the Extremely Large Telescope (ELT) becomes crucial. It is expected to offer higher spatial and spectral resolution than the JWST, enabling the study of scales less than 10 au with high spectral resolution for more than just the most extreme discs currently accessible with existing telescopes.

The observed lack of turbulence is puzzling because it fails to account for the known accretion rates and hints at the presence of another mechanism guiding the disc evolution. This alternate mechanism would need to transfer mass and angular momentum without substantially causing observable gas turbulence or major mixing of the dust within the mid-plane. Potential mechanisms for this behaviour could include magnetic phenomena such as magnetic winds \citep{bai2017global}.

\paragraph{Constraining the vertical thermal structure:}

%%%%%%%%%%%%%%%%%%%%%CS
Gradients in disc temperature, are expected both radially and vertically. The observational confirmation of vertical temperature gradients has significantly strengthened in recent times, yet extracting and interpreting firm numbers from these observations remains a complex task. This complexity arises from the intricate connection between vertical temperature gradients and numerous fundamental properties, like including the evolutionary state of the dust and the chemical composition of the disc, which affects heating and cooling agents \citep{gorti2004models,woitke2009radiation}. Further complicating matters are the disc inclination and associated projection effects, which cloud the direct observational constraints on temperature. Often, these require highly detailed spatial resolution data \citep{rosenfeld2013spatially,pinte2018direct}. These and other factors contribute to the intrinsic difficulty in observationally constraining vertical temperature gradients. 

Various models, both analytic and Monte Carlo-based, have been developed across the literature to anticipate the temperature profiles of irradiated accretion discs (e.g. \citealt{bruderer2012warm,woitke2019consistent}). These models exhibit different degrees of complexity and may encompass factors such as layered dust settling, accretion heating, thermochemistry, and interactions with the vertical gas density profile. A key aspect is the relationship between gas scale height and temperature, mediated by gas pressure and magnetism counteracting the gravitational forces from the star and disc \citep{bai2009heat}. For instance, the increased pressure support found in warmer discs leads to larger scale heights compared to cooler discs. Additionally, a more expanded vertical structure absorbs greater heating radiation from the star. This amplifies the pressure support, which is in turn balanced by an increased surface area for cooling.

Disc substructures, including features like inner holes or radial gaps, can alter the illumination of the disc, leading to localised changes in its vertical profile. For example, these modifications might create puffed-up inner rims \citep{natta2001reconsideration,bi2021puffed} or the outer edges of disc gaps \citep{jang2012gaps}. Such effects are critical to estimate the deviation from the standard and simplified assumption that a disc is locally vertically isothermal. This assumption often leads to a Gaussian vertical density profile and, despite its limitations, continues to be employed in models of disc structure due to its straightforward nature.

In the early stages, observational constraints on vertical temperature gradients within discs mainly concentrated on thermal continuum and optically thick gas. Detailed fitting of the SED indicated the necessity of a warm irradiated surface to reproduce the IR excess and the 10-micron silicate feature. Nonetheless, specific temperature details were at that times challenging to separate from other disc properties such as composition and structure \citep{woitke2011unusual,woitke2019consistent}. Optically thick CO emission has been a popular temperature tracer, utilising both high-J CO studies \citep{fedele2016probing} and $T_{gas}$ derived from multi-isotopologue analyses.

The advent of high-resolution observations of gas and dust by interferometers has led to an increasingly sophisticated understanding of temperature constraints in discs. These advanced tools provide the fine-grained resolution needed to map not only radial temperature gradients but also dissect the disc layers to more precisely determine local temperatures. The following paragraphs will further detail some of the current observational insights into the temperatures of both the elevated molecular layer and the disc midplane.

The molecular layer of the disc, a region where molecules are shielded from dissociating radiation yet remain warm enough to stay in the gas phase, has accumulated extensive temperature data, particularly in the outer disc at temperatures less than 100 K. This rich dataset is attributed to the layer temperature coinciding with the energy needed to stimulate rotational emission in several common interstellar molecules, detectable at submillimetre wavelengths. The temperature of this specific layer can be ascertained through a combination of techniques, based on chemistry, line excitation, and line optical depth.
\begin{itemize}
    \item \textbf{Chemistry Technique} This method uses the straightforward fact that molecules will freeze out onto dust grains at particular dust temperatures, though this can vary slightly depending on surface properties, presence of other ice species, local gas density or pressure, and gas temperature (e.g., \citealt{collings2004laboratory}). For more edge-on discs, a sharp decline in submillimetre emission might indicate a molecule freeze-out boundary.
    \item \textbf{Line Excitation Technique} This approach uses multiple transitions of optically thin lines to pin down the gas excitation temperature over the observed gas column. In discs, the excitation temperature is usually equivalent to the gas kinetic temperature, barring the upper thin layers or at extensive radii, depending on a molecule critical density. For example, \citet{loomis2018distribution} used the close spacing of specific CH$_3$CN transitions to estimate the rotational temperature and column density for that molecule. With the insight from astrochemical models about the molecule emitting layer, they suggested that the molecular layer temperature ($z/r \approx 0.3$) might be as low as 30 – 40 K
    \item \textbf{Line optical depth} When the spectral line emission is optically thick, it is possible to measure the brightness temperature at the location where $\tau = 1$, interpreting it as the gas temperature \citep{cleeves2018constraining}.   
\end{itemize}

Understanding the midplane of the disc is of great interest as it can influence both the composition of planetesimals and the specific mechanisms involved in the formation of planets. Some techniques for estimating the midplane temperature share similarities with those used for the molecular layer. These include locating snow lines and using emissions that become optically thick at heights corresponding to the midplane \citep{zhang2017mass}.

However, the concentration of large dust grains that settle in this region also provides some unique tools. For instance, millimetre-sized dust grains, emitting at centimetre to submillimetre wavelengths, can generate optically thick emissions, particularly in the inner disc. In the dense midplane, where frequent collisions cause an equilibrium between the two components, this is equivalent to the gas temperature \citep{2023ARA&A..61..287O}. 

Another approach to estimate the midplane temperature involves using highly spatially resolved observations of an optically thick line such as $^{12}$CO in an inclined disc. Through projection effects, the optically thick upper surface on the near side of the disc, along with the thick bottom surface on the far side, can offer constraints on the temperature difference between the layers \citep{dullemond2020midplane}. This method is mainly suitable for either the warmer regions of a disc surrounding a low-mass star, where tracers like CO do not freeze out, or for discs around more massive stars.

To summarise, there are only a few cases where diverse techniques have been employed on the same disc to validate a structure consistently, especially for estimating midplane and molecular layer temperatures among other factors. For some low-mass stars, it appears that the molecular layer may not heat as intensely as some thermochemical models propose, potentially alluding to variations in the existence or nonexistence of cooling agents, diverging from what standard interstellar cloud chemistry would anticipate. The comparative studies using various techniques highlighted here are crucial for a more refined understanding of disc vertical structures.

\subsection{Modeling Protoplanetary discs}\label{sec:ModelingProtoplanetaryDiscs}

Protoplanetary discs are essentially accretion discs, which supply their central stars with material at rates typically within the range of $10^{-10}$ and $10^{-7}$ solar masses per year \citep{2014A&A...570A..82V,hartmann2016accretion,2016A&A...591L...3M}. The formation, development, and eventual dispersal of these discs are intrinsically related to their angular momentum content. Given that angular momentum is conserved,
%CS ? only in the steady case!
the manner in which this angular momentum is transported becomes a critical factor.

\subsubsection{Physical properties}

%\paragraph{Physical processes:}

The evolution of a disc comes through the combined effects of internal processes, surface processes, as well as infall and mass loss. In the early stages, self-gravity \citep{kratter2016gravitational} and infall \citep{lesur2015spiral} are significant factors. However, in the subsequent evolution phase, the one we are interested in for this thesis, is mainly dictated by the disc net vertical magnetic field $B_z$ and its coupling to the disc matter due to sufficient ionisation. 

\paragraph{Plasma $\beta$ parameter}
The strength of the vertical magnetic field can be characterised by the plasma beta, which is the ratio of thermal pressure $P_{th}$ to the magnetic pressure $P_m$,
\begin{equation}
    \beta= \frac{P_{th}}{P_m}= \frac{8 \pi nkT}{B_z^2}= \frac{8 \pi \rho c_s^2}{B_z^2},
    \label{eq:definitionbetaplasma}
\end{equation}
where $n$ is the particle number density, $k$ the Boltzmann constant, $T$ the temperature, $\rho$ the mass density and $c_s$ the sound speed. 
In the case of very high plasma beta ($\beta$ $\gg 10^5$) the evolution is not driven by magnetic processes, but rather by other mechanisms like vertical shear instabilities \citep{nelson2013linear,flock2020gas}, hydrodynamic processes \citep{lyra2019initial}, and mass loss through photoevaporation \citep{2017RSOS....470114E}.

Weak yet non-zero $B_z$ fields, responsible for $\beta$ within the range of $10^3 - 10^5$, can induce levels of both turbulent and laminar MHD transport that overcome those found in non-magnetised discs \citep{2013ApJ...775...73S,2017A&A...600A..75B,2021JPlPh..87a2001P}. These fields are associated with mass and angular momentum loss via MHD winds \citep{2013ApJ...767...30B}. Lower $\beta$ values, which can still be plausible outcomes of star formation \citep{xu2021formation} are also an equilibrium state for simplified models of protoplanetary discs \citep{guilet2014global}.

Theoretically, discs where $\beta$ approaches infinity would undergo very slow evolution. Conversely, discs that are more strongly magnetised are anticipated to have shorter lifetimes due to the influence of magnetic braking. This theoretical insight leads to two central questions. The first concerns to whether disc evolution is primarily the result of turbulent transport or MHD winds. The second question concerns the source of turbulence, whether it is primarily generated by hydrodynamic or MHD processes, recognising that some level of turbulence must exist, even if it is not the principal cause of disc evolution.

Modern simulations enable the examination of hydrodynamic and MHD transport processes over short periods, using physical parameters like ambipolar diffusion strength which corresponds to what is expected in discs. However, connecting these individual simulations into a comprehensive, long-term evolution model demands extra, complex steps.

Basically, all MHD transport processes hinge on the parameter $\beta(r, t)$. One dimensional effective theories exist for this parameter (as found in works by \citealt{lubow1994magnetic,leung2019local}), but they still need to be fully validated by simulations. Adding to the complexity, as we showed in the previous section, there is no standard tracer that directly gives the gas surface density. Comparisons based on observations need intermediate steps using dust evolution \citep{birnstiel2012simple} or chemical models \citep{miotello2014protoplanetary,2016A&A...586A.103W}, and these models introduce further uncertainties.

Finally, the disc evolution is also affected by external factors, such as disc truncation in multiple systems and interference from massive stars, including fly-by or external photoevaporation (e.g., \citealt{winter2018protoplanetary}).

Protoplanetary discs display both inward (accretion) and outward (winds) motions. While these phenomena could potentially be attributed solely to thermal and hydrodynamic processes, various factors suggest the significant role of magnetic processes. The feasibility of hydrodynamic instabilities creating turbulent viscosity appears uncertain if not unlikely, except possibly for gravitational instabilities during the Class 0 phase \citep{2023ASPC..534..465L}. Also, propelling molecular outflows to velocities of several km s$^{-1}$ solely via thermal processes is challenging without raising the gas temperature to the point of molecular dissociation. However, magneto-centrifugal winds can theoretically expel the gas without any heating in ideal MHD conditions. MHD necessitates the gas ability to support electrical currents, demanding the presence of mobile charge carriers such as free electrons, atomic and molecular ions, and even small charged grains or polycyclic aromatic hydrocarbons (PAH).

\paragraph{Magnetic fields:}
There is not a lot of concrete observational evidence for magnetic fields in protoplanetary discs. The expected values generally range from approximately one Gauss at 1 au, declining to a few mG at several tens of au \citep{2007Ap&SS.311...35W}. Nevertheless, these theoretical values may deviate substantially across several orders of magnitude. It is not possible to use the Zeeman effect for measurements, except for the innermost part of the disc. In this region, toroidal magnetic fields of the order of one kG have been detected, but it remains uncertain whether these fields belong to the host star or the disc itself \citep{2005Natur.438..466D}. At larger distances tens of au, attempts to measure the field strength through Zeeman splitting in molecular lines have only provided upper limits, with $B_z < 0.8$ mG and $B < 30$ mG \citep{2019A&A...624L...7V}.

The field topology can also be inferred from polarisation in the continuum, i.e., dust thermal emission. It is assumed that dust grains generally align themselves perpendicular to the magnetic field lines, emitting thermal radiation with a specific polarisation, perpendicular to the orientation of the local field \citep{2007ApJ...669.1085C,2014Natur.514..597S}. However, polarisation in sub-millimetre radiation has been shown to be due in many cases to the scattering of dust thermal emission by dust grains themselves, a process called self-scattering \citep{2015ApJ...809...78K,2016MNRAS.456.2794Y}. Efforts using multi-wavelength observations have attempted to separate these two effects \citep{2017ApJ...851...55S}, yet the interpretation of results concerning magnetic topology remains uncertain, see also \citet{2023Natur.623..705S}.

Besides, the intensities of magnetic fields can be inferred from meteoritic and cometary evidence within our own solar system. If we assume that the field becomes embedded in the body during its formation in the parent disc, we can make such inferences. Field strengths on the order of $0.1$ G around 1 au are suggested from the residual magnetisation in meteorites based on this concept \citep{2014Sci...346.1089F}, while an upper limit of $B < 30$ mG in the region around 15-45 au is derived from the magnetisation of Comet 67P/Churyumov-Gerasimenko \citep{2019ApJ...875...39B}.

If the gas is sufficiently ionised and in a magnetised  medium, gas and magnetic field couple. Once gas becomes coupled with magnetic fields, the strength and topology of the magnetic fields start to play a crucial role in system dynamics. The dimensionless plasma parameter $\beta$ is typically used to quantify the magnetic field impact on the dynamics of plasmas. 

This parameter also connects with physical values of the disc. Given a "typical" T-Tauri disc with $\Sigma = 300 \, \text{g cm}^{-2} \, \text{R}^{-1} \, \text{au}$, one gets for the poloidal components (\citealt{2021JPlPh..87a2001P}, eq. 2.4):

\begin{equation}
    B_p \approx 10 \, \text{R}_{au}^{-11/8} \beta_{p}^{-1/2} \, \text{G}.
\end{equation}

Thus, the expected value in discs with $\beta_{p} \approx 10^{4}$ is $B_p \approx 5 \, \text{mG}$ at 10 au, consistent with the above-mentioned constrains. 

If the large-scale magnetic field primarily controls angular momentum transport, triggering an MHD wind, the total field B is closely related to the accretion rate. This allows the deduction of a minimum field strength for a given accretion rate \citep{bai2009heat}, 

\begin{equation}
    B \gtrsim 0.31 \, \text{R}_{AU}^{-5/4} (\dot{M}_{acc}/10^{-7} M_{\odot}.\text{yr}^{-1})^{1/2} \, \text{G}.
\end{equation}
Note that this lower limit requires careful interpretation, as in most MHD winds, $B$ is predominantly influenced by its toroidal component, which changes sign on both sides of the disc and is thus challenging to measure directly from optically thin tracers.

\paragraph{Ionisation:}

In a typical Class II low-mass system, only small radii ($\sim 0.1$ au) harbor temperatures high enough for partial thermal ionisation. It is however not clear if at these small radii, high thermal ionisation can be sustained deep in the disc or if it is just limited to the surface. The thermally ionised region expands farther from the star and deeper in the disc for high luminosity stars. In addition, highly accreting systems release heat that can further increase the temperature thus the ionisation of the disc \citep{mohanty2018inside}. In regions of the disc where thermal ionisation is not efficient, as in much of the interstellar medium, ionisation depends on non-thermal sources. These encompass X-rays \citep{2017MNRAS.472.2447G}, far-UV photons \citep{2011ApJ...735....8P} emitted by the protostar corona and accretion layers, cosmic rays \citep{2017A&A...603A..96R,2021ApJ...912..136S},  radionuclides like $^{26}$Al \citep{Umebayashi09}, and potentially powerful electric fields linked to MRI \citep{2019ApJ...878..133O}, or flare loops reaching the disc, which are the focus of this thesis.

Determining the ionisation degree is complicated due to the uncertain strengths of the above-listed ionising processes and the unclear abundance and size distribution of grains, which tend to capture free charges and facilitate recombination \citep{2016ApJ...833...92I}. However, in regions close to the star, where midplane temperatures range between 500-1500 K, thermal emission from grains might boost the abundance of free electrons \citep{2022MNRAS.509.5974J}. At even higher temperatures where grains sublime, thermal ionisation of alkali metals (K, Na, etc.), which possess low first-ionisation potentials, becomes the primary source. We dedicate Chapter \ref{C:ionisation} to the description of the ionisation of protoplanetary discs.

\subsubsection{Viscous thin discs}

Over the past five decades, the topic of angular momentum transport has received a lot of attention. Two main transport mechanisms have been proposed: turbulence and magnetised disc winds (MDW). The former has been widely explored due to the simplicity of its representation using the \citet{Shakura73} pseudo-viscosity framework. We present this model here. The structure of accretion discs of T Tauri stars is generally described by a geometrically thin ($H/R\ll1$) and optically thick disc model ($\tau_\nu\gg1$), composed of gas and dust (e.g. \citealt{Shakura73,1974MNRAS.168..603L,1999ApJ...527..893D}). The gas is mainly composed of molecular hydrogen (H$_2$), while the dust consists of sub-micron sized grains composed of silicates, carbonaceous materials and ices (e.g. \citealt{2003ARA&A..41..241D,2010ARA&A..48..205D}). 
%CS It works also with weakly ionised media and even fully ionised as long as B is weak (no current).

To describe how a disc evolves, we start by examining mass and angular momentum conservation for a differential ring at radius \( R \) and thickness \( dR \). Assuming an axisymmetric disc described by its surface density \( \Sigma(R) \), we let \( v_R(R) \) denote the radial velocity, where a positive value indicates an outward direction and \( v_R < 0 \) signifies accretion.

For a differential ring located at radius \( R \), the mass flow rate at the inner edge is,

\[
\dot{M}_{\text{inner}} = 2\pi R \Sigma(R) 
v_R(R)
%|v_R(R)|
.
\]

And at the outer edge,

\[
\dot{M}_{\text{outer}} = 2\pi (R + dR) \Sigma(R + dR) v_R(R + dR)
\]

The mass change rate within the ring is the difference between these two,

\[
2\pi R dR \frac{\partial \Sigma}{\partial t} = \dot{M}_{\text{inner}} - \dot{M}_{\text{outer}}
\]

By substituting these expressions, we obtain,

\begin{equation}
    R \frac{\partial \Sigma}{\partial t} = - \frac{(R + d R) \Sigma(R + d R) vR(R + d R) - R \Sigma(R) v_R(R)}{d R}  
\end{equation}

Taking the limit $dR\longrightarrow 0$ on the right hand side is the definition of the derivative of the function $R\Sigma v_R$ with relation to $R$. We deduce the following expression,

\begin{equation}
    R \frac{\partial \Sigma}{\partial t} = - \frac{\partial}{\partial R}(R \Sigma v_R)
\label{eq:MassContinuityThindisc}
\end{equation}

This is the equation for mass conservation in a thin disc.

For angular momentum conservation, if \( \Omega(R) \) represents the orbital frequency, the angular momentum per unit area is \( R^2 \Omega \Sigma \). Conservation of angular momentum also yields a corresponding equation, but here we must take into account the torques acting on the ring. Thus, the equation for angular momentum conservation incorporates these torque effects.

To further study how a disc evolves over time, we again start by considering a differential ring of radius \( R \) and thickness \( dR \). The equation for angular momentum conservation in such a thin disc, accounting the torque \( \tau(R) \), can be written as:

\begin{equation}
    R \frac{\partial}{\partial t}\left(R^2 \Omega \Sigma\right) + \frac{\partial}{\partial R}\left(R^3 \Omega \Sigma v_R\right) = \frac{1}{2\pi} \frac{\partial \tau}{\partial R}
\label{eq:AngularMomentumConservation}
\end{equation}

In this equation, a non-zero radial derivative of the torque \( \frac{\partial \tau}{\partial R} \) represents the differential torque across the ring. If this derivative is zero, it indicates that the angular momentum remains constant.

To make further progress, assumptions about the source of the torque are necessary. The simplest assumption is that the torque comes from viscous interactions within the disc. The torque \( \tau \) in a disc with viscous drag can be expressed as,

\begin{equation}
    \tau = 2\pi R \cdot R \nu \Sigma \frac{\partial \Omega}{\partial R} \cdot R.
\end{equation}

Here, \(\nu\) represents the turbulent viscosity. We decompose $\tau$ in three terms separated by dots. The first term arises from integrating around the annulus. The second term represents the viscous force per unit length and the third factor \(R\) corresponds to the lever arm of the torque. The fact that \(\tau \propto \frac{\partial \Omega}{\partial R}\) reflects that the viscous torque is non-zero only when the disc exhibits differential rotation. When \(\Omega\) is constant, the disc rotates as a solid body, and there are no viscous torques between adjacent annuli. We substitute this expression for \(\tau\) into Eq. \eqref{eq:AngularMomentumConservation} to obtain,

\begin{equation}
    \frac{\partial}{\partial t}(R^2 \Sigma \Omega) + \frac{1}{R}\frac{\partial}{\partial R}(R^3 \Sigma v_R \Omega) = \frac{1}{R}\frac{\partial}{\partial R}\left(R^3 \nu \Sigma \frac{d\Omega}{dR}\right).
\label{eq:ViscousAngularMomentumConservation}
\end{equation}

This equation describes angular momentum conservation in a viscous accretion disc.

Combining this equation with the equation for mass conservation allows us to eliminate the variable \( v_R \). After some manipulations, we get,

\begin{equation}
    \frac{\partial \Sigma}{\partial t} = -\frac{1}{R}\frac{\partial}{\partial R}\left[\frac{1}{\frac{\partial}{\partial R}\left(R^2 \Omega\right)}\frac{\partial}{\partial R}\left(R^2 \Omega\right) \frac{\partial}{\partial R}\left(R^3 \nu \Sigma \frac{d\Omega}{dR}\right)\right].
\label{eq:GeneralEvolutionDisc}
\end{equation}

In the specific case where the disc is in Keplerian rotation, the evolution equation simplifies to:

\begin{equation}
    \frac{\partial \Sigma}{\partial t} = \frac{3}{R}\frac{\partial}{\partial R}\left[R^{1/2}\frac{\partial}{\partial R}\left(\nu \Sigma R^{1/2}\right)\right].
\label{eq:KeplerianDiscEvolution}
\end{equation}

Equation \(\eqref{eq:KeplerianDiscEvolution}\) is typically nonlinear, but can become linear if the viscosity \( \nu \) is independent of \( \Sigma \). Analytic solutions are seldom available, making numerical methods the convenient approach for most problems.

When considering axially symmetric motion with rotation around the z-axis, the velocity components are \( v_r \), \( v_z \), and \( v_\phi \), with \( v_z = 0 \). Here, \( P \) denotes the gas pressure, and \( \rho \) the mass density. The gravitational potential \( \Phi(r, z) \) can be described as \( -\frac{GM_*}{\sqrt{R^2 + Z^2}} \), where \( M_* \) is the mass of the central star and \( G \) is the gravitational constant. \( Z \) is the distance from the midplane of the disc.

By coupling Eq \eqref{eq:KeplerianDiscEvolution} with the radial component of the momentum equation, we can derive the pressure profile of the disc. This can be expressed as,

\begin{equation}
    \frac{\partial v_R}{\partial t} + v_R \frac{\partial v_R}{\partial R}-\frac{v_\phi^2}{R}= -\frac{G M_*}{R^2}-\frac{1}{\rho}\frac{\partial P}{\partial R}
\label{eq:RadialPressureProfile}
\end{equation}

This simplified framework provides some advantages, primarily in that the accretion disc equations become a closed set once the turbulent viscosity \( \nu \) is specified. Although the underlying causes of this turbulence are still not entirely clear, it is possible to estimate \( \nu \) values from observational data.

We first explore the hydrostatic solutions that arise from the Keplerian disc evolution equation \(\eqref{eq:KeplerianDiscEvolution}\). Subsequently, we turn our attention to the origins and implications of the turbulent viscosity.

\subsubsection{Hydrostatic Structure}

%The radial profiles of surface density, $\Sigma(R)$, and temperature, $T(R)$, of the disc are often approximated by power-law relationships,

%\begin{equation}
%\Sigma(R)= \Sigma_0 \left(\frac{R}{1 \rm au }\right)^{-1},
%\end{equation}

%\begin{equation}
%T(R)= T_0 \left(\frac{R}{1 \rm au }\right)^{-1/2},
%\end{equation}
%
%where $R$ is the radial distance from the central star, and $\Sigma_0=300$ g cm$^{-2}$ and $T_0=280$ K (e.g., \citealt{2011ApJ...732...42A,2013ApJ...774...16R}).

All the work done in Chapter \ref{C:PublicationI} and \ref{C:PublicationII} of this thesis uses the disc structure generated by the thermochemical radiation code, {\tt ProDiMO} \citep[PROtostellar disc MOdel,][]{woitke2009radiation,kamp2010radiation,thi2011radiation,woitke2016consistent,Rab18}; we discuss it in details in Sec. \ref{sec:focusonProDiMO}.  While {\tt ProDiMO} assumes a hydrostatic disc structure, it is well established that T Tauri systems are magnetised and a plethora of evidence points towards the necessity of treating the dynamics of these systems within the MHD framework. Nevertheless, we use {\tt ProDiMO} because it can compute the abundances of hundreds of chemical species, a critical aspect of our work, especially to estimate the energy losses of particle in discs as discussed in Sect. \ref{sec:EnergyLossesDisc}. Thus, we introduce the simple
%I think in the vertical direction a static model is not that simplistic
hydrostatic thin disc model used by {\tt ProDiMO}. Besides describing the disc model that we use, providing an hydrostatic perspective of a disc offers a basic understanding of what a T Tauri disc structure might resemble to.

Considering the pressure structure Eq. \eqref{eq:RadialPressureProfile}, assuming $v_r = 0$ and $v_z = 0$,
\begin{equation}
\frac{v^2_\phi}{r} = \frac{1}{\rho} \frac{\partial P}{\partial r} + \frac{\partial \Phi}{\partial r} 
\label{eq:hydrostaticradialstructure}
\end{equation}
and
\begin{equation}
0 = \frac{1}{\rho} \frac{\partial P}{\partial z} + \frac{\partial \Phi}{\partial z}, 
\end{equation}
is derived under the "1+1D" modelling concept (see \citealt{d1998accretion,dullemond2002vertical}). This model suggests that the radial pressure gradient $\frac{1}{\rho} \frac{\partial p}{\partial r}$ in the first equation is small compared to the centrifugal acceleration and gravity. As a result, the radial and vertical components of the equations of motion are decoupled. The radial component leads to keplerian orbits, expressed by,
\begin{equation}
v_\phi = \sqrt{ \frac{r^2 GM_*}{(r^2 + z^2)^\frac{3}{2}} },
\label{eq:hydrostaticazimuthalvelocity}
\end{equation}
which leaves the radial distribution of matter undetermined, as it is primarily influenced by the actual angular momentum distribution in the disc.

The vertical component of the equation of motion can be independently solved for each vertical column in the disc as follows,
\begin{equation}
\frac{1}{\rho} \frac{dP}{dz} = - \frac{Z GM_*}{(R^2 + Z^2)^\frac{3}{2}} .
\end{equation}
Up to here we are in the standard Shakura-Sunyaev thin disc model in the steady approximation.

Integrating this equation from the midplane upwards by substituting the density for the pressure via $p = c_T^2 \rho$, where $c_T$ is the isothermal sound speed, computed from the thermal structure as,
\begin{equation}
    c_T^2=\frac{k T}{\mu m_p},
\end{equation}
where again $k$ is the Boltzman constant, $\mu$ the mean molecular weight, $m_p$ the proton mass and $T$ is the temperature.
We can then determine the surface density $\Sigma(r)$, using
\begin{equation}
\Sigma(R) = 2 \int_0^ {z_{max}(R)} \rho(R,Z) dZ,
\end{equation}
with the factor 2 accounting for the symmetry of the lower half of the disc.

In {\tt ProDiMO}, as the disc is steady, a power-law distribution for the surface density is assumed, 
\begin{equation}
    \Sigma(R) = \Sigma_0 R^{-\epsilon} ,
    \label{eq:columndensityradialpowerlaw}
\end{equation}

where $\Sigma_0$ is adjusted from the specified disc mass $M_{disc}$ as:
\begin{equation}
M_{disc} = 2\pi \int_{R_{in}}^{R_{out}} \Sigma(R) R dR,
\end{equation}
where $R_{in}$ and $R_{out}$ represent the inner and outer disc radii, respectively.

In summary, the structure of the disc is determined by the parameters $M_{disc}$, $M_*$, $R_{in}$, $R_{out}$, $\epsilon$ and the function $c_T^2 (r, z)$. The latest function is given by the thermal structure of {\tt ProDiMO} and is derived from the radiative and chemical equilibrium model, while we presented in the previous section how to constrain observationally the other parameters.

\paragraph{Hydrostatic inner rim:}

Using a radial surface density power-law equation like Eq. \eqref{eq:columndensityradialpowerlaw} to describe the disc between the inner radius ($R_{in}$) and the outer radius ($R_{out}$) is common but fundamentally artificial and not physically accurate. Equation \eqref{eq:hydrostaticradialstructure} reveals that a sharp radial cutoff would result in infinite force due to the radial pressure gradient $\partial p/\partial r$. This would push gas inwards at the inner radius and outwards at the outer radius, effectively smoothing out the radial density structure at these boundaries.

Let us assume an abrupt cut-off of the inner disc. We explore the gas motion as it is forced inwards due to the radial pressure gradient at the inner boundary. The conservation of specific angular momentum, $l_{in} = R_{in} v_{\phi}(R_{in}) = r v_{\phi}(r)$, during this motion implies that as the gas is pushed inwards, it will spin up. This spinning continues until the increased centrifugal force balances the radial pressure gradient, along with gravity. The equilibrium of forces in this relaxed state, according to Eq. \eqref{eq:hydrostaticazimuthalvelocity}, is expressed as,

\begin{equation}
\frac{l_{in}^2}{r^3} = \frac{1}{\rho} \frac{\partial p}{\partial r} + \frac{\partial \Phi}{\partial r}
\end{equation}

This provides us with an equation for $\rho(r)$. Using again $p(r) = c_T^2 \rho(r)$ and assuming $c_T^2 = const$, we have,

\begin{equation}
\ln \left(\frac{\rho(r_0)}{\rho(R_{in})} \right)= -\frac{1}{c_T^2} \left[ \frac{l_{in}^2}{2r^2} + \frac{GM_*}{r} \right]^{r_0}_{R_{in}} 
\label{eq:hydrostaticinnerrimdensity}
\end{equation}

In this equation, $r_0$ represents any arbitrary point within $R_{in}$. Expressing Eq. \eqref{eq:hydrostaticinnerrimdensity} in terms of column densities and estimating $c_T^2$ in the midplane, we derive,

\begin{equation}
\Sigma(r_0) \approx \Sigma(R_{in}) \exp \left(-\frac{1}{c_T^2} \left[ \frac{l^2_{in}}{2r^2} + \frac{GM_*}{r} \right]^{r_0}_{R_{in}}\right).
\label{eq:hydrostaticinnerrimcolumndensity}
\end{equation}

A similar formulation could be applied for the column density outside the outer boundary. CO observations by \citet{hughes2008resolved} suggest that such methods can enhance model fits. However, \citet{woitke2009radiation} opted to apply this approach for smoothed edges solely to the inner boundary of the {\tt ProDiMO} model. 
The reference disc model we will use throughout this thesis is generated by {\tt ProDiMO}, based on the hydrostatic structure we have just described. Although this is a simplified assumption, it allows us to model complex chemical reactions and radiative transfer, taking into account heating and cooling processes in protoplanetary discs within a manageable computational time. We will explore these features of {\tt ProDiMO} in more detail in Sec.\ref{sec:focusonProDiMO}. Acknowledging the limitations of using a hydrostatic structure, we now turn to an examination of the reasons for adopting a MHD approach to describe protoplanetary discs.

\paragraph{Steady-state accretion disc:}
In this section, we lift the assumption of zero radial velocity to explore the steady-state solutions. By focusing on Eq. \eqref{eq:ViscousAngularMomentumConservation} and setting the time-dependent terms to zero, we find that the accretion rate \( \dot{M} = -2\pi R \Sigma v_R \) remains constant in a steady state. Under the condition of no torque at the inner boundary (\( d\Omega/dR = 0 \) when \( R = R_{\text{in}} \)), integrating over \( R \) yields,

\begin{equation}
    -R^2 \Omega \dot{M} + R^2 \Omega_{\text{in}} \dot{M} = 2\pi R^3 \nu \Sigma \frac{d\Omega}{dR}
\end{equation}

Inserting the expression for \( \Omega = \sqrt{\frac{GM_*}{R^3}} \) results in,

\begin{equation}
\Sigma(R) = \frac{\dot{M}}{3\pi \nu(R)} \left(1 - \sqrt{\frac{R_{\text{in}}}{R}}\right).
\label{eq:SteadyStateAccretionRate}
\end{equation}

Assuming \( l \gg l_{\text{in}}\), (i.e \( R^{1/2} \gg R^{1/2}_{\text{in}}\) for Keplerian disc), the accretion rate simplifies to,

\begin{equation}
\dot{M} = 3\pi \nu \Sigma
\label{eq:FarFieldAccretionRate}
\end{equation}

This underscores the central role that the viscosity \( \nu(R) \) plays in both mass distribution and the accretion mechanisms within the disc. However, to have a more comprehensive understanding, we need to discuss the nature of this viscosity. We will address this now.

\subsubsection{Origin of thin disc viscosity}\label{sec:TurbulenceAngularMomentum}
\paragraph{Turbulence as driver of angular momentum transport:}
Until now, we have deliberately avoided discussing viscosity. Nonetheless, it is time to explore the physical processes behind the transport of angular momentum and observed accretion rates. It is clear that the turbulent viscosity, denoted by \( \nu \), cannot be simplified to fluid viscosity in this setting. A closer look at inter-molecular forces suggests that, at astronomical unit scales in protoplanetary discs, the typical time scale for viscosity surpasses \(10^{13}\) years. Therefore, another form of viscosity is essential to explain observed accretion rates in young stars and other accreting systems.

In a thin disc setting, the orbital speed is much faster than the speed of sound \( (v_\phi = (H/R)^{-1} c_s )\). With only molecular forces contributing to fluid viscosity, the flow would have an extraordinarily high Reynolds number, well above \(Re=L^2/\nu T > 10^{11}\), where L and T are the characteristic length and time scales of the flow.

In such a framework, it is necessary that turbulence drives angular momentum transport within accretion discs. This concept forms the basis of the disc model, initially developed by \citet{1973A&A....24..337S}. In this model, the characteristic scale of turbulent eddies is roughly \( H \) (the vertical height of the disc), and the characteristic speed is about \( c_s \) (the sound speed). Based on dimensional considerations, they proposed:

\begin{equation}
    \nu = \alpha c_s H.
\end{equation}

Here, \( \alpha \) is a dimensionless factor that indicates the turbulence intensity and is generally less than or equal to 1.

We can relate this to Eq. \eqref{eq:KeplerianDiscEvolution}, which ties a single time derivative of \( \Sigma \) to a second-order radial derivative, resembling a diffusion equation. By recasting Eq. \eqref{eq:KeplerianDiscEvolution} using variables \( X= 2 R^{1/2} \) and \( S=R^{1/2} \Sigma \), the diffusive nature becomes clearer,

\[
\frac{\partial S}{\partial t}=3 \frac{\partial^2 \nu S}{\partial X^2}.
\]

Here, the diffusion constant is linked to the viscosity \( \nu \). The typical diffusion time frame, neglecting constants of order unity, can be described as,

\[
t_\nu \approx \frac{R^2}{\nu}, \quad v_R \approx \frac{R}{t_\nu}\approx \frac{\nu}{R}.
\]

This characteristic time
%bof duration...
is more commonly known as the viscous time-scale, which quantifies the time needed for viscosity-induced accretion to significantly alter the local surface density.

Within this framework, if disc accretion is solely driven by turbulence, the accretion rate can be approximated from Eq. \eqref{eq:FarFieldAccretionRate} as,

\begin{equation}
    \dot{M}_{acc} \approx 3 \times 10^{-8} \left(\frac{\alpha}{10^{-2}}\right) \left(\frac{\epsilon}{0.1}\right)^2 \left(\frac{R}{10 \text{ au}}\right)^{1/2} \left(\frac{M_{\star}}{M_{\odot}}\right)^{1/2} \left(\frac{\Sigma}{10 \text{ g.cm}^{-2}}\right) M_{\odot} \text{yr}^{-1},
    \label{eq:massaccretionratelesur22}
\end{equation}
where the disc aspect ratio $\epsilon \equiv H/R$  and $\Sigma(R)$ is assumed to follow a shallow power law. Thus, typically $\alpha \approx 10^{-2}$ is required to match the observed accretion rates by viscosity alone.

Equation \ref{eq:KeplerianDiscEvolution} can also be derived from the more complex Navier-Stokes equation framework \citep{2008NewAR..52...21L}. This method, while more complex, associate the viscosity \( \nu \) with the viscous stress tensor. In classical shear viscosity, the only non-zero component of this tensor is the \( ({R,\phi}) \) component. Integrating this stress tensor component vertically reads,

\begin{equation}
    T_{R,\phi} = \nu R \Sigma \frac{d\Omega}{dR}.
    \label{eq:viscousstresstensor}
\end{equation}

We have already established that molecular fluid viscosity is inadequate for the observed accretion rates. However, Eq. \eqref{eq:viscousstresstensor} hints at an alternative. For non-viscous fluids, one can break down the equations into average and fluctuating components (refer to \citealt{1998RvMP...70....1B,2008NewAR..52...21L}). Identifying Eq. \eqref{eq:viscousstresstensor} with the \( \alpha \)-prescription allows to link the \( \alpha \) parameter with the parameters of a turbulent flow. When the fluid is both magnetised and self-gravitating, \( \alpha \) can be expressed in terms of the various stresses in turbulent flow as,

\begin{equation}
    \alpha = \left\langle \frac{\delta v_R \delta v_\phi}{c_s^2} - \frac{B_r B_\phi}{4 \pi c_s^2} - \frac{g_r g_\phi}{4 \pi G \rho c_s^2} \right\rangle.
    \label{eq:alphaNavierStokes}
\end{equation}

In this expression, the angle brackets indicate a time-averaged sum over the fluctuating flow fields. The equation incorporates hydrodynamic (Reynolds) stress, magnetic (Maxwell) stress, and gravitational stress, confirming the role of turbulence in enabling angular momentum transfer.

It is important to specify that the term "turbulent viscosity" is somewhat ambiguous. While it may behave like fluid viscosity at macroscopic scales, this approximation breaks down at lengths comparable or smaller than the scale of the turbulence itself ($\lesssim H$). However, this aspect may be overlooked at the scale of accretion disc, while it may becomes important when considering planetary formation. 

Additionally, we have reframed the meaning of the \( \alpha \) parameter. In classic accretion disc theory, \( \alpha \)-viscosity serves as a predictive tool based on assumed disc conditions; it is often treated as an input parameter in models. In contrast, Eq. \eqref{eq:alphaNavierStokes} describes \( \alpha \) as an efficiency indicator for how turbulence transports angular momentum. This changes the causal direction implied by the \( \alpha \)-prescription: \( \alpha \) becomes an outcome of local disc conditions rather than a determinant. This opens the door for \( \alpha \) to be directly quantified in numerical simulations of turbulent accretion discs. In order to discuss the origin of turbulence we need an MHD description of discs.

\paragraph{MRI-driven turbulence:}

Equation \eqref{eq:alphaNavierStokes} illustrates that the sources of turbulence in accretion discs could be hydrodynamic, MHD, or even gravitational. However, gravitational forces usually come into play only in especially massive discs. For less massive Class II phase protoplanetary discs, gravitational instability is generally not a concern.

In the framework of pure hydrodynamics, stability is guided by the Rayleigh criterion, which states \citep{1974IAUS...59..185Z}, 

\[
\frac{d}{dR}(R^2\Omega) < 0.
\]

Here, \( \Omega \) denotes angular velocity, and \( R \) the radius. Keplerian accretion discs however, increase in specific angular momentum as the radius increases, rendering them stable to hydrodynamic fluctuations. Consequently, hydrodynamic turbulence is unlikely to play a substantial role in angular momentum transport.

Conversely, for magnetised fluids, the most promising instability is the Magnetorotational Instability (MRI). From the perturbation of a rotating magnetised fluid, a different stability criterion emerges, which can be expressed as \citep{1994ApJ...421..163B}

\[
\frac{d\Omega^2}{d \ln R} < 0.
\]

Keplerian discs satisfy this condition, making them subject to this MHD instabilities. Although a complete explanation of MRI is beyond the scope of this section, Appendix \ref{app:Internshipreport} features Mathieu Venet's Master's internship report. Under the supervision of Alexandre Marcowith and myself, Mathieu focused on the linear development of MRI in partially ionised fluids.

In a disc influenced by a vertical magnetic field, MRI has solutions, where vertical disc layers alternately move inward or outward. These motions do not result in a net angular momentum transfer when considered in a vertically averaged sense. To understand the full complexity, one must engage in numerical simulations, an active research domain, see next section. Most models suggest that magnetic stresses are the primary drivers, with typical efficiencies for MRI-driven angular momentum transfer denoted by \( \alpha \) values ranging between \( 10^{-3} \) and \( 10^{-2} \).

Ideal MHD conditions are unlikely to be universally applicable across protoplanetary discs. Factors like disc ionisation levels, which are critical for magnetic field coupling, differ across the disc. For MRI to operate, the ionisation fraction must exceed \( 10^{-12} \). Although thermal ionisation are high enough in many astrophysical contexts, this is not always true for protoplanetary discs. 

In regions where the ionisation level is insufficient, so-called "dead zones" appear, where MRI is inactive. Thermal ionisation near the star is generally sufficient, and cosmic rays ionize at larger radii. However, at intermediate distances (around several au), the disc midplane is likely devoid of MRI activity. In these zones, accretion may proceed via a partially-ionised surface layer. Further details on this subject can be found in \citet{2019fpdp.book.....A,2021JPlPh..87a2001P}.
%%%%%%%%%%%CS
\paragraph{Viscosity inferred from MRI simulations:}
In the quasi-ideal MHD regime (i.e., when magnetic diffusive length-scales are much smaller than $H$) and in the presence of a net-flux vertical magnetic field, the magnetic field growth saturates, halting the MRI \citep{murphy2015anisotropic,hirai2018study}. Such saturated MRI activity conditions are present in the inner and outer regions of protoplanetary discs \citep{parkin2014global,suzuki2014magnetohydrodynamic,flock20173d,zhu2018global}. Fully saturated MRI turbulence shows an angular momentum transport coefficient $\alpha$ ranging between $0.01$ and $1.0$, depending on the initial magnetic field strength and topology. For example, \citet{salvesen2016accretion} found $\alpha \approx 11\beta^{-0.53}$ for $10 < \beta < 10^5$ in stratified shearing box simulations threaded by a vertical field, suggesting $\alpha$ is roughly proportional to $B_p$. 

While this result holds in "ideal" MHD, the magnetic Prandtl number $P_m$, the ratio between the viscous and magnetic diffusion rate, is expected to be very small in protoplanetary discs, even in regions where ideal MHD applies. This small $P_m$ regime was investigated using non-stratified high-resolution simulations, achieving convergence for the first time \citep{meheut2015angular} with a value of $\alpha = 3.2 \times 10^{-2}$ and with $\beta = 10^3$. We note that this value of $\beta$ is pretty small for standard disc \citep{2019A&A...632A..44T}. Indeed, MRI simulations in the low $P_m$ limit show turbulent decay \citep{mamatsashvili2020zero} and no sustained turbulence.

As we move further out or deeper in the protoplanetary disc, non-ideal effects such as Ohmic, Hall, and Ambipolar diffusion grow in importance. These reduce the intensity of MRI turbulence, $\alpha $ is reduced by four orders of magnitudes \citep{bai2014hall,simon2018origin}. Multiple disc factors become critical as these non-ideal aspects depend on the strength of the magnetic field, local gas and dust density, and the local high-energy X-ray, Cosmic ray, and even far ultraviolet (FUV) radiation field \citep{gole2018nature}.

\subsubsection{Outflows}
\begin{figure}
    \centering
    \includegraphics[width=0.5\linewidth]{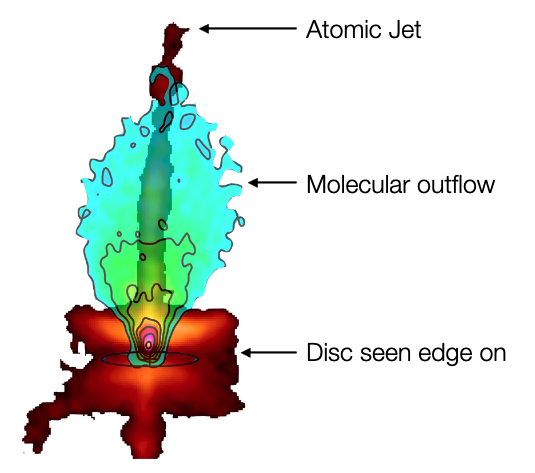}
    \caption{Observation of the disc and the atomic jet seen by the Hubble Space Telescope \citep{1996ApJ...473..437B} and the molecular wind observed in CO(2-1) by ALMA \citep{2018A&A...618A.120L} of HH30, a protoplanetary disc seen edge-on. Figure reproduced from \citep{2021JPlPh..87a2001P}.}
    \label{fig:OutflowsComponentObservationl}
\end{figure}

Since the 1990s, outflows have been thought to explain the atomic jets observed in a number of young stellar objects. In the literature, the distinction between jets and winds is somewhat ambiguous, but is often implicitly associated with the collimation and velocity of the outflow. Jets are typically viewed as fast ($100-1000$ km/s) and well-collimated, thereby appearing narrow at large distances, while winds are usually perceived as slow ($1-30$ km/s) conical outflows. These slow outflows are sometimes called "molecular" as they are detected in molecular lines. See Fig. \ref{fig:OutflowsComponentObservationl} for the distinction of the components. We distinguish two types of outflows: thermally driven outflows (photoevaporation), where the primary energy source is an external radiation field (which could be due to the central star or the immediate environment), and magnetised outflows, which are powered by accretion energy.

We have discussed in the previous section the standard model of viscous accretion discs. On the other hand, disc winds, initially proposed by \citet{Blandford82}, have seen a resurgence in popularity over the past decade as a means of transporting angular momentum in otherwise "dead" discs. In fact even with MRI the alpha parameter is usually too small, see \citet{2013ApJ...769...76B}, to drive the accretion in a "lively" disc.

In a similar way as we derived the evolution equation of a viscous accreting disc, an evolution equation can be derived accounting for mass losses due to winds (see \citealt{2021JPlPh..87a2001P}). The correlations between accretion flow, angular momentum transport, and the long-term evolution of discs are apparent in the mass and angular momentum conservation equations, averaged over the disc thickness,

\begin{equation}
    \frac{\partial \Sigma}{\partial t} + \frac{1}{2\pi R}\frac{\partial \dot{M}_{acc}}{\partial R} = -\zeta \Sigma \Omega_{K},
\end{equation}

and
\begin{equation}
    \frac{\dot{M}_{acc} \Omega_{K}}{4\pi} = \frac{1}{R}\frac{\partial}{\partial R}(R^2\alpha P) + \zeta(\lambda - 1)P \left(\frac{R}{H}\right)^2.
    \label{eq:accretionrateLesur22}
\end{equation}

Here we define the disc mass accretion rate as $\dot{M}_{acc}$, the surface density as $\Sigma$, and the pressure as $P$. The Keplerian angular velocity is again represented by $\Omega_K$ and the geometrical thickness by $H$. These equations also include three dimensionless coefficients: $\zeta$, the mass loss parameter due to a hypothetical outflow; $\alpha$, as defined by \citet{Shakura73}, which describes radial angular momentum transport within the disc, potentially caused by turbulence; and finally, $\lambda$, the outflow lever arm \citep{blandford1982hydromagnetic}, quantifying the specific angular momentum extracted vertically by an outflow. All three coefficients may vary with radius and time. 

Knowing these three coefficients allows for a complete prediction of the evolution of the system and accretion theory.

\paragraph{Thermal outflows:}\label{sec:thermaloutflows}
In a simplistic view, 
%\textcolor{magenta}{CS yes this is oversimplistic and it leads to wrong statements like, no thermal wind can exist in T Tauri stars but ok, I'll have to live with that}
gas can only be thermally unbounded if it is heated to temperatures that approach the local escape temperature, 
\begin{equation}
    T_{\text{esc}} = \frac{GM_{\star}\bar{m}}{R k_{\text{B}}},
\end{equation}
where $\bar{m}$ is the mean mass per particle. This is approximately $10^4$ K for $M_{\star} =M_{\odot}$, $R = 1$ au, and $\bar{m} = 0.68 m_{\text{p}}$ (ionised hydrogen at solar abundance). Depending on the irradiating spectrum, the gas is heated to different temperatures. For instance, Extreme-Ultraviolet (EUV) radiation ($13.6$ eV $< E < 100$ eV) primarily heats the gas through the photoionisation of hydrogen and helium, generally yielding to an almost isothermal gas with a temperature around $10^4$ K. 

For geometrically thin, non self-gravitating discs around a $1 M_{\odot}$ star, the radius at which the escape temperature of the gas equals $10^4$ K is approximately 1 au. The mass loss profile in this case is peaked around the so-called gravitational radius of the system, the cylindrical radius at which the sound speed of the heated gas equals the Keplerian orbital speed and has a total rate of about $10^{-10} M_{\odot}$/yr assuming an EUV flux of $10^{41}$ photons/s \citep{alexander2006photoevaporation}, as demonstrated by isothermal hydrodynamical simulations.

Soft X-ray radiation ($100$ eV $< E < 1$ keV) is not efficient at ionizing hydrogen directly, but it ejects inner-shell electrons from abundant metals (e.g. Carbon and Oxygen); the ejected photoelectrons have suprathermal energies and result in secondary ionisations. This leads to a more weakly ionised gas of a range of temperatures from a few thousand to $10^4$ K. Two-dimensional radiation-hydrodynamic models yield to mass loss profiles that are more extended with respect to the EUV case and total mass loss rates of the order of $10^{-8} M_{\odot}$/yr, assuming a soft X-ray flux of $10^{30}$ erg/s \citep{picogna2019dispersal}, or even higher for Carbon-depleted discs \citep{wolfer2019radiation}.

The dominant gas heating mechanism for FUV photons ($E < 13.6$ eV) is photoelectric heating from dust grains, which is proportional to the grain surface. Thus in the presence of abundant small grains or Polycyclic Aromatic Hydrocarbons (PAHs) in the disc atmosphere, FUVs can efficiently heat the gas and may also contribute to driving a thermal wind, as suggested by hydrostatic 1+1D models \citep{gorti2009time}. However, PAHs  are rarely observed in T-Tauri discs \citep{seok2017polycyclic} leading to large uncertainties in their atmospheric abundances. Also, accretion processes lead to a huge FUV flux variability. These reasons make it difficult to assess the role of FUV-driven winds on the final disc dispersal.

Recent calculations of thermal winds have combined two-dimensional (axisymmetric) hydrodynamics with treatments of the chemical and thermal state of the gas under the influence of these radiative processes. These works find that X-rays alone are less effective at driving photoevaporation than EUV \citep{wang2017wind}. \citet{2022A&A...668A.154R} find that for a number of sources, the photoevaporative disc-wind model is consistent with the observed signatures of the blueshifted narrow low-velocity component (NLVC), usually associated with slow disc winds. 
%\textcolor{magenta}{CS But in that case how do you explain the presence of coronae of 1 million degree around T Tauri stars? The Solar corona is not formed via radiation but due to magnetic turbulence and waves? A word of caution might be in order then.... After reading the end of the chapter I think you mean here the outer outflow so my remark is not valid... Then be more precise on your scales.}

\paragraph{MHD Outflows:}
The absence of any systematic method to promote turbulence in protoplanetary discs has reignited interest in the magnetised outflow (MO) model, especially in areas with weak ionisation. This concept, originally introduced by \citet{blandford1982hydromagnetic}, and later expanded by \citet{wardle1993structure} and \citet{konigl2010wind}, gained additional backing from numerical simulations. Initial confirmation came from shearing box models where \citet{bai2013wind} discovered that a magnetically inactive disc, mainly influenced by Ohmic and ambipolar diffusion, could still produce a MO due to the ionised layer situated at the disc surface. This finding was quickly extended to the more turbulent outer regions ($R > 30$ au) and to models incorporating the Hall effect \citep{lesur2014thanatology}.

The first generation of global simulations including non-ideal MHD effects \citep{gressel2015global,bai2017hall} indicated that these outflows are not merely an artifact of the shearing box model, and can indeed account for accretion rates around $10^{-8}$ M$_{\odot}$/yr, given a disc surface density $\Sigma \approx 10$ g/cm$^{2}$ at $R = 10$ au and a disc magnetisation $\beta_p \approx 10^{5}$. It was also discovered that the mass loss rate in the wind,
\begin{equation}
    \dot{M}_{\text{wind}} = 2\pi \int R dR \zeta \Sigma \Omega_K,
\end{equation}
 is roughly equal to the mass accretion rate, $\dot{M}_{\text{acc}}$ \citep{2023ASPC..534..465L}. 

A useful way to interpret these high mass ejection rates is to use the local ejection efficiency,
\begin{equation}
    \xi = \frac{1}{\dot{M}_{\text{acc}}} \frac{d \dot{M}_{\text{wind}}}{d \log R}
\end{equation}
to compare ejection to accretion rates. In a steady state, assuming a constant $\xi$, it follows that, 
\begin{equation}
    \dot{M}_{\text{wind}} = \dot{M}_{\text{acc}}(R_{\text{in}}) (( R_{\text{out}}/ R_{\text{in}})^{\xi}- 1)
    \,,
\end{equation}
for a wind-emitting disc region with inner radius $R_{\text{in}}$ and outer radius $R_{\text{out}}$.  

It can be demonstrated that if accretion is the only energy source propelling the outflow, then energy conservation demands $\xi < 1$ \citep{2021JPlPh..87a2001P}. This boundary is typically exceeded in MHD wind simulations (e.g. $\xi = 1.5$ in \citealt{bai2017hall}), indicating that these outflows cannot be powered solely by accretion energy. Thus another source of energy is necessary. As all of these initial models incorporated some form of prescribed coronal heating, it became evident that these outflows could be of magneto-thermal origin, i.e., the outflow is a blend of photo-evaporation and MHD wind. 

\paragraph{Inner region outflows:}
Jets, identified by their High Velocity Components (HVC), along with inner winds marked by the Low Velocity Component (LVC) and the Broad Component (BC), are generally considered to arise from disc radii of less than approximately \(0.5\) au. Multiple sources are likely responsible for the expelled gas in this complex region, including the central star, the interface between the star and its accretion disc, and the accretion disc itself \citep{2006A&A...453..785F}. This zone is also expected to exhibit magnetorotational instability (MRI) \citep{2011ARA&A..49..195A,flock20173d}.

The most central part of the expulsion process is attributed to winds from the central star. In addition to traditional stellar winds propelled by coronal gas pressure, these winds can also be driven by the combination of accretion energy, strong magnetic fields, and surface convection \citep{matt2005accretion,2008ApJ...681..391M,2011A&A...533A..46S}. The speeds of such accretion-powered winds usually meet or exceed the star's escape velocity, often reaching several hundred km/s. These winds are likely responsible for the noticeable blueshifted absorption features at He I and C II and are considered a mechanism for slowing down the accretion rate of the central star.

Near the co-rotation radius of about \(0.1\) au, the interaction between the star and the disc gets complex. Here, some stellar magnetic field lines stay closed, funneling disc material toward the star, while others open up, potentially initiating winds. Multiple scenarios could lead to material ejection in this region \citep{2006A&A...453..785F}.

The "X-wind" model suggests that open magnetic field lines from the star get anchored in the disc near the co-rotation radius, leading to a stable magnetocentrifugal wind (MCW) \citep{shu1994magnetocentrifugally}. These field lines then fan out, creating a magnetohydrodynamic (MHD) wind composed of a dense, axial jet and a broader, wide-angle wind, both moving at an average velocity of around \(150\) km/s \citep{shang1998synthetic}. This wide-angle component is thought to potentially sweep away ambient gas in slow molecular outflows. While the original X-wind model has faced substantial critique, current perspectives favor magnetospheric ejection mechanisms \citep{Ray21}.

MHD simulations concerning the interaction between a star and its disc suggest a distinct ejection geometry, primarily featuring an unstable "conical wind" launched at approximately \(45^\circ\) angles (\citealt{2009MNRAS.399.1802R,2013A&A...550A..99Z}). Yet, research by \citet{2022A&A...664A.176S} indicates that the potency of this conical wind may be significantly influenced by the stellar wind itself. Magnetic fields anchored at both the star and the disc stretch due to differential rotation between their anchor points, coiling the toroidal magnetic field and propelling mass outward at speeds greater than \(100\) km/s from about \(0.1\) au \citep{liffman2020infrared}. In cases where the disc possesses a net vertical magnetic field aligned with the stellar dipole, its reconnection with closed stellar field lines near the magnetopause can also result in potent, intermittent ejections at mid-latitudes \citep{ferreira2000reconnection}. The process of reconnection may also produce particles that add supplementary heat, aiding the launching of these outflows (see Chapter \ref{C:PublicationII}).

Centrifugal disc winds (MCW), originating from the co-rotation radius and extending outward, are other likely contributors to outflows within \(0.5\) au. These winds display a variety of speeds, mirroring the different radii from which they are launched, leading to a layered "onion-like" kinematic structure (\citealt{1999A&A...343L..61C}). The innermost portion of this type of disc wind is believed to cause the slow, narrow blueshifted absorption features observed at He I and C II, especially where the observer's line of sight passes through the disc wind. Additionally, the [O I] low-velocity component BC line width is thought to originate from a wind in the inner disc.

Within a few stellar radii, the concept of "failed winds" also comes into play. In this model, some vertical outflows lack the escape velocity and thus fall back onto the star, as illustrated in the 3D global simulations by \citealt{takasao2018three}. When the stellar magnetic field strength is below 1 kG, a well-developed magnetosphere around the star is unlikely \citep{takasao2018three}. These failed winds would then accrete onto the star at mid-latitudes at speeds close to the Keplerian velocity at the surface, between \(100-200\) km/s. Unlike magnetospheric accretion, which occurs in localized magnetic tubes, this form of accretion spans a broad latitude belt between \(15^\circ - 40^\circ\). No observations have been linked to MRI failed winds yet, but they could contribute to the observed variability in near-stellar emission lines. Furthermore, in conditions of failed MHD winds, thermal mechanisms alone might suffice to launch a wind.

It's anticipated that a combination of these various outflow processes, which are likely to vary due to changing accretion rates onto the star and occur within \(0.5\) au, jointly contribute to shaping, directing, and causing the variability of the jet.

\section{Conclusion}

In this chapter, we set the stage for our exploration of protoplanetary discs in T Tauri systems by identifying the critical components and observational techniques that enable our understanding of these objects. We began by focusing on the central star and its surrounding protoplanetary disc. We highlighted the significance of key observable parameters on the central star, such as luminosity and effective temperature, which serve as critical inputs for estimating the stellar mass and age via young star evolution models. Understanding these stellar properties is crucial for our subsequent investigation into protoplanetary discs as they control the gravitational potential, radiation field, and accretion processes within the disc.

We then shifted our focus more specifically towards the objects of interest of our thesis, protoplanetary discs. Specifically, we analysed the observational constraints that guide our understanding of these complex structures, including the mass distribution of gas and dust, as well as the disc radial and vertical thermodynamic structure. We provided an overview of the hydrostatic disc structure of the {\tt ProDiMO} code that will be the referential framework for our subsequent analyses.

Recognising the limitations of purely hydrostatic models, we also emphasised the necessity of incorporating Magnetohydrodynamic (MHD) theories into our analysis. We outlined that MHD mechanisms currently stand as the most plausible explanations for the accretion-ejection processes in T Tauri stars. However, we have also pointed out that the effectiveness of these MHD theories is closely related to the level of ionisation inside the disc, which paves the way for the study of disc ionisation in the next chapter.

As a conclusion, this chapter serves as a foundational cornerstone for our thesis, highlighting both the complexity and the interdisciplinary nature of the subject matter. It hopefully offers a comprehensive survey of the current knowledge and observational techniques, preparing the ground for our ensuing, in-depth examination of ionisation sources and their roles in protoplanetary discs surrounding T Tauri stars.

\chapter{Ionisation in discs}\label{C:ionisation}
%\textcolor{red}{il serait bon d'ajouter ici 2 references aux travaux de Paola Caselli qui a bcp contribué}
\section{Introduction}

In comparison to the parent molecular clouds, the disc gas is much denser, screening the ionizing radiation, leading to a lower ionisation degree. Even with this reduced ionisation degree, it is still expected to be high enough to allow partial coupling with the magnetic fields in discs. This coupling could instigate MHD instabilities, contribute to disc winds, and enable the necessary angular momentum transfer for mass accretion \citep{2013ApJ...767...30B,suzuki2014magnetohydrodynamic}. The ionisation rate, noted $\zeta$ in s$^{-1}$, is the production of electron per H$_2$ atoms per second and the ionisation degree noted $x_E=n_e/n_{n}$, is the relative abundance of electrons to neutral molecules. These are the key parameters determining the ionisation state of a disc. They influence both the physical and chemical dynamics of protoplanetary discs, which in turn shape the formation of planetary systems \citep{1988PThPS..96..151U}. 

The ionisation degree varies spatially within the disc, influencing its physical structure \citep{2017A&A...600A..75B}. This spatial variability also impacts ion-molecule reactions, as ion-molecular reaction rates are directly linked to ion levels. In addition, ionisation catalyses complex chemical interactions in T Tauri discs, leading to a multitude of molecular species that have an impact on the overall chemical composition of the disc. Knowledge of ionisation mechanisms is also essential for predicting the synthesis and distribution of organic molecules, potentially crucial for prebiotic chemistry in forming planetary systems. Furthermore, ionisation-induced heating modulates the thermal equilibrium in T Tauri discs, affecting their vertical structure and influencing the behavior of both dust and gas. This heat could also be one of the energy source for the launching of winds and jets. 

\begin{figure}[h!]
    \centering
    \includegraphics[width=1\linewidth]{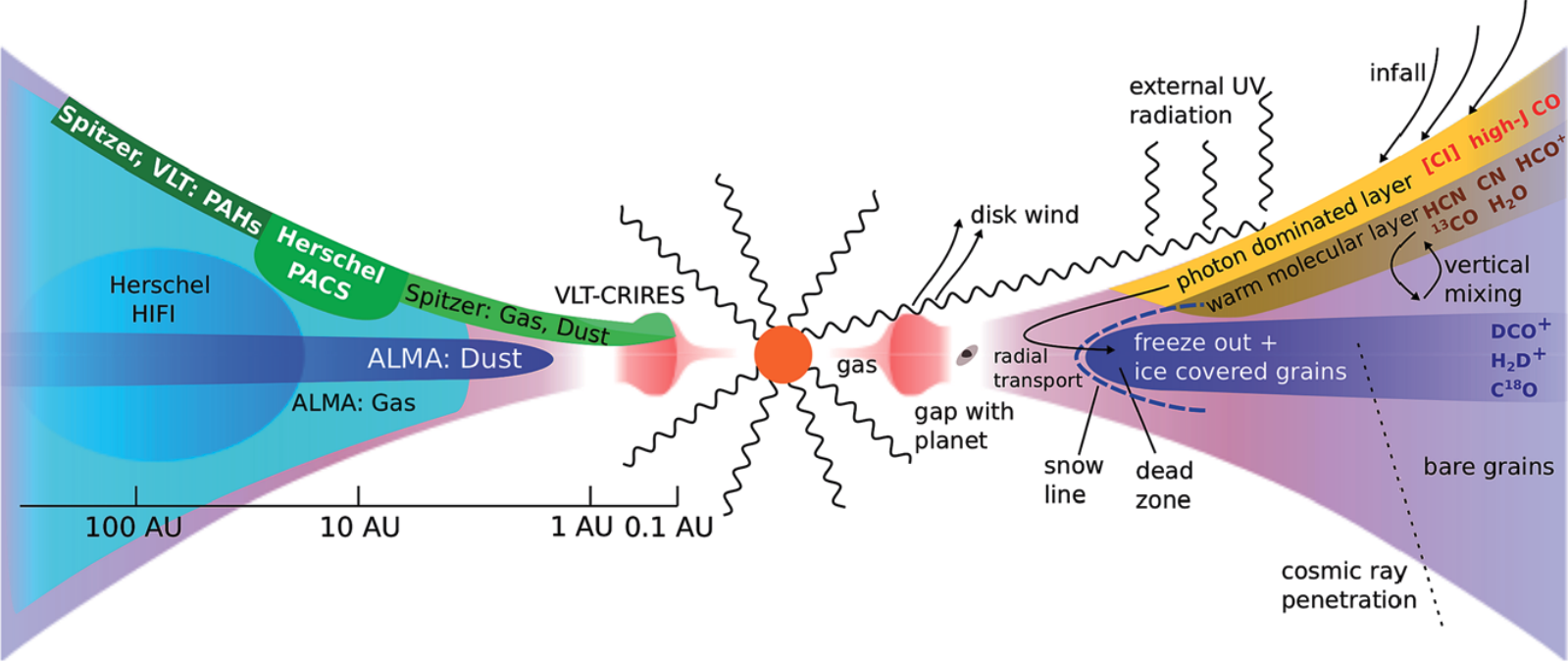}
    \caption{Sketch of different physical and chemical processes taking place in discs (right) that are probed by different observational facilities (left). The figure is reproduced from \citet{2014FaDi..168....9V}}
    \label{fig:ionisationTracersDistribution}.
    %\textcolor{magenta}{CS very nice plot indeed you could refer to it earlier.}
\end{figure}

Studies on disc ionisation typically aim to do two things: pinpoint the ionisation degree (see Sec. \ref{sec:ionisationConstrains}) and identify the primary ionisation sources (see Sec. \ref{Sec:ionisationSources}). This ionisation degree is influenced by a balance between ionisation and recombination rates, both of which depend on factors such as the disc chemical composition and the propagation of ionising radiation \citep{2004A&A...417...93S,woitke2009radiation,2019ApJ...872..107X}. Due to the chemical layering of these discs, the main carriers of positive charge differ between various disc regions (as shown by the coloured layer on the right hand side of Figure \ref{fig:ionisationTracersDistribution}). Consequently, the use of atomic and molecular ions as indicators of ionisation levels requires detailed numerical models and observations of the chemistry of the disc. On the computational front, simulations employing both radiative transfer and chemistry models like {\tt ProDiMO}, or radiative MHD simulations like the ones proposed by \citet{flock20173d}, generate observables needed to interpret the observational data, allowing to estimate the ionisation state of T Tauri discs.

%In recent years, observational methods and computational simulations have opened new ways for understanding ionisation processes within T Tauri discs. This progress has made possible to demonstrate a complex interplay between multiple sources of ionisation and their effects on the dynamics of the disc, structure, and evolution. Taking advantage of high-resolution spectroscopic and interferometric techniques, observers have successfully identified and characterised ionised species in T Tauri discs, thereby aiming to constrain ionisation rates and their spatial distributions.

%The critical role of ionisation in disc dynamics and structure cannot be overstated. Ionisation is pivotal for magnetic field coupling with the disc material, instigating various instabilities and turbulence. Specifically, ionisation is a fundamental factor in the development of magneto-rotational instability (MRI), a principal mechanism thought to driving angular momentum transport and accretion within discs.

The arising question is how to estimate the ionisation degree in order to evaluate its influence on the chemistry and dynamics of T Tauri discs.

This chapter is structured as follows. We first discuss current observational constraints and identify the most effective ionisation tracers relative to their positions within the disc in Sec. \ref{sec:ionisationConstrains}. Subsequently, we will delve into the commonly invoked ionisation sources for the modeling of ionisation rate distribution within the discs in Sec. \ref{Sec:ionisationSources}. Lastly, we will explore the impacts of ionisation on both the chemistry and dynamics of the discs in Sec. \ref{Sec:ionisationEffects}.

\section{Observational constraints on ionisation processes}\label{sec:ionisationConstrains}

\begin{figure}[h!]
    \centering
    \includegraphics[width=0.7\linewidth]{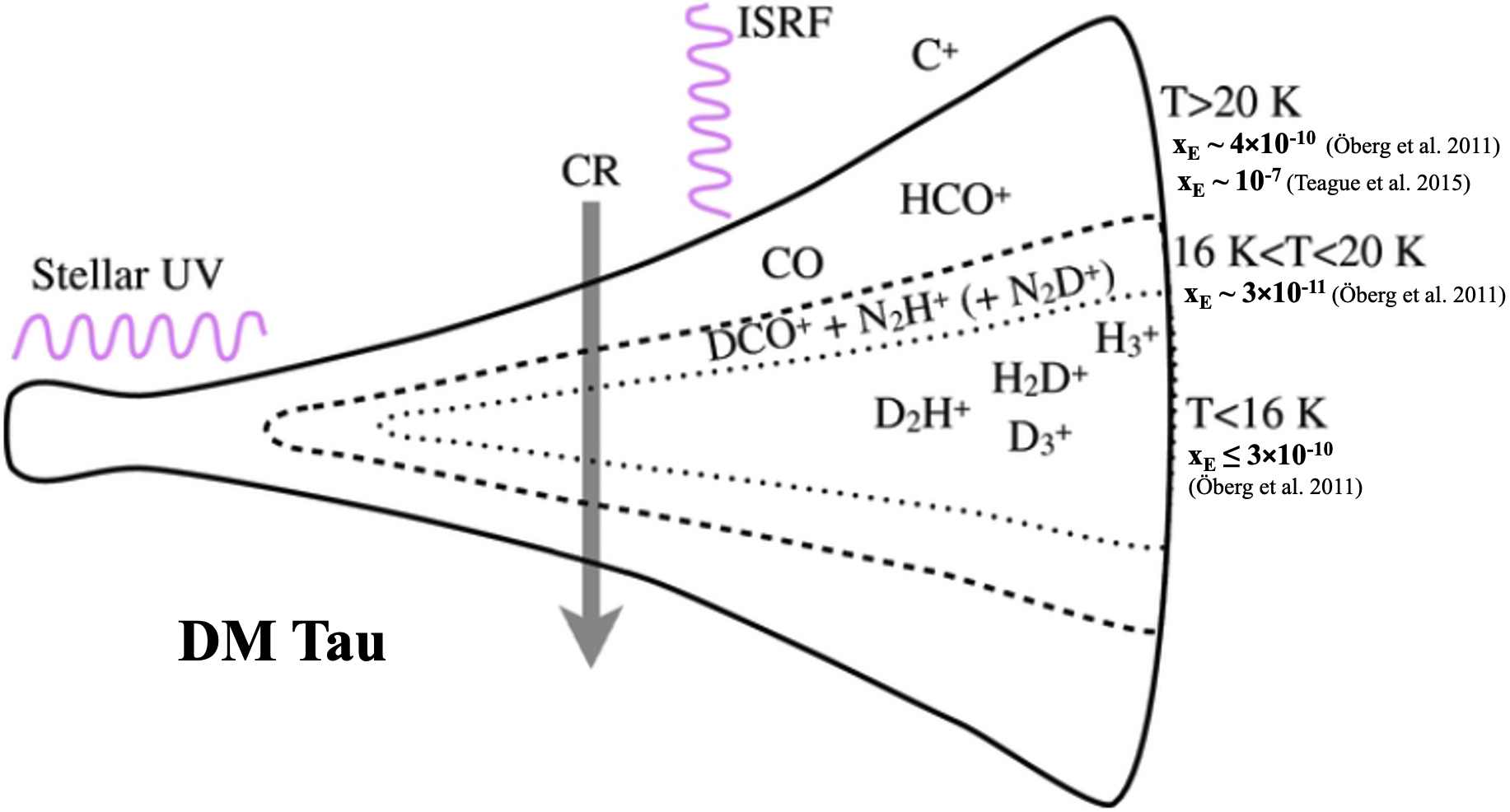}
    \caption{Schematic of the expected disc ion distribution as a function of disc layer temperature. The solid lines represent the CO photodissociation front, the dashed lines the CO snowline, and the dotted lines the transition region where the last heavy molecules freeze out. We added to the original figure of \citet{2011ApJ...743..152O} the ionisation degree observed in DM Tau at $R = 100$ au by \citet{2011ApJ...743..152O} and \citet{teague2015chemistry}.}
    \label{fig:DMTauionisationConstrains}
\end{figure}

Investigating ionisation processes is crucial, so the examination of molecular ions in protoplanetary discs is necessary. According to theoretical models, the key molecular ions tracers in discs are H$_3^+$, HCO$^+$, N$_2$H$^+$, along with their deuterated isotopologues (see \citealt{aikawa2001two,2007prpl.conf..751B,aikawa2015analytical,aikawa2018multiple,aikawa2021molecules}). The abundance of these molecular ions can vary both radially and vertically within the discs, see Fig. \ref{fig:ionisationTracersDistribution}. 
These ions are detected using millimetre and submillimetre telescopes \citep{2010ApJ...720..480O,2014ApJ...788...59W}. Additionally, atomic ions like C$^+$ and S$^+$ are observed in the FIR and submillimetre wavelength ranges \citep{2010A&A...518L.125T,2016A&A...588A.108K}. The observed abundances of these ionised species are related to ionisation rate, gas density, and the number and size distribution of dust grains. There are multiple ionisation sources in these discs, such as X-rays from the central star, Galactic CRs (GCR), stellar energetic particles (SEP), and short-lived radioactive nuclei (SLRs) \citep{1981PASJ...33..617U,glassgold1997x,Rab17}. These sources are represented on the right hand side of Fig. \ref{fig:ionisationTracersDistribution}. The ionisation rates of these sources are determined theoretically, and their impact on the disc is calculated using conjointly, chemical models and observational analyses \citep{2014ApJ...790L...1C,2014Sci...345.1590C}. By comparing the observed abundances of ionised species with the predictions of chemical models and ionisation rates, it is possible to constrain the ionisation processes in T Tauri discs \citep{2014A&A...563A..33W,2014Sci...345.1590C,2014ApJ...790L...1C}. For example, different species may trace different ionisation mechanisms and disc regions \citep{2011ApJ...740..109O,2018A&A...609A..93C}. For instance, C$^+$ observations are used to probe the ionisation of the disc surface layers by UV radiation \citep{2016A&A...592A..83K}, while N$_2$H$^+$ detection trace ionisation by X-rays and cosmic rays in the disc midplane \citep{2012ApJ...747..114W}, see Fig \ref{fig:DMTauionisationConstrains}.

Due to the complexity of the structure of the atomic and molecular ions in the discs and the small number of studies devoted to characterising the ionisation of the discs, there is no consensus on the typical ionisation degree and ionisation rate.

\begin{itemize}
    \item At the disc surface, known as the photon-dominated region (PDR), light yellow area of Fig. \ref{fig:ionisationTracersDistribution}, atomic ions like C$^+$ and S$^+$ are the primary charge carriers due to the process of photoionisation.
    X-rays predominantly ionize the surface layers of the disc. However, their penetration is considerably limited before reaching the disc midplane. For example, at 1 keV, the attenuation length for X-rays corresponds to a hydrogen column density of \(N_H \approx 10^{22} \, \text{cm}^{-2}\), while the actual column densities of disc are around \(10^{23}-10^{25} \, \text{cm}^{-2}\) at a radius of 100 au (Zhang et al. 2021). Although X-rays with higher energies have larger attenuation lengths, their ionisation rates in the midplane are still expected to be less than \(10^{-18} \, \text{s}^{-1}\) \citep{Rab18}. 
    
    \item HCO$^+$ takes over as the dominant ion when the ratio $n$(CO)$/n(e)$ exceeds approximately $10^3$, where $n(i)$ is the number density of species $i$ \citep{aikawa2015analytical}, see the shaded yellow area of Fig. \ref{fig:ionisationTracersDistribution}. CRs produce an ionisation rate of approximately \(5 \times 10^{-17} \, \text{s}^{-1}\) in molecular clouds \citep{2006PNAS..10312269D}, and they have a much larger attenuation length compared to X-rays. The exact length has been reassessed to be even larger in recent studies \citep{Padovani18}. However, GCR can be deflected by magnetic fields in the disc or by magnetised stellar winds \citep{1981PASJ...33..617U,cleeves2014ancient,Padovani18}. 

    \item In regions where the temperature falls below 20 K, corresponding to below the CO snow surface, CO stick to grains, causing an increase in the abundance of N$_2$H$^+$. As we move deeper into the disc and the temperature decreases, the dominant ion transitions are from N$_2$H$^+$ to H$_3^+$ and its deuterated isotopologues, see the deep blue area of Fig. \ref{fig:ionisationTracersDistribution}. Unfortunately, H$_3^+$ is not observable at millimetre wavelengths, and so far, its deuterated isotopologues have not been detected in discs. Instead, N$_2$D$^+$ is considered a viable alternative for estimating the ionisation degree in the cold midplane \citep{cleeves2014ancient}. At these locations, ionisation is due to the decay of \(^{26}\text{Al}\) and is estimated to be \(10^{-18} \, \text{s}^{-1}\) in the primordial solar system, based on meteorite evidence \citep{2009ApJ...690...69U}. Although the abundance of SLRs could differ across star-forming regions, recent models suggest that the \(^{26}\text{Al}\) abundance in the primordial solar system may be representative \citep{fujimoto2018short}.
\end{itemize}

In summary, both the main source of ionisation and the degree of ionisation are subject to spatial variations within the disc; see Fig. \ref{fig:DMTauionisationConstrains} for a schematic view of the layers with their associated ionisation tracers discussed above.

Although the rotational transitions of molecules like HCO$^+$, N$_2$H$^+$, and their isotopologues have been observed in multiple discs, only a few studies have conducted a comprehensive evaluation of their column densities and the associated disc ionisation degrees. These studies and their results are summarised in Tab. \ref{tab:ionisationConstrains}.

\begin{table}[]
    \small
    \centering
    \begin{tabular}{|c|c|c|c|c|}
    \hline
    Object (Authors) & Tracer & Disc layer & $x_E$ & $\zeta$ (s$^{-1}$)\\
    \hline\hline
     DM Tau   & HCO$^+$ & warmer molecular & $4 \times 10^{-10}$& - \\
    \citep{2011ApJ...743..152O} & DCO$^+$,N$_2$H$^+$ & cold molecular  & $3 \times 10^{-11}$& -\\   
      & H$_2$D$^+$ & midplane  & $\lesssim 3 \times 10^{-10}$&  -\\
    \hline
    DM Tau  & HCO$^+/$ DCO$^+$ & warm molecular  & $ 10^{-7}$ & - \\
    \citep{teague2015chemistry} & & & & \\
    \hline
    TW Hya & HCO$^+$ & midplane  &  $10^{-11} - 10^{-10} $ & $< 10^{-19} $\\
    \citep{2015ApJ...799..204C}  & & & & \\
    \hline
    IM Lup, AS 209, HD163296  & N$_2$H$^+$& midplane  & $3\times 10^{-11} - 3\times 10^{-10} $ & $> 10^{-18}$ \\
    \citep{2021ApJS..257...13A} & & & & \\
    \hline
    GM Aur, MWC 480 & N$_2$H$^+$& midplane  & $3\times 10^{-11} - 3\times 10^{-10} $ & $< 10^{-18}$\\
    \citep{2021ApJS..257...13A} & & & & \\
    \hline
    \end{tabular}
    \caption{Summary of the observational constrains on ionisation for different objects, all evaluated at R=100 au.  }
    \label{tab:ionisationConstrains}
\end{table}

For instance, \citet{2011ApJ...743..152O} used the IRAM 30 m telescope to study the H$^{13}$CO$^+$ J = 3 - 2 transition in the DM Tau disc. They combined their findings with prior Submillimetre Array (SMA) data of other molecules to estimate the ionisation degree, \( x_E \), in three thermal zones within the disc. In the warmer molecular layer where the temperature is above 20 K, they estimated \( x_E \approx 4 \times 10^{-10}\) based on HCO$^+$ data. In the moderately warm layer with temperatures between 16-20 K, where N$_2$H$^+$ and DCO$^+$ are prevalent, \( x_E \approx3 \times 10^{-11}\). Meanwhile, in the coldest and densest region of the disc ($T < 16$ K), the absence of H$_2$D$^+$ detection led to an upper limit for \( x_E \) of \(3 \times 10^{-10}\).

\citet{teague2015chemistry} extended these observations using the Plateau de Bure Interferometer to investigate HCO$^+$ J = 1 - 0 and J = 3 - 2 as well as DCO$^+$ J = 3 - 2 in DM Tau, with a resolution of approximately \(1.5"\). They determined the column densities for HCO$^+$ and DCO$^+$ at a 100 au radius to be \(9.8 \times 10^{12} \, \text{cm}^{-2}\) and \(1.2 \times 10^{12} \, \text{cm}^{-2}\), respectively. From these data, an ionisation degree around \(10^{-7}\) was inferred in the molecular layer, based on the DCO$^+$/HCO$^+$ abundance ratio. This calculation assumed a steady-state equilibrium between specific chemical reactions involving H$_3^+ +$ HD $\longrightarrow $ H$_2$D$^+ + $ H$_2$, H$_2$D$^+ +$ CO $\longrightarrow $ DCO$^+ +$  H$_2$, and DCO$^+$ destruction \citep{caselli2002molecular}.

In another study by \citet{2015ApJ...799..204C}, chemical models of the TW Hya disc were created, accounting for various CR ionisation rates (\( \zeta_{\text{CR}} \)) and X-ray spectra. Rather than directly measuring molecular column densities from observations, they computed the disc-integrated fluxes of molecular lines based on these models. These were then compared with their observations of HCO$^+$ and H$^{13}$CO$^+$ J = 3 - 2 in TW Hya, as well as existing data on HCO$^+$, H$^{13}$CO$^+$, and N$_2$H$^+$ from the literature (e.g., \citealt{qi2013h2co}). Their findings suggested that the model most consistent with observations featured a low CR ionisation rate (\( \zeta_{\text{CR}} \leq 10^{-19} \, \text{s}^{-1} \)) and moderate X-ray radiation. In this scenario, they estimated the ionisation degree to be in the range of \(10^{-11} \) to \( 10^{-10} \) near the midplane of the disc, beyond a radius of approximately 100 au. The study emphasized that since the ionisation degree can vary both spatially within a single disc and between different discs, acquiring observations with greater spatial resolution and covering more targets is crucial for advancing the field. TW Hya %\textcolor{magenta}{CS Is this not one of the nice ALMA image with the disc and the planets I sent you?} 
stands out as an optimal observational target due to its relative closeness. However, it is relatively old as a disc, and the total disc mass of $0.02 M_\odot $ derived from dust observations (which assumes a specific dust-to-gas mass ratio) is relatively large compared to the various observed young star-forming regions \citep{2016ApJ...831..125P}. As a result, the low ionisation rate observed by \citet{2015ApJ...799..204C} in the disc of TW Hya may not accurately represent the majority of discs present in star-forming regions.\\

Due to the complex nature of atomic and molecular ion configurations in discs, and the limited studies aimed at constraining disc ionisation, there is no clear agreement on what constitutes a typical ionisation degree and ionisation rate. We have seen that, based on their analysis of HCO$^+$ and N$_2$H$^+$ emissions in the disc of TW Hya, \citet{2015ApJ...799..204C} identified a low cosmic-ray ionisation rate ($<10^{-19}$ s$^{-1}$), which is significantly below the interstellar standard value ($\sim 10^{-17}$ s$^{-1}$). In contrast, \citet{aikawa2021molecules} reported relatively high ionisation rates in both the warm molecular layer and the midplane of other discs, consistent with common assumptions about X-ray and cosmic-ray ionisation. In addition \citet{2021ApJ...912..136S} noted a variation in the cosmic-ray ionisation rate across the IM Lup disc, suggesting ionisation due to particles emitted from the central star.

Since the ionisation degree can differ both spatially and between various discs, there is a high demand for observations at a greater spatial resolution and targeting more sources. Despite these challenges, significant progress has been made in recent years to improve the accuracy and precision of ionisation rate measurements in T Tauri discs. Advances in observational facilities, such as ALMA, Herschel, SOFIA, and JWST and the upcoming ELT, have enabled (or will) more detailed studies of ionisation tracers and their spatial distributions \citep{2014Sci...345.1590C,2021ApJ...912..136S,2022ApJ...932....6V}. 

Additionally, improvements in chemical models, radiative transfer codes, and rate coefficient databases have led to more accurate predictions of ionisation tracer abundances and line emission \citep{2018A&A...609A..91R,2022A&A...668A.164W}. These progresses have allowed to better constrain ionisation rates in T Tauri discs and to disentangle the contributions of different ionisation mechanisms. Recently, \citet{2022A&A...658A.189P} introduced a method to estimate ionisation rates using observations of near-infrared rovibrational transitions of H$_2$ primarily excited by secondary CR electrons. This method was tested by \citet{2022A&A...658L..13B} to obtain upper limits of the ionisation rate. In theory, the JWST should be able to detect these near-infrared H$_2$ lines, enabling the first-ever direct determination of spatial variation of the ionisation rate in dense gas. This will enable competing models of CR propagation and attenuation to be tested \citep{2011ApJ...739...60E,2019ApJ...879...14S,2018A&A...614A.111P,2021ApJ...917L..39G}.\\

We have seen that the dominant ionisation sources are expected to differ, within individual discs \citep{2015ApJ...799..204C,Rab17}, among discs surrounding different stars \citep{2015A&A...582A..88W}, and under different radiation conditions \citep{2013ApJ...766L..23W}. Theoretically, the significance of different ionisation sources could be determined through ion observations in discs, as each source affects ion abundances in distinct disc regions. For instance, \citet{2015ApJ...799..204C} found that cosmic-ray ionisation played a less significant role compared to an unexpectedly high contribution from X-rays in a specific disc, hinting at a flaring X-ray phase. Such X-ray flares could also affect ionisation levels over specific periods, as observed by \citet{2017ApJ...843L...3C} and \citet{2022ApJ...928...46W}. 
We end this section by emphasizing that the ionisation degrees we have discussed, are not only weakly constrained by observations but also heavily dependent on disc models. This underscores the need for a thorough theoretical understanding of ionisation sources within discs. This is crucial both for establishing consistent observational constraints and because, as of now, ionisation rates and degrees in the inner parts of discs are not observable. In the next section, we focus on the ionisation source models that have been developed to estimate the ionisation rate in T Tauri discs.

\section{Ionisation Sources in Protoplanetary Discs}\label{Sec:ionisationSources}

\begin{figure}[h!]
    \centering
    \includegraphics[width=0.7\linewidth]{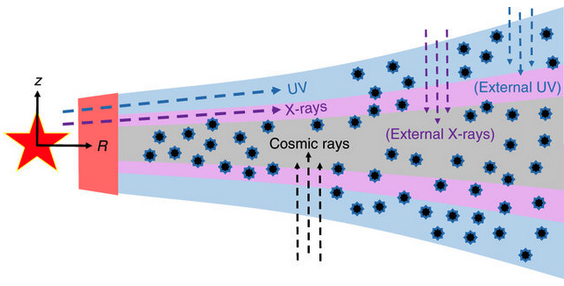}
    \caption{The figure represents the zones affected by different ionisation sources. These zones are color-coded: blue for UV-dominated areas, violet for X-ray dominated regions, and grey for cosmic-ray influenced regions. The central star is depicted on the left, with an adjacent red zone indicating a high-temperature area where thermal ionisation is dominant. Ice-coated grains are represented by black and blue star symbols. The boundaries separating these different zones, as well as the "snow surface" where ices begin to form, vary depending on the specific disc. The outer portion of the disc may be exposed to external UV radiation from the interstellar field, as well as external X-rays. The figure is reproduced from \citet{dupuy2018x}.}
    \label{fig:DominantionisationSource}
\end{figure}

Astrochemical models, which interpret observed molecular abundances, and non-ideal MHD codes, which simulate circumstellar discs, both use ionisation rates $\zeta$ as a core parameter. This rate measures the number of ionisation per unit of time and is determined by the interaction between ionisation sources (UV, X-rays, CR) and hydrogen molecules within the disc. Understanding the spatial distribution of the ionisation rates in T Tauri discs is crucial, as it directly impacts the disc chemistry, thermal balance, and dynamics. The ionisation rates vary significantly with the radial distance from the central star, as well as with the vertical height above the disc midplane, due to the varying contributions of the different ionisation sources; see Fig. \ref{fig:DominantionisationSource} and \citet{2004A&A...417...93S,2013ApJ...772....5C}.

As we will see, in the outer regions of the disc (beyond a few au), the ionisation rates are generally lower, as the contributions of internal sources decrease with distance from the star. Instead, the ionisation rates in the outer disc are mainly dominated by external sources, such as cosmic rays and interstellar radiation \citep{1981PASJ...33..617U,2009A&A...501..619P}. In the most embedded cold regions of the outer disc, ionisation could be dominated by Short-Lived Radionucleides (SLR). The ionisation rates in these regions is typically within the range of $10^{-17}$ to $10^{-15}$ s$^{-1}$ \citep{2013ApJ...772....5C,2016ApJ...833...92I}.

In the inner regions of the disc (within a few au), the ionisation rates are primarily determined by the internal sources, such as stellar radiation, accretion shocks, and flares. The ionisation rates in these regions reach values of $10^{-15}$ to $10^{-10}$ s$^{-1}$, depending on the specific ionisation source and the local disc properties \citep{1999A&A...351..233A,2017A&A...603A..96R}.

There is also a vertical distribution of ionisation rates. In the upper layers of the disc, where the gas is optically thin, the ionisation rates are higher due to the direct exposure to ionising radiation and energetic particles from both internal and external sources. Conversely, in the midplane region of the disc, where the gas is optically thick, the ionisation rates are significantly lower, as the ionising radiation is efficiently absorbed and scattered by the disc dust and gas \citep{2011ApJ...739...50B, 2016ApJ...833...92I}. In these regions, ionisation is dominated by high energy CR or radioactive decay of short lived isotopologues.

In this section, we present the various ionisation sources in protoplanetary discs that have been both theoretically and observationally investigated over the last decade. We first elucidate the origin of these sources and outline the constraints on the typical ionisation rates derived from the literature. Subsequently, we study the spatial distribution of these ionisation rates inferred from disc models. Finally, we present comparative maps of the dominant source of ionisation, depending on the specific region within the disc.

%To effectively study and model these processes, research relies on observational data, laboratory experiments, and theoretical calculations. By comparing the results obtained from these methods, one refine our understanding of the role played by CRs in the complex interplay between molecular clouds, star formation, and the interstellar medium.

%Each method for estimating the CR ionisation rate has its own degree of uncertainty, which impacts the accuracy of the resulting estimates. For instance, H$_3^+$ observations can only be conducted towards specific lines of sight in the direction of early-type background stars. Furthermore, $\zeta_{\rm ion,H_2}$ derived from H$_3^+$ is influenced by factors such as gas density, local ionisation fraction, and the attenuation of the interstellar ultraviolet field, all of which affect the final ionisation rate estimate.

%In denser regions, the primary challenge is the increased chemical complexity compared to diffuse clouds. This necessitates up-to-date, comprehensive reaction networks. \textcolor{magenta}{CS : I do not understand the previous sentence.} The main uncertainties stem from the unknown destruction and formation rates of various species, as well as the indeterminate amount of carbon and oxygen depletion on dust grains.

%\subsection{External ionisation sources}

\subsection{Radiation ionisation}

\subsubsection{The radiative environment of protoplanetary discs}
The radiation emitted by the central star plays a crucial role in shaping the physical and chemical properties of a protoplanetary disc, especially in the areas of heating, ionisation, and chemical transformations. Dust within the disc is of particular importance because it contributes to the disc opacity, affecting how radiation interacts with the material. A dynamic balance between various heating methods, such as chemical reactions and photoionisation, and cooling mechanisms, like line radiation from gas and continuum radiation from dust, dictates the thermal structure in the disc upper layer. Near the midplane, the temperatures of gas and dust couples due to interactions between them, as indicated by \citet{chiang1997spectral}. Both the gas chemical composition and the size diversity of the dust grains influence this heating and cooling equilibrium, necessitating intricate radiation thermochemical simulations for accurate modeling. Typically, the disc midplane remains relatively cool. However, high-energy radiation can cause extreme ionisation and heating in the surface layers of the disc, where the densities are low enough for gas and dust to thermally decouple. These processes can contribute to the photoevaporation of the gas disc and modify the chemical composition of its constituents. Consequently, our focus will primarily be on the impact of high-energy radiation on the gaseous component of the disc.

Circumstellar discs are exposed to a range of high-energy radiation, primarily from the stellar hot corona, across a broad range of wavelengths. This flux is typically divided into three categories: Far UltraViolet (FUV; 6 eV$< h\nu < 13.6 $ eV), Extreme Ultraviolet (EUV; 13.6 eV$ < h\nu < 100 $ eV), and X-ray ($h\nu > 100$ eV). Some focused jets also exhibit notable X-ray emissions, as observed by \citet{2008A&A...478..797G}. Stellar flares contribute to additional hard X-ray and gamma radiation, along with an increased flux of high-energy particles, often referred to as stellar cosmic rays, see Sec. \ref{sec:stellarEP}. These forms of high-energy radiation serve as ionising agents for the gas in the disc, and they can heat the gas to temperatures of several thousand Kelvin, catalysing chemical reactions. The ionisation and heating have additional consequences for the disc stability and longevity. Ionisation may trigger magnetorotational instability (MRI) by allowing ionised gas to couple with magnetic fields, while heating can lead to the photoevaporation of gas from the disc surface layers. Consequently, the lifespan of the disc could be significantly influenced by these high-energy processes originating from or near the central star.\\

Here we first provide an overview of the primary sources of radiation impacting the disc from outside-in. 

\begin{itemize}
    \item \textbf{Interstellar Radiation Field (ISRF) – } Most of the star formation occurs within embedded clusters in our galaxy. ISRF includes UV and X-ray photons emitted by nearby stars and other astrophysical sources \citep{2009ApJ...690.1539G}. For an isolated disc or a low-mass star-forming region, the ISRF delivers an omnidirectional incident flux. The UV flux is typically $\int F_\nu d\nu = 1.6 \times 10^{-3}\ \text{erg cm}^{-2} \text{s}^{-1}$ between 912 and 2000 Å. This corresponds to one Habing unit ($1 G_0$, see Eq. \eqref{eq:Habing} for the definition). As a comparison, the integrated flux from the central star at a distance of 100 au typically falls within the range of $G_0 = 240-1500$, decreasing proportionally to the inverse square of the distance ($\propto r^{-2}$). However, a disc within a stellar cluster can experience significantly higher values of $G_0$, ranging from 300 to 30,000, with a typical value of 3000. This can be attributed to the combined effect of many massive stars in the cluster, all contributing to the ISRF. Therefore, the ISRF within a cluster can overcome the FUV radiation effect in the outermost regions of the disc \citep{Cleeves13}.

    \item \textbf{Stellar coronal radiation – } X-ray and extreme-ultraviolet radiation from T Tauri stars primarily come from a magnetised corona, similar to the Sun. However, the intensity of X-ray radiation in these stars is significantly higher than that of the Sun, reaching up to $10^{-3}$ times the stellar bolometric luminosity, which translates to energy levels between $10^{29} $ and $10^{31}$ erg$/$s \citep{telleschi2007x}. This characteristic is also observed in Class I protostars. For instance, in the Orion region, the X-ray luminosities of protostars grow by about an order of magnitude when they evolve into T Tauri stars, although the exact evolution at energies below 1-2 keV remains unclear \citep{prisinzano2008x}.
    
    \item \textbf{Stellar X-ray flares – } X-ray flares are observed in T Tauri stars, capable of generating plasma temperatures as high as around $10^8$ K \citep{stelzer2007statistical,imanishi2001chandra}. They may originate from magnetic reconnection either in the stellar magnetosphere, at the star-disc interface, or above the circumstellar disc \citep{Feigelson99}. X-ray flares are the central point of this thesis, we dedicate Chapter \ref{C:Reconnection} to model the radiation and particle emission from these flares and Chapters \ref{C:PublicationI} and \ref{C:PublicationII} to their effects on T Tauri disc.

    \item \textbf{X-ray fluorescence – } Another interesting phenomenon is the photoionisation of the cooler disc gas by X-rays exceeding the Fe K edge at 7.11 keV, which leads to a distinct fluorescence line feature at 6.4 keV. This has been observed in several classical T Tauri stars, especially following intense X-ray flaring events \citep{tsujimoto2005iron}. In at least one case observed by \citet{2005ApJS..160..469F}, this strong fluorescence appeared to be relatively constant, rather than being associated with strong X-ray flares. This suggests that the 6.4 keV line could be stimulated by non-thermal electron impacts within densely packed, magnetically accreting loops.

    \item \textbf{Accretion-induced UV/X-rays – } T Tauri stars exhibit an ultraviolet flux that significantly exceeds what would be anticipated based on their effective temperature of approximately 4000 K. This excess in UV radiation is believed to stem from the accretion shock occurring on the surface of the star, as suggested by studies such as \citet{1998ApJ...509..802C}. Streams of material accreting from the circumstellar disc towards the central star can achieve speeds approximately equal to the free-fall velocity, typically around \(5 \times 10^2\) km/s \citep{1998ApJ...509..802C}. When this material reaches the stellar surface, it generates a shock with temperatures on the order of a few million kelvin. The X-rays generated from this high-temperature shock can be absorbed back into the shocked gas, further contributing to its heating. There is compelling evidence to suggest that such X-ray production related to accretion exists, particularly in the very soft X-ray spectrum. For example, there have been observations of excess flux in the O VII and Ne IX line triplets as reported by \citet{kastner2002evidence,gudel2007x}.
       
\end{itemize}

In the subsequent section, we focus on the impacts of the aforementioned radiation mechanisms on protoplanetary discs. However, we will omit discussing the ISRF ionisation, as we have previously noted our assumption that the T Tauri system is isolated and remains unaffected by its immediate surroundings. For the effects of the ISRF on ionization see \citet{Cleeves13,2019ApJ...883..121O}. Moreover, we do not explore the X-ray fluorescence line production induced by GeV particles since this has not been explored in the context of a T Tauri disc. Research such as that by \citet{2012A&A...546A..88T} has probed this phenomenon in the galactic centre.

\subsubsection{Stellar Photoionisation}\label{S:Photoion}
%\textcolor{red}{[la présentaiton ne va pas: présente les processus, la physique des processus et les résultats spécifiques pour chaque partie du disque, il manque la partie Xsection que tu devras à la première mention aux X et UV.]}

Stellar radiation is a crucial internal ionisation source in T Tauri discs, as it ionises the disc material and influences the disc chemistry, the thermal balance and the dynamics of the surface layers of discs \citep{2011ApJ...735...90G,2013MNRAS.436.3446E}. 
The ionisation by stellar radiation, at the disc surface depends on the central star spectral energy distribution and distance to the disc, but it is typically on the order of $\zeta=10^{-15} - 10^{-13}$ s$^{-1}$ \citep{2011ApJ...739...78B}. The ionisation rates due to stellar X and UV radiation are very sensitive to the disc structure, especially on the presence of dust, with higher ionisation rate in regions of lower dust opacity \citep{2004ApJ...615..972G}.

Stellar radiation efficiently ionise atomic hydrogen. They create a hot, ionised region at the disc surface, where the free electron density becomes greater than the neutrals density \citep{2009ApJ...705.1237G}. This layer is determinant in the global dynamics of the system. The ionised layer could accrete more efficiently and could be a driver in the launching of photoevaporative winds and jets allowing angular momentum transport from the disc \citep{2021MNRAS.508.3710R,2022EPJP..137.1357E}.

\paragraph{X-ray ionisation:}
The X-ray spectrum emitted by the star is continually absorbed as it travels from the star to the surface of the disc, primarily by the disc upper, less dense gas layers and potentially also by any existing disc wind. In the context of a geometrically thin, flat disc, \citet{glassgold1997x} provide an equation to describe the dependency of the photon flux spectrum \(F0\) on both the photon energy \(\epsilon\) and the position, which is defined by \(r\), the radius of the disc segment. The equation is:

\begin{equation}
    F_0(\epsilon, r) = \frac{R}{r} \times \frac{1}{2} \times \frac{1}{4\pi r^2} \times \frac{L_X}{kT} \times \frac{1}{\epsilon} \times e^{-\epsilon/kT}, \quad \epsilon > \epsilon_0
\end{equation}

In this equation, \(R/r\) represents the angle at which the X-rays hit the disc, \(1/\epsilon\) and \(1/(kT)\) are factors that convert the energy flux to photon flux and normalise the total luminosity to \(L_X\), respectively. The term \(\epsilon_0\) accounts for the low-energy cutoff due to absorption during the X-rays journey from the star to the disc.

\begin{figure}[h!]
    \centering
    \includegraphics[width=0.75\linewidth]{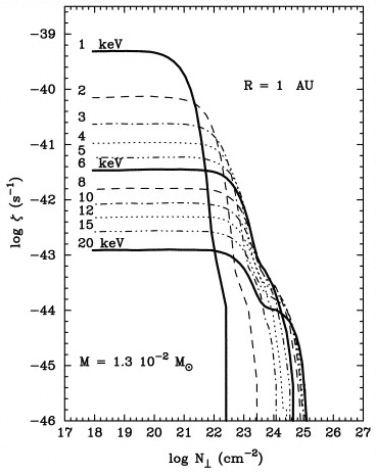}
    \caption{Ionisation rates corresponding to monochromatic X-rays (hence the total ionisation rate is obtained by integrating over the whole X-ray spectrum) as function of column density in a vertical direction through a disc of mass 0.013 $M_\odot$ situated at a radius of 1 au, assuming the gas has a typical interstellar composition. The X-ray luminosity from the source is normalised to 1 erg/s. The figure is reproduced from \citet{1999ApJ...518..848I}.}
    \label{fig:MonochromaticXrayionisationRate}
\end{figure}

As these X-rays reach the disc surface, their flux experiences further attenuation based on the photoionisation cross-section. Essentially, nearly all the initial energy from the X-rays is consumed for ionisation, using roughly \( \Delta \epsilon \approx 37 \) eV for each ionised pair. \citet{glassgold1997x} and \citet{1999ApJ...518..848I} carried out numerical analyses to determine ionisation levels for realistic disc profiles and shapes, also taking into account the Compton scattering of high-energy X-rays within the disc. They essentially find as can be seen in Fig. \ref{fig:MonochromaticXrayionisationRate} that at the disc surface, where the optical depth is $\ll 1$ or the column density is $\lesssim 10^{20}$ cm$^{-2}$, the ionisation rate by the softest X-rays remains roughly constant due to the spectral cutoff \( \epsilon_0 \). The ionisation rate then quickly diminishes as we move to greater column depths, where the harder photons primarily ionise heavier elements. These harder spectra also penetrate further into the disc, ionising its deeper layers. For a typical stellar coronal temperature of \( kT = 1 \) keV, the ionisation rate matches the cosmic-ray ionisation rate of \( 2 \times 10^{-17} \) s\(^{-1}\) at column depths between \( N_H \approx 10^{23}-10^{25} \) cm\(^{-2}\) and disc radii around 0.1 to 10 au. Beyond column densities greater than \( N_H \approx 1.5 \times 10^{24} \) cm\(^{-2}\), ionising X-rays with energies above 10 keV come into the game, where the Compton cross-section surpasses the absorption cross-section, leading to changes in the ionisation rate due to Compton scattering. 

\begin{figure}[h!]
    \centering
    \includegraphics[width=0.75\linewidth]{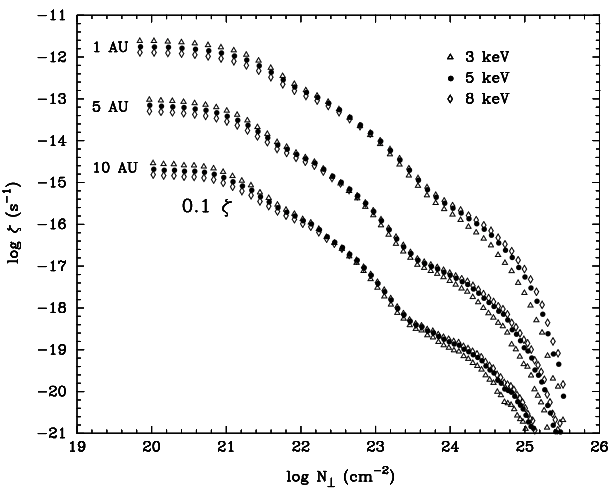}
    \caption{The ionisation rate of thermal X-ray spectra depends on vertical column density in a disc, considering an X-ray luminosity of \(10^{29} \text{ ergs/s}\), a minimum energy $E_0=1$ keV, and a disc mass of 0.013 $M_\odot$. These rates are derived at three radial distances: 1, 5, and 10 au. For each of these positions, three different X-ray temperatures are evaluated: 3 keV (open triangles), 5 keV (filled circles), and 8 keV (open diamonds). The figure is reproduced from \citet{1999ApJ...518..848I}.}
    \label{fig:XrayionisationRate}
\end{figure}
A parametric equation for the X-ray ionisation rate is extracted from the results of \citet{1999ApJ...518..848I}, plotted in Fig. \ref{fig:XrayionisationRate} and is given by \citet{bai2009heat},

\begin{equation}
    \zeta_X = L_{X,29} \left(\frac{r}{1 \rm au}\right)^{-2.2} \left[ \zeta_1 \left(e^{-\left(\frac{N_{Ha}}{N_a}\right)^{\alpha_1} }+e^{-\left(\frac{N_{Hb}}{N_b}\right)^{\alpha_1}}\right) + \zeta_2 \left( e^{-\left(\frac{N_{Ha}}{N_a}\right)^{\alpha_2} }+e^{-\left(\frac{N_{Hb}}{N_b}\right)^{\alpha_2}} \right) \right],
    \label{eq:parametricXrayionisationrate}
\end{equation}
where the constants and variables are defined as follows: \( L_{X,29} = \frac{L_X}{10^{29}\, \text{erg}\, \text{s}^{-1}} \) is the normalized X-ray luminosity, \( \zeta_1 = 6 \times 10^{-12}\, \text{s}^{-1} \)and \( \zeta_2 = 10^{-15}\, \text{s}^{-1} \) are constants. \( N_{a} = 1.5 \times 10^{21}\, \text{cm}^{-2} \), \( N_{b} = 7 \times 10^{23}\, \text{cm}^{-2} \) are constants representing reference column densities. \( N_{Ha} \), \( N_{Hb} \) are the column density of hydrogen nuclei above and below a given location in the disc, respectively. \( T_x = 3\, \text{keV} \) is the X-ray energy.
And \( \alpha_1 = 0.4 \), \( \alpha_2 = 0.65 \) are some additional parameters. 

\paragraph{UV ionisation:}
FUV radiation ionisation is significant at the disc surface, but it attenuates rapidly with increasing column density. UV photons are absorbed at column densities around $10^{19-20}$ cm$^{-2}$, and they heat the gas to temperatures ranging from $\sim 8 000$ to $12 000$ K.  Hence, ionisation due to FUV irradiation is limited to the disc surface layers.

For low-mass T Tauri stars, the strength of the UV radiation field is often parameterised in units of the local interstellar radiation field, called a Habing. The Habing unit is equal to $1.2 \times 10^{-4}~\rm erg ~ cm^{-2}~  s^{-1} sr^{-1} = 1.6 \times 10^{-3}~ cm^{-2}~  s^{-1} = 10^8$ photons cm$^{-2}$ s$^{-1}$. One Habing is the average FUV radiation field in the ISM. So the local radiation field $G_{ISM}\equiv 1$ Habing in the ISM.

T Tauri stars exhibit standard value for the FUV luminosity \( L_{FUV} = 10^{30} \, \text{erg s}^{-1}\) \citep{2019ApJ...883..121O}. The local radiation field $G_0$ T Tauri stars impose in the disc can be computed as a function of disc radius 
\begin{equation}
    G_0= 1.4 \times 10^4 \left( \frac{L_{FUV}}{10^{30} \rm ~ erg ~ s^{-1}}\right)\left( \frac{r}{3 ~ \rm au }\right)^{-2} \quad \rm Habing.
\label{eq:Habing}
\end{equation}

\begin{figure}
    \centering\includegraphics[width=\linewidth]{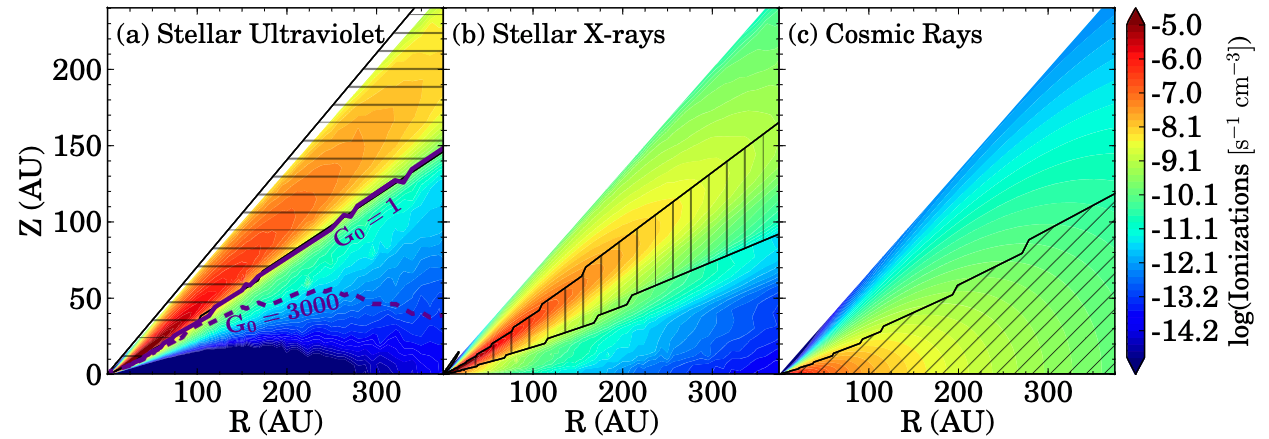}
    \caption{Relative contribution of stellar UV, X-rays, and cosmic rays to the overall ionisation rate as computed by \citet{Cleeves13}. Coloured plots display the volumetric ionisation rate due to each source on the same scale. The crosshatched area marks the region of the disc where each particular ionisation source supplies more than 30\% of the total ionisations per unit of time and volume. Panel (a) represents FUV ionisation of C, $n_C \zeta_{UV}$;  panel (b) represents X-ray ionisation of H$_{2}$, $n_{H_2} \zeta_X$;  panel (c) represents galactic cosmic ray ionisation of H$_2$, $n_{H_2} \zeta_{CR}$, for a standard ISM cosmic ray ionisation rate. The figure is reproduced from \citet{Cleeves13}.}
    \label{fig:VolumetricIonisationDistribution}
\end{figure}

The ionisation rate resulting from FUV photons is described by the following equation,

\begin{equation}
    \zeta = G_0  a_X  \exp\left(-b_X  A_v \right),
    \label{eq:uvionsationrate}
\end{equation}
where \( G_0 \) is the intensity of the radiation field in Habing units. The constants \( a_X \) and \( b_X \) are species-specific ionisation rate coefficients, and \( A_v \) is the dust extinction. In the given context, the coefficients for atomic carbon are specified, with \( a_C = 2.1 \times 10^{-10} \, \text{s}^{-1} \) and \( b_C = 1.7 \) \citep{mcelroy2013umist}. The reactions associated with FUV are included in the UMIST database. The UV volumetric ionisation rate ($\zeta_{UV}$ multiplied by the number density of carbon atoms), in units of $s^{-1}$ cm$^{-3}$ , is shown in Fig. \ref{fig:VolumetricIonisationDistribution}(a). The volumetric ionisation rate is a rough proxy for the electron production rate per unit volume.

These ionisation rates are then integrated into more detailed models of protoplanetary discs to understand the physical and chemical processes that they can trigger \citep{Cleeves13,Rab17,2019ApJ...883..121O}.

We conclude this section by highlighting the different regions where each ionisation source is dominant as per \citet{Cleeves13}. X-ray photons penetrate deeper than UV, mainly ionising dense regions of H$_2$ and He. While the gas and metals in the dust are strong X-ray photon absorber, H$_2$ ionisation predominantly contributes to the electron/ion population, with He contributing but to a lesser degree. Figure \ref{fig:VolumetricIonisationDistribution}(b) depicts the volumetric ionisation rate from X-ray ionisation of H$_2$, based on the X-ray ionisation cross sections by \citet{1999ApJ...518..848I}, Eq. \eqref{eq:parametricXrayionisationrate}, with $L_{X} = 10^{29.5} $erg s$^{-1}$. Using the FUV ionization Eq. \ref{eq:uvionsationrate} with $L_{FUV}= 2.9 \times 10^{31}$ erg s$^{-1}$ and a galactic cosmic ray ionisation rate $\zeta_{CR} = 5 \times 10^{-17} \exp [-\Sigma/(96 ~\rm g ~ cm^{-2})]$ (refer to Sect. \ref{sec:galacticCRionization} for GCR ionisation), the volumetric ionisation rates from the key ionisation sources are displayed, on the same scale, in Figure \ref{fig:VolumetricIonisationDistribution}. The surface layer is primarily influenced by UV ionisation of carbon (left panel), whereas X-rays are more inferential in the deeper layers (centre panel). The black crosshatched areas highlight regions where each ionisation source contributes to at least 30\% of the total ionisation. This plot does not include ionisation from stellar-origin energetic particles, which we discuss in Sec. \ref{sec:stellarEP}. By the end of the section about stellar energetic particles, we present a similar comparative graph illustrating the primary ionisation source, taking into account stellar energetic particles based on the models developed by \citet{Rab17}.

\subsection{Energetic particle ionisation}
\subsubsection{Galactic Cosmic Rays}\label{sec:galacticCRionization}
%\paragraph{Ionisation of the outer disc molecular layer}

GCRs are energetic particles, primarily electrons and protons, that originate from sources within our galaxy, such as supernova remnants \citep{1987PhR...154....1B,2002cra..book.....S}.
Several key questions, such as the mechanisms governing molecular cloud collapse, the fundamental processes controlling dust grain growth, the origins of chemical complexity in discs, and the sources of energetic events in T Tauri systems - all these problems share a connection to GCRs (see \citealt{2020SSRv..216...29P} for a review). Specifically, low-energy CRs ($E < 1$ GeV) are relevant for addressing these questions because the main processes in discs (e.g., ionisation, dissociation, and H$_2$ excitation by CR protons and electrons) have peaks of their cross sections between roughly 10 eV and 10 keV (see, e.g., \citealt{2009A&A...501..619P,2018A&A...614A.111P}).
 
GCRs penetrate T Tauri discs and initiate ionisation processes by colliding with gas molecules, primarily molecular hydrogen (H$_2$) and helium (He). These collisions create secondary particles, such as electrons and ions, that further interact with the disc gas, resulting in a cascade of ionising events \citep{2009A&A...501..619P,2015ApJ...812..135I}.

Although the ionisation of the disc surface layers is primarily dominated by far ultraviolet and X-ray photons \citep{2011ApJ...735....8P,2012ApJ...756..157G} and Sec. \ref{S:Photoion}, CRs are expected to play an essential role in the chemistry and dynamics of molecular layer in outer discs. With their ability to reach surface densities greater than $\Sigma > 10^{2}$ g  cm$^{-2}$ \citep{1981PASJ...33..617U,2018A&A...614A.111P}, CRs offer a substantial contribution to the overall ionisation processes deep in the disc, see Fig. \ref{fig:DominantionisationSource}.

\citet{2018A&A...614A.111P} recently developed a comprehensive ionisation model that is particularly applicable to high-density environments, like the inner areas of collapsing clouds and circumstellar discs. The model emphasises the importance of accurately selecting the transport regime for CR protons as well as robust models for the generation and transport of secondary CRs for precise ionisation calculations. This work outlines how the ionisation rate of molecular hydrogen ($\zeta_{H_2}$) varies with the column density $N$, considering multiple CR energy spectra parameters. Except for an extreme case involving elevated flux of stellar protons, these dependency models are quite universal and can be applied to various relevant settings. This extreme case of stellar protons is of central interest and has been of great inspiration for this thesis; we further discuss the results of this model in Sec. \ref{sec:stellarEP}.

In their study, \citet{2018A&A...614A.111P} used an analytical formula to represent the interstellar spectra of CR electrons and protons, and also accounted for heavier atomic nuclei with the same energy dependency. The total ionisation rate is then computed by summing contributions from both primary CRs like protons, heavier nuclei, and electrons—and secondary CR species, including electron-positron pairs and photons.

\begin{figure}[h!]
    \centering
    \includegraphics[width=0.6\linewidth]{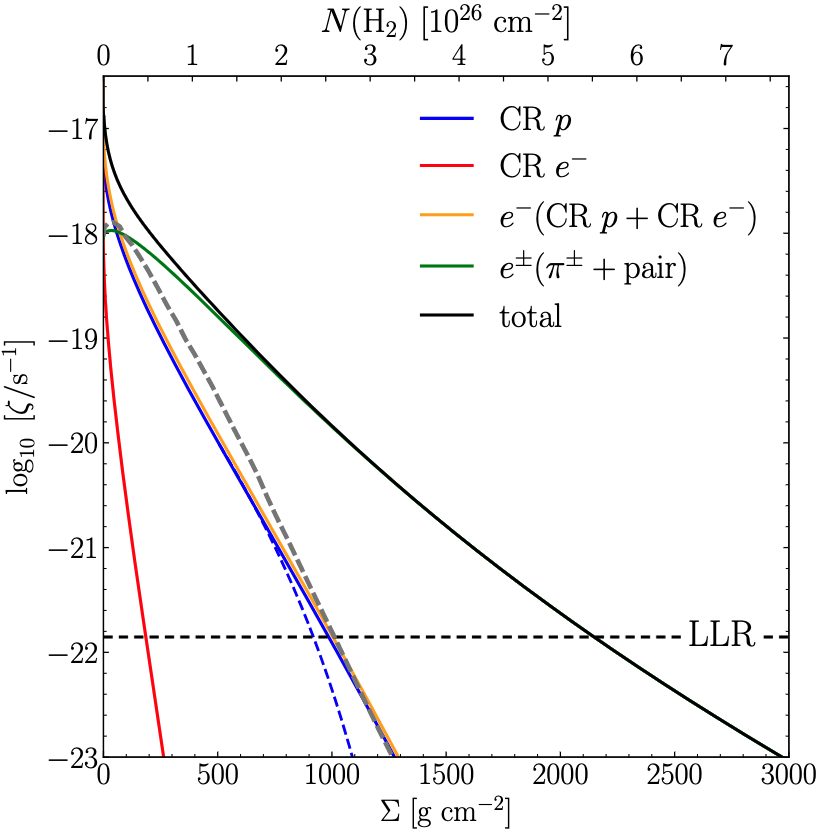}
    \caption{The graph depicts the production rate of molecular hydrogen ions, $\zeta_{H_2}$, as a function of both surface density $\Sigma$ (on the bottom scale) and column density $N$ (on the top scale). The black solid line represents the total ionisation rate. Individual contributors to this rate include ionisation from primary CR protons and electrons (shown as blue and red solid lines, respectively), ionisation from secondary electrons generated by primary CRs (indicated by an orange solid line), and ionisation resulting from electrons and positrons created through charged pion decay and pair production (represented by a green solid line). 
    An horizontal dashed line at $1.4 \times 10^{-22}$ s$^{-1}$ marks the total ionisation rate set by long-lived radioactive nuclei (LLR). The grey dashed line presents the $\zeta_{H_2}(N)$ rate as determined in the older study by \citet{1981PASJ...33..617U} for comparison.
    Additionally, the total electron production rate, arising from CR ionisation of heavier gas elements, is approximately $1.11 \times\zeta_{H_2}$. This figure is reproduced from \citet{2018A&A...614A.111P}.}
    \label{fig:GCRionisationRate}
\end{figure}
 
The total ionisation rate and contributions from various species are displayed in Fig. \ref{fig:GCRionisationRate}. The ionisation rate $\zeta_{H_2}$ is plotted against both column density ($N$) and surface density ($\Sigma$). The numerical relationship between $N$ (cm$^{-2}$) and $\Sigma$ (g cm$^{-2}$) is defined as $N = 2.55 \times 10^{23} \Sigma$, based on an average molecular weight of 2.35 in the interstellar medium (ISM).

Figure \ref{fig:GCRionisationRate} also reveals that below a specific transition surface density, denoted as $\Sigma_{tr}$ approximately equal to 130 g cm$^{-2}$, CR protons and their secondary electrons are the main drivers of ionisation. However, as surface density increases, electron-positron pairs formed by photon decay gradually take over this role. When surface density reaches or exceeds 600 g cm$^{-2}$, these pairs become the exclusive agents of ionisation, with a contribution about ten times greater than that of CR protons.

In conditions where the surface density is less than or equal to $\Sigma_{tr}$, ionisation is influenced by the effective surface density through which CRs travel along magnetic field lines. This effective surface density can vary significantly depending on the magnetic field configuration, and it is generally much larger than the line-of-sight surface density. On the other hand, when surface density is above $\Sigma_{tr}$, ionisation becomes independent of magnetic field effects and is dictated solely by line-of-sight factors.

These findings represent a significant departure from older models, such as those presented by \citet{1981PASJ...33..617U}. In their work, the total ionisation rate showed an exponential decrease with a constant characteristic attenuation scale—around 115 g cm$^{-2}$ for surface densities between 100 and 500 g cm$^{-2}$, and about 96 g cm$^{-2}$ for higher densities. In contrast, \citet{2018A&A...614A.111P} report on a characteristic scale that increases continuously with surface density, ranging from approximately 112 g cm$^{-2}$ to 285 g cm$^{-2}$ for surface densities between 100 and 2100 g cm$^{-2}$, with an error margin of less than 10\%. This divergence is mainly attributed to the more sophisticated transport models for primary CR protons and secondary CR photons that \citet{2018A&A...614A.111P} used in their work.

\citet{2013ApJ...772....5C} highlighted the importance of stellar winds and magnetospheres in the modulation of GCR ionisation rates. This effect may strongly attenuate the GCR flux that reaches the disc \citep{2018ApJ...853..112F,2022ApJ...937L..37F}. Consequently, this leads to spatial and temporal variations in the GCR ionisation rates.

%Furthermore, GCRs may play a role in triggering \textcolor{magenta}{magneto-rotational instabilities (MRI) in T Tauri discs, which constitute a key mechanism for angular momentum transport and accretion in the disc (\citep{1994ApJ...421..163B}. The efficiency of the MRI} depends on the ionisation fraction of the gas, as sufficient ionisation is required for the magnetic field to couple effectively with the disc material \citep{1999MNRAS.307..849W}. As GCRs contribute significantly to \textcolor{magenta}{the} ionisation in T Tauri discs, they are a crucial factor in determining the MRI-active regions and overall disc dynamics \citep{2011ApJ...739...50B,2021MNRAS.506.1128S}.

\paragraph{Exclusion of Galactic Cosmic Rays:}\label{sec:galacticCRExclusion}
Cosmic rays are considered as a significant source of ionisation in the protoplanetary discs, though their actual presence is yet to be fully understood. As seen in our solar system, stellar winds can limit the propagation of CRs within the circumstellar environment, thereby affecting their entry into the disc. \citet{2013ApJ...772....5C} first studied the influences of both CR modulation by stellar winds and magnetic field structures, and investigate how these processes contribute to the reduction of ionisation rates within discs.

Our Sun expels the low energy end of the GCR spectrum within a region known as the heliosphere. These low energy CRs are largely responsible for ionising H$_2$ in the interstellar medium (ISM) \citep{Padovani2009}. The degree to which the solar wind modulates these CRs varies across the solar magnetic activity cycle, especially for energies below 100 MeV. During solar minimum and maximum, the CR flux observed at Earth can vary more than an order of magnitude \citep{2013ApJ...772....5C}.

The dense, slow solar wind arises from the hot ($\sim 1-2$ MK) solar corona carrying magnetised plasma from the solar surface. The density of the wind and magnetic field strength decrease with distance from the Sun until pressure from the ISM exceeds that of the solar wind, causing a magnetic "pile-up" that blocks low energy CRs from freely streaming through the solar system \citep{weymann1960coronal,opher2011magnetic}.

Young T-Tauri stars with strong magnetic activity, are expected to exhibit similar stellar winds, driving out low energy CRs within an equivalent "T-Tauriosphere" \citep{guenther1997spectrophotometry,vidotto2009simulations}. 

Although the magnetic fields generation differs between T-Tauri stars and main sequence dwarfs like our Sun, the properties and ability to drive a stellar wind primarily depend on the presence of a corona, the star mass, stellar rotation, and the general magnetic topology on the surface. T-Tauri stars display strong X-ray emissions thought to originate from both the stellar corona and accretion shocks \citep{kastner2002evidence,brickhouse2010deep}.

From our understanding of the CR modulation in the solar system, we can extrapolate our understanding on how stellar wind modulation may operate in other systems. To study the relationship between solar activity and cosmic ray modulation influenced by stellar winds, \citet{2013ApJ...772....5C} employ a parametric form of the cosmic ray energy spectrum, \( J_{CR}(E) \), at a distance of 1 au from the star. This approach is often referred to as the "force-field" approximation. It allows us to represent the observed variations in the CR spectrum over the solar cycle using just one variable, the modulation potential \( \phi \). This method has been shown to be accurate at heliocentric distances near 1 au, according to studies by \citet{usoskin2005heliospheric}. However, it is worth noting that this approximation has its limitations, which are discussed in \citet{2013ApJ...772....5C}.

The modulated cosmic ray proton spectrum \( J_{CR}(E, \phi) \) is given by the equation:

\begin{equation}
    J_{CR}(E, \phi) = J_{LIS,CR}(E + \phi) \frac{ E (E + 2E_r)}{(E + \phi)(E + \phi + 2E_r)},
    \label{eq:ModulatedCRSpectrum}
\end{equation}
where \( J_{LIS,CR}(E) \) is defined as,

\begin{equation}
    J_{LIS,CR}(E) = \frac{1.9 \times 10^{-9} P(E)^{-2.78}}{1 + 0.4866 P(E)^{-2.51}} .
    \label{eq:LISSpectrum}
\end{equation}
In these equations, \( P(E) = \sqrt{E(E + 2E_r)} \) and the proton rest mass energy \( E_r \) is $0.938$ GeV. Both the energy \( E \) and the modulation potential \( \phi \) are measured in GeV per nucleon. Equation \ref{eq:LISSpectrum} gives the Local Interstellar Spectrum (LIS) \citep{burger2000rigidity} that is used by \citet{2013ApJ...772....5C}. It is important to note that the wind modulation efficiency is not uniform but varies with distance from the Sun. For instance, the cosmic ray flux at an energy \( E_{CR} = 300 \) MeV changes by a factor of approximately 6 when going from \( D = 1 \) au to \( D = 80 \) au, as indicated by \citet{caballero2004limitations}. While the force-field approximation offers a simplified model, it tends to overestimate the cosmic ray flux at low energies and large heliocentric distances.

To generalise the findings from our Sun to more magnetically active T-Tauri stars, \citet{2013ApJ...772....5C} have correlated the temporal evolution of the modulation potential \( \phi(t) \) with other time-resolved solar parameters. These parameters include the Sun average magnetic field strength from the Wilcox Solar Observatory, the sunspot count from the SPIDR5 database \citep{o1997spidr}, and the fractional area covered by sunspots \citep{balmaceda2009homogeneous}. Given that the solar wind mass loss rate is connected to the area covered by open magnetic field lines \citep{cohen2011independency}, a correlation between \( \phi \) is correlated to the the magnitude of the open \( |B| \)-field component. Such open field lines are often found in solar coronal holes, which allow plasma to escape freely, as opposed to X-ray bright regions where plasma is confined. 

By correlating solar magnetic activity with the modulation parameter \( \phi \), \citet{2013ApJ...772....5C} made preliminary estimates for the cosmic ray modulation around more magnetically active young stars. Measurements like magnetic field strength and spot coverage are useful because they can also be measured for other stars. Counting spots is less useful for T-Tauri stars, which are thought to have single, large-area spots \citep{donati2007magnetic,donati2011large,donati2011close}. Magnetic field characteristics on T-Tauri stars are intricate, consisting of multiple components and varying significantly in strength \citep{hartmann2016accretion}.

Spot coverage fraction varies over time and can range from 3\% to 17\% \citep{bouvier1989spots}. \citet{2013ApJ...772....5C} suggests modulation parameters \( \phi \) of 4.8 GeV, 9.2 GeV, and 18 GeV for spot coverage of 2\%, 4\%, and 8\%, respectively. According to the force-field approximation, these \( \phi \) values sufficiently characterise the shape of the cosmic ray energy spectrum \( J_{CR}(E, \phi) \), as described in Eq. \eqref{eq:ModulatedCRSpectrum}. 

\citet{2013ApJ...772....5C} predicted that modulation by stellar winds can reduce the CR flux in the circumstellar environment by many orders of magnitude, resulting in CR ionisation rates substantially lower ($\zeta_{GCR}< 10^{-20}$ s$^{-1}$ ) than typically assumed in models of MRI turbulence and circumstellar chemistry. When spot coverage is at least 10\% and the stellar X-ray luminosity \( L_{XR} \) exceeds \( 10^{29} \) erg/s, the cosmic ray flux becomes less significant than the star ionising X-ray flux, rendering CR ionisation negligible throughout the disc \citep{2013ApJ...772....5C}. 

With the exclusion of GCRs by the T-Tauriosphere, it is crucial to understand and identify alternative sources of ionisation within the system. This could include CRs generated by the star itself or by processes within the protoplanetary disc. These sources can have a significant impact on the disc physical and chemical evolution. Therefore, the study of internal sources of ionisation becomes a key area of research for understanding the environments of young stars.

\paragraph{Ionisation of T Tauri discs by different proton components:}
As we just seen, low-energy CRs from the galaxy might be blocked from entering the extended heliosphere, or "T-Tauriosphere", that surrounds a young star, due to the star stronger winds and elevated magnetic activity \citep{2013ApJ...772....5C}. Our understanding of the scope and form of this cosmic ray exclusion zone is limited to extrapolating from what we know about the Sun heliosphere. \citet{2013ApJ...772....5C} proposed that the T-Tauriosphere could potentially encompass the entire disc. However, the CR particles mainly responsible for ionisation at column densities higher than approximately 100 g/cm$^2$ possess energies above a few GeV. Despite the above analysis, the impact of stellar wind modulation on cosmic rays at these energies remains unclear. For T-Tauri stars, \citet{2013ApJ...772....5C} calculated modulation potential values (\( \phi \)) ranging from 4.8 to 18 GeV at a distance of 1 au. This results in a drop in the cosmic ray flux at 10 GeV by factors of roughly 6 to 100. The modulation potential (\( \phi \)) could vary with distance, a detail that might be clarified by more advanced magnetospheric models. Additionally, the presence of a magnetically active young star could increase the generation of stellar cosmic rays (see Sect. \ref{sec:stellarEP}). Let us describe a first model which accounts of different energetic particle component below.

\citet{2018A&A...614A.111P} employed their CR propagation model to estimate the ionisation at 1 au from a protostar using two distinct proton spectra. One spectrum was modulated by T-Tauri stellar winds and the other represented an enhanced proton flux generated by flares from an active T-Tauri star. Their findings markedly differed from earlier research.

\begin{figure}[h!]
    \centering
    \includegraphics[width=0.8\linewidth]{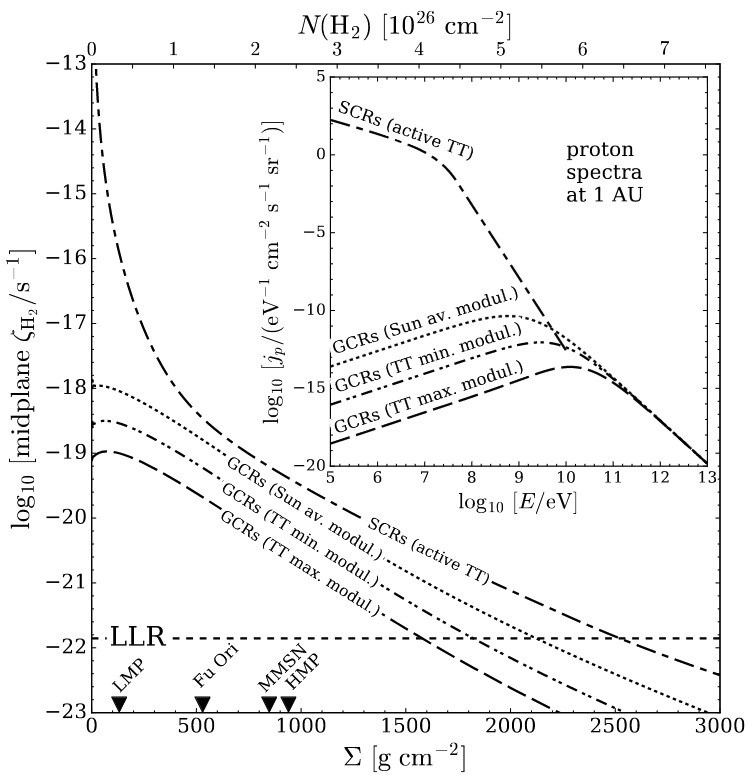}
    \caption{The ionisation rate at the disc's mid-plane ($\zeta_{H_2}$) is plotted against surface density ($\Sigma$) on the bottom scale and column density ($N$) on the top scale. Various types of proton spectra are represented at a distance of 1 au from the central star: Galactic cosmic rays with average solar modulation are shown as a dotted line; T-Tauri wind-induced minimum and maximum modulations are depicted by a dash-dot-dotted line and a long dashed line, respectively; and cosmic rays from an active T-Tauri star are indicated by a dash-dotted line. The horizontal dashed line at $1.4 \times 10^{-22}$ s$^{-1}$ indicates the total ionisation rate set by long-lived radioactive nuclei (LLR).
    For reference, black triangles on the horizontal axis mark characteristic mid-plane disc densities ($\Sigma_{disc}/2$) at a distance of 1 au for different types of stars: low-mass protostars (LMP), high-mass protostars (HMP), FU Orionis stars (FU Ori), and the minimum-mass solar nebula (MMSN). The figure is reproduced from \citet{2018A&A...614A.111P}.}
    \label{fig:ModulatedGCRionisationRate}
\end{figure}

Figure \ref{fig:ModulatedGCRionisationRate} illustrates the CR ionisation rates at 1 au for both maximum and minimum modulation by a T-Tauri stellar wind, corresponding to \( \phi = 18 \) GeV and \( \phi = 4.8 \) GeV, respectively. These are labeled as "GCRs (max. modulation)" and "GCRs (min. modulation)." For completeness, the figure also displays the ionisation rate from galactic cosmic rays influenced by an average solar wind (\( \phi = 1 \) GeV), denoted as "GCRs (Sun avg. modulation)."

In contrast to the findings of \citet{2013ApJ...772....5C}, \citet{2018A&A...614A.111P} report significantly different ionisation rates under both minimum and maximum modulation scenarios. Specifically, their model predicts an ionisation rate that is about 30 times higher at surface densities below approximately 100 g/cm$^2$ for the minimum modulation model. The rate then drops off much more sharply above surface densities of around 1200 g/cm$^2$. The disparity is even more pronounced for the maximum modulation model, where the ionisation rate is higher by a factor of roughly 260 at surface densities below 100 g/cm$^2$ and declines more quickly above about 1400 g/cm$^2$.

These variations are attributed to several factors. \citet{2018A&A...614A.111P} incorporated electron-positron pair creation through photon decay and also considered the relativistic behaviour of ionisation cross-sections for protons. The more rapid decrease in their results at higher surface densities is due to losses from heavy elements in the medium (see Sect. \ref{sec:EnergyLossesDisc}). It is worth noting that in both minimum and maximum modulation scenarios, the ionisation is almost solely due to relativistic protons that propagate diffusively (see Sect. \ref{sec:TransportMechanisms}).

For ionisation rates related to proton fluxes originating from T-Tauri flares, \citet{2018A&A...614A.111P} results align closely with those from \citet{Rab17}, differing by less than 5\% for surface densities up to around 300 g/cm$^2$. We go further into the subject of Stellar energetic particles in Sect. \ref{sec:stellarEP}. At larger surface densities, the ionisation rate decreases gradually, as electron-positron pair creation significantly boosts ionisation levels. However, it remains uncertain what proportion of cosmic rays generated during a flare event can be funnelled into the disc along magnetic field lines without traversing turbulent regions, as opposed to those that may escape along open field lines perpendicular to the disc.

\subsubsection{Radioactive decay of short-lived isotopes}

One source that could ionise T Tauri discs even deeper than GCR are short-lived radionucleides (SLR), such as $^{60}$Fe, $^{41}$Ca, $^{36}$Cl, $^{26}$Al, and $^{10}$Be \citep{1998ApJ...506..898L,2006NuPhA.777....5W}. These isotopes are synthesised in massive stars and supernovae and are subsequently incorporated into the molecular cloud from which the T Tauri disc forms \citep{1995ApJ...449..204T}.

The decay of short-lived isotopes produces energetic particles, such as alpha particles and gamma-ray photons, that ionise the surrounding gas in the disc. 

\citet{cleeves2013radionuclide} introduced a straightforward analytical solution to calculate the ionisation rate, \(\zeta_{\text{SLR}}\), due to the decay of SLRs in protoplanetary discs. They address the radiative transfer equations for the decay products inside the disc, accounting for the energy loss in regions with low surface density. This loss of radiation becomes significant beyond 30 au for discs with typical masses around \(0.04 M_{\odot}\).

Many earlier studies in disc chemistry and physics have often overlooked the role of SLRs, focusing instead on CRs because of their abundant presence in the interstellar medium. However, since stellar winds could significantly attenuate the CR flux in the circumstellar environment, this leads to GCR ionisation rates, \(\zeta_{\text{GCR}}\), that could be either comparable to or much lower than those caused by SLRs (around \(10^{-18} \text{s}^{-1}\)).

\citet{cleeves2013radionuclide} also calculate the total ionising particle flux and the associated ionisation rates across different positions in the disc, applying these findings to a variety of disc models. A power-law approximation for the ionisation rate in the disc midplane is provided, as function of the gas surface density and time.

The ionisation rate of SLR is shown to depend weakly on both time $t$ and surface density $\Sigma$,
\begin{equation}
    \zeta_{SLR}= 2.5\times 10^{-19} \left(\frac{1}{2}\right)^{1.04 t } \left(\frac{\Sigma}{1 ~\rm g ~ cm^{-2}}\right)^{0.27} \quad \rm s^{-1}.
\end{equation}

For a typical $t=1$ Myr disc, $\zeta$ ranges beteween $10^{-19} - 10^{-20} \rm s^{-1}$. These values establish a minimum for the ionisation rate within the disc. The ionisation process is further described as being dominated by short-lived radionuclides in the outer disc and close to the midplane, where other ionisation sources are not present. Long-lived radionuclides also contribute to ionisation, but only at a much lower level, approximately a few \( 10^{-22}\, \text{s}^{-1} \), \citep{Umebayashi09,2018A&A...614A.111P}.

\subsubsection{Locally produced Energetic Particles}\label{sec:stellarEP}
Despite the low ionisation rate estimated by \citet{cleeves2015constraining} in the outer disc see Tab. \ref{tab:ionisationConstrains}, \citet{2018ApJ...856...85S} finds using CO chemical models that some unidentified physical process must be taking place in these discs. This study suggests the presence of Stellar Energetic Particles (SEP) ionisation affecting the chemistry at $R=19$ au. In recent years, the concept of local cosmic ray production within star forming regions has introduced a new field of study. Multiple research groups are now focusing on developing theories and examining observational implications of these locally-produced cosmic rays. The work presented in this thesis is part of this field of research. The community studied the impact of locally produced CRs from M-dwarfs on Earth-like exoplanetary atmospheres (e.g., \citealt{2016A&A...585A..96T}), their propagation through T Tauri winds (e.g., \citealt{2018ApJ...853..112F}), their influence on protoplanetary discs (e.g., \citealt{2017MNRAS.472...26R,2019ApJ...883..121O}), and their effects on protostellar clusters \citep{2018ApJ...861...87G, 2019ApJ...878..105G}. Investigations in this emerging field will benefit from existing and upcoming telescopes, such as, SKA and its precursors (e.g., MeerKAT, LOFAR), which will study Synchrotron emissions at various scales. CTA, which will analyse $\gamma$-ray emissions from high-mass protostars and H II regions. 

As already discussed, CRs are capable of penetrating high column densities, however, the influence of external CRs might be negligible due to shielding by the T-Tauriosphere (although see the conclusion by \citet{2018A&A...614A.111P}). Unlike the external CRs, those that originate from the source itself, such as those accelerated at the stellar surface because of accretion shocks or flares, are not subject to this limitation \citep{Rodgers-Lee17}. They could act as a considerable source of ionisation \citep{Padovani18,2018ApJ...861...87G}.

\paragraph{Particles accelerated at Accretion Shocks \citep{2019ApJ...883..121O}}
As the material falls towards the star, it is accelerated by the stellar gravitational potential and eventually collides with the stellar surface, leading to the formation of a shock front. This shock front heats and ionises the infalling gas, producing high-energy radiation, primarily in the form of UV and X-ray photons \citep{1998ARep...42..322L,2007A&A...474L..25G}. In addition to the ionising radiation produced by accretion shock, the shock generates energetic particles, which contributes to the ionisation of the surrounding gas \citep{2016A&A...590A...8P}. These particles are accelerated at the shock front by first order Fermi process, gaining sufficient energy to ionise the gas as they collide with other particles in the disc. 

\citet{2016A&A...590A...8P} conducted a preliminary study investigating the role of energetic particles produced by accretion shocks in the ionisation of T Tauri discs. This process is similar to the acceleration of cosmic rays in supernova remnant shocks. They developed a detailed model of the interaction between the accelerated particles and the disc material, taking into account the physics of particle acceleration and propagation in the magnetised, partially ionised environment of the disc. 

\cite{2018ApJ...861...87G} demonstrated that accreting protostars produce a robust source of CRs that affect the ionisation of both the dense core and the more extensive molecular cloud. The influence of CRs accelerated by accretion shocks on the chemistry and dynamics of circumstellar discs is being further explored by \citet{2019ApJ...883..121O}.

CRs and photons, are attenuated geometrically as they move away from the central star. They also undergo energy losses through interactions with the gas along their path. The exact propagation regime of CRs remains unknown since it is influenced by factors as the magnetic field geometry, the gas density, and the turbulence. We study in more details the question of the transport of CR in disc in Chapter \ref{C:Propagation}.  Generally speaking, the flux of CRs decreases as $r^{-a}$, where two limit cases are free-streaming ($a = 2$) and diffusion ($a = 1$). 
\citet{2018ApJ...861...87G} estimated that in dense cores, the high CR fluxes accelerated by the accretion shock require an attenuation of $r^{-2}$ or steeper to maintain the observed low gas temperatures.\\

\begin{figure}[h!]
    \centering
    \includegraphics[width=\linewidth]{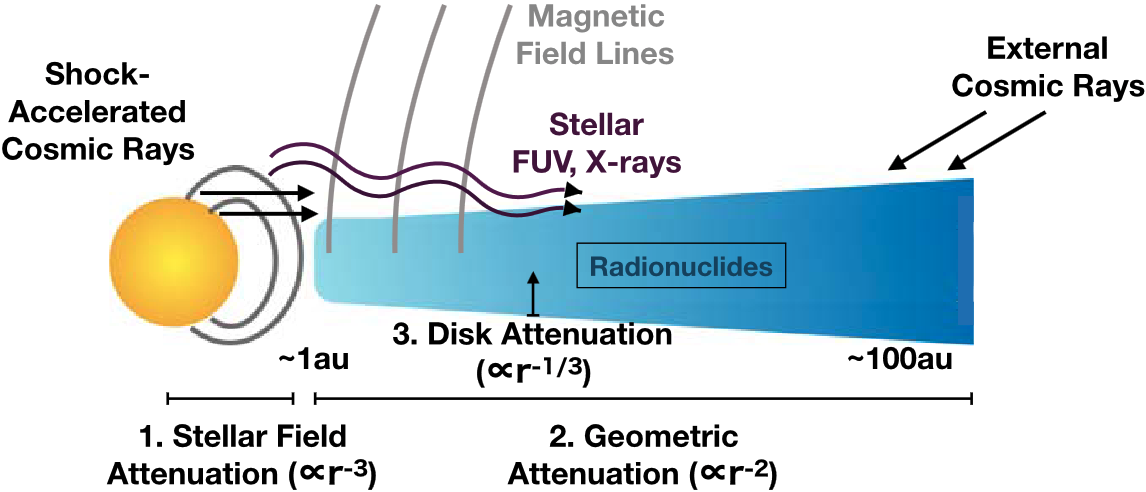}
    \caption{Schematic of the sources of ionisation and the associated regions. Here we consider only local sources of ionisation, namely those due to the central star, which generally dominate over external sources such as the Galactic X-ray, CR, and FUV backgrounds.}
    \label{fig:StellarEPOffner}
\end{figure}

To calculate the local CR ionisation rate, \citet{2019ApJ...883..121O} starts with the initial spectrum of shock-accelerated CRs at the stellar surface. This spectrum is then propagated through three distinct regions, considering both geometric attenuation and energy losses. These regions are represented in Fig. \ref{fig:StellarEPOffner}, 1. the accretion columns between the star and the inner edge of the accretion disc, 2. from the inner disc radius to the disc  surface, and 3. within the disc itself.

\begin{itemize}
    \item \textbf{Losses in accretion funnels} — The CR spectrum undergoes attenuation through inverse funnelling, along with energy losses resulting from interactions with the gas upstream of the shock. The energy losses considered make the assumption of completely ionised gas and combines the effects of Coulomb and pion-production losses \citep{2002cra..book.....S}.
    Also, in this region, the CR flux is further reduced due to inverse funnelling. A specific term, referred to as the funnelling attenuation factor, is defined as $f_{fun}=\frac{B_*}{B_d}$, where $B_*$ and $B_d$ are the magnetic fields at the star surface and at the disc surface, respectively. This effect is further developed in Sec. \ref{sec:mirroringfocussing}.
    The situation described implies that when CRs are tightly coupled to the magnetic field, the spectrum   suppression is steeper than the commonly assumed \( r^{-2} \) of the free-streaming limit. \citet{2019ApJ...883..121O} estimated that the inverse funnelling typically reduces the flux by a factor of around 10 to 100 between the star and the inner disc.

    \item \textbf{Losses between the inner disc radius and the disc surface} — In this region, it is expected that the particle flux encounter negligible energy losses due to interactions with neutral material above the disc. Thus they undergoing geometric attenuation (\( a = 2 \)). This is valid for hydrogen column density less than \( 10^{20} \, \text{cm}^{-2} \). 
    Across most radii, the ionisation rates at the disc surface \( \zeta_0 \approx 10^{-16} \, \text{s}^{-1} \) are noticeably higher than the GCR background ionisation rate of \citet{2013ApJ...772....5C}, \( \zeta_{GCR} < 10^{-20} \, \text{s}^{-1} \), assuming that GCRs are effectively blocked by the T-Tauriosphere \citep{2013ApJ...772....5C}.

    \item \textbf{Losses inside the disc} — Within the disc, the expression from \citet{Padovani18} can be used to estimate how the ionisation rate $\zeta$ decreases with disc column density, represented by \( N_H \)
    
    \begin{equation}
        \zeta(N_H) = \zeta_0 \left(\frac{N_H}{N_0}\right)^{-\gamma},
    \end{equation}
    where $N_0$ depends essentially on the ionising CR and $\gamma$ on the shape of the CR energy distribution. $\zeta_0$ is the ionisation rate at the disc surface. 

    It is important to note that the parameters of this attenuation process depend on multiple factors, such as the spectral shape, the interactions processes that are treated, and the incorporation of secondary particles, which refer to additional high-energy particles generated by CR interactions. While \citet{Padovani18} have used a cosmic ray spectrum that reflects the cosmic-ray background of the interstellar medium. \citet{2019ApJ...883..121O} suggest, that for shock accelerated particles, $\zeta_0=10^{-16} \, \text{s}^{-1}$, $N_0=10^{18}$ cm$^{-2}$ and $\gamma=0.36$. In Chapter \ref{C:Propagation}, we estimate these parameter values for other acceleration processes and for electron and proton CRs.  
\end{itemize}

Another variable to consider is the turbulence in the magnetic field. \citet{Rodgers-Lee17} introduce a turbulent diffusion parameter, denoted by $D$, to investigate variations in cosmic-ray propagation, ranging from 3 to 380 times the Larmor radius ($r_L$). They demonstrate that the attenuation of the cosmic-ray spectrum can vary between $\propto r^{-2}$ in the case of free-streaming, where $D\gg r_L$ and $\propto r^{-1}$ in the case of Bohmian diffusion $D= 3 r_L$. The diffusion coefficient depends on the the assumed turbulence model. In Chapter \ref{C:Propagation} we estimate the diffusion coefficients and deduce $D$ based on the effects of different turbulence models.

\begin{figure}[h!]
    \centering
    \includegraphics[width=0.6\linewidth]{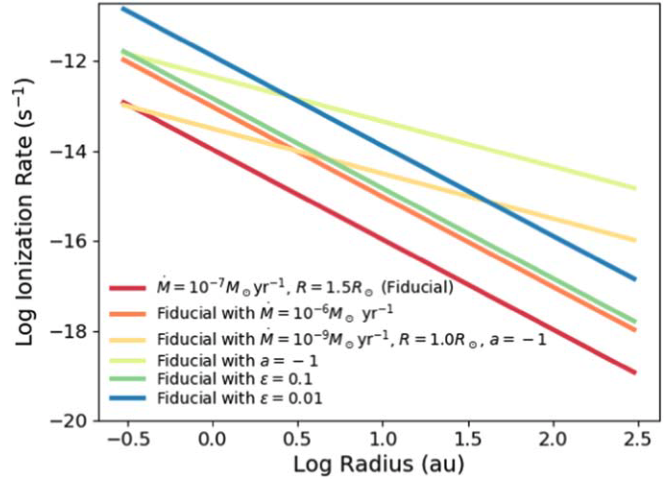}
    \caption{Ionisation by CRs produced by diffusive shock acceleration from the accretion shock at the stellar surface and attenuated as described in the text. See \citet{2018ApJ...861...87G} for details on the various model parameters. The figure is reproduced from \citet{2019ApJ...883..121O}.}
    \label{fig:AccretionShockionisationRateOffner}
\end{figure}

Stellar energetic particles ionisation rates can be significantly high, particularly in the inner regions of the disc (within a few au), where they can reach values of $10^{-19}$ to $10^{-13}$ s$^{-1}$ for the fiducial model of \citet{2019ApJ...883..121O}, see Fig. \ref{fig:AccretionShockionisationRateOffner}. 

%\citep{2019ApJ...883..121O} showed that these high ionisation rates influences the disc thermal balance and trigger MRI like instability, chemical reactions, leading to an increase in molecular ion abundances and a more complex molecular inventory. 

\paragraph{Energetic particles from stellar flares \citep{Rab17}}

Since the precise particle spectra and fluxes of young stars are unknown, the information must be derived from what is known about our Sun. Solar energetic particles likely originate from flares or shock waves driven by coronal-mass ejections (CMEs), as detailed by \citet{2015SSRv..194..303R}. It is important to note that particle fluxes are not continuous. They are produced in events that last from several hours to days.

Based on the observed X-ray luminosities of solar analogues in the Orion nebula, \citet{2002ApJ...574..258F} estimated that SEP fluxes in young stars could be around $10^5$ times greater than in the current Sun. T Tauri stars, being very active, experience X-ray flares several times a day, potentially even below the detection limit \citep{getman2021a}. This means that the X-ray flares, and consequently SEP emission events, can overlap, resulting in a continuous, enhanced SEP flux. 

\begin{itemize}
    \item \textbf{Injection Model –  } \citet{2017A&A...603A..96R} assume a continuous and increased SEP flux for young T Tauri stars, consistent with the assumption of powerful and overlapping flare and CME events particular to these stars.
    
    Using measurements from \citet{mewaldt2005proton} associated to five different solar particle events, \citet{2017A&A...603A..96R} derived SEP spectra, focusing on protons. Their spectra extend to energies usual for GCRs, up to approximately 10 GeV. This is similar to the maximum energy of about 30 GeV for particles accelerated on protostellar surfaces, as suggested by \citet{2015A&A...582L..13P,2016A&A...590A...8P}.

    \citet{reedy2012update} reported proton fluxes for the Sun spanning five solar cycles, with average cycle fluxes at 1 au ranging from \(59-213 \, \text{protons cm}^{-2} \text{s}^{-1}\) for \(E_p > 10 \, \text{MeV}\). \citet{Rab17} focus on one specific event observed by \citet{reedy2012update} and referred to as the "active Sun" spectrum, with proton flux of \(151 \, \text{protons cm}^{-2} \text{s}^{-1}\) at 1 au for \(Ep > 10 \, \text{MeV}\).

    To model the "active T Tauri" spectrum, \citet{Rab17} amplified the active Sun spectrum by approximately a factor of \(10^5\), consistently with the observations of \citet{2002ApJ...574..258F}. This scaling yielded an SP flux of \(1.51 \times 10^7 \, \text{protons cm}^{-2} \text{s}^{-1}\) for \(Ep > 10 \, \text{MeV}\). This is consistent with the estimation of roughly \(10^7 \, \text{protons cm}^{-2} \text{s}^{-1}\) for young solar analogues in the Orion Nebula Cluster by \citet{2002ApJ...574..258F}. Figure \ref{fig:SPandSunSpectraRab} illustrates both the active Sun and active T Tauri SP spectra, and also includes a comparison with two Galactic cosmic-ray spectra that \citet{Rab17} are using to estimate the GCR ionisation rate.

    For comparison they also show the model of \citet{2009ApJ...703.2152T} that adopt the same methodology based on an enhanced solar energetic particles flux. In their model, referenced as the Turner model, \citet{2009ApJ...703.2152T} assumed that SEPs exhibit behaviour similar of GCRs, particularly in terms of particle energies. The differences between the approach of Rab et al and Turner et al, is that in \citet{Rab17}, SEPs can only penetrate the disc radially and therefore do not reach the disc midplane. In the Turner model however, the particles also penetrate the disc vertically down. Allowing them to reach the midplane.

    \begin{figure}[h!]
        \centering
        \captionsetup{width=.8\linewidth}
        \includegraphics[width=0.6\linewidth]{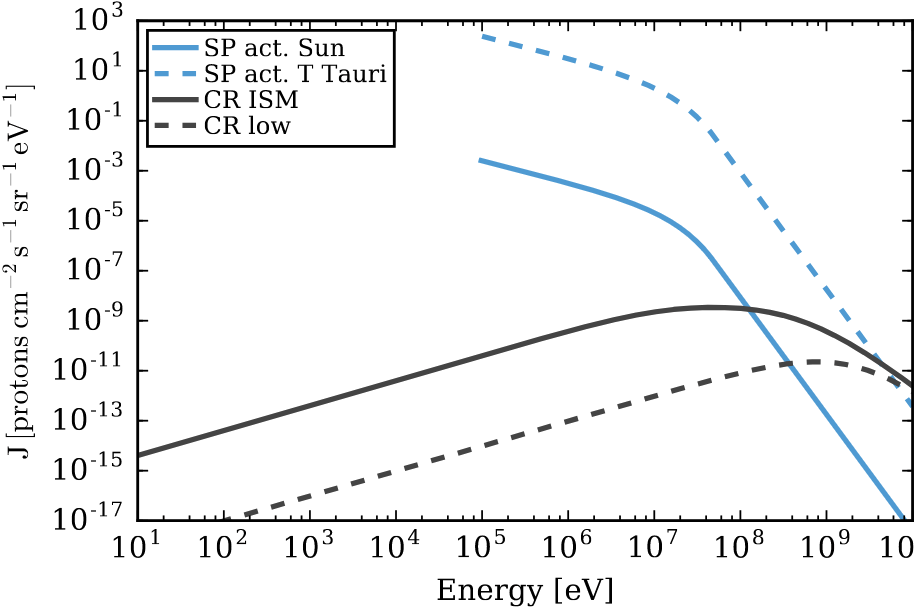}
        \caption{Stellar energetic proton (SP in the inlet) and cosmic-ray (CR) input spectra. The blue solid and dashed lines show the active Sun and active T Tauri SP spectrum, respectively. The black solid line shows the “LIS W98” CR spectrum from \citet{1998ApJ...506..329W} and the dashed black line the attenuated “Solar Max” CR spectrum from \citet{2013ApJ...772....5C}. The figure is reproduced from \citet{Rab17}.}
        \label{fig:SPandSunSpectraRab}
    \end{figure}

    \item \textbf{Transport Model –} In their modelling of energetic particle transport through the disc gas, \citet{Rab17} used the continuous slowing down approximation. We further present this particle transport model in Sec. \ref{sec:FreeStreaming}. \citet{2017A&A...603A..96R} further assumed that stellar particles (SPs) travel in straight lines without scattering. This assumption is also discussed in Sec. \ref{sec:TransportMechanisms}. Their model assumes that particles originate from a point source on the stellar surface. The effect of magnetic fields, which could either increase or decrease the ionisation rate (discussed in Sect. \ref{sec:mirroringfocussing}), was disregarded for this model. But they considered the effect of geometric dilution. Hence, they use an ionisation rate scaling as \(1/r^2\), where \(r\) is the distance to the star in au. 

    The ionisation rate is computed using the disc structure and the radiative transfer and chemical model of {\tt ProDiMO} \citep{2009A&A...501..383W}. We discuss this model in details below in Sec. \ref{sec:focusonProDiMO}. Figure \ref{fig:ionisationRateAllSourcesRab} plots the ionisation rate dependence with column density for the different sources presented above and accounting for the additional ionisation rate from stellar flares as computed by \citet{2017A&A...603A..96R}.

    \item \textbf{Ionisation Rate – } We examine here the ionisation rates, as computed by \citet{Rab17}, attributed to Stellar Energetic Particles (SEP), X-rays, and Cosmic Rays (CR) as a function of the total column density of hydrogen, denoted as \(N_{\langle H \rangle}= N_{\text{H}} + 2N_{\text{H}_2}\) . Figure \ref{fig:ionisationRateAllSourcesRab} presents this comparative analysis based on different energy spectra discussed.

    Examining Fig. \ref{fig:ionisationRateAllSourcesRab}, the ionisation of SEP ionisation similar to that of our active Sun, does not compete with the X-ray ionisation, assuming standard T Tauri X-ray luminosities. However, SEP ionisation becomes comparable to, or even surpasses, X-ray ionisation when the column density \(N\langle H \rangle\) ranges between \(10^{24}\) and \(10^{25} \, \text{cm}^{-2}\) in the case of an active T Tauri SEP spectrum. For \(N_{\text{H}} < 10^{23} \, \text{cm}^{-2}\), the ionisation rate \( \zeta_{\text{SP}} \) is primarily influenced by particles with energies \(E_p < 5 \times 10^7 \, \text{eV}\). Above this column density, higher energy particles become predominant. A noticeable drop in the ionisation rate occurs at \(N\langle H \rangle \approx 2 \times 10^{25} \, \text{cm}^{-2}\), attributable to the strong energy loss of SEPs at these high densities.

    For X-rays, Fig. \ref{fig:ionisationRateAllSourcesRab} distinguish typical and elevated X-ray spectra. The ionisation rate is greater for the high X-ray spectrum due to its elevated luminosity. Moreover, higher-energy X-ray photons not only penetrate deeper but are also scattered more efficiently compared to their lower-energy counterparts. As a result, the ionisation rate of X-rays increases by several orders of magnitude for \(N\langle H \rangle \geq 10^{24} \, \text{cm}^{-2}\) in the high X-ray scenario.

    Among the different ionisation sources considered by \citet{Rab17}, GCR include the highest energetic particle component, with proton flux peaking at around \(10^8 - 10^9 \, \text{eV}\). The cosmic ray ionisation rate, \( \zeta_{\text{CR}} \), largely remains constant and only diminishes for \(N\langle H \rangle \geq 10^{25} \, \text{cm}^{-2}\), where pion losses becomes notably effective.

    \begin{figure}[h!]
        \centering
        \captionsetup{width=.8\linewidth}
        \includegraphics[width=0.6\linewidth]{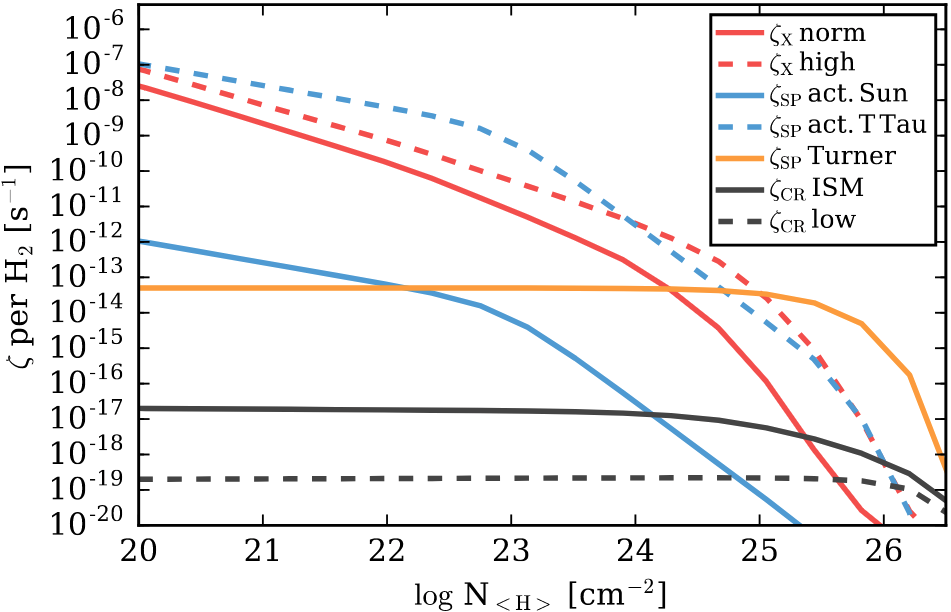}
        \caption{SP, CR, and X-ray ionisation rates \(\zeta\) are presented as functions of the hydrogen column density \(N_{\langle H\rangle_{ver}}\). The solid red line plots the ionisation rate derived from an X-ray "normal" luminosity \(L_X=10^{30}\) erg s\(^{-1}\), whereas the dashed red line displays the ionisation rate from a "high" X-ray luminosity \(L_X=5\times 10^{30}\) erg s\(^{-1}\). The solid blue line plots the ionisation rate due to an "active" Sun particle flux, while the dashed blue line shows the ionisation from the anticipated stellar particle flux of T Tauri stars, \(10^5\) times the active solar flux. The orange line represents the ionisation rate according to the Turner model. The solid black line plots the ionisation rate due to GCR assuming the ISM flux, while the dashed black line presents the ionisation rate by GCR accounting for the particle exclusion by the T Tauriosphere, as determined by \citet{Cleeves13}.}
        \label{fig:ionisationRateAllSourcesRab}
    \end{figure}

    \item \textbf{Ionisation rates over the disc –} Figure \ref{fig:ionisationSources} illustrates the dominant ionisation sources in the disc for various models of radiation computed by \citet{2017A&A...603A..96R}. An ionisation source is considered dominant if its value exceeds the combined values of the other two ionisation sources.
    
    CI\_XN\_SP indicates a scenario where the X-ray flux is normal, while CI\_XH\_SP represents a comparatively high X-ray flux scenario. In both cases, the GCR flux is the particle flux of the ISM. According to these models, SEPs are the principal ionisation source in the upper layers of the disc (above the white solid contour line representing $N_{\langle H \rangle,rad} = 10^{25}$ cm$^{-2}$), whereas in the midplane, CRs or X-rays dominate.
    The picture is quite different for the Turner model, CI\_T and CL\_T plots. In these models, SPs penetrate the disc vertically. So they reach higher vertical column densities before they are completely attenuated. Consequently, these SEP become the dominant ionisation source in the midplane of the disc. According to this models, X-rays dominate in the upper layers as $\zeta_{\text{SP,Turner}} < \zeta_{\text{X}}$ for low column densities, see Fig. \ref{fig:ionisationRateAllSourcesRab}.

    \begin{figure}[h!]
        \centering
        \captionsetup{width=.8\linewidth}
        \includegraphics[width=0.9\linewidth]{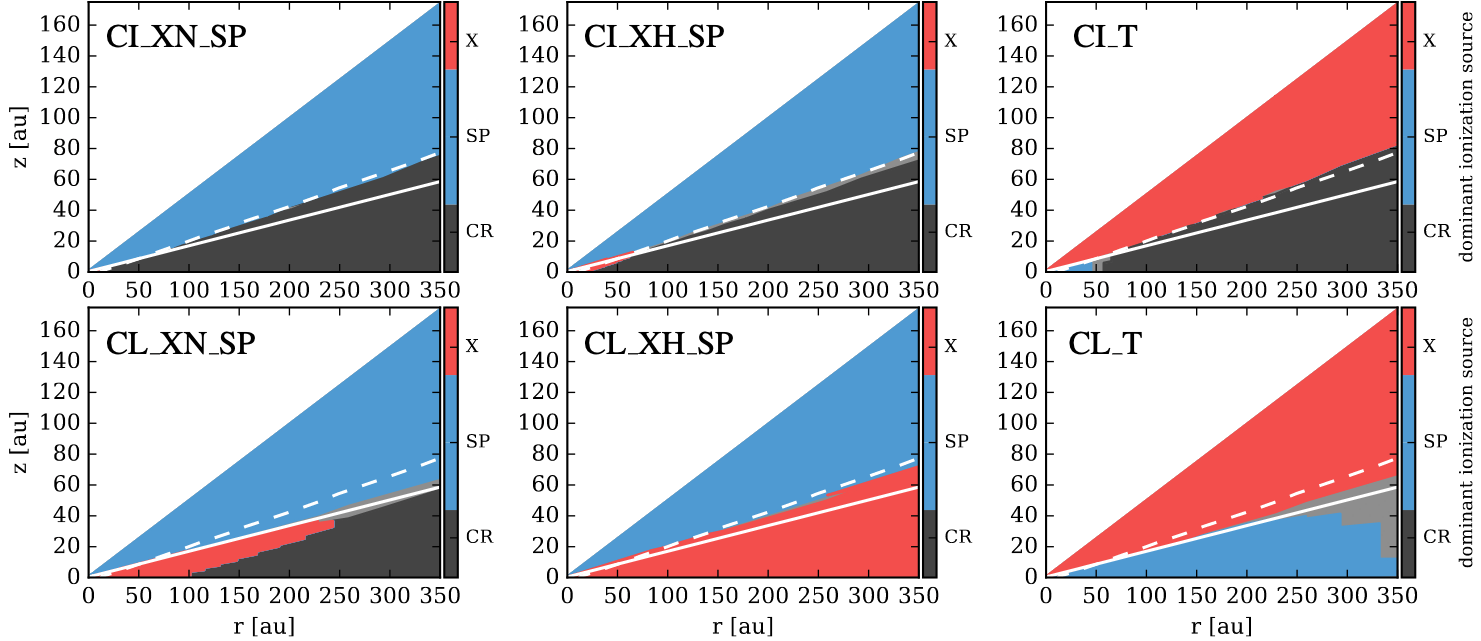}
        \caption{The dominant ionisation sources are indicated by different colors (see the color bar). Each color representing an ionisation source, X-rays, Stellar Energetic Particles (SEPs) and Cosmic Rays (CRs). The light grey area indicates a region without a dominant ionisation source. The white solid contour line shows $N_{\langle H \rangle, rad} = 10^{25}$ cm$^{-2}$, while the white dashed line represents the CO ice line. The model names are located at the top left of each panel. The top row named CI depicts models with an Interstellar Medium (ISM) CR ionisation rate, while the bottom row illustrates models with a low CR ionisation rate (CL).
        The first column shows normal X-ray models (XN), the second column presents high X-ray models (XH), and the third column presents the Turner models (T). This plot is adapted from \citet{2017A&A...603A..96R}
    }
        \label{fig:ionisationSources}
    \end{figure}

\end{itemize}

Considering varying properties of the competing high-energy disc ionisation sources, X-rays and Galactic cosmic rays, \citet{Rab17} studied the interplay of the different ionisation sources and identified possible observational tracers of SP ionisation. The main conclusions drawn from their study are as follows.

At hydrogen column densities \(N_{\langle H\rangle}> 10^{25}\, \text{cm}^{-2}\), even the most energetic particles are stopped, and the SEP ionisation rate drops. Given that the radial hydrogen column densities for full T Tauri discs are typically \(N_{\langle H\rangle} \gg 10^{25}\, \text{cm}^{-2}\), the mid-plane SEP ionisation rate is \(\zeta_{\text{SP}} \ll 10^{-20}\, \text{s}^{-1}\) already at a distance of 1 au from the star. For the assumed SP flux (see above), SEPs become the dominant H$_2$ ionisation source in the warm molecular layer of the disc above the CO ice line, provided that SEPs are not shielded by magnetic fields. This is even true for enhanced X-ray luminosities (i.e., \(L_X = 5 \times 10^{30}\, \text{erg s}^{-1}\)). SEP ionisation can increase the observational tracers, HCO\(^+\) and N\(_2\)H\(^+\) column densities by factors of between approximately three and ten for disc radii \(r \lesssim 200\, \text{au}\). The impact is more significant in models with low GCR ionisation rates (i.e., \(\zeta_{\text{GCR}} \approx 10^{-19}\, \text{s}^{-1}\)). SEP ionisation becomes insignificant for an SEP flux one order of magnitude lower than the proposed value for T Tauri stars. In such a case, \(H_2\) ionisation is solely dominated by X-rays and GCRs. As SEPs cannot penetrate the deep layers of the disc, X-rays and/or CRs usually remain the dominant H$_2$ ionisation source in the cold disc layers (i.e., below the CO ice line). Therefore, HCO\(^+\), which traces the warm molecular layer, is more sensitive to SP ionisation than N\(_2\)H\(^+\) , which resides in the cold molecular layer. Simultaneous modelling of spatially resolved radial intensity profiles of molecular ions tracing different vertical layers of the disc, allows to disentangle the contributions of the competing high-energy ionisation sources to the total H$_2$ ionisation rate. Consequently, such observations would allow to constrain the SP flux in discs. However, it is worth noting that this method is likely to be model-dependent \citep{Rab17}.\\

In this section we have demonstrated the importance of stellar energetic particles as a key ionisation factor in disc. To gain a more comprehensive understanding of the extent to which stellar particles reach the disc, further enhancements to the existing models are necessary, particularly in terms of modelling the transport of the energetic particles, including their interactions with magnetic fields. We tackle this aspect in Chapter \ref{C:Propagation}. In addition, the results of vertically propagating particles presented in \citet{Rab17}, becoming a dominant source of ionisation are noteworthy and have inspired the model presented in Chapter \ref{C:PublicationI}. In this chapter, we explore the effect of particles produced during flares that propagate along the magnetic field lines vertically into the disc. According to our model, the particles produced by flares, may be a dominant source of ionisation deep in the disc. However, according to our model (at least at its current stage of development), flares inject particles only close to the star, in the inner disc.

\section{Effects of ionisation on T Tauri Disc Dynamics}\label{Sec:ionisationEffects}
\subsection{Influence of ionisation on the disc chemistry}

\subsubsection{Ionisation balance equations and chemical networks}
Understanding the ionisation processes in T Tauri discs requires the development of theoretical and numerical models that capture the complex interplay between ionisation sources, chemical reactions, and the physical conditions in the disc. To quantify the effect of ionisation on the disc chemistry, detailed chemical models account for the various ionisation processes presented above and their impact on the reaction rates like {\tt ProDiMO}. One of the main objective of these models is to derive the ionization fraction $x_E$, the ratio of the densities of electron to neutral molecules.

Before presenting such numerical thermochemical models, to fix a framework, we discuss a simple model to compute the ionisation balance. We illustrate such ionisation balance as proposed by \citet{2002MNRAS.329...18F}. The authors used this model with the main aim to investigate the triggering of some MHD instabilities, but their calculation uses a framework which will be useful for our purposes, as it provides insights into the temporal evolution of electron density that could be induced by temporally variable ionisation sources.

 \citet{2002MNRAS.329...18F} propose to evaluate the ionisation state by solving ionisation balance equations, which describe the rate at which species are ionised and recombine, and the chemical networks that determine the production and destruction of molecules. The next equations is a simplified system as the one used in \citet{2002MNRAS.329...18F}. 

Let us consider \(n_n\), \( n_N \), \( n_m \), and \( n_M \) as the densities of neutral molecules, of neutral metals, of molecular ions, and metal ions, respectively. The rate equations governing the density of electrons (\( n_e \)) and molecular ions (\( n_m \)) can be expressed as:

\begin{equation}
    \frac{{dn_e}}{{dt}} = \zeta n_n - \beta_h n_e n_m - \beta_r n_e n_M 
\end{equation}
\begin{equation}
    \frac{{dn_m}}{{dt}} = \zeta n_n - \beta_h n_e n_ m - \beta_t n_N n_m. 
\end{equation}
The ionisation rate $\zeta$ depends on the ionisation sources considered e.g., cosmic rays, stellar radiation, accretion shocks, as we saw above. Here, \( \beta_h \) represents the dissociative recombination rate coefficient for molecular ions, \( \beta_r \) is the radiative recombination rate coefficient for metal atoms, and \( \beta_t \) is the charge transfer rate coefficient from molecular ions to metal atoms. From numerical calculations, based on prior studies \citep{oppenheimer1974fractional,spitzer1978review}, we use:

\begin{equation}
    \beta_r = 3 \times 10^{-11} T^{-1/2} \, \text{cm}^3 \, \text{s}^{-1}
    \label{eq:metalrecombrate}
\end{equation}
\begin{equation}
    \beta_h = 3 \times 10^{-6} T^{-1/2} \, \text{cm}^3 \, \text{s}^{-1}
    \label{eq:nometalrecombrate}
\end{equation}
\begin{equation}
    \beta_t = 3 \times 10^{-9} \, \text{cm}^3 \, \text{s}^{-1} 
    \label{eq:chargetransferrate}
\end{equation}

Charge neutrality is maintained by, $n_e = n_M + n_m$. In steady-state, the above system yields,

\begin{equation}
    x_E^3 - \frac{{\beta_t}}{{\beta_h}} x_M x_E^2 - \frac{\zeta}{\beta n_n} x_E^2 + \frac{{\zeta \beta_t}}{{\beta_h \beta_r n_n}} x_M = 0
    \label{eq:steadystateFromang}
\end{equation}

where \( x_M = n_M/n_n \) and \(x_E = n_E/n_n\). If no metals are present (\( x_M = 0 \)), then Eq. \eqref{eq:steadystateFromang} simplifies to,

\begin{equation}
    x_E = \sqrt{\frac{{\zeta}}{{\beta_h n_n}}} 
\end{equation}

This would correspond to a situation where all metals are sequestered in sedimented grains.

While the effect of the ionisation sources on the disc chemistry via the chemical model of {\tt ProDiMO} is reserved for Sec. \ref{sec:focusonProDiMO}, it remains interesting to compute the electron density via the simple chemical model presented above. In particular because temporal dynamics remain absent from the {\tt ProDiMO} framework. This simple model provides insights into the temporal evolution of electron density modulations induced by non-thermal ionisation processes. This simple model is used in Chapter \ref{C:PublicationII} to study the temporal effects of energetic particles produced by flares on electron density.

To go beyond the ionisation balance equations, one should also consider a chemical model that govern the formation and destruction of molecules in the disc. These models include gas-phase reactions, such as ion-neutral reactions and dissociative recombination, as well as surface reactions on dust grains (e.g. \citealt{2009A&A...501..383W}). 

Recent advances in computational chemistry and astrochemical databases have enabled the development of comprehensive chemical models for T Tauri discs, which include thousands of reactions and hundreds of species. These models have been used in combination with theoretical ionisation rates to investigate the impact of ionisation processes on the disc chemical composition and the formation of complex molecules (e.g. \citealt{2014Sci...345.1590C, 2017A&A...603A..96R,2019ApJ...883..121O,2022A&A...668A.164W,2023A&A...672A..92A}).

%The study \textcolor{magenta}{which study?} \textcolor{magenta}{confirms the results of} \citet{2011ApJ...735....8P} that ionised carbon and sulfur are abundant in the disc surface layers due to the influence of the stellar Far-UV field. However, the inclusion of heavier elements in the network results in magnesium \textcolor{magenta}{coming over} sulfur as the dominant ion at $\Sigma \approx 10^{-2}\, \text{g cm}^{-2}$, causing the ion fraction to increase by a factor of a few. Furthermore, reactions involving magnesium and nitrogen replace the formation of complex hydrocarbon radicals near the disc surface. In a general sense, the ionisation induced by X-rays and cosmic rays causes the ion/neutral transition region to shift towards slightly lower column densities.

%\textcolor{magenta}{CS The following paragraph sound very general. I guess it should be in the introduction of this chapter or you are repeating youself.}ionisation processes in T Tauri discs play a significant role in driving the disc chemistry, shaping the molecular complexity, and influencing the formation of organic molecules and also  prebiotic species like amino acids \citep{1999A&A...351..233A,2004A&A...417...93S}. Ionisation directly impact the abundance and distribution of charged particles, free electrons, and reactive radicals, initiating a rich network of chemical reactions and affecting the overall chemical composition of the disc \citep{2011ARA&A..49...67W,2018IAUS..332....3V}.

%%%%%%%%%%%%%%%%%%CS
\subsubsection{Ionisation-induced heating}\label{ionisationinducedheating}

\paragraph{Heat deposition:}
Ionisation processes in T Tauri discs contribute to the heating of the disc material which is necessary to determine the disc thermal structure \citep{1998ApJ...500..411D,2009ApJ...699.1639E}. Various ionisation mechanisms presented in Sec. \ref{Sec:ionisationSources}, including cosmic rays, X-rays, and UV radiation, deposit energy into the disc gas, which in turn raises the gas temperature and affects the thermal equilibrium \citep{1999ApJ...518..848I,2017A&A...603A..96R}.

Ionisation-induced heating is particularly significant in the upper layers of the disc, where ionisation rates are higher due to the penetration of external radiation \citep{2004ApJ...615..972G}. For instance, when an electron is ionised by X-ray photons, it loses its remaining kinetic energy through inelastic collisions with the atoms and molecules in the disc. These collisions deposit heat in the disc. \citet{glassgold2012cosmic} showed that $\sim 1/2$ of the X-ray and CR energy goes into heat.

Ionisation-induced heating competes with other sources of heating and cooling in the disc, such as viscous heating, cooling by molecular line emission and thermal Bremsstrahlung. The balance between these various heating and cooling mechanisms determines the disc temperature structure and thermal equilibrium \citep{1998ApJ...500..411D,2004ApJ...615..972G,2004ApJ...615..991K}.

The first estimate of heat deposition by ionisation in a hydrogen-helium gas mixture was made by \citet{1999ApJS..125..237D}. They focused on the heat generated from elastic scattering and collisional de-excitation of rotationally excited hydrogen molecules. \citet{2012ApJ...756..157G} further explored the heating that results from all ionisation and excitation processes, emphasising the interactions between cosmic-ray and X-ray generated ions and heavy neutral species, referred to as chemical heating. In molecular regions such as dense core or protoplanetary discs, chemical heating is the dominant factor, contributing to 50\% of the energy used in ion pair production. The heating per pairs of ions varies, ranging from approximately $E_\mathrm{ion}= 4.3$ eV in diffuse atomic gas to around 13 eV in moderately dense molecular cloud regions and up to 18 eV in highly dense protoplanetary disc regions. Cosmic-ray and X-ray heating rely on the medium physical properties like molecular and electron fractions, total hydrogen nuclei density, and temperature. Additionally, chemical heating, the primary mechanism for cosmic-ray and X-ray heating, plays a part in UV irradiated molecular gas.

The energy deposition from ionisation processes can be described by the volumetric heating rate $\Gamma_\mathrm{ion}$ in erg cm$^{-3}$ s$^{-1}$,
\begin{equation}
\Gamma_\mathrm{ion}= n_\mathrm{gas} Q_\mathrm{ion} \zeta 
,
\end{equation}
where $n_\mathrm{gas}$ is the number density of the gas, $Q_\mathrm{ion}$ is the average energy per ionisation and $\zeta$ is the ionisation rate.

In Chapter \ref{C:PublicationII}, we evaluate the heating rate generated by ionisation due to CR in the disc inner region. A short term increase in this ionisation and subsequent heating rate, potentially caused by an event like a flare, can have long-lasting impacts on gas ionisation and material accretion rates. In other words, if the disc experiences a temporary boost in ionisation and heating from cosmic rays that push temperatures above approximately 1000 K, this activates the thermal ionisation of alkali metals. The enhanced ionisation allows the gas to couple more effectively with magnetic fields, thereby initiating MHD instabilities such as MRI. This, in turn, spurs the accretion of material, introducing an extra heat source in the form of viscous accretion heating. This sustained heat is sufficient to keep alkali metals ionised thermally, thereby maintaining efficient accretion processes.

\paragraph{Thermal ionisation and ionisation fraction:}
We have established in that high-energy radiation, particularly from stellar UV and X-ray photons, plays a significant role in altering the surface layers of protoplanetary discs due to the intense photoionisation and heating that they produce in these low density regions. While dust and chemical reactions mainly determine the temperature profile in deeper regions, the upper atmospheres of discs are mainly heated by radiation. Specifically, X-rays can heat these upper layers to temperatures reaching several thousand Kelvin at a distance of 1 au from the star. This results in a strong temperature inversion in the vertical profile of the disc. Near the star, the high temperature may even produce regions dominated by ionisation due to matter collisions. We call this process thermal ionisation. Indeed, thermal ionisation of alkali metals seems to be the dominant source of ionisation in the innermost regions of the disc, typically at $\sim 0.1$ au. In thermal equilibrium, the ionisation state of a single species with an ionisation potential $\chi$ follows the Saha equation,

\begin{equation}
\frac{n_{\text{ion}} n_e}{n} = 2 \frac{U_{\text{ion}}}{U} \left(\frac{2\pi m_e k_B T}{h^2}\right)^{3/2} \exp\left[-\frac{\chi}{k_B T}\right].
\end{equation}

In this equation, $n_{\text{ion}}$ and $n$ denote the number densities of the ionised and neutral species, respectively, and $n_e(=n_{\text{ion}})$ is the electron number density. The partition functions for the ions and neutrals are $U_{\text{ion}}$ and $U$, respectively, and $m_e$ is the electron mass. The temperature dependence is not strictly the typical exponential Boltzmann factor, because the ionised state is favored over the neutral state due to entropy effects \citep{2011ARA&A..49..195A}.

In protoplanetary discs, thermal ionisation becomes significant when the temperature rises enough to start ionising alkali metals. For potassium, with an ionisation potential $\chi = 4.34$ eV, we denote the abundance of potassium relative to all other neutral species as $f = n_K/n_n$, and the ionisation degree  $x_E \equiv n_e/n_n$.

As long as potassium remains weakly ionised, the Saha equation yields,

\begin{equation}
x_E \approx  10^{-12} \left(\frac{f}{10^{-7}}\right)^{1/2} \left(\frac{n_n}{10^{15} \text{cm}^{-3}}\right)^{-1/2} \left(\frac{T}{10^3 \text{K}}\right)^{3/4} \exp\left[-\frac{2.52 \times 10^4}{T}\right],
\end{equation}
where the final numerical factor in the denominator is the exponent value at $10^3$ K. 
\begin{figure}
    \centering
    \includegraphics[width=0.7\linewidth]{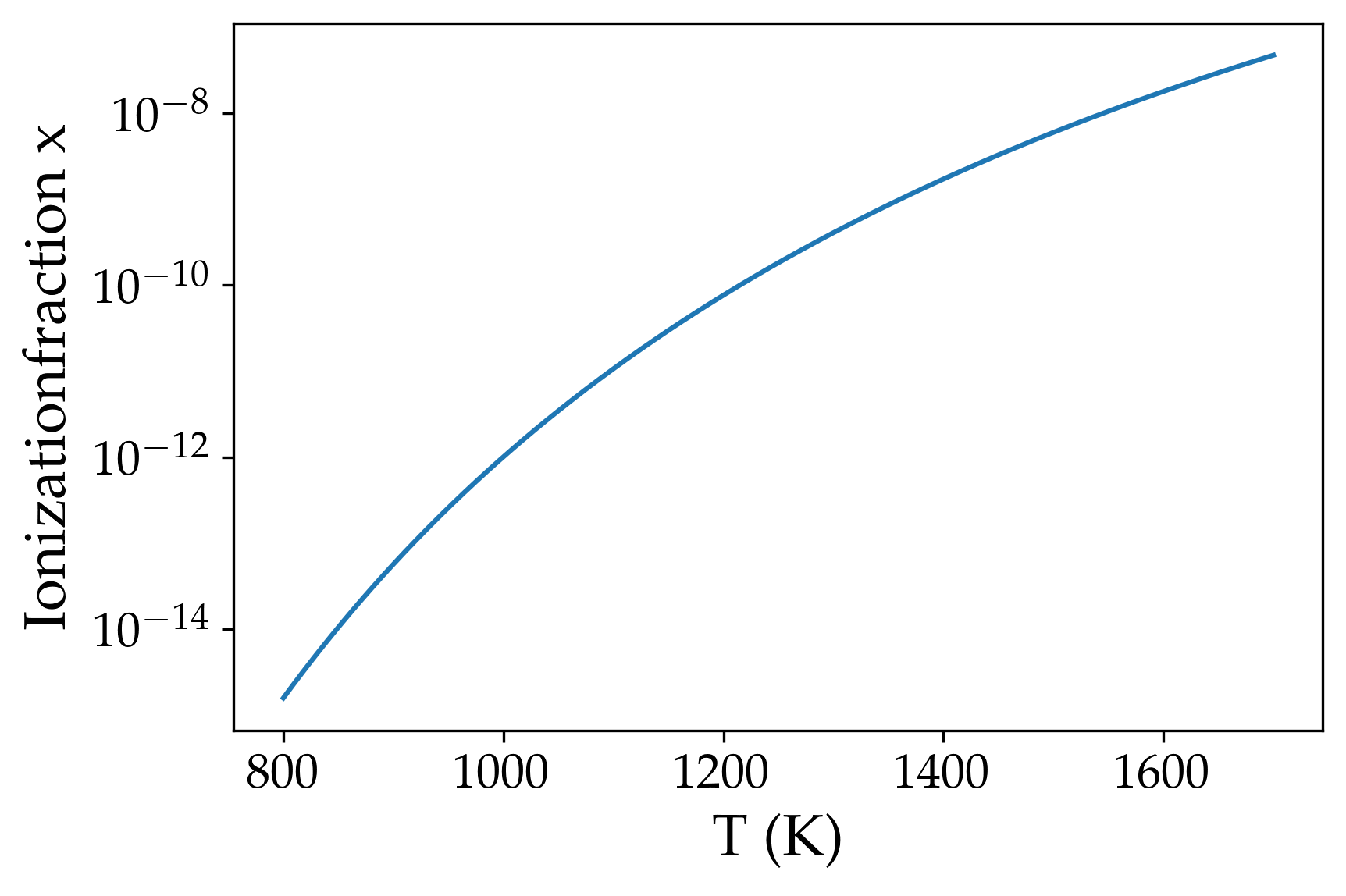}
    \caption{The figure illustrates the thermal ionisation fraction as a function of temperature, as predicted by the Saha's equation for the inner disc. Here, we operate under the assumption that potassium, with an ionisation potential $\chi = 4.34$ eV and a fractional abundance $f = 10^{-7}$, is the sole element contributing to ionisation. The number density of neutrals is considered to be $n_n = 10^{15}$ cm$^{-3}$.}
    \label{fig:ThermalionisationFraction}
\end{figure}

The ionisation fraction at different temperatures is depicted in Fig. \ref{fig:ThermalionisationFraction}. Ionisation fractions that are large enough to be relevant for studies of magnetic field coupling are reached at temperatures of $T \sim 10^3$ K, though the numbers remain exceptionally small, of the order of $x \sim 10^{-12}$ for these parameters. In the outer part of protoplanetary discs ($r \gtrsim 1$ au), the gas is mostly cold with $T < 500$ K. This implies that thermal ionisation due to collision between molecules is inefficient.

To summerize, the additional heating rate from locally produced CR, may play a role in our understanding of the launching mechanisms of winds and jets as we discussed in Sect. \ref{sec:thermaloutflows}. The energy needed to propel outflows is expected to arise from the inner parts of the protoplanetary disc. The elevated temperatures and the corresponding pressure gradient created by heating mechanisms act as the driver that make the disc matter gravitationally unbound, propelling it outwards, initiating the launching of thermal winds. On the other hand, the heating can also lead to increased ionisation of the gas, which makes it easier for the gas to couple to the magnetic fields, an essential mechanism for the acceleration and collimation of MHD winds and jets. Therefore, the heating rates and the consequent thermal structure of the inner disc influence directly the power, direction, and stability of the launched outflows.

\subsubsection{Modeling the disc chemistry with the Protoplanetary Disc Model {\tt ProDiMO}} \label{sec:focusonProDiMO}
\paragraph{The necessity of radiative thermo-chemical numerical model for protoplanetary discs:}
Observational tracers like molecular and atomic species provide critical insights into the disc key physical properties, such as mass, temperature, structural layout, and turbulence. Interpreting these observations necessitates a comprehensive understanding of the disc chemical processes. This is not easy, as discs are chemically complex structures with temperatures ranging from around 10 to several thousand Kelvin and densities varying between approximately \(10^4\) and \(10^{16} \, \text{cm}^{-3}\). We presented ionisation sources, including stellar radiation like UV and X-rays as well as cosmic rays. Numerical simulations are necessary to estimate there influence on the disc chemical structure and complexity.

%Additionally, the role of dust is not to be underestimated; it acts as a shield against stellar radiation, facilitates the formation of ice in colder regions, and even serves as a catalytic surface for molecular formation. On top of these factors, dynamic processes such as viscosity changes, turbulence, and interactions between the disc and emerging planets—as well as the star own evolution complicate the chemical landscape of the disc. This makes protoplanetary discs incredibly complex systems, both physically and chemically. 

Incorporating all pertinent processes in chemical disc modelling is a complex task. Many models tend to rely on a static two-dimensional disc structure and constant stellar parameters. However, advanced radiation thermo-chemical disc models take into account variables like gas temperature and dust radiative transfer, often going into fine details. These models can also explore the effects of turbulent mixing and the evolution of dust particles. Comprehensive reviews of these types of models are available in works by \citet{2014prpl.conf..317D} and \citet{2023ARA&A..61..287O}. These high-computational-cost simulations often require simplifications, particularly in radiative transfer and chemistry, but they are valuable in providing insights into how dynamical processes affect the chemical evolution of discs.

In this section, we are not delving into the specificities of chemical models but rather highlighting the overall chemical structure of the disc based on {\tt ProDiMO} simulations.

\paragraph{Overview of {\tt ProDiMO} features:}
In order to evaluate the disc thermal and chemical characteristics, all along this thesis, we consider a model generated by \textsc{{\tt ProDiMO}}\footnote{\url{https://www.astro.rug.nl/~ProDiMO}~(Version: 1.0 7e3ecc64)} (Protoplanetary Disc Model; \citealt{2009A&A...501..383W,2010A&A...510A..18K,2016A&A...586A.103W,2018A&A...609A..91R}). This code computes the radiation field within a 2D disc structure, accounting for the absorption, scattering, and re-emission of photons from both external and internal sources. {\tt ProDiMO} has been extensively used to model the ionisation structure and chemistry of T Tauri discs by computing emission lines. The emission lines emanating from protoplanetary discs are primarily produced in the irradiated surface layers, where the gas typically exhibits higher temperatures than the dust. {\tt ProDiMO} comprehensive thermo-chemical model enables the interpretation of these emission lines. Subsequently, by comparing these emission lines with observational data, constraints can be established on various disc physical parameters. These parameters include, but are not limited to, the disc mass, the position of the inner rim, and the flux of stellar energetic particles, among others \citep{woitke2009radiation, 2011MNRAS.412..711T, 2017A&A...603A..96R}.  

{\tt ProDiMO} has been used to model the spectral energy distributions of the disc (SEDs, \citealt{thi2011radiation}), the chemistry of water deuteration \citep{2010MNRAS.407..232T}, UV-fluorescence including CO rovibrational emissions \citep{2013A&A...551A..49T}, and multiple observations from the GASPS program using Herschel. The X-ray physics is also implemented in {\tt ProDiMO} \citep{2011A&A...526A.163A,2012A&A...547A..69A,2012A&A...547A..68M}. The {\tt ProDiMO} code is able of executing within a few CPU-hours per model. It can perform automated fittings of the observed continuum emission and thousands of gas lines. This process requires the execution of thousands of models, an operation that would be excessively time-consuming with a complete 3D non-ideal MHD radiation chemical code. In return, {\tt ProDiMO} does not address the hydrodynamic equations of the gas. Instead, the disc density structures are parameterised. Only the vertical hydrostatic structure can be self-consistently readjusted with the gas temperature. The main physical parameters of the disc model (kept fixed) are listed in Tab. \ref{table:ProDiMOparameters}.\\

\begin{table}
\caption{Parameters of the disc model \citep{2017A&A...603A..96R}. 
}
\label{table:discmodel}
\centering
\begin{tabular}{l|c|c}
\hline\hline
Quantity & Symbol & Value  \\
\hline
stellar mass                          & $M_*$                    & $0.7~ M_{\odot}$\\
stellar effective temp.               & $T_{\mathrm{*}}$                  & 4000~K\\
stellar luminosity                    & $L_{\mathrm{*}}$                  & $1.0~\mathrm{L_{\odot}}$\\
FUV excess                            & $L_{\mathrm{FUV}}/L_{\mathrm{*}}$ & 0.01\\
FUV power law index                   & $p_{\mathrm{UV}}$                 & 1.3\\
\hline

disc gas mass                         & $M_{\mathrm{disc}}$               & $0.01~\mathrm{M_{\odot}}$\\
dust/gas mass ratio                   & $d/g$                             & 0.01\\
inner disc radius                     & $R_{\mathrm{in}}$                 & 0.07~au\\
tapering-off radius                   & $R_{\mathrm{tap}}$                & 100~au\\
reference scale height                & $H(R=100\;\mathrm{au})$             & 10 au\\
                    
\hline

\end{tabular}

\label{table:ProDiMOparameters}
\end{table}

We now examine the various physical and chemical processes highlighted by the disc chemical modelling.

\paragraph{Chemical structure:}
To explore the disc chemical structure, we categorise the disc into radial regions and vertical layers. Chemically, the disc midplane is divided into three specific areas:
\begin{itemize}
    \item \textbf{The inner zone – } It is reaching up to a few au from the star, is mostly shielded from radiation and has temperatures exceeding 100 K.
    \item \textbf{The middle zone – } It extends from a few au up to 100 au, with temperatures ranging between 20 K and 100 K and most chemical species adhering to dust grains, like water ice.
    \item \textbf{The outer zone – } It has temperatures below 20 K, where most species are frozen onto dust grains and are only partially shielded from radiation.
\end{itemize}

Figure \ref{fig:DustGasStructure} illustrates this by showing the abundance of CO and water, in both gaseous and solid phases, as a function of radius within the midplane. Notable are the distinct locations of the "ice-lines" due to varying adsorption energies for CO and water, as well as the resurgence of gaseous-phase abundance in the outer disc, particularly for CO.

Vertically, the disc is typically divided into three separate layers, also represented Fig. \ref{fig:ionisationTracersDistribution},
\begin{itemize}
    \item \textbf{The midplane – } it corresponds to the zones described above, it is characterised by ice-based chemistry involving interactions between dust and gas, as well as surface chemistry. 
    \item \textbf{The molecular layer – } it has temperatures between 20 and 100 K, where most molecules are in gaseous form. This layer is somewhat protected from radiation and exhibits a wide variety of ion-neutral chemical interactions.
    \item \textbf{The photon-dominated layer – } it primarily involves photochemical processes like ionisation and photo-dissociation, with most species present in an atomic and ionised state.
    
\end{itemize}
\begin{figure}
    \centering
    \includegraphics[width=0.9\linewidth]{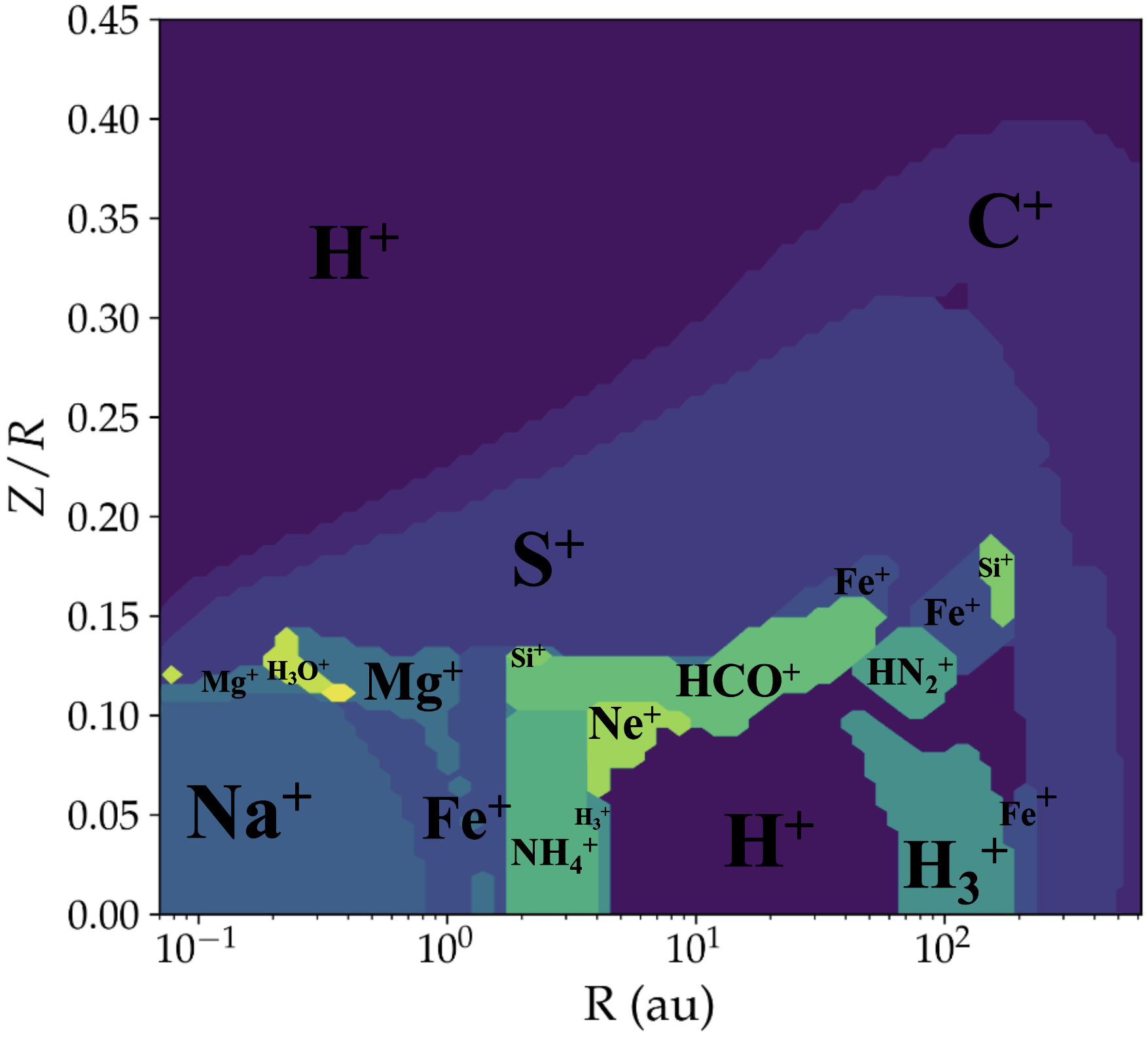}
    \caption{The figure produced from {\tt ProDiMO} results, identifies the dominant ions in various regions of the disc. The surface layer resembles a Photo-Dissociation Region (PDR) and is primarily occupied by atomic ions such as H\(^+\), C\(^+\), S\(^+\), and Mg\(^+\). Below 1 au, in the disc inner region, radiative and accretion heating elevate temperatures, leading to a prevalence of Na\(^+\) ions. The warm molecular layer is rich in molecules containing carbon, nitrogen, and oxygen, which react with H\(_3^+\) generated by non-thermal ionisation sources. Here, molecular ions like HCO\(^+\), NH\(_4^+\), and HN\(_2^+\) are predominant. Conversely, in the deeply embedded cold midplane, where most species are in a frozen state on grains, H\(^+\) and H\(_3^+\) ions are the most prevalent.}
    \label{fig:DominantCationAllDisc}
\end{figure}

Varying physical conditions in these zones result in diverse chemical behaviours. In the inner disc region, chemical reactions with high activation barriers become relevant, unlike in the generally colder interstellar environments \citep{2010ApJ...721.1570H}. This results in a different chemistry than in the well studied ISM. This difference in chemistry has been observed, as noted by \citet{2014FaDi..169...49P}, but current thermo-chemical models still struggle to explain these observed molecular abundances. One possible explanation could be the neglect of radial migration processes, such as the inward movement of large dust particles. These migrating particles are likely coated with water ice, which evaporates in the warmer inner zone, increasing the oxygen levels and subsequently affecting the local chemistry. However, some models like that of \citet{2014A&A...563A..33W} do account for accretion flows and indicate that complex molecules, such as methanol, can see a significant increase in abundance due to the migration of ice-coated dust particles into the inner zone. Figure \ref{fig:DominantCationAllDisc} shows the distribution of the most abundant cation as computed by {\tt ProDiMO}. At the disc surface, the ionising flux arising from FUV photons is predominant, and the gas layer essentially acts as a photodissociation region (PDR). In this region, atomic ions like  H$^+$, C$^+$, S$^+$, and Mg$^+$ are most abundant. There, the main chemical activities are driven by the direct ionisation of atomic species and the breaking apart of molecules due to X-ray and UV photons.  \\

The region below the PDR, known as the warm molecular layer, harbours diverse chemical reactions. As the column density increases, the abundance of atomic ions such as carbon decreases, and they become neutral. The edge of the PDR is encountered around a column density, \(N \sim 10^{22}\, \text{g cm}^{-2}\), beyond which other ion become the most abundant. This layer is influenced by high temperatures and both UV and X-ray radiation \citep{2004A&A...417...93S,2023ARA&A..61..287O}. Accurate representation of chemical interactions in this area is crucial for determining the gas temperature, which in turn is important for processes like photo-evaporation in discs. X-rays also play a significant role in this layer, contributing to the ionisation of noble gases like Ne and Ar. In the inner region ($R\lesssim 1$ au), it can heat the gas to temperatures of thousands of Kelvins. In this region, Na$^+$ is the most abundant cation. With its low ionisation potential the high temperatures of the inner disc are sufficient for thermal ionisation. 

In our standard T Tauri disc model, the chemical richness of this layer is evident, particularly in the context of CO. However, the specific location and thickness of this layer can vary from one disc to another and are also affected by the properties of the dust. In this region, the most dominant ions are molecular like HCO$^+$, HN$_2^+$ and NH$_4^+$. The presence of H$_3^+$ ions, byproduct of the ionisation of H$_2$ by CRs, gives rise to various ion-neutral reactions, facilitating more complex chemistry like the formation of hydrocarbon radicals initiated by H$_3^+$ \citep{2005pcim.book.....T}. For example, H$_3^+ + $C $\rightarrow$ CH$^+ + $H$_2$ is a reaction illustrating the formation of hydrocarbon radicals. Subsequent reactions with molecular hydrogen lead to hydrocarbon ions like CH$_3^+$ and CH$_5^+$. These reactions contribute to an increasing molecular complexity. 

 \citet{2011ApJ...727...76V} revealed that when large dust grains are included in chemical models, this layer becomes more extended, reaching deeper into the disc compared to models that only consider typical ISM dust properties, which usually involve smaller grains. Similar findings were reported by \citet{2013ApJ...766....8A}. They utilised a comprehensive chemical model that included dust evolution and found that, on average, column densities of neutral molecules like CO and H$_2$O are greater in models that account for dust evolution.

Another crucial chemical phenomenon in this warm molecular layer involves CO isotopologue chemistry, especially isotopologue-selective photo-dissociation \citep{2009A&A...503..323V}. This particular process might account for the unusual ratios of $^{17}$O and $^{18}$O isotopes found in meteorites. When utilising CO isotopologue line ratios to gauge disc gas masses, this selective photo-dissociation must also be taken into account. Thermo-chemical disc models by \citet{2018A&A...619A.113M} reveal that relying on uniform CO isotopologue ratios across the disc can result in significant underestimation of disc masses by as much as an order of magnitude. Additionally, the potential reduction of gas-phase CO in discs, as suggested by \citet{2012A&A...541A..91B} and \citet{2013ApJ...776L..38F}, should also be considered. While this depletion might be attributed to isotope-selective photo-dissociation, especially of C$^{18}$O, other near-midplane chemical processes like the dissociation of CO by He$^{+}$ \citep{2014FaDi..168...61B,2014ApJ...790...97F} or surface chemistry mechanisms \citep{2015A&A...579A..82R} must not be overlooked if CO and its isotopologues are being used as indicators of disc mass.\\

In the more radiation-protected and cooler middle zone of the disc midplane, CRs and SLR ionisation, along with potentially high-energy X-rays, are the key chemical influences. The ionisation rates from these processes are relatively low, around \(10^{-20} - 10^{-17} s^{-1}\) \citep{cleeves2015constraining} compared to those in the disc upper layers. Secondary electrons and UV radiation generated by CR also plays a significant role in this embedded region \citep{Padovani18}. In this region where most of the Carbon and Nitrogen are freezed-out on grains, the most abundant species are H$^+$ and H$_3^+$.

On the other hand in the cold outer zone ($R\gtrsim100$ au), non-thermal desorption mechanisms, like photo-desorption, gain importance. Observations of the DCO$^+$ molecule in the IM Lup system \citep{teague2015chemistry} hint at the effectiveness of non-thermal desorption processes. The interstellar background radiation may also contribute to DCO$^+$ abundances. Surface chemistry, such as the hydrogenation of CO to form H$_2$CO, can create complex molecules that can then be photo-desorbed. Methanol serves as a key example of this formation pathway \citep{2010ApJ...722.1607W,2014prpl.conf..317D}, although current observations, like those in the TW Hya disc \citep{2016ApJ...831..200W}, show methanol abundances up to 100 times lower than model predictions. This discrepancy likely suggests that existing models overestimate either the efficiency of methanol formation or the effectiveness of various desorption mechanisms. Vertical turbulent mixing can move ice-coated dust particles to higher layers where the efficiency of photo-desorption is greater. This results in elevated gas-phase abundances of complex molecules that originally formed on dust grains, as noted by \citet{2011ApJS..196...25S}. However, this vertical mixing can also have the opposite effect, it can diminish the abundance of complex organic molecules. Transporting atomic hydrogen to deeper layers can substantially inhibit the formation of these complex molecules, according to \citet{2014ApJ...790...97F}.

The resulting chemical structure presented in this section is accounting for the standard non-thermal ionisation sources usually considered, i.e, stellar UV and X-ray radiation aswell as the ISRF, GCR and SEP. In Chapter \ref{C:PublicationI} we consider an additional non-thermal source of ionisation, energetic particles produced by flares. And in Chapter \ref{C:PublicationII} we include this additional ionisation source in {\tt ProDiMO}, the outgoing disc structure is discussed there.

\subsection{Influence of ionisation on the disc viscosity and the accretion processes}\label{sect:role of mri}
%\subsection{Magnetorotational instability}

In the previous section, we presented the effect of ionisation on the disc chemistry. In this section we will discuss the effect of ionisation on the disc dynamics, focusing on the viscosity and accretion rate.
In Sec. \ref{sec:ModelingProtoplanetaryDiscs} we showed that the MRI \citep{1991ApJ...376..214B} plays a crucial role in the accretion processes and overall evolution of protoplanetary discs. The MRI has been extensively studied in the context of protoplanetary discs through theoretical models \citep{2014ApJ...780...42T} and laboratory experiments \citep{2018PhR...741....1R}. The conditions for the MRI to be triggered, are basically a weak magnetic field $\beta \geq 1$ (\citealt{2021JPlPh..87a2001P}, eq. 6.64), shear and sufficiently high ionisation. The current understanding of T Tauri discs suggests that these conditions are met in the inner ($T > 1000$ K) and outer protoplanetary disc regions ($\Sigma \leq 15$ g cm$^{-2}$). 
If the disc accretion is solely driven by turbulence of either hydrodynamical or MHD origin, the accretion rate can be calculated using Eq. \eqref{eq:accretionrateLesur22} and is approximately given by \citep{hartmann1998accretion},

\begin{equation}
\dot{M}_{\text{acc}} \approx 3 \times 10^{-8} \left(\frac{\alpha}{10^{-2}}\right) \left(\frac{\epsilon}{0.1}\right)^2 \left(\frac{R}{10\, \text{AU}}\right)^{1/2} \left(\frac{M_{\star}}{M}\right)^{1/2} \left(\frac{\Sigma}{10\, \text{g cm}^{-2}}\right)
\label{eq:MassaccretionRatehartman}
\end{equation}

In this equation, we define the disc aspect ratio as $\epsilon \equiv H/R$, and $\Sigma(R)$ the surface density. Consequently, it is typically necessary for the viscous $\alpha$ parameter to be around $10^{-2}$ to match the observed accretion rates based solely on turbulence. 

In this section we first review the values of $\alpha$ obtained from numerical simulations. Then we introduce the parameters that define the MRI active region, the region where $\alpha$ values are sufficiently high to sustain efficient accretion. And finally based on a simple model we compute the expected value of $\alpha$ in a disc as computed by {\tt ProDiMO}. We show that $\alpha$ is directly linked to the ionisation degree. From this, we infer a rough estimation of the mass accretion rate expected for standard {\tt ProDiMO} discs. 

\subsubsection{MRI Active zones}\label{sec:MRIActiveRegion}
\paragraph{Impact of radiative environment on MRI:}
Recent local simulations using radiative 3D MHD codes have studied the combined effect of radiative transfer and non-ideal MHD, key heating and cooling processes and MRI in the inner disc \citep{2017ApJ...835..230F}. In optically thin zones of protoplanetary discs, irradiation heating primarily defines the temperature, and the MRI strength conforms to local temperature and pressure profiles, as demonstrated by \citet{2013A&A...560A..43F} in their global radiation MHD simulations.

In optically thick regimes \citep{scepi2018turbulent}, MRI-induced heating elevates the disc temperature, thereby increasing disc scale height, enhancing turbulence and causing convection. The bulk of MHD heating occurs in the magnetic reconnection layers \citep{ross2018dissipative}, which typically require high resolution to resolve the heating rate \citep{mcnally2014temperature}.

In protoplanetary discs, the inner gas disc is largely thought to remain optically thin to its own thermal radiation \citep{flock20173d}, yet a detailed understanding of gas opacities is still a topic of investigation. Particularly, the inner gas disc, spanning the silicate sublimation line to the co-rotation radius, is heavily reliant on heating and cooling processes. Therefore, gas opacity exhibits various molecular lines \citep{malygin2014mean}.

MRI activity in this area governs the final accretion process stage of T Tauri stars, transferring matter onto the star while also fuelling magnetically driven winds. The gas opacity has a strong influence on thermodynamics and the MRI, so we emphasise on this point in the next section.  It is particularly important for the evolution of the MRI-active, inner disc.

\paragraph{MRI active zones from MHD simulation:}
The transition area in protoplanetary discs, spanning from MRI-active to MRI-damped regions cause variations in the viscous $\alpha$ parameter. These may modify $\alpha$ by factors ranging from 10 to 1000 based on magnetic field strength, ionisation degree, and hydrodynamical turbulence level \citep{2014prpl.conf..411T,flock20173d}. 

The dead-zone inner edge typically lies between 0.05 and 0.3 au for a young T Tauri-like star \citep{mohanty2018inside,flock2019planet}, reaching up to 1 au for a Herbig-Haro star due to higher stellar luminosity \citep{flock20173d}. However, in highly optically thick discs, the dead-zone inner edge may be even further out due to large accretion rate and substantial accretion heating \citep{mohanty2018inside,jankovic2021mri}.

In the inner disc, the presence of dust grains turns out to be the determining factor in the local intensity of X-ray and UV radiation. 
%For example, \citet{flock20173d} conducted a radiative MHD simulation of the inner disc. Because of the strong radiation field they are considering, for exemple at ($R=0.1$, $Z/R=0.1$), the temperature is found to be much higher than computed by {\tt ProDiMO}. , \(T\gtrsim1200\) K, which allows the triggering of MRI at this position. The main difference that leads to this discrepancy between the thermal structure computed by {\tt ProDiMO} and \citet{flock20173d} is the location of the dust sublimation radius. 
Inside the dust sublimation radius, the distance from the central star where the temperature is high enough to cause dust grains to change from a solid to a gas phase, radiation from the central star cause the dust to sublimate. This leaves a region essentially free of solid dust particles. Beyond this radius, the temperature is low enough for dust to remain in solid form. Self shielding itself from the heating of X-rays and UV, the presence of dust lead to a sharp drop in the temperature, killing thermal ionisation.

The dust sublimation radius \( R_{\text{sub}} \) can be estimated from the star  luminosity \( L_* \) and effective temperature \( T_{\text{eff}} \):
\begin{equation}
    R_{\text{sub}} = 0.03 \sqrt{Q_R} \sqrt{\frac{L_*}{L_\odot}}\left(\frac{1500\, K}{T_{\text{sub}}}\right)^2
    \label{eq:sublimationradius}
\end{equation}
with \( Q_R \) being the ratio of the absorption efficiencies of the dust for the incident field and the reemitted field, and \( T_{\text{sub}} \) being the dust sublimation temperature. As emphasised in \citet{2002ApJ...579..694M}, \( Q_R \) depends on the dust properties and on the central star  effective temperature. For grain radii ranging from \( 0.03 \, \mu \text{m} \) to \( 1 \, \mu \text{m} \), \( Q_R \) ranges from 1 to 4 for stars with \( T_{\text{eff}} = 4000\, K \); \( Q_R \) equals 6 for \( T_{\text{eff}} = 5200\, K \), and 8 for \( T_{\text{eff}} = 5770\, K \). 

From observations of \( T_{\text{eff}} \) and \( L_* \), sublimation radii lie between \( 0.05\, \text{au} \lesssim R_{\text{sub}} \lesssim 0.3\, \text{au} \), with median values at \( R_{\text{sub}} \approx 0.1\, \text{au} \) (see \citet{2021A&A...655A..73G} and references therein). {\tt ProDiMO}, with values of \( T_{\text{eff}} = 4000\, K \) and \( L_* = L_\odot \), has a sublimation radius in the standard observed range. Conversely, the simulation of \citet{flock20173d} uses \( T_{\text{eff}} = 10\,000\, K \), corresponding to a Herbig star. With such a high surface temperature, the dust sublimation radius is naturally pushed farther from the star to \( R \approx 0.55\, \text{au} \).

For the study and delimitation of the MRI active region in the inner disc, the position of the sublimation radius is a determining parameter. In the case of {\tt ProDiMO}, beyond \( R_{\text{sub}} = 0.07\, \text{au} \), the dust screens all X and UV rays, suppressing all sources of heat and making the temperature drop to \( T < 1000\, K \).

\paragraph{MRI-active region in hydrostatic simulation}

In a simplified vision, the development of the MRI can be characterised by two dimensionless parameters. The magnetic Reynolds number, $Rm$, which quantifies the coupling between the plasma and the field, and the ambipolar diffusion number, $Am$, which quantifies the coupling between the neutral gas and the ionised plasma. 

The Magnetic Reynolds number is defined as, 

\begin{equation}
Rm  \equiv \left(\frac{c_s h}{\eta}\right)
\end{equation}
where $c_s$ is the local sound speed, $h = c_s / \Omega_K$ is the scale height, $\Omega_K=\sqrt{\frac{GM}{R^3}}$ is the Keplerian orbital frequency, and $\eta$ is the magnetic atomic diffusivity \citep{1994ApJ...421..163B}.

\begin{equation}
\eta = 234 \left(\frac{T}{K}\right)^{\frac{1}{2}} x_E^{-1} \, \rm cm^2 s^{-1},
\end{equation}

where $x_E = n_e/n_n$ is the fractional abundance of electrons whereas $c_s$ can be expressed as,
\begin{equation}
    c^2_s=\frac{k_B T}{\mu m_p}
\end{equation}
where $k_B$ is the Boltzmann constant, $m_p$ the proton mass, $T$ the temperature and $\mu=3/2$ is the mean molecular weight. So the magnetic Reynolds number can be expressed as function of the disc parameters,
\begin{equation}
    Rm\approx 10^5 \left(\frac{T}{100 \rm K}\right)^{\frac{1}{2}} \left(\frac{R}{1 \rm au}\right)^{\frac{3}{2}} \left(\frac{x_E}{10^{-8}}\right).
\end{equation}

The Elsasser number for ambipolar diffusion $Am$ represents the frequency at which neutral particles collides with ions normalised to the orbital frequency. %$\nu_{in}=n_{charge}\beta_{in}$.

\begin{equation}
Am \equiv \frac{\nu_{in}} {\Omega_K} = \frac{\beta_{in} n_{\rm e}}{\Omega_K}= \frac{x_E \beta_{in} n_{\rm n}}{\Omega_K} ,
\end{equation}

where $n_{\rm e} = x_{E}n_{\rm n}$ is the total number density of singly charged species and $\beta_{in} = 2 \times 10^{-9} \, \rm cm^3 s^{-1}$ is the collisional rate coefficient for singly charged species to distribute their momentum to neutrals. Thus,
\begin{equation}
    Am\approx 1\left(\frac{x_E}{10^{-8}}\right)\left(\frac{n_n}{10^{10}~\rm cm^{-3}} \right) \left(\frac{R}{1 \rm{au} } \right)^{3/2}
\end{equation}

Numerical simulations suggest that the MRI is active for magnetic Reynolds numbers of order $10^2-10^4$ \citep{2007ApJ...659..729T}. \citet{2013ApJ...772....5C} noted that constraints based on $Am$ also depend on the magnetic field strength and the ratio of thermal to magnetic pressure, $\beta$. Following \cite{2019ApJ...883..121O}, one define the MRI-active region as $Re > 3000$ and $Am > 0.1$.

\begin{figure}
    \centering
    \includegraphics[width=0.6\linewidth]{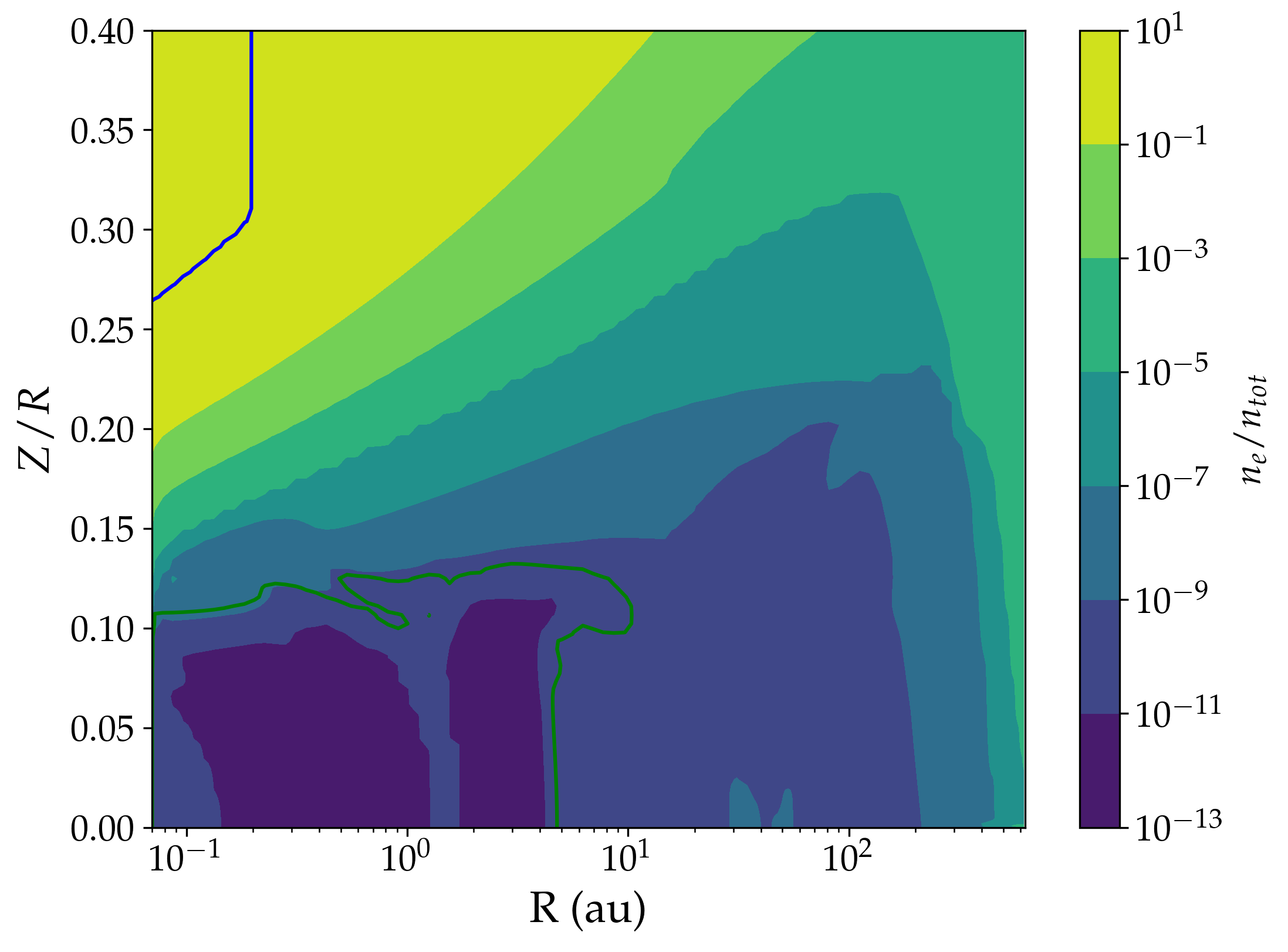}
    \caption{Two-dimensional structure of the accretion disc model. The height of the disc $Z$ is scaled to its radius ($Z/R$). The background colour plot represents the ionisation fraction $n_e/n_{tot}$ distribution as per {\tt ProDiMO}. The green solid line marks $z_{inf}(R)$, denoting the region within which the disc remains MRI inactive, where $Rm<3000$. The blue solid line marks $z_{sup}(R)$, where $Am<0.1$}
    \label{fig:MRIActiveRegionProDiMO}
\end{figure}

In Figure \ref{fig:MRIActiveRegionProDiMO}, we determine the MRI active region, using the thermal and chemical disc structure computed by {\tt ProDiMO}, based on the standard parameters provided in Tab. \ref{table:ProDiMOparameters}.

\subsubsection{Mass accretion rate}\label{sec:Derivationofaccretionrate}
We present, as a proof of concept, a simple model for the determination of the accretion rate. This model allows to highlight the close link between accretion rate and ionisation degree. We showed Eq. \eqref{eq:MassaccretionRatehartman}, that assuming efficient angular momentum transport via the MRI, the disc mass accretion rate can be written in terms of an $\alpha$-disc model. We rewrite Eq. \eqref{eq:MassaccretionRatehartman} in order to express it in terms of parameters that {\tt ProDiMO} is computing,

\begin{equation}
\dot{M} = 2\times 3\pi \mu m_p \nu  N_{H,MRI} = 6\pi N_{H,MRI} \alpha \frac{kT}{\Omega}, 
\label{eq:MRIAccretionRate}
\end{equation}
where the effective turbulent viscosity is given by $\nu = \alpha c_s h$ and $\mu$ is the mean molecular weight. The factor of 2 accounts for accretion along the top and bottom disc surfaces and 

\begin{equation}
    N_{H,MRI}(R)= \int_{z_{inf}}^{z_{sup}} n_H(R,z) dz,
\end{equation}

is the hydrogen number column density in the height range $z_{inf}-z_{sup}$ where the MRI is active.

This formula assumes very efficient angular momentum transport namely $l_{in}\ll l$ where $l=R^2\Omega$ is the specific angular momentum. So it means that it applies far from the inner disc radius.

To compute the value of $\alpha$, we adopt the relation between $\alpha$ and $Am$ at the MRI saturated state from numerical simulations \citep{2011ApJ...736..144B},

\begin{equation}
\alpha =\frac{1}{2}\left[\left(\frac{50}{Am^{1.2}}\right)^2+\left(\frac{8}{Am^{0.3}}+1\right)^2\right]^{-1/2}.
\label{eq:AmbipolarAlphaParameter}
\end{equation}

\begin{figure}
    \centering
    \includegraphics[width=0.6\linewidth]{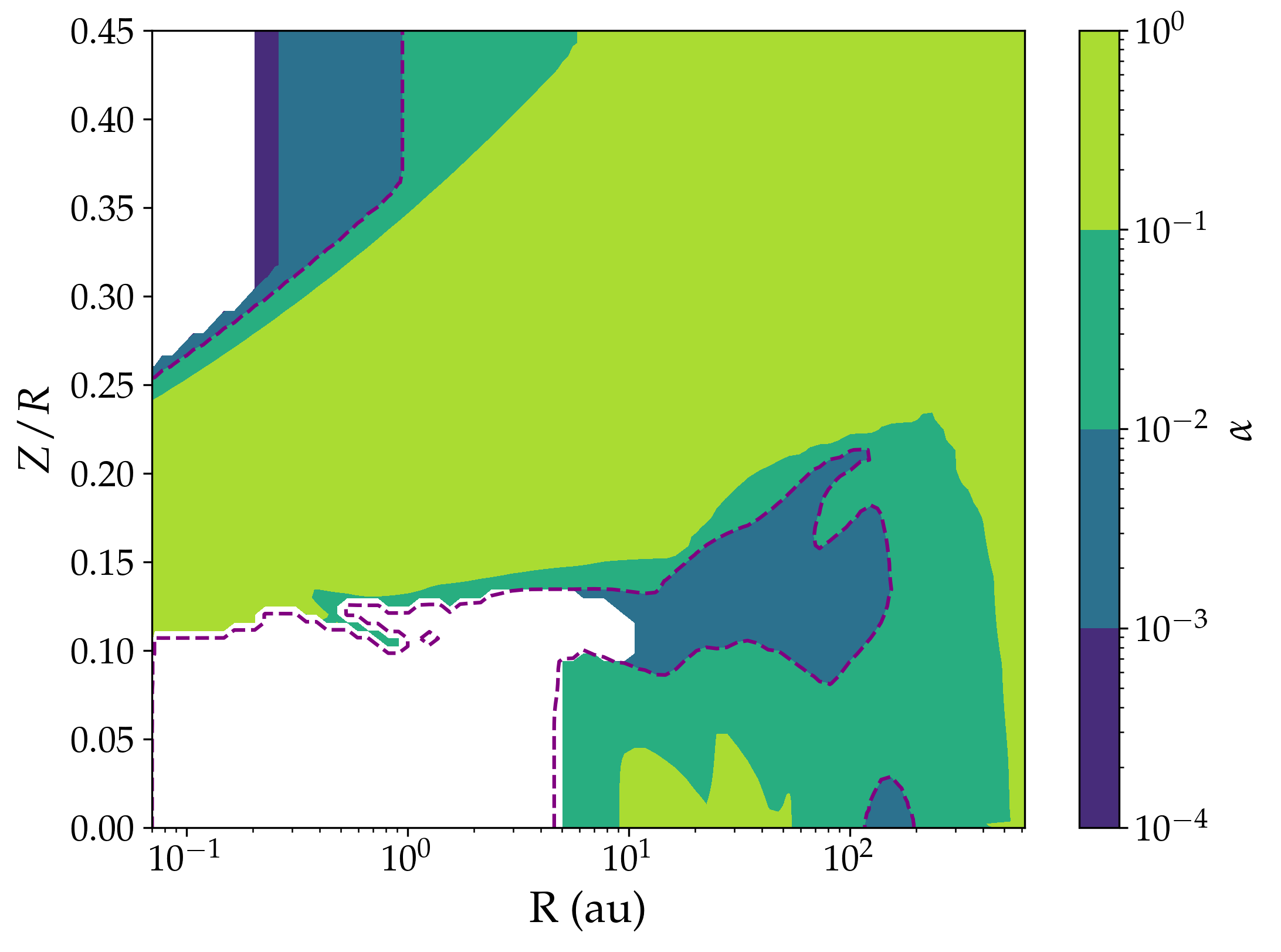}
    \caption{Two-dimensional structure of the accretion disc model. The height of the disc $Z$ is scaled to its radius ($Z/R$). The figure shows the viscosity parameter Eq. \eqref{eq:AmbipolarAlphaParameter}. The dashed lines delimit the region where $\alpha_{\rm eff}= 10^{-2}$. The white area is the MRI inactive region, where $Rm<3000$ and $Am<0.1$.}
    \label{fig:AlphaViscAmbiProDiMO}
\end{figure}

These equations demonstrate the direct link between ionisation degree and disc viscosity. Figure \ref{fig:AlphaViscAmbiProDiMO} plots the values of the viscosity parameter $\alpha$ assuming the disc structure of {\tt ProDiMO}.

By controlling the disc viscosity, ionisation processes play a crucial role in regulating the mass accretion rate onto the central star and the disc overall evolution. 

The ionisation rates and their spatial distribution in the disc have a direct influence on the disc viscosity and accretion processes. In regions with higher ionisation rates, the coupling between the gas and the magnetic field is stronger, leading to more efficient MRI-driven turbulence and angular momentum transport \citep{2007ApJ...659..729T}. This results in higher disc viscosity and accretion rates \citep{2019SAAS...45....1A}. Conversely, in regions with lower ionisation degree, the MRI may be suppressed or significantly weakened, leading to lower disc viscosity and accretion rates \citep{2011ApJ...739...50B}.

Using Eq. \eqref{eq:MRIAccretionRate}, we estimate the average accretion rate of the {\tt ProDiMO} disc to be $\dot{M}\approx9.3 \times 10^{-9} ~ \rm M_\odot /yr$. This rate is within the range of typical observed values in T Tauri stars, given the parameters (as seen in Tab. \ref{table:ProDiMOparameters}) used for the disc structure computation. In Chapter \ref{C:PublicationII}, we introduce a more sophisticated model for to compute the turbulence parameter $\alpha_{\text{eff}}$. We rely on the model proposed by \citet{2019A&A...632A..44T}, which introduces an empirical non-ideal MHD MRI-driven equation for \(\alpha_{\text{eff}}\) accounting for Ohmic diffusivity. In Chapter \ref{C:PublicationII}, we compute the viscosity in different disc solutions, to compare the effect of an additional ionisation source by flares. Our findings will show that by incorporating the additional ionisation due to flares, the MRI inactive zone is spatially reduced which lead to an increase of the accretion rate in the inner disc by an order of magnitude in comparison to a flare-absent disc model. The accretion rate could be increased up to distances of $R\lesssim 1$ au, depending on the flare geometry. Details of the flare geometry will be developed in Sec. \ref{sec:TTauriFlareGeometry}.

\section{Conclusion}
This chapter presents the basis needed for our understanding of ionisation processes in the protoplanetary discs surrounding Tauri stars. We initiated the discussion by elucidating the observational techniques employed to constrain ionisation in these discs. By examining a variety of tracers, which can be either molecular and atomic ions or neutral species, we have established that the choice of the best tracer depends on the specific region of the disc being examined. To quantify the full extent of ionisation, we demonstrated the necessity of employing chemical disc models, as these enable the estimation of the abundance of unobservable species.

Our study progressed into an exhaustive review of the sources of ionisation that have been modelled to date. In the inner disc, ionisation is primarily governed by the star UV and X-ray radiation, which not only directly ionises but also heats the disc, thereby sustaining a high level of thermal ionisation. In contrast, the outer disc has often been considered to be predominantly ionised by GCR, though this assumption needs to be further investigated due to the potential exclusion of GCR by the winds of the T Tauriosphere. We also introduced recent studies that explore the impact of locally-produced energetic particles within the T Tauri system, setting the stage for our pioneering research, as presented in Chapters \ref{C:PublicationI} and \ref{C:PublicationII}, which will focus on flare-induced ionisation effects in the inner disc.

We have demonstrated the critical role of ionisation in determining the dynamics and chemistry of both the disc and associated jets. Importantly, we showed that ionisation is a pivotal factor in triggering the accretion processes, and we provided a formula to calculate mass accretion rates based on ionisation fractions. This is essential for understanding mechanisms such as magneto-rotational instability, which can only occur in sufficiently ionised regions. Moreover, we discussed the potential heating effects of non-thermal ionisation sources, a feature that could be significant in explaining the mechanics behind the observed speeds of jets and winds. These non-thermal sources also have repercussions on the disc chemical profile, influencing observational tracers and potentially triggering the formation of complex molecules such as amino acids, which are interesting from a prebiotic chemistry point of view.

To sum up, constraining the spatial distribution of ionisation is crucial for comprehending its myriad effects on disc dynamics, chemistry, and associated processes. 

As we look forward to the next chapter, we will introduce an alternative ionisation source through a particle acceleration model generated by magnetic reconnection in flares. This upcoming model targets the inner disc region, where flare activity is high and can be constrained observationally. %In future studies, we intend to explore weaker, currently undetectable magnetic reconnection events, which could serve as additional ionisation sources in the mid-range of the disc (1-10 au).

\chapter{Magnetic reconnection and Particle Acceleration}\label{C:Reconnection}

%\textcolor{red}{remarque générale : la lecture est toujours difficile parce que tu mets des sections les unes après les autres de manière descriptive mais sans but précis. Il faut que tu te fixes un cap dans chaque chapitre et que tu amènes le lecture à ce que tu veux décrire. La description doit être progressive du général au particulier. Dans le lay-out cette démarche doit être explicitée clairement. En l'état le chapitre est encore trop brouillon. }

\section{Introduction}
In this chapter, we study in detail the critical subject of particle acceleration, an ubiquitous process in space plasma. Various mechanisms, such as stochastic acceleration in turbulent electric fields or diffusive shock acceleration facilitate this energy gain under differing conditions and settings.

One essential mechanism for particle acceleration is magnetic reconnection, an ubiquitous process in plasmas during which magnetic field lines undergo abrupt reconfiguration. This leads to a local conversion of magnetic energy into kinetic and thermal energy. Occurring in environments ranging from the Earth magnetosphere to solar flares, accretion discs and jets, magnetic reconnection plays a crucial role in plasma dynamics, leading to phenomena like coronal mass ejection and acceleration of plasma jets.

The astrophysical relevance of these processes is particularly striking in T Tauri stars, which exhibit heightened activity in the X-ray wavelength spectrum. These X-ray emissions are conjectured to arise from up-scaled solar-like flares produced due to the star strong magnetic activity. Notably, these flares may have magnetic loops that extend beyond the disc truncation radius, anchoring them into the T Tauri circumstellar disc.

The arising question is how to model particle acceleration by magnetic reconnection to be able to study their impact over the chemical and dynamical properties of the inner disc surrounding T Tauri stars. 

We start by outlining the existing theoretical reconnection model within the MHD framework in Sec. \ref{sect:ClassicalMHDTheories}. Following that, we explore various particle acceleration mechanisms particularly efficient in turbulent reconnection regions in Sec. \ref{sec:kineticapproach}. We then consider how observational data from solar flares can be employed to constrain these mechanisms in Sec. \ref{sec:solarflares}. Subsequently, we extend these findings to construct a model for T Tauri flares in Sec. \ref{sec:ttauriflares}, to finally model supra-thermal particle distribution based on X-ray observations of these T Tauri flares in Sec. \ref{sec:ParticleEmission}.

\section{The theoretical framework of magnetic reconnection}\label{sec:reconnectiontheoreticalframework}
\subsection{Classical MHD theories}\label{sect:ClassicalMHDTheories}
\subsubsection{Magnetic reconnection: a non-ideal MHD process}

\paragraph{Ideal MHD and frozen-in conditions:} The MHD theory, despite its simplifications, provides a good model for understanding large-scale plasma dynamics. This theory overlooks kinetic processes in the plasma but still effectively describes the dynamics of the plasma at large scales compared to thermal particle mean free path.

To be illustrative, we restrict ourselves for now to plasmas that are magnetised and fully ionised, consisting of protons and electrons. Within the framework of ideal MHD, the equation that describes the evolution of the magnetic field is,
\begin{equation}
  \frac{\partial \mathbf{B}}{\partial t}= \nabla \times (\mathbf{u} \times \mathbf{B})
  ,
  \label{eq:idealinduction} 
\end{equation}
where, $\mathbf{B}$ is the magnetic field and $\mathbf{u}$ the plasma bulk velocity.

The velocity $\mathbf{u}$ is given by the momentum-weighted sum of the velocities of the protons ($\mathbf{u}_p$) and electrons ($\mathbf{u}_e$), with respective masses $m_p$ and $m_e$. Hence, $\mathbf{u}$ can be written as,
\begin{equation}
   \mathbf{u}= \frac{m_p \mathbf{u_p}+ m_e \mathbf{u_e}}{m_p+m_e} 
\end{equation}

Equation \eqref{eq:idealinduction} is derived by combining Faraday's Law, which describes the time evolution of the magnetic field, with the ideal form of Ohm's Law, which relates the electric field ($\mathbf{E}$), the magnetic field, and the bulk velocity of the plasma as,
\begin{equation}
    \mathbf{E}+\frac{\mathbf{u}\times \mathbf{B}}{c}=0
    \label{eq:idealohm}
\end{equation}

%According to this law, in a frame of reference moving with the bulk plasma velocity, the electric field and the magnetic field are perpendicular to each other.

The frozen-in condition is a useful concept in ideal plasma physics to have a qualitative understanding of magnetic reconnection. In the state where the plasma is frozen to the magnetic field, there is a relationship between plasma fluid elements and the field lines over time. If at an initial time, an infinitesimal distance ($\delta r$) between two plasma fluid elements is tangential to a field line, it will remain so at any subsequent time. This condition is represented by the following equation,
\begin{equation}
    \frac{d}{d t} (\delta \mathbf{r} \times \mathbf{B})=0
\end{equation}

The frozen-in condition implies that the magnetic field lines are unbreakable, introducing key constraints on the magnetic field topology of the system. According to Eq. \eqref{eq:idealinduction}, the plasma, viewed as a single fluid, is frozen-in. The frozen-in condition can also apply to a specific species or even a particular population of a species, which is collectively frozen-in. This scenario occurs in the ion diffusion region, where electrons remain coupled to the magnetic field, whereas ions are decoupled.

Moreover, in the range of application of ideal MHD, the magnetic flux $\Phi_B$ through a surface $S(t)$ remains constant over time,

\begin{equation}\label{eq:dphidt}
    \frac{d \Phi }{dt }= \frac{d}{dt} \left( \int_{\Omega(t)} d\mathbf{S} \cdot \mathbf{B} \right) =0
\end{equation}

Here, the integration is done over the surface $\Omega(t)$, enclosed by the closed curve $\gamma(t)$, which moves with the plasma velocity field.

The derivation of Eq. \eqref{eq:dphidt} takes into account that the differential surface area $d\mathbf{S}$ can be expressed as $d\mathbf{S} = \mathbf{u} dt \times d\mathbf{l}$, where $d\mathbf{l}$ is tangential to $\gamma(t)$, and also includes the ideal Ohm’s law and  Eq. \eqref{eq:idealinduction}. Thus, this property is equivalent to the frozen-in concept.

When the plasma is frozen to the magnetic field, the motion of the field lines corresponds to the movement of the plasma fluid elements, and they share the same velocity. This velocity, called the drift velocity $u_{E \times B}$, is derived by taking the cross product of $\mathbf{B}$ and the electric field $\mathbf{E}$ from Eq. \eqref{eq:idealohm}, yielding,

\begin{equation}
    \mathbf{u}_{\mathbf{E}\times \mathbf{B}} = c \frac{\mathbf{E}\times \mathbf{B}}{\mathbf{B}^2}
    \label{eq:DriftVelocity}
\end{equation}

Consequently, the motion of the plasma and magnetic field lines are coupled.

\paragraph{Breaking the ideal MHD:}
The theoretical framework of ideal MHD, as discussed in the preceding paragraph, can model large-scale plasma dynamics. These dynamics are characterised by a length scale $L$ and time scale $T$, where kinetic effects can be neglected. However, there are scenarios where the characteristic scale diminishes to the scale of the particles. Such reductions lead to the formation of regions where the conditions of ideal MHD are locally broken. In these regions, the system can modify its magnetic topology and transition to a lower energy state, often through explosive energy conversion. The formation of such regions is typically associated with the presence of current sheets, which are regions of strong magnetic field gradient. The process responsible for such energy release and magnetic field topology modification is called magnetic reconnection.

Magnetic reconnection is a multi-scale process. Large scale systems, which can be modelled by ideal MHD, are influenced by particle-scale processes that occur in regions of local ideal MHD violation. These small scales, localised processes eventually lead to a reconfiguration of the global topology and connectivity of field lines, enabling plasma regions, initially disconnected, to connect each other. This reconfiguration affects the paths of fast particles and induces heat by conduction. 

\begin{figure}
    \centering
    \includegraphics[width=\linewidth]{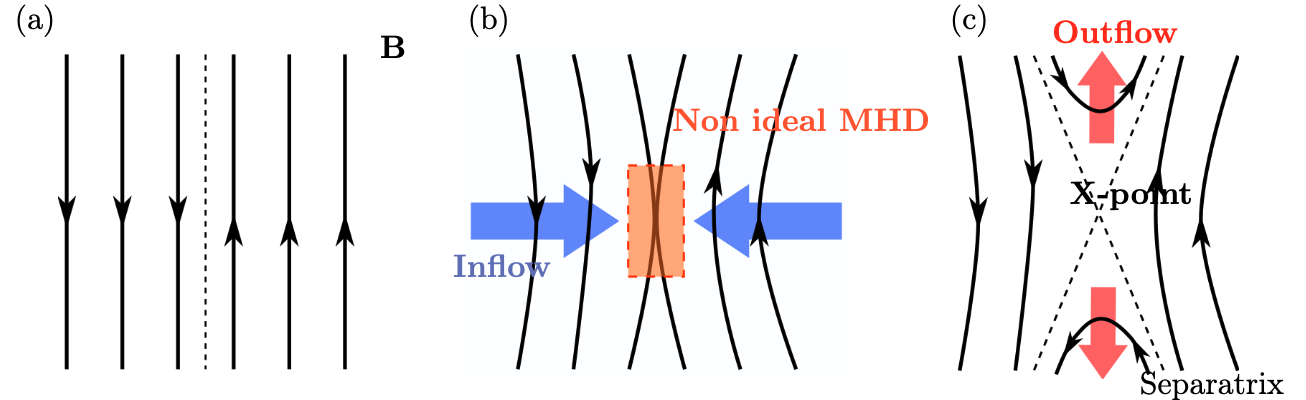}
    \caption{Schematic picture of magnetic field merging. Black lines
represent magnetic field lines.}
    \label{fig:2DReconnection}
\end{figure}

We here present simple illustration of the magnetic reconnection processes, we consider a 2D configuration and describe the system evolution from a large-scale MHD perspective \citep{2000mare.book.....P}. This involves considering two neighbouring regions where magnetic field lines are oriented in opposite directions. There is a current sheet (high $\mathbf{\nabla} B$) corresponding to the change in the magnetic field direction (see Fig. \ref{fig:2DReconnection}a). As the system evolves it enhances the magnetic intensity gradient across this boundary, the current sheet narrows, and the configuration becomes unstable. This intensification of gradients leads to the formation of a non-ideal region where the magnetic field vanishes, see Fig. \ref{fig:2DReconnection}(b). Within this non-ideal region, Eq. \eqref{eq:idealinduction} is no longer applicable. This region is the diffusion region, where the magnetic field can eventually break the freeze-in condition with the plasma. Finally, the magnetic topology is reconfigured, see Fig. \ref{fig:2DReconnection}(c).

In the diffusion region, an additional term in the induction equation needs to be taken into account,

\begin{equation}
  \frac{\partial \mathbf{B}}{\partial t}= \nabla \times (\mathbf{u} \times \mathbf{B}) - \nabla \times (\eta \nabla \times \mathbf{B})
  \label{eq:resistiveinduction} 
\end{equation}

Here, $\eta = c^2 / (4\pi\sigma)$ is the resistive diffusion coefficient (dimensionally it scales as $L^2/T$).  Note that here $\eta$ is the atomic diffusivity due to collision, not to be confused with the anomalous coefficient due to turbulence. Resistivity illustrates the dissipation thus the {\it irreversibility} required for reconnection. Note that Eq. \eqref{eq:idealinduction} can be viewed as the $\eta \rightarrow 0$ limit of the equation for magnetic field evolution in a resistive plasma.

\begin{figure}[h!]
    \centering
    \includegraphics[scale=0.4]{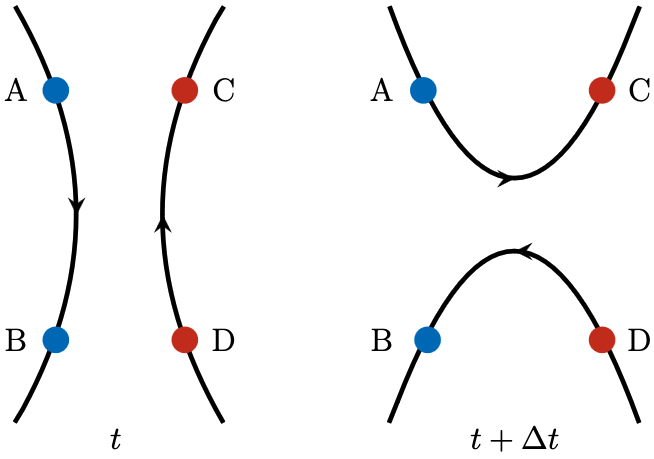}
    \caption{Schematic representation of the change in connectivity operated by the reconnection process. At time $t$, prior to reconnection, the fluid element A and B (C and D) are frozen to the same field line (left panel). After reconnection (right panel), the connectivity has changed and now A and C (B and D) are connected.}
    \label{fig:ReconnectionofDisconnectedRegion}
\end{figure}

The processes within the diffusion region lead to breaking and reconnecting of magnetic field lines, leading to the release of magnetic energy that is converted into kinetic energy of the plasma. Specifically, the plasma accelerates and heats up in the outflow region (Fig. \ref{fig:2DReconnection}c).

Considering now a collisional plasma with uniform resistivity $\eta$, the induction equation and Ohm's law can be expressed as follows,

\begin{equation}
\frac{\partial \mathbf{B}}{\partial t} = \nabla \times (\mathbf{u} \times \mathbf{B}) + \eta \nabla^2 \mathbf{B}, 
\label{eq:inductionequnifresistivity}
\end{equation}
and 
\begin{equation}
    \mathbf{E} + \frac{\mathbf{u}}{c} \times \mathbf{B} = \frac{\mathbf{J}}{\sigma}.
\end{equation}

The non-ideal terms consist of the resistive diffusivity $\eta$ in the induction equation and the resistive current in Ohm's law, both expressed in terms of the electrical conductivity $\sigma = c^2/ 4\pi \eta$.

The non-ideal induction equation highlights the competition between the diffusion of the magnetic field and the ideal evolution. This balance is represented by the magnetic Reynolds number, $R_m \equiv \ell \tilde{u} / \eta$, where $\ell$ is a characteristic length scale and $\tilde{u}$ is a typical velocity of the system. In ideal MHD, $R_m \gg 1$ and reconnection is suppressed. Where $R_m \ll 1$, field diffusivity prevails, making reconnection possible but not mandatory \citep{2000mare.book.....P}.

The MHD description of magnetic reconnection may not provide detailed insights into the kinetic processes within the diffusion region, but it helps in understanding several critical large-scale properties of this process. \\

In summary, magnetic reconnection is a phenomenon induced by the presence of strong gradients and narrow current sheets within a plasma environment. This process is inherently multi-scale, from the macroscopic dynamics of plasma down to the microscopic interactions between individual particles. During reconnection, magnetic energy is converted into both thermal and kinetic energy of the plasma. Furthermore, this process ends up in a reorganization of the magnetic field topology, thereby allowing the mixing of plasma populations that were initially isolated from each other, see Fig. \ref{fig:ReconnectionofDisconnectedRegion}. Understanding magnetic reconnection is of critical importance to the field of particle acceleration. In the context of solar physics, for instance, magnetic reconnection has been identified as a crucial process driving particle acceleration during solar flares \citep{2009ApJ...700L..16D,2022NatRP...4..263J}. Therefore, exploring the complexities of magnetic reconnection in detail is essential for understanding particle acceleration in a variety of astrophysical contexts. Before moving to particle acceleration, in the following sections, we first present the main MHD models developed to explain the process of magnetic reconnection. 

\subsubsection{The Sweet-Parker Model}

\begin{figure}[h!]
    \centering
    \includegraphics[width=0.7\linewidth]{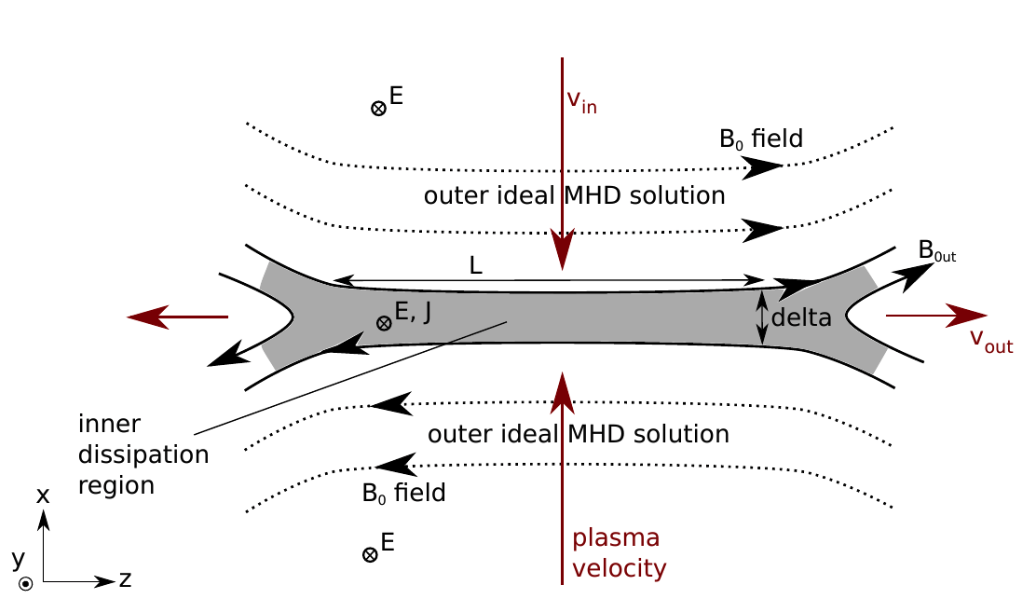}
    \caption{The Sweet-Parker configuration. The plasma from an outer ideal region flows toward a dissipation region where the resistivity is finite, then decouples from the magnetic field, and is expelled from this inner region by the tension force to form outflows. The figure is reproduced from \citet{melzani:tel-01126912}}
    \label{fig:SPReconnection}
\end{figure}

The Sweet-Parker model, is a foundational theory of magnetic reconnection in collisional plasmas initially introduced by \citet{1957PhRv..107..830P}  and subsequently refined by \citet{1958NCim....8S.188S}. In a 2D steady-state scenario summarised in Fig.\ref{fig:SPReconnection}, the plasma from an outer ideal region flows parallel to the x-direction toward a dissipation region with length scale $L$ and thickness $\delta$ \citep{2000mrp..book.....B}. The inflow velocity is determined by the $E \times B$-drift velocity in the plasma, illustrated by the $v_{in}$ red arrows of Fig. \ref{fig:SPReconnection}. In the outer ideal region ($R_m \gg 1$), the plasma is frozen-in to the magnetic flux. However, in the diffusion region where resistivity dominates, the plasma decouples from the magnetic field, enabling the field to reconnect and expel plasma in the z-direction, forming exhausts \citep{2000mrp..book.....B}, illustrated by the $v_{out}$ red arrows of Fig. \ref{fig:SPReconnection} .

By applying mass and energy conservation, neglecting the gas pressure, assuming steady-state and that the field energy dominates at the inflow and the kinetic energy of the particles at the outflow, two important relations can be derived,

\begin{equation}
u_{\text{out}} = \sqrt{2}\frac{ B_0}{\sqrt{4\pi m n_{\text{in}}}}= \sqrt{2} u_{A, \text{in}}
\end{equation}
\begin{equation}
\frac{\delta}{L} = \frac{u_{\text{in}}}{u_{A, \text{in}}} = M_{A, \text{in}},
\end{equation}
where $u_{A, \text{in}}$ is the Alfvén speed in the ideal zone and $M_{A,\text{in}}$ is the Alfvénic Mach number of the inflow. The outflow speed is of the order of the Alfvénic speed of the inflow. The inflow velocity is given by the $E \times B$-drift, $u_{\text{in}} = cE_y/B_0 = \eta/\delta$, and is generally sub-Alfvénic. The Lundquist number, $S_L$, and the reconnection rate, $R$, are defined as
\begin{equation}
S_L \equiv \frac{L u_{A, \text{in}}}{\eta} \sim \left(\frac{L}{\delta}\right)^2 \sim \left(\frac{u_{A, \text{in}}}{u_{\text{in}}}\right)^2 \sim M_{A, \text{in}}^2,
\label{eq:LundquistNumber}
\end{equation}
\begin{equation}
R \equiv \frac{u_{\text{in}}}{u_{\text{out}}} \sim \frac{\delta}{L} \sim S_L^{-1/2}.
\label{eq:sweetParkerReconnectionRate}
\end{equation}
The Lundquist number $S_L$ is equal to the magnetic Reynolds number, $R_m$, when the typical velocity is equal to the Alfvénic speed of the inflow. Highly conducting plasmas have high Lundquist numbers. Laboratory plasma experiments typically have Lundquist numbers between $10^2$ and $10^8$. In astrophysical plasma, they are higher, up to $10^{20}$, resulting in very low Sweet-Parker reconnection rates.

Despite its fundamental role in understanding magnetic reconnection, the Sweet-Parker model has some limitations. One major issue is the slow reconnection rate that it predicts, which is inconsistent with the fast reconnection rates observed in space and astrophysical phenomena such as solar flares and magnetospheric substorms \citep{1997PhPl....4.1936Y,2001ApJ...559..452Z}. 
For example, in solar corona flares, the value of $S_L$ is around \(10^8\), while $u_A$ is approximately \(100\) km/s, and L is about \(10^4\) km. As a result, the time scale according to the Sweet–Parker model is on the order of several tens of days. However, observations indicate that the release of magnetic energy actually occurs within a few minutes to an hour. This inconsistency is referred to as the fastness problem of the Sweet–Parker reconnection.

Additionally, the model assumes steady-state conditions and does not account for non-ideal MHD effects like Hall MHD, two-fluid effects, and plasma pressure anisotropy, which  influence the reconnection process and the reconnection rate \citep{1998JGR...103.9165S, 1999PhPl....6.1781H}.

To address these limitations, several extensions and modifications to the Sweet-Parker model have been proposed. One such extension is the Petschek model, which introduces a new geometry for the reconnection region and predicts faster reconnection rates.

\subsubsection{Petschek Model}

\begin{figure}
    \centering
    \includegraphics[width=0.7\linewidth]{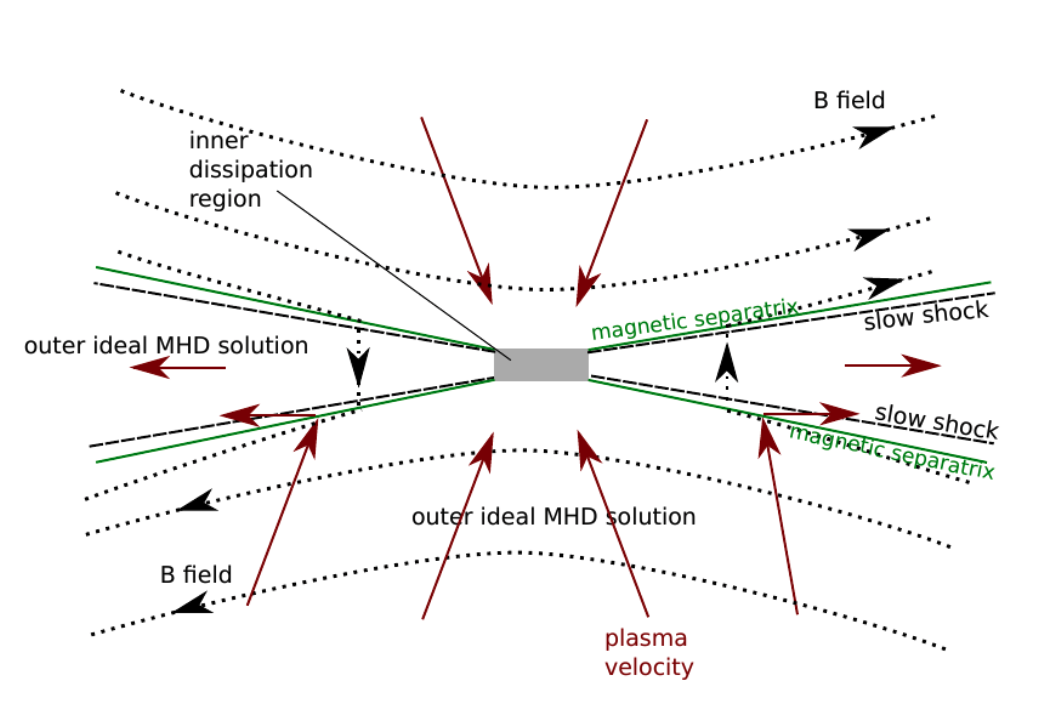}
    \caption{Petschek configuration. The magnetic field is shown as dotted lines. The plasma flows (red arrows) across the magnetic separatrices (green), inside which there are the standing slow-mode shocks. The shocks are switch-off shocks, which means that the tangential component of the magnetic field vanishes after the shock. The figure is reproduced from \citet{melzani:tel-01126912}}
    \label{fig:PetschekReconnection}
\end{figure}

The Petschek model for magnetic reconnection is a fundamental theoretical framework that was developed to address the slow reconnection rates of the Sweet-Parker model. The Petschek model \citep{1964NASSP..50..425P}, introduces a new geometry for the reconnection region, including the presence of a pair of slow-mode shocks, emanating from the diffusion region and separating the inflow and outflow regions \citep{2000mare.book.....P}, see the green lines in Fig. \ref{fig:PetschekReconnection}. Particles can thus be accelerated without having to pass through the inner dissipation region. Instead, magnetic energy can be converted to kinetic particle energy in the shocks. This shock configuration allows a more efficient energy conversion from magnetic to kinetic energy, leading to higher reconnection rates. The reconnection rate in the Petschek model, is expressed as \citep{2000mare.book.....P},

\begin{equation}
    R \equiv \frac{u_{\text{in}}}{u_{\text{out}}} \sim  \frac{1}{\ln{S_L}},
\end{equation}
where again $S_L$ is the Lundquist number defined above. This expression shows that the reconnection rate in the Petschek model is significantly faster than the Sweet-Parker model, as it depends logarithmically on the Lundquist number rather than linearly. Petschek reconnection is thus qualified as fast reconnection.\\

Despite the success of the Petschek model in predicting faster reconnection rates, it still has some limitations. One of the main issues is the difficulty to maintain the assumed steady-state configuration of the slow-mode shocks in the presence of plasma resistivity \citep{1997PhPl....4.1964B}. Some studies have shown that the Petschek model can be unstable to small perturbations or not applicable in certain plasma conditions, such as high Lundquist number plasmas \citep{2009PhPl...16l0702C}. Moreover, numerical simulations conducted years later revealed that for the solutions of Petschek to be physically feasible, they required localised anomalously large resistivity profiles, which might be unphysical. The diffusion region seems to extend and rather adopts a Sweet-Parker like form. This observation is consistent with both theoretical predictions and laboratory experiments. Although Petschek-type solutions continue to raise interest in the field of solar physics, the required very high resistivity has not been detected in laboratory experiments, observations of the magnetosphere or kinetic simulations.

To address these limitations, several extensions and modifications of the Petschek model have been proposed, incorporating additional physical effects such as Hall MHD \citep{2001JGR...106.3759S} or multi-fluid effects \citep{1997PhPl....4.1964B,2023ApJ...946..115W}. These extended models have provided a more comprehensive understanding of the magnetic reconnection process and the factors that influence the reconnection rate, such as the plasma beta \citep{2009ApJ...700L..16D}, collisionality \citep{2011NatPh...7..539D}, and the presence of guide fields \citep{2017PhPl...24b2124S}.

\subsubsection{Turbulent reconnection}

The turbulent reconnection model was first proposed by \citet{1999ApJ...517..700L}, who argued that the reconnection rate should depend on the properties of the turbulence, such as the energy spectrum or the correlation length. They introduced the idea of "reconnection diffusion", where the reconnection process is described as a diffusive process driven by the turbulent motions in the plasma.
\begin{figure}[h!]
    \centering
    \includegraphics[width=0.6\linewidth]{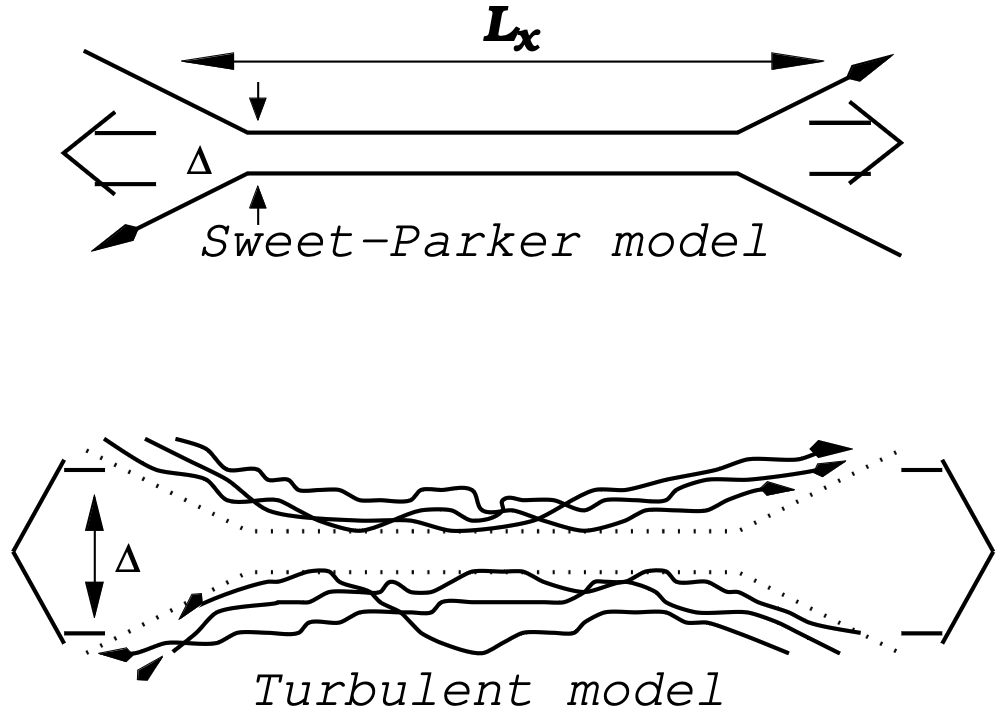}
    \caption {Upper diagram: The Sweet-Parker model represents magnetic reconnection. In this configuration, the outflow thickness ($\Delta$) is determined by microscopic Ohmic diffusivity. The ratio between this microscopic scale and the astrophysical large scale ($L_x \gg \Delta$) determines the reconnection rate. 
    Lower diagram: The \citet{1999ApJ...517..700L} model corresponds to a turbulent reconnection scenario. In this case, the outflow width ($\Delta$) is governed by macroscopic turbulence of the magnetic field lines. This width can be equivalent to the scale $L_x$ in trans-Alfv\'enic turbulence. The figure is reproduced from \citet{Lazarian_2020}.}
    \label{fig:TurbulentReconnection}
\end{figure}

The basic principles of the turbulent reconnection model arise from the interplay between turbulent motions and magnetic fields in plasmas. Turbulence in a plasma leads to the entanglement and stretching of magnetic field lines, which enhances the reconnection process by increasing the effective turbulent resistivity and creating a broadened current sheet see Fig. \ref{fig:TurbulentReconnection}. Furthermore, turbulence forms multiple small-scale diffusion regions that contribute to the overall reconnection rate \citep{2009ApJ...700...63K}.

The presence of turbulence significantly enhances the reconnection rate compared to the classical Sweet-Parker model, which is limited by the plasma resistivity. The reconnection rate is basically enhanced by reducing the magnetic Reynolds number switching from the Ohmic to the turbulent one. In Eq. \eqref{eq:inductionequnifresistivity}, $\eta$ should be replaced by the effective turbulent resistivity $\eta_{T}$, with $\eta_T \gg \eta$. Namely, in the turbulent reconnection model, the reconnection rate is primarily determined by the properties of the turbulence and is independent of the plasma Ohmic resistivity. As a result, the reconnection rate in the turbulent reconnection model can be significantly higher than the Sweet-Parker model prediction, even in highly conducting plasmas.
%\textcolor{magenta}{CS : Do you mention sliding reconnection at some point? I think you mention 3D reconnection below. If you have a guide field, sliding is anavoidable... so probably you do not have to mention it here but you should be more precise on stating that you are still in 2D reconnection?}

The scaling laws in the turbulent reconnection model relate the reconnection rate to the properties of the turbulence. \citet{1999ApJ...517..700L} derived a scaling law for the reconnection rate as follows,

\begin{equation}
R=\frac{u_{in}}{u_{out}} \propto \frac{l_{in}}{L } \frac{u_l}{ u_A},
\end{equation}

where $R$ is the reconnection rate, $u_{in}$ is the inflow velocity, $u_A$ is the Alfvén velocity, $l_{in}$ is the scale of the inflow region, $L$ is the global reconnection scale, and $u_l$ is the turbulent velocity at the scale $l_{in}$. This scaling law demonstrates that the reconnection rate in the turbulent reconnection model is independent of the plasma resistivity and depends solely on the properties of the turbulence.

Recent research has further investigated the scaling laws in the turbulent reconnection model by exploring the dependence of the reconnection rate on various plasma parameters. The plasma parameters studied were the magnetic field strength, plasma density, and the amplitude of the turbulent fluctuations, see \citet{2020PhPl...27a2305L} and references therein for a review. These studies have provided a more comprehensive understanding of the scaling laws in the turbulent reconnection model, revealing a complex interplay between turbulence and magnetic reconnection in various plasma regimes. 

Numerical simulations incorporating turbulence have shown that the reconnection rates are indeed significantly enhanced, reaching values of up to $0.1$ or higher, which is an order of magnitude faster than the Sweet-Parker model prediction \citep{2012NPGeo..19..297K}. These simulations also demonstrated that the presence of turbulence affects the overall dynamics of the reconnection process, leading to the development of instabilities forming plasmoids, secondary magnetic islands, and turbulent fluctuations in the current sheet \citep{2013ApJ...771..124Z}. We discuss now magnetic reconnection where the current sheet develop these instabilities.

\subsubsection{Multiple X-line reconnection}
\begin{figure}[h!]
    \centering
    \includegraphics[width=0.85\linewidth]{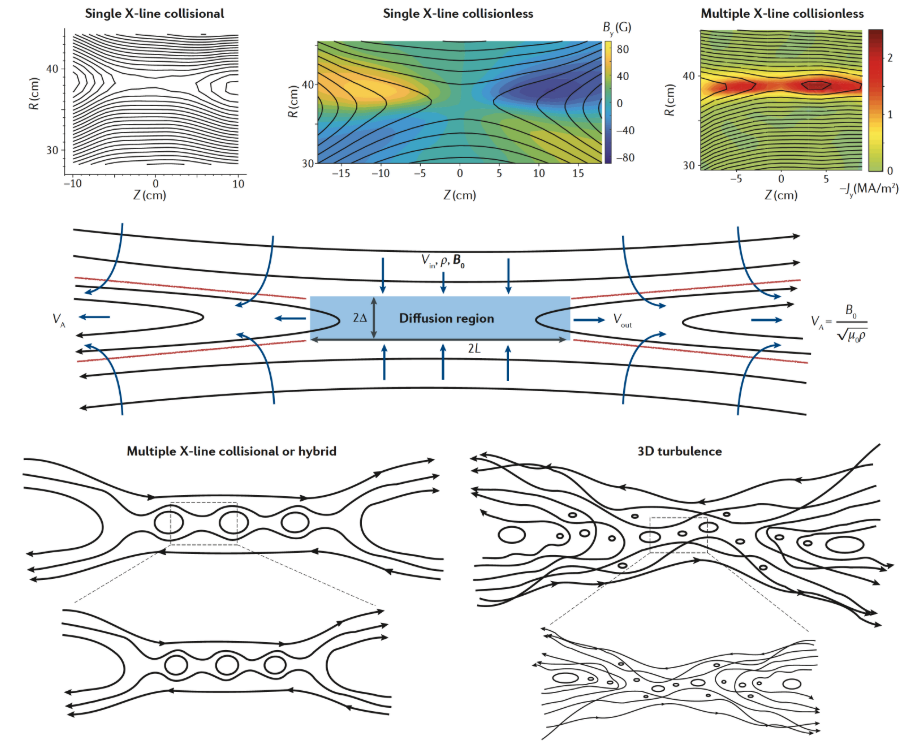}
    \caption{The top panels illustrate phases of magnetic reconnection confirmed through laboratory experiments: a single X-line in a collisional setting, a single X-line in a collisionless environment, and multiple X-lines in a collisionless context. These data were captured using the Magnetic Reconnection Experiment (MRX) and are presented in radial (R) and axial (Z) coordinates. The out-of-the-plane magnetic field (B$_Y$ ) and current density (J$_Y $) are also depicted. Bottom panels provide conceptual illustrations of plasmoid mediated reconnection and 3D turbulent reconnection, both demonstrating some degree of self-similarity. This figure is reproduced from \citet{2022NatRP...4..263J}}
    \label{fig:ReconnectionRegime}
\end{figure}

The next step in the understanding of magnetic reconnection was enabled by the increased scale of simulations. These simulations demonstrated that the Sweet-Parker model was unstable on astrophysical scales. And that such instabilities allow for multi-scale couplings \citep{shibata2001plasmoid}. Specifically, linear tearing instability theory indicates that Sweet-Parker layers are inherently prone to develop secondary magnetic structures, or plasmoids, when the Lundquist number \(S_L\) exceeds a critical value \(S_c\) \citep{comisso2016general}. This critical value is influenced by flow shear and convection time through the layer. While linear theory did not predict this, numerical simulations began to show plasmoid formation at \(S_c \approx 10^4\) \citep{bhattacharjee2009fast,samtaney2009formation}, corresponding to diffusion regions where \(L/\Delta \approx \sqrt{S_c} \approx 100\).

The tearing instability (also called plasmoid instability) has been studied in earlier work \citep{biskamp1982effect} but its importance was overlooked due to the low Lundquist numbers in initial simulations. As the Lundquist numbers grow larger in simulations, the tearing instability grow stronger \citep{shibata2001plasmoid}, resulting in the fragmentation of the layer into numerous plasmoids, each separated by new current sheets, see Fig. \ref{fig:ReconnectionRegime} bottom left panel. These new sheets can also be unstable to plasmoid formation if \(S_L > S_c\), based on the new local parameters. 

In the plasmoid-dominated regime, both MHD simulations \citep{comisso2016general} and theoretical models \citep{uzdensky2010fast} suggest that the reconnection rate \(R\) becomes independent of \(S_L\), with \(R \approx 0.01 \approx 1/\sqrt{S_c}\). In this context, the thickness of each newly-formed layer scales as \(\Delta \propto \sqrt{\eta L}\), where \(L\) is the length of the original current sheet. This results in a quick cascade to smaller scales, which will stop either when the new layer is stable or when its thickness reaches the ion kinetic scale, then triggers kinetic reconnection.

This has been verified through kinetic simulations, both in 2D \citep{daughton2009transition} and 3D \citep{stanier2019influence}, which show that plasmoids can evolve into extended 3D flux ropes that can engage in more complex interactions that are absent in 2D. Both kinetic and MHD simulations indicate that turbulence within 3D reconnection layers is intrinsically self-generated due to the dynamics of these flux ropes.

\subsection{Magnetic reconnection in the collisionless regime and particle acceleration}\label{sec:kineticapproach}

In the preceding section, we discussed how the collisional MHD approach provides some understanding of magnetic reconnection. However, this method neglects the kinetic processes taking place within the reconnection zone. In this section, we start by outlining the criteria that establish which magnetic reconnection regime is applicable based on the properties of the plasma. Our aim is to pinpoint the regime relevant to reconnection events in T Tauri flares. Following this, we do a concise overview of the collisionless kinetic approach, which helps to elucidate some of the mechanisms responsible for particle acceleration to supra-thermal energies.

\subsubsection{Reconnection regime criteria}\label{sect:ReconnectionRegime}
We showed in Sect. \ref{sect:ClassicalMHDTheories} that magnetic reconnection can be described by resistive MHD. Resistive MHD reconnection models are parameterised solely by the dimensionless Lundquist number $S_L$ defined in Eq. \eqref{eq:LundquistNumber}. We recall it here,
\[
    S_L= \frac{L_{CS} V_A}{\eta},
\]
where $L_{CS}$ is the half-length of the reconnection current sheet, defined as $L_{CS} = \epsilon L$, where $L$ is the plasma size and $0 \le \epsilon \le 1/2$, $V_A$ is the Alfvén velocity as a function of the reconnecting magnetic field component (in the 2D plane), and $\eta$ is the plasma resistivity due to coulomb collisions.

As seen in Sect. \ref{sect:ClassicalMHDTheories}, the Sweet-Parker and Petschek models predict reconnection rates as an explicit function of $S_L$. When collisions are sufficiently rare or $S_L$ is large, physics beyond resistive MHD becomes indispensable. In this regime, reconnection rates are high and almost independent of $S_L$. This is why most of the works in recent decades have focused on reconnection in the collisional or collisionless limit. The term collisionless refers to the reconnection process dominated by ion and electron kinetic effects.

The collisional MHD description provides a satisfactory explanation for the magnetic reconnection in plasmas where all the resistive layers remain larger than the kinetic scale of the ions. \citet{2005PhRvL..95w5002C,2007ApJ...671.2139U} proposed the following condition to determine if the MHD description of reconnection is valid or if kinetic processes should be accounted for. They proposed that the MHD description of reconnection is valid when the diffusion region thickness is larger than the kinetic scale. The diffusion region is coloured in grey in Fig. \ref{fig:SPReconnection}, its thickness is called $\delta$. Where $\delta$ can be expressed in the Sweet-Parker model from Eq. \eqref{eq:sweetParkerReconnectionRate}, 

\begin{equation}
    \delta=\frac{L_{CS}}{\sqrt{S_L}}.
\end{equation}
The relevant kinetic scale depends on the presence of a guiding magnetic field or not (i.e. the part of the magnetic field which does not enter in the reconnection process, see Sec. \ref{sect:constrainingmaxenergy} for a discussion about this notion). In case no guiding field is present the relevant 
kinetic scale is the ion skin depth, 
\begin{equation}
d_i = {c \over \omega_{pi}} \ ,   
\end{equation}
where $\omega_{pi}= 4\pi q_i^2 n_i / m_i$ is the ion plasma frequency.
On the other hand, in case a guiding field is present the relevant kinetic scale is the kinetic scale is the ion sound radius,
\begin{equation}
    \rho_s\equiv{\frac{\sqrt{k(T_i+T_e)m_i c^2}}{q_i B}},
\end{equation}
where $T_e$ and $T_i$ are the electron and ion temperatures, $B$ the magnetic field, and $n_i$, $m_i$ and $q_i$ are the ion density, mass and charge, respectively. Let us move with the guiding field case hereafter. 
%Or the kinetic scale can be the ion skin depth, the depth in a plasma to which low-frequency electromagnetic radiation can penetrate, 
%\begin{equation}
 %   d_i\equiv c/\omega_{pi}
%\end{equation}
%where $c$ is speed of light and $\omega_{pi} $ is the ion plasma frequency. Note that \( \rho_i \approx d_i (\beta/2)^{1/2} \), where \( \beta = \frac{8\pi n_o (T_i + T_e)}{B_{tot}^2} \). In the case of a weak guide field, there is no out of plane component of the magnetic field, the force balance across the layer dictates that \( \beta = 1 + \beta_{\text{up}} \), where \( \beta_{\text{up}} \) is the beta parameter just upstream of the layer. For most scenarios of interest, \( \beta_{\text{up}} \approx 1 \), which implies that \( \rho_i \approx d_i \) when guide fields are weak.

The predicted transition criteria is the following: if $\delta<\rho_s$ then reconnection should be treated with the collisionless kinetic approach while if $\delta>\rho_s$ then the collisional MHD approach is valid. This transition from collisional to collisionless leads to significant differences in the structure of the reconnection layer, as well as changes in the magnitude and scale of the reconnection rate. 

This qualitative change can be characterised by the effective plasma size, 
\begin{equation}
    \lambda = \frac{L}{\rho_s}.
\end{equation}
So the Lundquist number can be rewritten as,
\begin{equation}
    S_L = \epsilon ^2 \lambda^2.
    \label{eq:colisionalreconnectioncondition}
\end{equation}

%If in the upstream plasma $\beta_{up} \ll 1$, then $\rho_S$ is equal to $\delta$ due to the balance of forces across the current sheet. The boundary between collisional and collisionless reconnection is defined by Eq.\eqref{eq:colisionalreconnectioncondition}, where the effective plasma size is

\begin{figure}
    \centering
    \includegraphics[width=0.6\linewidth]{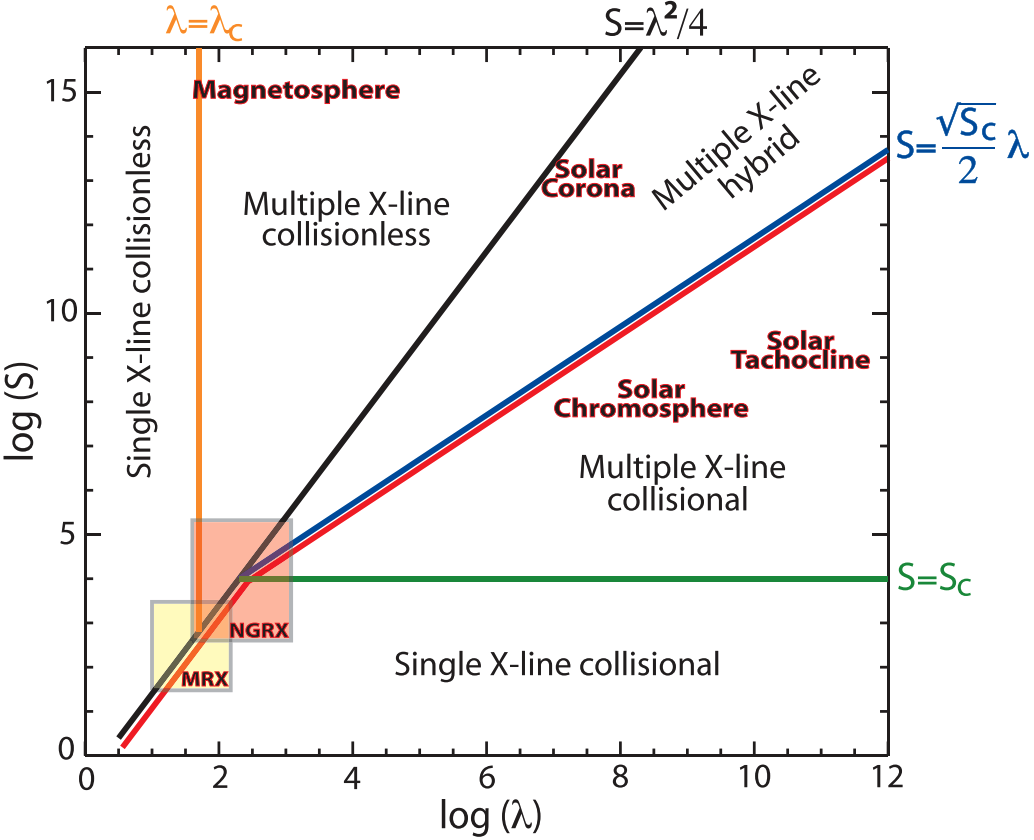}
    \caption{Regimes of magnetic reconnection, as established by \citet{2011PhPl...18k1207J}. $S$ refers to the Lundquist number (called $S_L$ in the text), see Eq. \eqref{eq:LundquistNumber}, $\lambda$ represents the normalised size of the macroscopic system. The red curve is calculated with a plasma beta $\beta=0.2$ and a magnetic reconnection rate of $R = 0.05$. The figure is reproduced from \citet{2011PhPl...18k1207J}.}
    \label{fig:ReconnectionParameterSpace}
\end{figure}

Thus, collisional and non-collisional reconnection regimes are delimited by the black line in the parameter space of $(\lambda, S_L)$ in Fig. \ref{fig:ReconnectionParameterSpace}. This phase diagram of magnetic reconnection illustrates different regimes of reconnection as a function of two dimensionless parameters, the Lundquist number $S_L$ and the normalised size $\lambda$. If $S_L$ (resp.  $\lambda$) is small, the reconnection regime is collisional (resp. collisionless) with a single X-line (as in Sweet-Parker reconnection, see Sect. \ref{sect:ClassicalMHDTheories}).
%\textcolor{red}{cela veut dire quoi (Petschek) ?}.
In addition, the phase diagram, plotted in Fig. \ref{fig:ReconnectionParameterSpace}, also shows the different regions corresponding to reconnection scenarios in different astrophysical environments. It shows reconnection regions in the Earth  magnetosphere, the solar corona, the solar chromosphere and the solar tachocline. 

When $S_L$ and $\lambda$ are large, the scenario changes. Three new phases of multiple reconnection of X-line appear, characterised by the presence of magnetic islands, see Fig \ref{fig:ReconnectionRegime}. The dynamics of the new current sheets between these islands is determined either by collisional or non-collisional physics. This underlines the importance of understanding both regimes and the transition between them.

In section \ref{sec:ReconnectionRegimeinTTauri}, we will use this diagram to determine the reconnection regime during T Tauri flares. However, to that aim we must first constrain the plasma properties and reconnection parameters of T Tauri flares.

\subsubsection{Maxwell-Vlasov equations}

We do now a short overview of the kinetic approach to magnetic reconnection, which will allow us to have an understanding of the microphysical process at play in the diffusion region. The strength of microphysical studies lies in their foundation on first principle physics, allowing them to sidestep challenging questions, such as: which equations from the MHD family are the best for modelling reconnection events? - or - what values should be assigned to transport coefficients like resistivity, viscosity, and Hall parameters? These considerations are naturally accounted for by solving the kinetic Vlasov–Maxwell equations. At lengthscale smaller than $\rho_i$, in the collisionless regime, ions decouple from electrons, causing the magnetic field to be "frozen" into the electron fluid rather than the overall plasma. This leads to parameters other than just resistivity influencing the Ohm's Law. A more complex version of the Ohm's Law for electrons in the context of a two-fluid, non-relativistic plasma model can found in \citet{melzani:tel-01126912},

\begin{align}
    \textbf{E} + \frac{\textbf{v}_i}{c} \times \textbf{B} &= \frac{1}{n_e e} \textbf{J} \times \textbf{B} - \frac{m_e}{e} \left( \frac{\partial \textbf{v}_e}{\partial t} + \textbf{v}_e \cdot \nabla \textbf{v}_e \right) - \frac{1}{n_e e} \nabla \cdot P_e \\
    &+ \frac{\chi}{(n_e e)^2} \textbf{J} + \frac{\chi_e}{n_e e} \nabla^2 \textbf{v}_e.
\end{align}

Where \( n_e \) denotes the electron number density, \( \textbf{v}_i \) and \( \textbf{v}_e \) are the ion and electron velocities, \( c \) is the speed of light, and \( e \) is the elementary charge. The term \( \chi \) accounts for the effect of collisions between ions and electrons, which can be anisotropic and dependent on the magnetic field orientation. \( \chi_e \nabla^2 \textbf{v}_e \) describes the electron viscosity due to electron-electron collisions. \( P_e \) is the pressure tensor, defined as:

\[
P_e = \int d^3 \textbf{v} \, m_e (v_a - \bar{v}_a)(v_b - \bar{v}_b)
\]
where \( a, b = x, y, z \), and \( \bar{v} \) is the mean particle velocity.

Both thermal and bulk inertia of electrons contribute to the non-ideal terms in this formulation. Importantly, when the plasma is entirely collisionless (\( \chi = \chi_e = 0 \)), these inertia terms are the sole contributors to the non-ideality of the plasma.

\begin{figure}
    \centering
    \includegraphics[width=\linewidth]{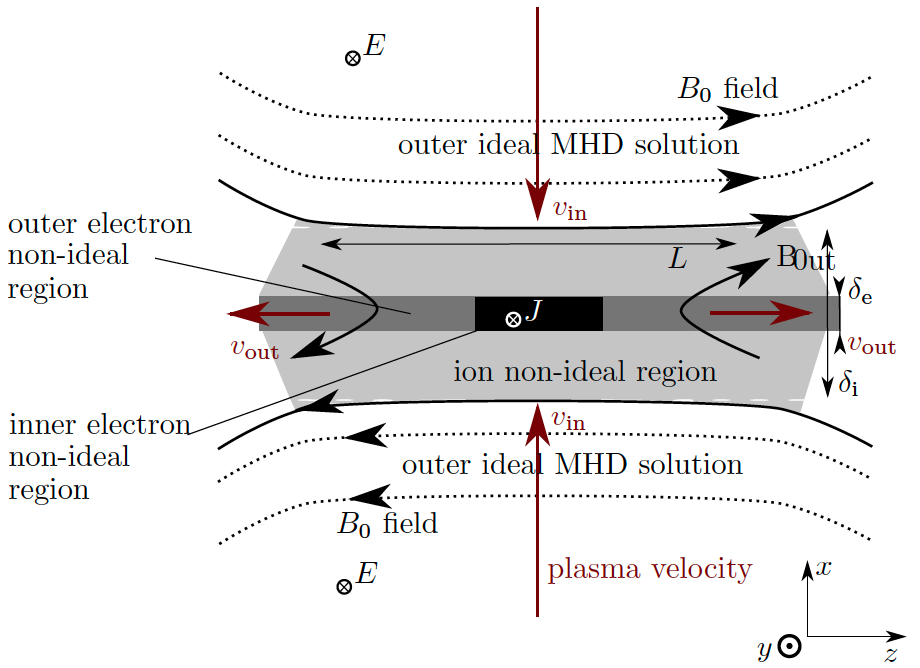}
    \caption{Collisionless Magnetic Reconnection in an electron-ion plasma. The figure illustrates reconnection on scales smaller than the ion inertial length. In this case, the diffusion region for ions are considerably larger than those for electrons. The structure of the current sheet resembles an X-point more than a double Y-point. Typically, ion trajectories do not intersect the electron diffusion region. The figure is reproduced from \citet{melzani:tel-01126912}.}
    \label{fig:ionelectronnonidealregion}
\end{figure}

Figure \ref{fig:ionelectronnonidealregion} reveals that the diffusion region is now divided into two regions: a larger ion diffusion region with dimensions denoted by \( \delta_i \), and a smaller electron diffusion region \( \delta_e \). In this context, \( \delta_{i,e} \) represents the ion and electron inertial lengths. At these scales, the Hall effect becomes significant. Specifically, the magnetic field lines are now carried along by the electron motion, while the ions no longer follow this movement. Although the Hall term does not directly cause magnetic reconnection, it is a subject of ongoing debate whether it may contribute to the efficiency of the reconnection process. This is because the Hall effect has the potential to accelerate electrons to higher speeds, thereby increasing their bulk inertia.

The Particle-in-Cell (PIC) method is the numerical approach most commonly used in astrophysics for this purpose. However, this approach comes at a cost: the computational requirements for solving kinetic equations are far greater than for MHD models \citep{2020LRCA....6....1M}. Kinetic simulations, despite their computational intensity, are limited to examining localised phenomena on scales up to a few thousand electron inertial lengths or about a hundred proton inertial lengths. While this is sufficient for studying specific features like a single X-point or a small chain of plasmoids, it is inadequate for exploring how these reconnection sites fit into a larger-scale context or how they originated. The ways to address this problem involve hybrid approaches that combine MHD models with localised particle models or improved MHD solutions through the propagation of test particles.

\subsubsection{Acceleration Mechanisms}\label{Sect:AccelerationSites}

A large amount of evidence points towards the role of magnetic reconnection in the process of non-thermal particle acceleration and the creation of power-law energy spectra observed in space and solar plasmas \citep{2006Natur.443..553D,2018SSRv..214...82O}. Several studies, both analytical and numerical, shed light on the underlying mechanisms that govern particle acceleration in magnetic reconnection \citep{2001JGR...106.3715B,2011NatPh...7..539D,2021PhPl...28e2905L}. 

Understanding the mechanisms and spatial transport of particle acceleration in the context of 3D reconnection is currently a very active research topic \citep{2014PhPl...21l2902S,2020PhPl...27j0601D}. More specifically, the aim is to understand the processes that allow magnetic reconnection to enable particles to transition from thermal to non-thermal energies \citep{2022PhPl...29d2902H}. As of now, there is no definitive understanding of the specific physical mechanism responsible for the acceleration of non-thermal particles during magnetic reconnection. While various plausible theories have been proposed, and some mechanisms have been identified as being effective, multiple processes are likely at play simultaneously. We here outline the most commonly discussed acceleration processes. We aim to offer a general overview of these mechanisms. We will then select and discuss in Sec. \ref{sec:magnrecTTauriFlare} the mechanisms expected to be specifically relevant in T Tauri flares. \\ 

Three primary particle acceleration processes dominate \citep{2022NatRP...4..263J},

\begin{itemize}
    \item Fermi acceleration, which occurs as particles stream along and drift in relaxing curved magnetic field lines.
    \item Localised electric fields $E_{\parallel}$ parallel to the magnetic field directly accelerating particles.
    \item Betatron heating, which happens as particles drift into regions of stronger magnetic fields while conserving their first adiabatic moment $\mu = \frac{mv_{\perp}^2}{B}$.
\end{itemize}

These mechanisms are described by guiding centre equations of motion for the particles. In this reference frame, the energisation rate in the non-relativistic limit can be approximately by the following equation \citep{2022NatRP...4..263J},

\begin{equation}
\frac{d{\cal E}}{dt} = q E_{\parallel} v_{\parallel} + \mu\frac{dB}{dt} + q\mathbf{E} \cdot \mathbf{u} + \frac{1}{2} m\frac{d}{dt} |\mathbf{u}_{\mathbf{E}\times \mathbf{B}} |^2,
\label{eq:EnergyRateReconnection}
\end{equation}
where ${\cal E}$ is the energy of a particle, $\textbf{u}_{\textbf{E}\times \textbf{B}}$ is the $\textbf{E} \times \textbf{B}$ drift velocity defined in Eq. \eqref{eq:DriftVelocity}, the total derivative is $\frac{d}{dt} = \frac{\partial}{\partial t} + \textbf{u}_{\textbf{E}\times \textbf{B}} \cdot \nabla$, $E_{\parallel}$ is the parallel component of the electric field, and $v_{\parallel}$ is the drift-corrected guiding center parallel velocity.\\

\begin{figure}[h!]
    \centering
    \includegraphics[width=\linewidth]{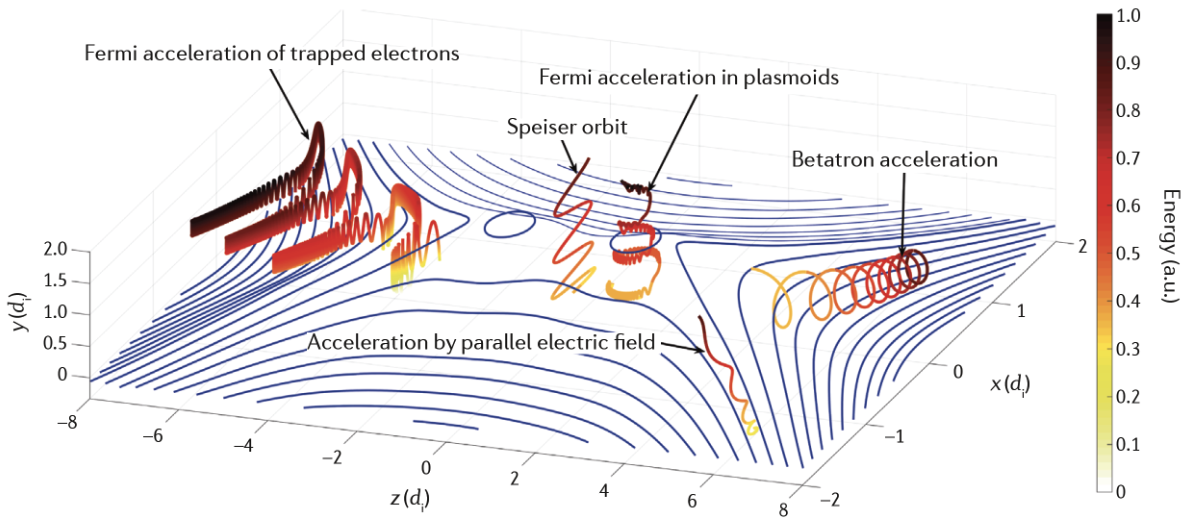}
    \caption{Illustration of particle acceleration mechanisms. The reconnection electric ﬁeld is along the $y$ direction. Figure is reproduced from \citet{2022NatRP...4..263J}.}
    \label{fig:AccelerationMechanismReconnection}
\end{figure}

Figure \ref{fig:AccelerationMechanismReconnection} details the mechanisms of particle acceleration. The acceleration caused by the reconnection electric field takes place near the X-line due to the parallel electric field near the separatrices. Electrons (resp. ions) are propelled in the direction opposite to (resp. along) that of the electric field. This generates a positive direct acceleration term, $qE_\parallel v_\parallel$, which stands as the first term on the right-hand side of Eq. \eqref{eq:EnergyRateReconnection}. 

The betatron acceleration mechanism occurs where particles move into the outflow regions with increasing magnetic fields intensity, see Fig. \ref{fig:AccelerationMechanismReconnection}. The particles travel in the positive z direction due to the $\textbf{E}\times \textbf{B}$ drift. In this direction, the magnetic field lines come closer together, so the magnetic field increases, i.e. $\nabla B$ is in the direction of positive z. Thus the second term on the right-hand side of Eq. \eqref{eq:EnergyRateReconnection} is positive.

Also, the inflowing field components drift toward the reconnection region due to the $\textbf{E}\times \textbf{B}$-drift. \citet{lazarian2005production} suggest that this motion can potentially serve as a mechanism for particle acceleration through a first-order Fermi process. This would occur if particles could traverse the reconnection region and oscillate between the lower and upper inflow areas into the sheet. They propose that this acceleration mechanism would yield a particle energy spectrum of \(N(E) \propto E^{-5/2}\).

Fermi acceleration occurs in areas with highly curved magnetic field lines. In Fig. \ref{fig:AccelerationMechanismReconnection}, plasmoids stretch along the z direction, leading to the highest curvature of field lines at $x = 0$. Inside the plasmoid, the particles experience a significant velocity boost in the $y$ direction due to a strong curvature drift, and gain energy from the reconnection electric field, $q \mathbf{E} \cdot \mathbf{u} > 0$. In the outflow area, the third term on the right-hand side of Eq. \eqref{eq:EnergyRateReconnection} is positive. The trapped particles gain energy each time they traverse the regions of high curvature located at $x = 0$.\\

We now turn our attention on extended current sheets that break into plasmoids, thereby enabling a diverse range of potential mechanisms for particle acceleration. In a sufficiently long sheet, multiple X-points form due to the tearing instability, each with their own exhaust regions. These exhausts from neighbouring X-points collide and give rise to plasmoids, areas of plasma enclosed by a strong, closed magnetic field loop (see Fig. \ref{fig:ReconnectionRegime}). The formation of these plasmoids, particularly at the interfaces between adjacent exhausts serves for Fermi II particle acceleration due to the random approach of magnetic mirrors, as described, for example, in \citet{lapenta2015separatrices}. 

Inside a plasmoid, the magnetic field tends to zero, and thermal plasma is confined within, unable to escape due to the strong, surrounding magnetic field that redirects particles back into the plasmoid interior. As reconnection continues, these plasma "islands" may enlarge, and their enclosing magnetic fields may intensify. Plasmoids can also merge and expand in size, and post-merger contraction can be a significant mechanism for particle acceleration.

Additionally, as outlined in Sec. \ref{sect:ClassicalMHDTheories}, the fracturing of current sheets due to tearing mode instabilities, or a mixture of tearing and kink modes in three spatial dimensions, creates turbulence in the region where magnetic reconnection occurs. This induces turbulence, along with the existing turbulence in the inflows, consistent with the observed speed of collisional reconnection \citep{2009ApJ...700...63K,2012NPGeo..19..251L}. In the context of test particles, it has been demonstrated that both MHD fast and slow compressible modes can accelerate energetic particles via second-order Fermi acceleration. \citet{zhdankin2018system} explored forced turbulence through PIC simulations and found that particles can indeed achieve extremely high energies even in scenarios involving collisionless turbulence. \citet{zhdankin2018system} compute the temporal evolution of the accelerated particle spectrum, which eventually forms a power law in the energy distribution of particles, scaling with the Lorentz factor as \( \gamma^{-3} \).\\

The recent reviews by \citet{2020LRCA....6....1M} and \citet{2022NatRP...4..263J} provide deeper discussions of these processes. Additionally, numerical simulations based on first principles provide insights into the current understanding of electron acceleration during magnetic reconnection in the non-relativistic regime \citep{2020PhPl...27j0601D,2021PhPl...28e2905L}. They suggest that power-law non-thermal particle populations are efficiently produced in 2D and 3D. \\

The study of particle acceleration and transport in large-scale reconnection layers, such as solar flares, presents considerable challenges, such as anticipating the impacts of plasma parameters and simulation setup on the results or understanding the properties of turbulence related to particle acceleration and transport in 3D reconnection, as well as understanding how ions are accelerated. 

\section{Reconnection in Solar Flares}\label{sec:solarflares}
Solar flares provide information about high-energy processes in otherwise inaccessible stellar atmospheres. More importantly in our present context, the study of solar flares provides an essential basis for understanding the flaring activity of T Tauri stars.

\subsection{Multiwavelength observations of solar flares }
\begin{figure}
    \centering
    \includegraphics[width=0.8\linewidth]{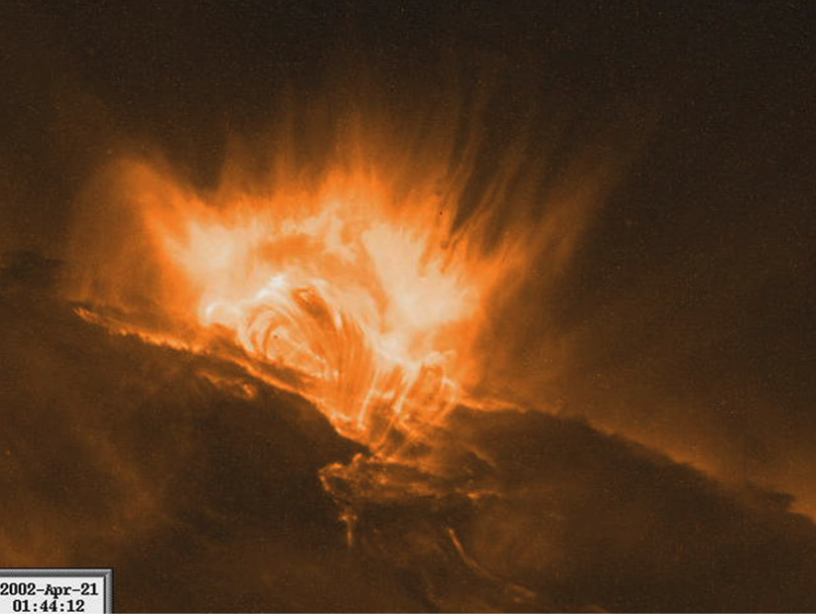}
    \caption{A flare observed in the Fe XII line at 195 \AA (sensitive to 1.5 MK plasma) by the TRACE satellite. (credit: TRACE Project, NASA)}
    \label{fig:SolarCoronalFlareObservation}
\end{figure}

Solar flares are transient phenomena triggered by a sudden release of magnetic energy on the solar surface. They emit a wide range of photons from radio waves to high-energy gamma rays (refer to reviews by \citealt{2011LRSP....8....6S,2017LRSP...14....2B}). This occurs through multiple emission processes. The main origin of these processes, magnetic reconnection, heats up corona and chromosphere plasma, resulting in bright optical, UV, and soft X-ray emissions, see Fig \ref{fig:SolarCoronalFlareObservation}. 

\begin{figure}[h!]
    \centering
    \includegraphics[width=0.6\linewidth]{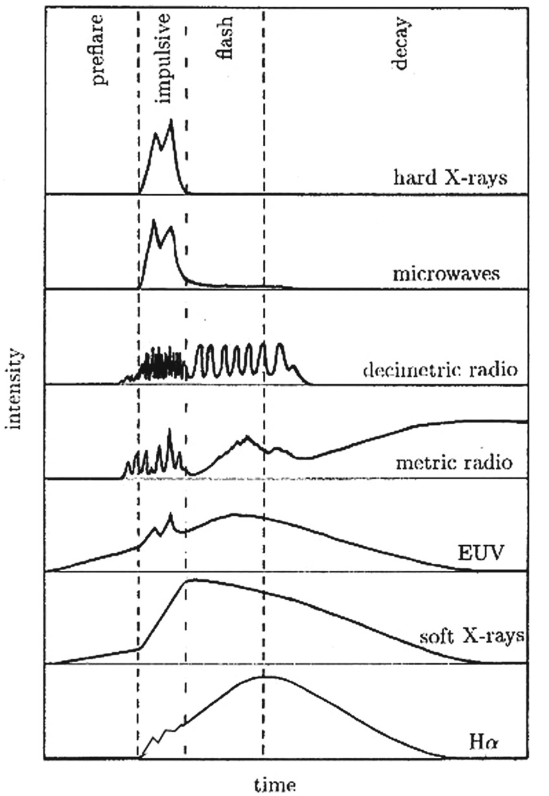}
    \caption{A schematic profile of the flare intensity at several wavelengths. The various phases indicated at the top greatly vary in duration. In a large event, the preflare phase typically lasts a few minutes, the impulsive phase 3–10 min, the flash phase 5–20 min, and the decay one to several hours. Image reproduced from \citet{benz2002kinetic}.}
    \label{fig:FlarePhase}
\end{figure}

Magnetic reconnection also accelerates non-thermal electrons, leading to the emission of hard X-ray and radio signals \citep{1994Natur.371..495M,2015Sci...350.1238C,2018SSRv..214...82O}. The Fermi satellite has also observed high-energy gamma rays ($E_\gamma > 100$ MeV) from solar flares \citep{2021ApJS..252...13A}. This emission happens in two phases: an impulsive phase that lasts around 3–10 minutes during which the soft X-ray luminosity is increasing, followed by a phase of gradual decrease in soft X-ray luminosity lasting roughly 2–20 hours. See Fig. \ref{fig:FlarePhase} for a schematic representation of these phases.

Fig. \ref{fig:FlarePhase} shows the timing of various emissions during a solar flare. In the preflare phase, the flare region coronal plasma slowly heats up and is visible in soft X-rays and EUV. However, non-thermal emissions sometimes occur in different locations from later phases.

During the impulsive phase, large numbers of energetic electrons and sometimes ions get accelerated. This phase occurs when most of the energy gets released. The appearance of hard X-ray footpoint sources at chromospheric altitude is a feature of this phase. Some high-energy particles get trapped and emit intensively in the radio band.

In the post impulsive phase, thermal soft X-ray and H$_\alpha$ emissions reach their maximum. Energy is gradually released during this period and further distributed. The rapid increase in H$_\alpha$ intensity and line width is known as the flash phase. It coincides with the impulsive phase, although H$_\alpha$ may peak later.

In the decay phase, the coronal plasma returns nearly to its original state, except in the high corona where magnetic reconnection, plasma ejections, and shock waves continue. These processes accelerate particles, resulting in meter wave radio bursts and interplanetary particle ejections. 
%Figure \ref{fig:FlarePhase} outlines these various phases.

In both phases, the high-energy gamma-ray flux surpasses the extrapolation of the hard X-ray power-law spectrum (e.g., \citealt{2017ApJ...835..219A}). This suggests that gamma rays and hard X-rays have different production mechanisms. The creation of gamma rays might be due to hadronic emission by high-energy protons \citep{2023ApJ...944..192K}. This implies that solar flares might boost protons to GeV energies.

However, the exact mechanisms for proton acceleration and resulting gamma-ray emission are not yet clear. The extended duration of the gradual phase could indicate ongoing proton acceleration, potentially linked to coronal mass ejection and the interaction with surrounding plasma \citep{2014ApJ...789...20A,2016ApJ...830...28P}. A recent multi-wavelength analysis points to multiple particle acceleration sites during solar flares, with the termination shock at the loop top possibly acting as a proton acceleration site during the impulsive phase \citep{2021ApJ...915...12K}. Non-thermal electron production at the termination shock is widely discussed \citep{1994Natur.371..495M,2015Sci...350.1238C}, but detailed investigation of proton acceleration and gamma-ray emission there is still required.

\subsection{Solar flares standard model}

\subsubsection{Solar flare geometry}
\begin{figure}[h!]
    \centering
    \includegraphics[width=1\linewidth]{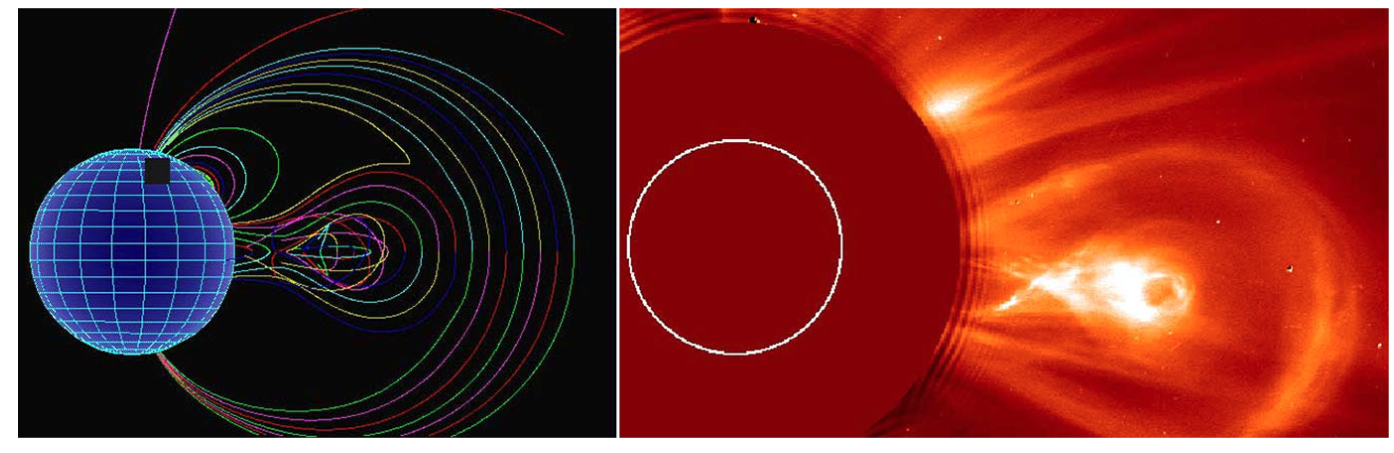}
    \caption{Left: a schematic drawing of the one-loop flare model. Right: observation of an apparent X-point behind a coronal mass ejection observed by LASCO/SOHO in white light (copyright by NASA)}
    \label{fig:OneLoopFlareModelObservation}
\end{figure}
\begin{figure}[h!]
    \centering
    \includegraphics[width=1.\linewidth]{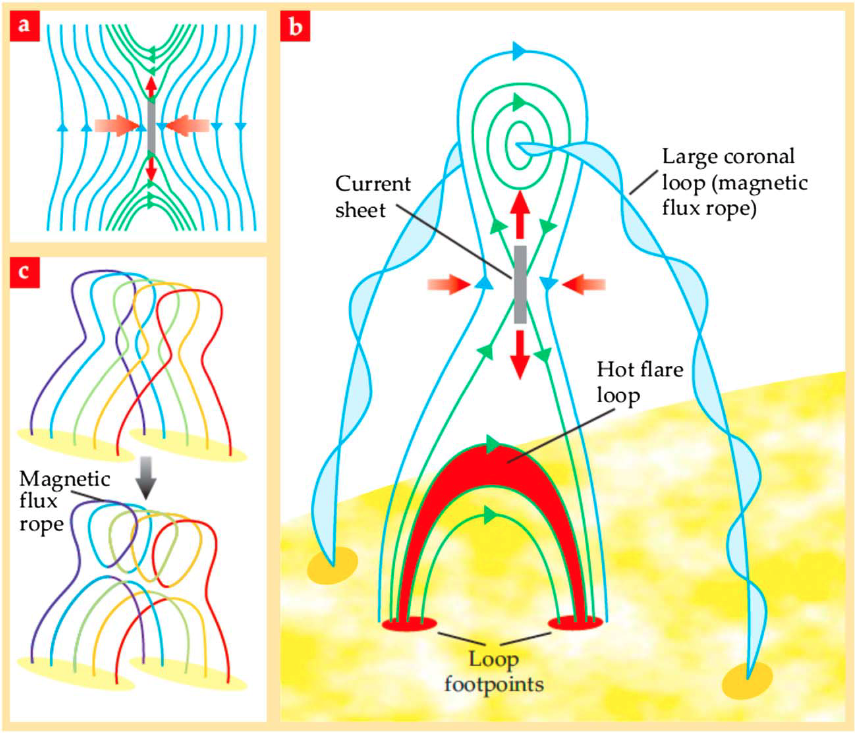}
    \caption{The Standard Model illustrates the magnetic progression of a Solar Eruptive Event (SEE). (a) It depicts the reconnection process where the inflowing magnetic field is shown in blue and the outflowing field in green. (b) This reconnection process leads to the formation of the new flare loop below and the magnetic flux rope above. (c) This shows how the sheared loops undergoe reconnection to produce the flux rope and a less sheared flare loop. Reproduced from \citet{10.1002/2016JA022651}.}
    \label{fig:SolarFlareStandardModel}
\end{figure}

The general consensus is that the energy released in an impulsive reconnection during a solar flare originates from sheared or anti-parallel magnetic fields. However, the geometry of these magnetic fields on a large scale in the corona remains elusive. The dominant theory is depicted in Figs. \ref{fig:OneLoopFlareModelObservation} (left) and \ref{fig:SolarFlareStandardModel} and illustrated by a coronagraph image in Fig. \ref{fig:OneLoopFlareModelObservation} (right). It is called the CSHKP scenario, named after the scientists who developed it \citep{1964NASSP..50..451C,1966Natur.211..695S,1974SoPh...34..323H,1976SoPh...50...85K}.

The CSHKP model is a simplified two-dimensional representation primarily based on the concept of a magnetic loop pinched at its ends. This loop can vary in size and may be moving outward, characterised by its legs having anti-parallel magnetic fields. When reconnection occurs, a plasmoid is ejected from the top of the loop. Key observational evidence supporting this model includes vertical cusp-like structures observed in soft X-rays following a solar flare, and these cusps increase in size over time, with higher loops displaying higher temperatures. 

Additional supporting observations consist of horizontal inflows of colder material and two sources of hot plasma that diverge from the reconnection point, moving both upward and downward. Although the latter has been observed more frequently, both types of movements have been reported in scientific literature. Further evidence includes the appearance of drifting pulsating structures in decimetric radio emission and hard X-rays, suggesting the presence of non-thermal and mildly relativistic electrons within the plasmoids.

There is also sporadic evidence of downward reconnection outflow in addition to the typically observed upward motion of the soft X-ray source. Long-standing evidence also includes the presence of two ribbons at the footpoints of an arcade of loops, visible in H$\alpha$, EUV, and X-ray emissions. This supports the concept of the one-loop model extended into three dimensions. It is worth noting that these specific elements of the CSHKP model are not frequently observed, suggesting that this type of solar flare might be less common than its theoretical prominence might indicate.

There are alternative flare geometry models involving more loops \citep{heyvaerts1977emerging,2017LRSP...14....2B}, but for simplicity, we keep these geometries outside the scope of our study. 

Magnetic confinement is essential in shaping the typical light curve of a flare, which exhibits a rapid increase in emission due to the rapid heating of the plasma, followed by a slow decay due to cooling by thermal conduction and radiation. \citet{2007A&A...471..271R} has shown that under very general conditions, which most observed flares meet, the time evolution of emission and maximum temperature can provide an estimate of the half-length ($L_f$) of the flare magnetic structure.

\begin{equation}
    L_f=\frac{\tau_d T^{1/2}}{\alpha F(\epsilon)},
    \label{eq:lengthsolarflare}
\end{equation}
where $\alpha \simeq 3.7~10^{-4}$ is generally valid for solar-type stars, $T$ is the temperature of the flare at the peak of soft X-ray luminosity and $\tau_d$ is the characteristic decay time of the soft X-ray emission. The confinement factor $F(\epsilon)$ allows to include the effect of a progressive heating in the rising phase of the flare. The function has an empirical expression, $F(\epsilon) = {0.63 \over \epsilon-0.32} +1.41 > 1$.  The value of $\epsilon$ varies from one flare to the other but is generally close to 1. In Sect. \ref{sec:TTauriFlareGeometry} we will discuss the values of $\epsilon$ in the specific case of T Tauri stars.

Due to the differential rotation of the Sun, the loops described above lengthens, eventually becoming unstable and collapsing inwards to reconnect with adjacent loops (as shown in Fig. \ref{fig:SolarFlareStandardModel}c). The result is a new, less twisted magnetic field loop below the reconnection region, and a twisted magnetic flux rope above it. The yellow ovals in the diagram represent the flare ribbons, the photospheric intersections of the ends of the loop.

In the reconnection region shown in Fig. \ref{fig:SolarFlareStandardModel}a, the blue lines symbolise the oppositely directed incoming magnetic field lines that reconnect in the thin current sheet (grey area). According to models such as Petschek's, the thickness of the current sheet is similar to its width. After reconnection, the green lines represent the ejected magnetic field flowing both upwards and downwards. 

Figure \ref{fig:SolarFlareStandardModel}b shows a cross-section of the loop reconnection on a larger scale. As long as the reconnection continues above, the downward ejected field helps to form the new arch loop (in red). Meanwhile, the upward ejected field contributes to the formation of the magnetic flux rope above the reconnection site. If the flux rope becomes unstable, it moves away from the Sun in the form of a coronal mass ejection (CME). This scenario is supported by observations, see Fig. \ref{fig:OneLoopFlareModelObservation}.

\subsubsection{Reconnection dominating the impulsive phase}
Observations of X-class flares, defined by the luminosity in the GOES band ($1.55 - 12.4$ keV; $L_X \sim 3 \times 10^{26}$ erg s$^{-1}$ to $3 \times 10^{27}$ erg s$^{-1}$), revealed that the size of the flare loop and the temperature of evaporation plasma are typically $l_{loop} \sim 10^9 - 10^{10}$ cm and $T_{flare} \sim 10$ to $40$ MK, respectively. The duration of the impulsive and gradual phases are typically $t_{imp} \sim 10^2 - 10^{3}$ s and $t_{grad} \sim 10^3 – 10^{4}$ s, respectively (e.g., \citealt{2017LRSP...14....2B}). The magnetic field strength and number density of the coronal region can be $B_{rec} \sim 30 - 300~ \text{G}$ and $n_{rec} \sim 10^{8} - 10^{9}$ cm$^{-3}$, respectively (e.g., \citealt{2011LRSP....8....6S}). We search for the appropriate values of $B_{rec}$ and $n_{rec}$ so that the resulting quantities are in agreement with the observations. 

We are setting up a system of normalised parameters for notation purposes. The numerical subscript indicates the power of ten by which the physical parameter has been normalised in the cgs unit system. For example $B_{rec,2}=\left(\frac{B_{rec}}{10^2 G}\right)$.

Magnetic reconnection produces bipolar outflows which velocity is roughly equal to the Alfv\'{e}n velocity around the reconnection region, the so-called inflow region,
\begin{equation}
V_{out} \approx V_{A} = \sqrt{\frac{B_{rec}^{2}}{4\pi m_{p}n_{rec}}} \approx 6.9 \times 10^{6} B_{rec,2} n^{-1/2}_{rec,9} \, \text{cm s}^{-1},
\end{equation}
where $m_p$ is the proton mass. 

The width of the sunward reconnection outflow is estimated to be $l_{out} \sim \eta_{rec}l_{loop} \sim 1.0 \times 10^{8}l_{loop,9.5}\eta_{rec,-1.5}$ cm, where $\eta_{rec} = V_{in}/V_{A}$ is the reconnection rate and $V_{in}$ is the reconnection velocity. In solar flares, $\eta_{rec} \sim 0.01–0.1$ is observationally estimated (e.g., \citealt{2005ApJ...632.1184I,2006ApJ...637.1122N,2012ApJ...745L...6T}). 

The duration of the reconnection event is given by,
\begin{equation}
t_{rec} = \frac{l_{loop}}{V_{in}} \approx 1.4 \times 10^{2} l_{loop,9.5} \sqrt{\frac{n_{rec,9}}{B_{rec,2}}} \eta_{rec,-1.5} \, \text{s}.
\end{equation}
This is the typical timescale over which the plasma around the reconnection region brings the magnetic field to it. This timescale corresponds to the duration of the impulsive phase, i.e., $t_{imp} \approx t_{rec}$. This supports that magnetic reconnection is dominating the impulsive phase.

\subsection{X-ray emission from Solar flares}
\subsubsection{X-ray emission model}\label{sect:Xrayemissionscenario}
Flare observations can be interpreted as a sequence of phenomena. This starts with energy being released in the corona due to magnetic field reconnection, heating the plasma in the reconnection region to extreme temperatures and accelerating electrons to super-thermal energies, see Fig. \ref{fig:SolarCoronalFlareObservation}. Flares sudden emergence suggests a slow accumulation of magnetic energy, which is abruptly released similarly to a blockage breaking event. 

In plasmas, this accumulation can happen with low magnetic diffusion, followed by a drastic increase in the diffusion rate during the flare, releasing the magnetic energy. Diffusion may surge in contracting current sheets when they become thinner than the particle gyroradius, causing particles to decouple from the magnetic field, as seen in Sect. \ref{sect:ClassicalMHDTheories}. High-frequency turbulence waves may also abruptly increase resistivity, accelerating magnetic diffusion and rearranging the magnetic field. Energy storage and rapid release typically mean small collision rates or low density, which is why impulsive flare reconnection is generally believed to happen in the corona. This is consistent with the position of the reconnection regime of coronal flares Fig. \ref{fig:ReconnectionParameterSpace}. 

Then, the energy moves from the corona into the dense chromosphere along a magnetic loop. Energetic electrons, and potentially ions, precipitating from the coronal acceleration site shed their energy in the dense, underlying chromosphere via Coulomb collisions, inciting a dynamic response in the plasma. This reaction can also result from heat conduction, when thermal particles transfer the energy released in the corona. 

The chromosphere temperature rises and the resulting pressure surpasses the surrounding chromospheric pressure. As the excess pressure builds up rapidly, the heated plasma expands explosively along the magnetic field in both directions. This plasma expansion into the corona was first reported by \citet{1980ApJ...239..725D} and \citet{1980ApJ...241.1175F}. \citet{Milligan_2009} identified plasma in the temperature range of $T = 2–16$ MK, with velocities between 300 to 400 km/s. Chromospheric evaporation is further discussed in \citet{2015SoPh..290.3399M}.

The evaporated hot plasma seems to be the main source of the soft X-ray emission \citep{1968ApJ...153L..59N,Silva_1997}. As a result, the soft X-ray emitting plasma absorbs a significant portion of the energy from the precipitating non-thermal electrons. Thus, the soft X-ray emission is proportional to the integrated preceding hard X-ray flux.

When the magnetic containment of a substantial part of the corona is disrupted, it expands and is ejected by magnetic forces in a coronal mass ejection. The shock front associated with this motion is also a site of particle acceleration, especially for high-energy solar cosmic rays observed near Earth.

The hypothesis of a direct causal relationship between soft X-ray emissions and hard X-rays produced by energetic electrons has given rise to a straightforward model. This model asserts that the energy release in a flare involves particle acceleration. However, the actual acceleration process is not included in this scenario.

\subsubsection{X-ray emission dominating the decaying phase}
The reconnection process induces heating within the outflow plasma. However, this heat dissipates through thermal conduction. Therefore, considering the balance between the heating induced by reconnection and cooling caused by conduction, the temperature within the outflow can be approximated. This has been explored in the works of \citet{2001ApJ...549.1160Y} and \citet{2002ApJ...577..422S}.

\begin{align*}
T_{out} \approx \left(\frac{B_{rec}^3 l_{loop}}{2 \kappa_0 \sqrt{4\pi m_{p} n_{rec}}}\right) ^{2/7}  \\
\end{align*}

The Spitzer thermal conduction coefficient is $\kappa = \kappa_0 T^{5/2}$, with $\kappa_0 \approx 10^{-6}$ cm$^2$/s. Given this temperature, the outflow is supersonic and forms a termination shock when it collides with the reconnected field lines if the guide field is weaker than the reconnecting field. The 
%acoustic : I guess it is either acoustic or sonic not both? 
sonic Mach number of the termination shock is given by \citep{2009ApJ...701..348S,2016ApJ...823..150T},

\begin{equation}
    M_{\text{{f}}} \approx \frac{V_{out}}{C_s} \approx 5.8 l^{-1/7}_{loop,9.5} B^{4/7}_{{rec,2}} n^{-3/7}_{{rec,9}}
\end{equation}
where $c_{s} = \sqrt{\gamma k_{\text{{B}}} T_{\text{{out}}}/m_{p}}$ is the adiabatic sound speed, $\gamma = 5/3$ is the specific heat ratio and $k_{\text{{B}}}$ is the Boltzmann constant.

The evaporated plasma fills the flare loop, and its temperature is estimated from MHD simulations with thermal conduction \citep{2001ApJ...549.1160Y} as

\begin{equation}
    T_{\text{{evap}}}  \approx \frac{T_{out}}{3}  \approx 3.4 \times 10^7 l^{2/7}_{loop,9.5} B^{6/7}_{{rec,2}} n^{-1/7}_{{rec,9}} \quad \text{K}
\end{equation}

The density of the evaporated plasma is given by equating the thermal pressure of the evaporating plasma to the its magnetic pressure \citep{2002ApJ...577..422S}. The density is,

\begin{equation}
    n_{\text{{evap}}} \approx \frac{B_{rec}^2}{16 \pi k_B T_{evap}} \approx 4.2 \times 10^{10}  l^{-2/7}_{loop,9.5} B^{8/7}_{{rec,2}} n^{1/7}_{{rec,9}}  \quad \text{{cm}}^{-3} 
\end{equation}

The evaporated  plasma emits soft X-rays by thermal Bremsstrahlung as observed in the soft X-ray band. The thermal free–free luminosity from the evaporated plasma in per unit volume \citep{rybicki1991radiative} is:

\begin{equation}
    L_{ff} \approx 1.7 \times 10^{27} n_{\text{{evap}}}^2 T^{1/2}_{\text{{evap}}} ~\rm{erg/s}
\end{equation}

The cooling timescale is the free-free cooling time scale of the evaporated plasma. We derive it by equating the Bremsstrahlung energy density $\sim L_{ff} t_{ff}$ and the thermal energy, which gives,

\begin{equation}
    t_{\text{{cool}}} \approx 10^4  l^{1/7}_{loop,9.5} B^{-5/7}_{{rec,2}} n^{-3/14}_{{rec,9}} \quad \text{{s}} 
\end{equation}

The evaporated plasma falls back to the stellar surface in $t_{ff}$, and this timescale corresponds to the duration of the soft X-ray decaying phase, i.e., $t_{decay}\approx t_{ff}$. This suggest that the decay phase is dominated by bremstrahlung cooling.

\subsection{Non-thermal Emission in Solar flares}
\subsubsection{Particle acceleration}

A significant portion of the total energy released during a flare goes initially to kinetic energy of non-thermal electrons and protons. Observations in white-light suggest that a significant amount of energy is transported to the chromosphere and the photosphere, in addition to the particles seen in X-rays and gamma-rays \citep{lin1976non}. Hence, understanding the acceleration process is a very problematic aspect of flare physics and is at the heart of the flare enigma. Next paragraph presents the main current issues.

\paragraph{Acceleration mechanisms:}
The conversion of magnetic energy into kinetic particle energy is divided observationally in two stages. During the impulsive flare phase, there is a "bulk energisation", which results in an over 100-fold increase in electron energy, from coronal thermal energy levels (around 0.1 keV) in less than 1 second, as shown by hard X-ray observations \citep{1984ApJ...287L.105K}. The other type of acceleration takes place in a secondary phase, possibly resulting from the first phase, such as a shock wave triggered by a flare or an associated coronal mass ejection. 

Given the free motion of charged particles in dilute plasmas and the difference in inertia between electrons and ions, electrons should be accelerated during any significant impulsive event in the corona. Waves of various types, from MHD to collisionless shock waves, are anticipated to be triggered by the flare. They may also resonate, which would accelerate particles via stochastic Fermi acceleration. From a plasma physics perspective, acceleration is expected. However, there is debate about which process is dominant.

Several mechanisms are capable of converting magnetic energy into accelerated particles. More than one of those mechanisms is likely to occur during a flare, having secondary effects. The most widely discussed mechanisms can be categorised into three types (e.g., \citealt{melrose1990particle,benz2002kinetic}),
\begin{itemize}
    \item stochastic Fermi acceleration,
    \item acceleration by an electric field parallel to the magnetic field,
    \item acceleration by perpendicular and parallel shocks (first and second order Fermi acceleration).
\end{itemize}
These processes have been discussed from the kinetic point of view in Sect. \ref{Sect:AccelerationSites}.

%The stochastic acceleration by the magnetic field component of low-frequency waves in solar flares \citep{miller1997critical,schlickeiser1998quasi,petrosian2006damping} is the most widly accepted. Yet there is not a complete agreement on this mechanism. This process is known as "transit-time damping" of waves.
%\textcolor{red}{et donc ? pourquoi cela s'arrête là, qu'est ce que le TTD ?}

\paragraph{Current observational constraints for particle acceleration mechanisms:}
Particle acceleration during flares might be linked to magnetic reconnection in low-density plasma, though this remains controversial. Present flare observations hold abundant information that helps constraining possible scenarios for particle acceleration. The three types of acceleration mechanims mentioned earlier have been refuted more often than they have been confirmed by observations. Specifically,
\begin{itemize}
    \item High-frequency waves close to the plasma frequency, like Langmuir waves, can not drive stochastic acceleration. If they were driving the acceleration, we should observe radio decimeter waves, which would be present in every flare. However, acceleration by transit-time damping aligns with the frequent lack of radio emission, as the predicted turbulence frequency is significantly lower than the plasma frequency and does not trigger radio emission.
    \item The presence of large-scale, stationary electric fields would result in anisotropy in the velocity phase space, making them subject to velocity-space instabilities. These would be observable in the radio band. This kind of radio emission should have a strong correlation with hard X-ray emission. However, such a correlation is not seen in every flare.
    \item While shock acceleration plays a key role in interplanetary space, they seem to have a weak impact on the acceleration of particles in solar flares. Shocks are possibly produced by the reconnection process. Radio signatures of such shocks have been reported \citep{aurass2002shock,mann2006electron}, but these are still disputed and, if confirmed, extremely rare. Still, slow shocks are anticipated at the interface between inflow and outflow regions of reconnection, as well as fast shocks at the bow of the reconnection outflows. High energies could be easily achieved, 100 MeV ions in less than 1 s \citep{1985ApJ...298..400E,tsuneta1998fermi}.
\end{itemize}
There are, however, other plausible scenarios, such as fluctuating low-frequency electric fields parallel to the magnetic field. Such fields could arise from high-amplitude, low-frequency turbulence like kinetic Alfvén waves. They can accelerate and decelerate electrons, resulting in net diffusion in energy space \citep{arzner2004particle}. But none of which are firmly supported by observations.

\subsubsection{Constraining non-thermal particles spectra}\label{eq:constrainingnonthparticle}

\paragraph{Constraining the power-law index with Hard X-rays:}
Imaging spectroscopy of hard X-ray sources during solar flares has provided insights into the spatial distribution and spectral properties of the accelerated particles produced by the magnetic reconnection events. Observations from instruments such as RHESSI have revealed the presence of distinct hard X-ray sources in the solar corona \citep{2010ApJ...714.1108K}. 

\begin{figure}[h!]
    \centering
    \includegraphics[width=0.9\linewidth]{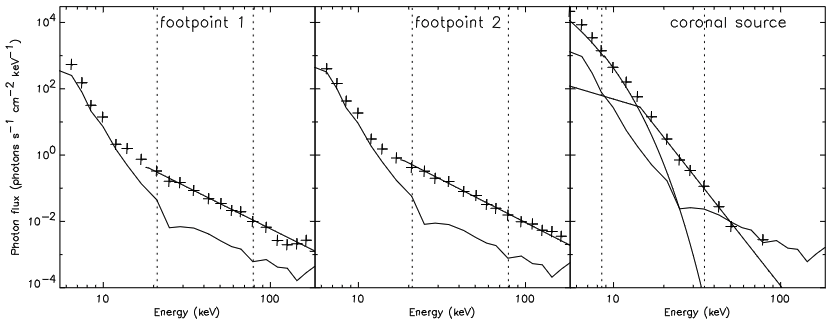}
    \caption{The spectra of the sources at three different locations of a flare indicated on top of each panel, as observed by RHESSI at peak time. The left and middle spectra correspond to footpoints where a power-law was fitted within the energy range delimited by the dotted lines. The spectrum on the right panel is the spectrum of the coronal source, where a fit was established for a power-law and a thermal population between the dashed lines. This figure is reproduced from \citet{battaglia2006relations}.}
    \label{fig:HardXRPowerlawspectra}
\end{figure}
Fig. \ref{fig:HardXRPowerlawspectra} presents the spectrum in photon/s$/$cm$^{2}/$keV of the two footpoints as observed in hard X-rays. There is no evidence of thermal emission within the RHESSI energy range. The non-thermal power-law indices for these are identical at $2.7 \pm 0.1$. The power-law of the coronal X-ray non-thermal emission (see right panel of Fig. \ref{fig:HardXRPowerlawspectra}) is at $5.6 \pm 0.1$. This discrepancy, larger than 2 in the index values, could be interpreted as an effect of the return current electric field causing the flattening of the footpoint spectra. Assuming X rays are emitted by the Bremstrahlung emission of a non thermal distribution of electron, \citet{1971SoPh...18..489B} has found an analytical relationship between the non-thermal X-ray power law index $\gamma_X$ and the power-law index $\delta_e$ of the non-thermal electron distribution emitting it: $\gamma_X=\delta_e-1$. The inferred electron energy spectra observed in solar flares exhibit a power-law shape with indices in the range of $3 \leq \delta_e \leq 6$. These are consistent with the predictions of magnetic reconnection-driven particle acceleration models presented in \citep{2020PhPl...27j0601D,2021PhPl...28e2905L}. 

In Chapter \ref{C:PublicationI}, we study the effects of energetic particles produced by flares of T Tauri stars. We assume that the acceleration processes are the same as in the Sun so the particles will have power-law indices in the same range.

\paragraph{Constraining maximal energy with X-rays and gamma-rays:}\label{sect:constrainingmaxenergy}

The presence of non-thermal electrons in solar flares has been confirmed, as demonstrated by various studies \citep{2011SSRv..159..107H,oka2018electron}, although the signals detected in the hard X-ray bands are generated by nonrelativistic electrons. These relativistic electrons are confined within the flare loop for a duration $t_{esc}\sim 1 s$ (e.g., \citealt{2008ApJ...681.1725S}). $t_{esc}$ is considerably shorter than all cooling timescales, including Bremsstrahlung, Synchrotron and Inverse Compton processes. Hence, the maximum energy of electrons should match that of protons, $E_{e,max} = E_{p,max}$. 

Current hard X-ray and soft gamma-ray observations cannot constrain the maximum energy of electrons due to the lack of observable cutoff features in the photon spectrum below the MeV photon range \citep{2023ApJ...944..192K}.

The emissions from relativistic electrons within the flare loops are found to be difficult to detect. Both Bremsstrahlung and Inverse Compton processes primarily yield GeV gamma rays, but these are less bright than hadronic gamma rays. The synchrotron radiation spectrum peaks in the soft X-ray band, but it is easily overshadowed by the thermal Bremsstrahlung emission from the evaporation plasma.

Nonthermal electrons, produced in the solar corona, travel to the chromosphere and dissipate their energies via Bremsstrahlung and Coulomb interactions. This scenario is consistent with the hard X-ray observations at the footpoints of the flare loop (for example, \citealt{2011SSRv..159..107H}). However, it is important to note that Bremsstrahlung photons produced by the relativistic electrons cannot be detected due to the relativistic beaming effect. Almost all photons are beamed towards the Sun because of their inefficient isotropisation.

\paragraph{Constraining the location of the acceleration region and the effect of the guide field from microwave emissions \citep{2022Natur.606..674F}:}
Decimetric and microwave radio emissions observed during solar flares are primarily generated by the gyrosynchrotron emission associated to non-thermal, high-energy electrons in strong magnetic fields \citep{2007ApJ...666.1256B}. The spectral properties of radio emissions offer constraints on the power law index of the non-thermal electrons \citep{2018ApJ...859...17F}. 
\citet{2022Natur.606..674F} presents a microwave analysis using imaging spectroscopy data from the Expanded Owens Valley Solar Array (EOVSA) to study the behaviour of suprathermal and thermal electrons during a solar flare event that took place on September 2017 10$^{th}$. These observations also constrain the theoretical results of energetic particles of \citet{2021PhRvL.126m5101A}, further discussed in Sect. \ref{sect:collisionnalmultiXline}, on the effect of the guide magnetic field (the field component that is not involved in the reconnection process).

\begin{figure}[h!]
    \centering
    \captionsetup{width=\linewidth}
    \includegraphics[width=\linewidth]{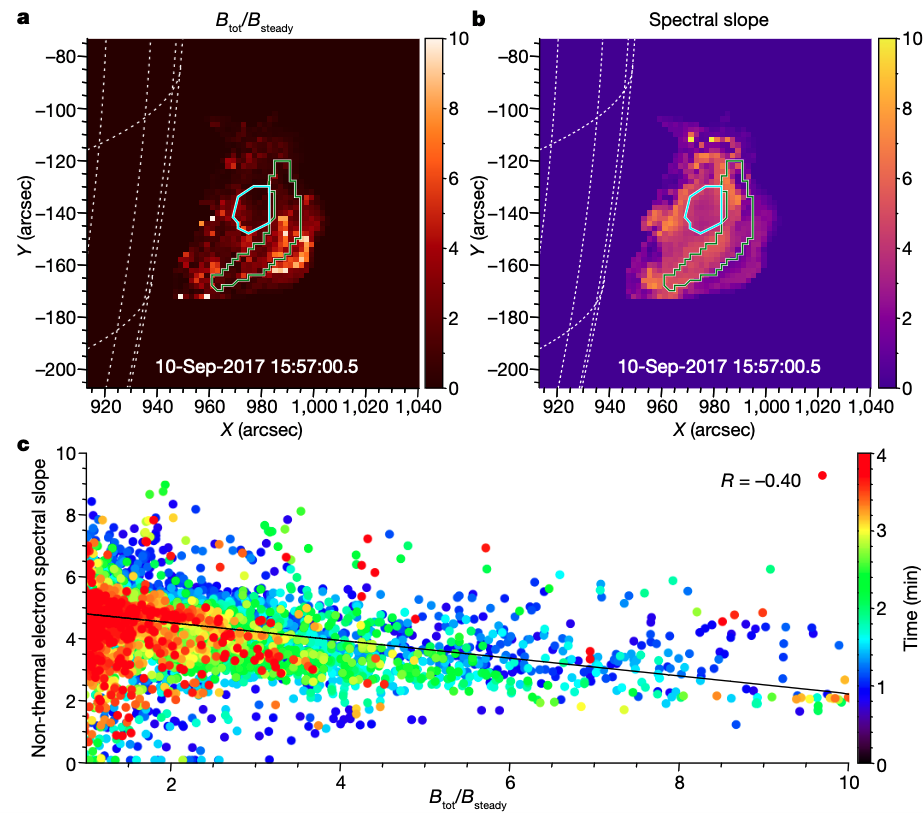}
    \caption{Relationship between the spectral index of the suprathermal component as a function of the reconnecting magnetic field, $b_g^{-1} = B_{tot} (t)/B_{steady}$. The regions measured are delimited with green and cyan contours. (a) Maps of the inferred $b_g^{-1}$ ratio and (b) the suprathermal electron energy spectral slope. In the bottom panel (c), the solid line is the linear fit to the data over the initial two minutes. The figure is reproduced from \citet{2022Natur.606..674F}.}
    \label{fig:RadioObservationSpectralIdex}
\end{figure}

\begin{itemize}
    \item \textit{Location of the acceleration region –} The observed spatio-temporal patterns reveal fundamentally different characteristics between the green and cyan region plotted Fig. \ref{fig:RadioObservationSpectralIdex} throughout the observed period by \citet{2022Natur.606..674F}. In the green region, suprathermal electrons are predominant, whereas thermal electrons dominate the distribution in the cyan region. It appears that the suprathermal electrons in the green are accelerated directly at their current location, rather than being transported from a different region. This green region is characterised by three main features: rapid magnetic energy release, reduced presence of thermal plasma, and a high density of suprathermal electrons, likely accelerated by the aforementioned release of magnetic energy. These features suggest that the observations of \citet{2022Natur.606..674F} identified the region of the solar flare, where the mechanism responsible for electron acceleration is at place. Any process that generates a suprathermal particle population must draw a subset of charged particles from the thermal reservoir and increase their energies locally. Consequently, as electrons are accelerated, the number of suprathermal particles increases, while the count of thermal particles diminishes accordingly. In the green region, the thermal plasma content is highly reduced, possibly below 10\%. This implies that the majority, if not all, of the initial thermal electrons within this volume have transitioned to a suprathermal state during this energy discharge. In conclusion, the energy release from the solar flare drives particle acceleration, efficiently transitioning nearly all local electrons with thermal energies (e.g., below approximately 1 keV) into a suprathermal group with energies surpassing 20 keV (see \citet{2022Natur.606..674F} for a further discussion).
    \item \textit{Effect of the guide field –} In their study, \citet{2022Natur.606..674F} compare the observed total magnetic field, $B_{tot} (t)$, at a specific time and location in the flare region, with its value $B_{steady}$ when it reaches a steady state after the flare. $B_{tot}$ is a measure of both reconnecting and the guiding field, $B_{tot}=B_{rec}+B_{g}$. As these two components are indistinguishable observationally, the guide field, which does not undergo reconnection and thus remains post-flare, can be estimated using $B_{steady}$. The guide field parameter is defined as the ratio of the non reconnecting magnetic field to the total magnetic field,  $b_g \equiv B_g / B_{tot}(t) \approx B_{steady} / B_{tot}(t)$. 
    They investigate the relation between the power-law index of the suprathermal electron distribution $\delta_e$ and the ratio $b_g(t)$. \citet{2022Natur.606..674F} found that the power law spectral index $\delta_e$ is correlated to $b_g$ with $\delta_e \propto b_g^{0.40}$. Large values of $b_g$ produce large spectral index and thus a softer energy spectrum, see Fig. \ref{fig:RadioObservationSpectralIdex}. These results validate the theoretical results of the guide field suppressing the efficient particle acceleration of \citet{2021PhRvL.126m5101A} presented in Sect. \ref{sect:collisionnalmultiXline}. These observations constrain the power-law index range used to build our particle emission model in T Tauri flares, see Sect. \ref{sec:ParticleEmission}. 
\end{itemize}

To summarise the study of \citet{2022Natur.606..674F}, the hard X-rays emitted by high energy electrons, are detected in high density region. As a result, the areas observed might not represent the full range of accelerated electrons. In this study, they also observe the spatial distribution of both thermal and non-thermal electrons during a solar flare using microwave data, revealing the true size of the acceleration region. These data illustrate a region predominantly filled with non-thermal electrons, suggesting extensive acceleration. In contrast, the surrounding region primarily consists of thermal particles with fewer non-thermal electrons. This efficient acceleration occurs precisely where magnetic energy is liberated.

\section{T Tauri Flares }\label{sec:ttauriflares}
Young stellar objects are known to exhibit flaring activity in X-ray bands throughout all stages of their evolution, ranging from the Class 0/Class I protostar phase to the Class II/Class III pre-main-sequence phase \citep{1996PASJ...48L..87K,Feigelson99,getman2021a}. YSO flares, in contrast to solar flares, are both more energetic and more frequent. The total X-ray energy of these flares can reach between $10^{35}$ and $10^{37}$ erg, which is probably only a small fraction of the total radiated and magnetic energy released (see \citealt{2012ApJ...759...71E}). These observations suggest that protostellar flares are approximately $10^{4}$ to $10^{6}$ times more energetic than the largest recorded solar flares (see \citealt{2013JSWSC...3A..31C}). Yet, the mechanisms that trigger these YSO flares, especially during the T Tauri phase, are still a matter of debate.

T Tauri flares have a significant impact on the thermochemical structure of its surrounding disc by producing powerful X-rays. T Tauri flares may also accelerate CRs, which could further affect the chemical structure through gas ionisation \citep{2023MNRAS.519.5673B}. Studying these effects is the goal of this thesis and is presented in Chapter \ref{C:PublicationI} and \ref{C:PublicationII}.

However, the CR production efficiency in YSO flares remains uncertain. Given that YSO flares are far more energetic than solar flares, it is anticipated that CR particles could be accelerated to much higher energies. CR protons emit gamma rays via hadronuclear interactions, whereas CR electrons emit millimetre/submillimetre signals via synchrotron radiation. These signals can penetrate even dense, cold medium without attenuation, making them a valuable tool for investigating flaring activities and CR productions. Theoretical modelling of non-thermal signatures from YSOs, including jet and magnetosphere emissions, have been already addressed in a few publications (e.g., \citealt{2007A&A...476.1289A,2011ApJ...738..115D,2021MNRAS.504.2405A}), but these mostly focus on a specific T Tauri star or emission from protostellar jets. The general detectability of nonthermal signatures from YSO flares has just recently started to be explored. \citet{2019MNRAS.483..917W} modelled the radio and X-ray emission from T Tauri flares, while \citet{2023ApJ...944..192K} modelled the gamma ray emission.

In order to gain insights into the properties and behaviour of T Tauri flares, we turn to observations and models of solar flares presented in the previous section, which have been extensively studied and well-documented.  Solar flares, being more easily observable due to the proximity of the Sun, serve as excellent templates for studying T Tauri flares. The similarities in eruption mechanisms, magnetic reconnection, and particle acceleration processes between these two types of flares make the study of solar flares a useful tool for constraining the characteristics of T Tauri flares.

While comparing solar flares to T Tauri flares is valuable, it is important to acknowledge the differences between these two types of flares. T Tauri stars possess strong magnetic fields and rapid rotation, which lead to more complex and luminous flare behaviour compared to solar flares. Additionally, the presence of a circumstellar disc around T Tauri stars can influence the dynamics and characteristics of the flares.

\subsection{T Tauri flare geometry}\label{sec:TTauriFlareGeometry}
%\citet{2019MNRAS.483..917W} studied these flares to calculate X-ray and radio emissions. Their model used a multipolar magnetic field and varied parameters to estimate possible luminosities. Their model found luminosities much higher than those of solar flares, which is consistent with the active nature of T-Tauri stars.

%\citet{2020MNRAS.496.2715W} then used a time-dependent 3D MHD model to simulate a flare atmosphere and an accretion disc system. The model includes a star undergoing accretion triggered by a flare event, leading to the development and evolution of a flux tube filled with non-thermal electrons.

%Previous studies have examined accretion and flaring \textcolor{magenta}{CS what do you mean here by flaring?} in young stars using MHD modelling. The properties of the magnetic field were found to have a significant impact on certain \textcolor{magenta}{certain is inappropriate... which ones BTW} properties of the plasma. Some models have explored different configurations of the magnetic field. For example, \citet{2017ApJ...838..100R} compared purely dipolar fields with multipolar field configurations. 

%%%%%%%%%%%%%%%%%%%CS
Flares, as described by the standard one loop solar flare model, are essentially sudden energy releases that occur in a diffuse plasma confined in a "magnetic bottle". Energy is lost through optically thin radiation and efficient heat conduction to the chromosphere of the star (see Sect. \ref{sect:Xrayemissionscenario}). 

Magnetic confinement is essential in shaping the typical light curve of a flare, which exhibits a rapid increase in emission due to the rapid heating of the plasma, followed by a slower, almost exponential decay due to cooling by thermal conduction and radiation. \citet{2007A&A...471..271R} has shown, in the context of solar flares, that under very general conditions, which most observed solar flares meet, that the characteristic decaying time of emission $\tau_d$ and the maximum temperature $T$ can provide an estimate of the half-length ($L_f$) of the flare magnetic structure. \citet{2011AJ....141..201M} propose a modification of \citet{Serio91} relation. Assuming this relation holds for T Tauri flares, the half-length of a T Tauri flare loop can be estimated analogously as Eq. \eqref{eq:lengthsolarflare}, 

\begin{equation}
    L_f=\frac{\tau_d T^{1/2}}{\alpha F(\epsilon)} ,
    \label{eq:Halflength3}
\end{equation}
where $\alpha \simeq 3.7~10^{-4}$ is largely valid for solar-type stars and the empirical function $F(\epsilon) = {0.63 \over \epsilon-0.32} +1.41 > 1$. The introduction of the function $F(\epsilon)$ includes the effect of a progressive heating in the rise phase of the flare. Conversely, \citet{Serio91} assumed an impulsive heating with $F(\epsilon)=1$. $\epsilon$ depends on the star and varies for a given star from one flare to the other. \citet{2011AJ....141..201M} derived this parameter for a sample of flaring class I to III young stellar objects. The mean value of $\epsilon$ depends of the YSO class, namely $\langle \epsilon \rangle  = 1.085$ and $ F(\langle \epsilon \rangle )  \simeq 2.23$ for class I YSOs and  $\langle \epsilon \rangle  = 1.43$ and $ F(\langle \epsilon \rangle )  \simeq 1.98$ for class II YSOs. For a fiducial calculation we assume $F(\epsilon) = 2$ hereafter.

Following this interpretation, \citet{2005ApJS..160..469F} analysed a set of large flares observed in the Orion cloud thanks to the continuous Chandra observation of the Orion nebula (the COUP data set). They concluded that in about 10 of these flares, the length of the eruptive structure is between 3 and 5 stellar radii ($R_*$). Structures of similar length have never been observed in older stars. The authors noted that such extensive structures, if anchored to the stellar surface, would face major stability problems due to the centrifugal force, given that young stellar objects are fast rotators, with a rotation period of around 3 to 6 days. They therefore suggested that these loops probably connect the star and the disc at the corotation radius.

\begin{figure}
    \centering
    \includegraphics[width=0.7\linewidth]{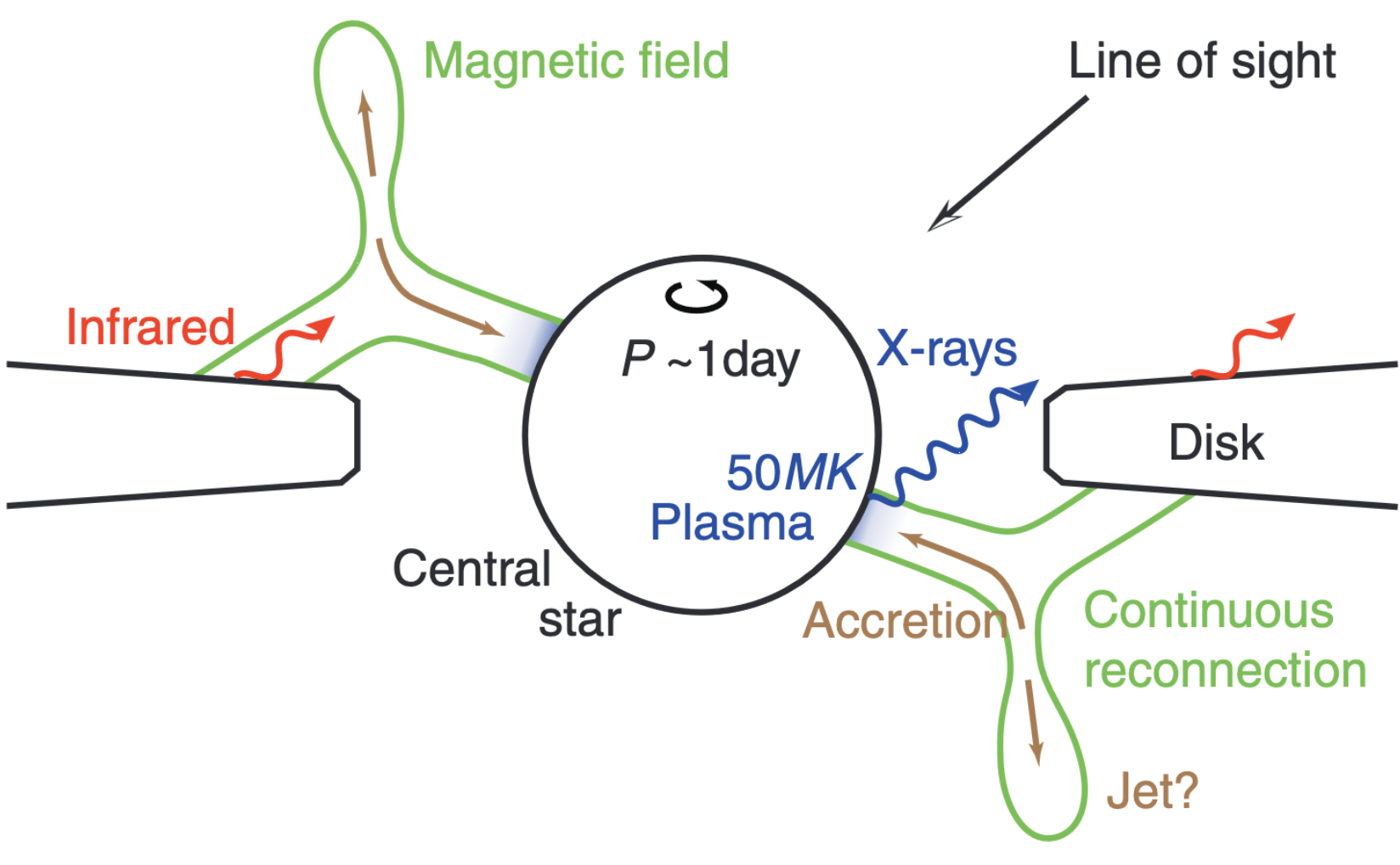}
    \caption{Star-disc flare geometry, possible geometry of a T Tauri flare reproduced from \citet{2012ApJ...754...32H}. Differential rotation between the star and the disc shears the stellar bipolar magnetic fields, and the magnetic fields twist and continuously reconnect. Matter, accelerated by the magnetic reconnection, collides with the stellar surface, thermalizes and emits hard X-rays.}
    \label{fig:StarDiscFlareLoop}
\end{figure}

A similar analysis has been carried out for several dozen of young stellar objects in the $\rho$ Oph region using data from a large XMM-Newton program. And the rate of very long flare structures in $\rho$ Oph is similar to that observed in Orion. Through analysing X-ray light curves from the Chandra, XMM-Newton and Suzaku observatories, \citet{2012ApJ...754...32H} used the rotational modulation in the X-ray emission of Class I protostar to propose a magnetic geometry of the X-ray emitting structure. The geometry is reproduced in Fig. \ref{fig:StarDiscFlareLoop}. The light curve can be reproduced by the emission from two hot spots located at opposite poles on the stellar surface. These hot spots, which likely cover significant portions of the stellar surface, suggest high plasma density ($\ge 5\times 10^{10}$ cm$^{-3}$). 

\begin{figure}
    \centering
    \includegraphics[width=0.7\linewidth]{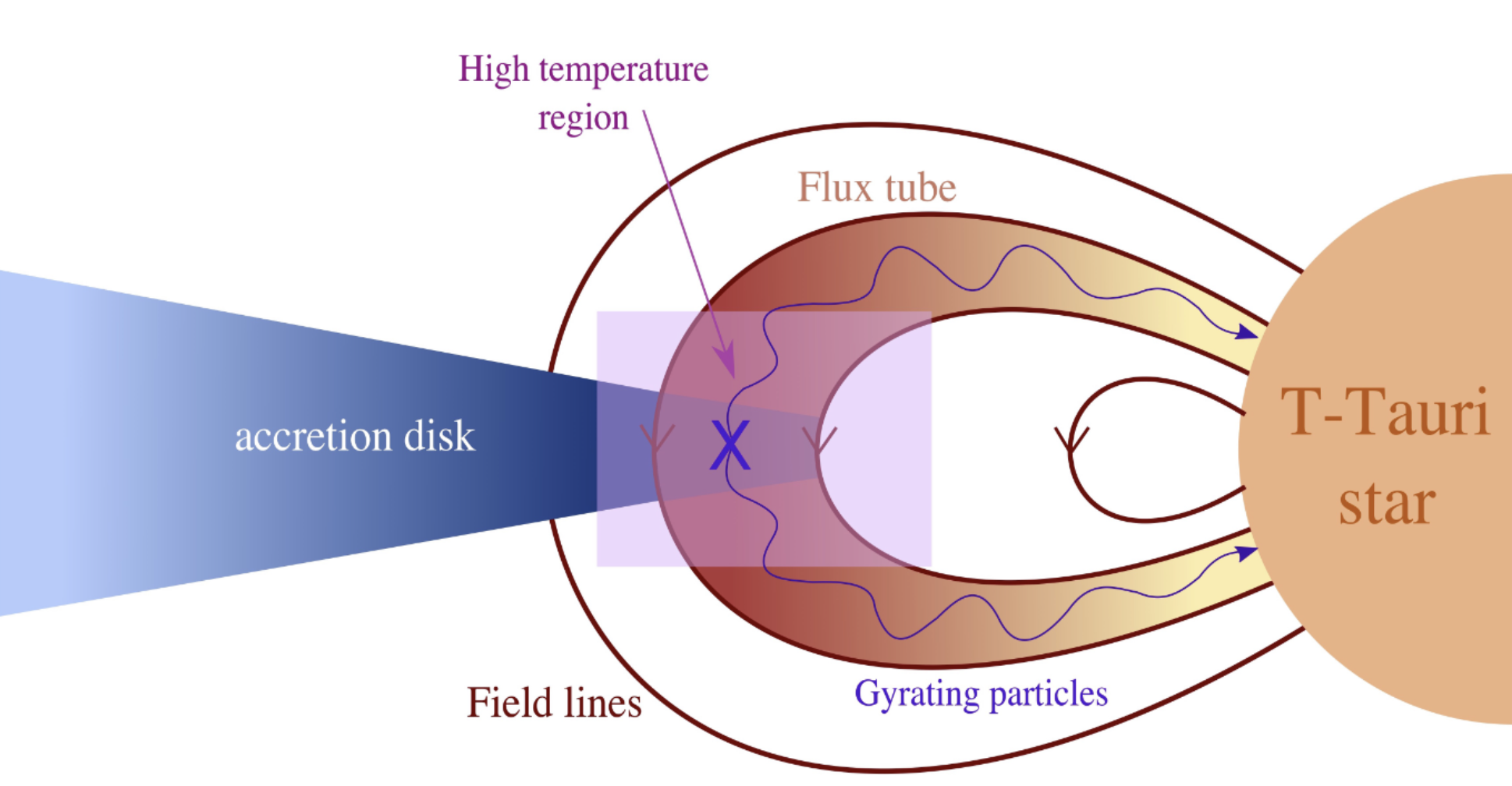}
    \caption{Coronal flare geometry. Magnetic reconnection takes place at the marked X-point, in the interaction region between the star and disc   magnetic field lines. Non-thermal particles move in a circular motion from this point, traveling along the field lines towards the surface of the star emitting X-rays. In Chapter \ref{C:PublicationI} we study the effects of particles entering the disc.  This figure is reproduced from \citet{waterfall2020predicting}.}
    \label{fig:TTauriCoronalFlare}
\end{figure}

\citet{2021ApJ...920..154G,2021ApJ...916...32G} performed an all sky survey in X rays, MYStIX and SFiNCs. It led to several conclusions. In particular, circumstellar discs do not affect flare morphology, they are not related to flare energy, and very hot flares occur in accreting young stellar objects. Consistently with this scenario, another flare geometry is proposed, see Fig. \ref{fig:TTauriCoronalFlare}. This geometry is presented in \citet{2020MNRAS.496.2715W}. They used a time-dependent 3D MHD model to simulate a flare atmosphere and an accretion disc system and studied the emission produced by a flare with the two foot-points anchored to the surface of the star. In this configuration, the top of the loop would be in contact with the inner disc. The area with the highest particle density is at the top of this loop, which also coincides with the region of highest temperature.

All in all, the existence of flare loops interconnecting stellar discs has been the subject of debate over the last decade. The main impact that each of these models has on our results is the position at which the energetic particles produced in the reconnection region enter into the disc. Assuming the coronal flare geometry, with the two foot points anchored to the stellar surface, the particles are penetrating at $R\approx L_f$. On the other hand assuming the star-disc flare model, with the loop linking the stellar surface with the disc, particles are penetrating at $R\approx 2 L_f$. The star-disc flare model injects particles farther in the disc. We show that this has a major impact on the ionisation distribution in the inner disc thus potentially to the size of the MRI active region. Depending on the actual occurrence rate of these large flares, they could influence the early evolution of circumstellar discs.

%As mentionned, the X-ray emmission is supposed to be generated by bremsstrahlung emission of energetic electrons. We present now how we construct a model of particle emission based on the the X-ray observation. 

%Despite the considerable progress made, many challenges remain. The physical scales involved in T Tauri stars flares are much larger, and the environment, including the presence of an accretion disc, is considerably more complex. Solar flare models must therefore be extended and modified to take account of these differences. High-resolution, multi-wavelength observations of T Tauri stars flares, taking advantage of advances in instrumentation, are needed to test and refine these models. Combining these observations with advanced computer models will provide a better understanding of T Tauri stars flare activity.

%\textcolor{magenta}{CS you never mention why and how you will use thoses results} I guess this is the reason Alexandre as put a similar comment further down

\subsubsection{Constraining T Tauri flare thermal structure}\label{sec:ConstrainingPlasmaDensity}

The X-ray flux depends on various physical parameters, such as the size of the reconnection site and the local plasma density. By using COUP data, we can correlate both the emission size and gas density with the local gas temperature, thereby deriving an expression for the total plasma density $n_k=n_e$ that depends solely on temperature. This quantity is important to constrain as it is necessary to estimate the distribution of energetic particle produced by the flare. The method is further discussed in Chapter \ref{C:PublicationI} but we summarise it here. We define the normalised quantity,
\[
L_{10}=\frac{L}{10^{10} \rm cm} , \quad n_{e,10}=\frac{n_e}{10^{10} \rm cm^{-3}}, \quad T_{6}=\frac{T}{10^6 \rm K}. 
\]

We derive a relationship between the flare loop length and temperature averaged over the entire flare loop $T$ using data from \citet{getman2008a,getman2008b},

\begin{equation}
    L_{10}(T)=7.61 \times T_6^{0.45}
\end{equation}
And electron density $n_e$ can be expressed as a function of $(L/R_*)$, where $R_*$ denotes the stellar radius, based on the linear regression in Fig. 11 from \citet{getman2008b},

\begin{equation}
n_{e,10}(L/R_*)=4.60\times \left(\frac{L}{R_*}\right)^{-1.32}
\label{eq:nedelR}
\end{equation}

Again based on data from \citet{getman2008a}, see in Chapter \ref{C:PublicationI}, Fig. 5 of \citet{2023MNRAS.519.5673B}, we deduce a linear regression-based relationship between $L/R_*$ and $L_{10}$,

\begin{equation}
\frac{L}{R_*}=0.03 \times L_{10}^{1.17}
\label{eq:LRdeL}
\end{equation}

Combining these equations allows for the electron density in the flare area to be represented as a function of temperature,

\begin{equation}
n_{e,10}(T)=20.5 \times T_{6}^{-0.69}
\label{eq:obsdensity}
\end{equation}

From this estimation of the electron density in the flare, we estimate a normalisation of the non-thermal electron distribution, see Sec. \ref{sec:parametricinjectionmodel}.

\subsection{Magnetic reconnection in T Tauri flares}\label{sec:magnrecTTauriFlare}

\subsubsection{Identifying the reconnection regime in T Tauri flares}\label{sec:ReconnectionRegimeinTTauri}
To determine the regime of magnetic reconnection anticipated over the circumstellar disc of a T Tauri star, we refer to the phase diagram in Fig.\ref{fig:ReconnectionParameterSpace} \citep{2011PhPl...18k1207J}, reproduced in Sect. \ref{sect:ReconnectionRegime}. This diagram is used to determine whether the reconnection will be non-collisional, collisional MHD with plasmoids, or collisional following the Petsheck model. Given that each reconnection type has distinct acceleration mechanisms, it is crucial to forecast the specific type that is likely to occur above protoplanetary discs.

As mentioned in Sect. \ref{sect:ReconnectionRegime}, there are two primary parameters that control the position in the phase diagram in Fig.\ref{fig:ReconnectionParameterSpace}. The first one is the system size, denoted as $L_0$, the second is the global Lundquist number, $S_L$ defined by the Eq.\eqref{eq:LundquistNumber}.

These parameters can be described by the temperature at the location of the flare, the particle number density and the magnetic field strength. We assume a standard flare size of $L = 10^{10}$ cm, and the typical density and temperature values of the plasma at the flare site $n_e = 10^{9} ~ \rm cm^{-3}$ and $T = 1$ MK respectively. Based on these parameters and the phase diagram Fig.\ref{fig:ReconnectionParameterSpace}, we conclude that the magnetic reconnection should occur in the "Multiple X-line collisional" regime. We want now to constrain the energetic particle distribution in this regime. So we focus now on simulations of acceleration of particles in the "Multiple X-line collisional" regime.

%The location in $S-\lambda$ parameter space of the reconnection regime occurring above the disc is approximately identical to that of the reconnection regime in the solar chromosphere. Therefore, observing reconnection processes in the solar chromosphere will allow us to constrain the particle acceleration process in the region above the disc of T Tauri stars.

\subsubsection{Particle acceleration in the collisional multi X-line regime}\label{sect:collisionnalmultiXline}
\begin{figure}
    \centering
    \includegraphics[width=0.7\linewidth]{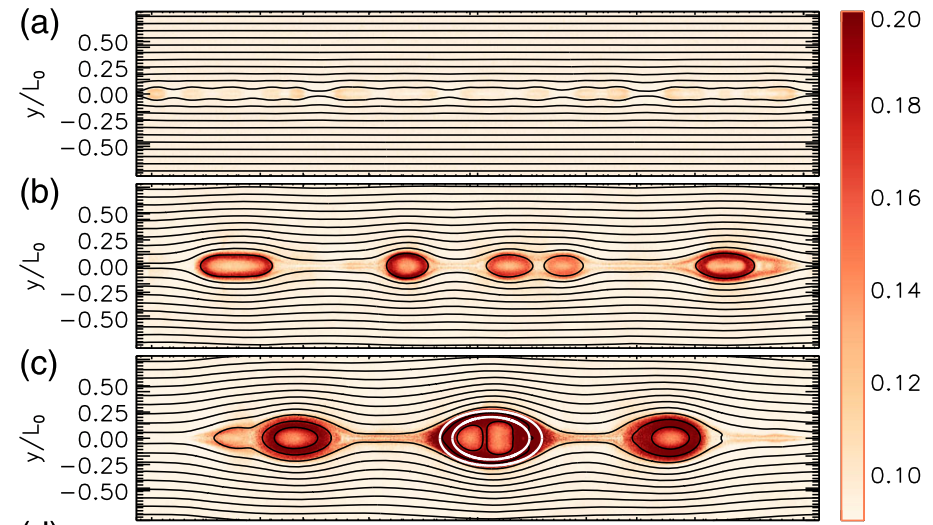}
    \caption{Normalised mean electron energy for a simulation with $B_g/B = 0.25$ at $t/\tau_A = 2.5, 5$, and 8,  (a),(b), and (c) respectively. With field lines overplotted. The figure is reproduced from \citet{arnold2021electron}}
    \label{fig:MultiXlineArnold}
\end{figure}

\citet{2021PhRvL.126m5101A} explored the spectral index $\delta_e$ of accelerated electrons in the collisional regime using 2D simulations that account for accelerated particle feedback. This model studied the reconnection regime relevant in the physical condition assumed above T Tauri discs deduced from the parameter space of \citet{2011PhPl...18k1207J}. This is the model for the particle acceleration in magnetic reconnection that we use in Chapter \ref{C:PublicationI}.
The simulations of \citet{2021PhRvL.126m5101A} predict a value of the power-law index of the non-thermal particles energy distribution. It also accounts for the effects of the surrounding guiding field. The model incorporates the convective loss of electrons that are inserted into large flux ropes. The model is based on the increase in electron energy as the merged field lines change from a figure-of-eight formation to a circular formation, see Fig. \ref{fig:MultiXlineArnold}. 

This change in energy can be determined by analysing the geometry of the magnetic field before and after the merger. The calculation of \citet{2019PhPl...26j2115Z} reveals the rate of energy gain,
\begin{equation}
    \frac{dE}{dt}= E \frac{g}{\tau_r}
\end{equation}

The model estimates $\tau_r$, the merging time of a flux rope of initial radius $r_i$, where $R$ is the merging rate, about 0.1. The factor, $g$, resulting from the influence of the out of the reconnecting plane component guide field of the magnetic field $B_g$ is also taken into account,
\begin{equation}
    g=\left(1+2\frac{B_g^2}{B_x^2}\right)^{-1}
\end{equation}
The model provides an equation for determining the energy distribution function of electrons undergoing acceleration due to reconnection in a one-dimensional current layer while experiencing convective losses. The model includes a simple constant diffusion in the current layer and a convective loss mechanism. Electrons are introduced with a initial Maxwellian distribution at low temperature while the resulting spectrum of energetic electrons is a power-law,
\begin{equation}
    F(E)\sim E ^{-(1+\frac{r_i}{g R L})}
    \,,
\end{equation}
with a spectral index $\delta$ determined by the reconnection rate $R\approx0.1$, the ratio between the characteristic radius $r_i$ and the half-width $L$ of the current layer, and the strength of $B_g$ through g. 
The scaling relationship of this model is fitted with the simulation of \citet{2021PhRvL.126m5101A} giving $r_i/L\approx 0.22$. All together the power-index of the energy distribution of the non-thermal electrons $\delta_e$ defined as $F(E)\sim E^{-\delta_e}$ is,
\begin{equation}
    \delta_e \approx 3+4 b_g^2,
    \label{eq:relationindicechampguidearnold}
\end{equation}
where $b_g=B_g/B_x$ is the guide field parameter. In the absence of a guide field, the power law index $\delta_e=3$. This value is going to be our reference value for the power-law index of the non-thermal particle energy distribution. According to these results, the presence of a guide field softens the non-thermal distribution considerably. Although the dependence of the power law index \(\delta_e\) on \(b_g\) differs from what one might observe in Fig. \ref{fig:RadioObservationSpectralIdex}, in both scenarios the power law index increases with the increasing guiding field. The discrepancy between Eq. \eqref{eq:relationindicechampguidearnold} and the regression in Fig. \ref{fig:RadioObservationSpectralIdex} arises because the multi X-line collisional reconnection regime studied by \citet{arnold2021electron} is not the expected regime in the Solar corona. The regime examined by \citet{arnold2021electron} giving Eq. \eqref{eq:relationindicechampguidearnold} corresponds to the multi X-line collisional regime anticipated above T Tauri discs.

\subsubsection{Plasma properties and magnetic reconnection in T Tauri flares : connection to accretion discs}

In this section, we discuss the conditions for magnetic reconnection to occur in circumstellar discs, focusing on the physical properties of circumstellar discs.

\paragraph{Magnetic field configuration:}
High-resolution MHD simulations have been able to reproduce the complex magnetic field structure and the dynamics of the disc, revealing the formation of current sheets and the occurrence of magnetic reconnection in various magnetic field configurations \citep{2013A&A...550A..99Z,2019A&A...624A..50C,2023arXiv230413746L}.

Moreover, observational studies have also provided important constraints on the magnetic field strength and configuration in circumstellar discs using for example polarimetry or Zeeman splitting \citep{2009ARA&A..47..333D,2014Natur.514..597S,2017A&A...607A.104B}. These observations have revealed the presence of strong magnetic fields and complex field topologies in the disc, supporting the occurrence of magnetic reconnection and its role in regulating the disc dynamics.

\begin{figure}
        \centering
        \includegraphics[width=\linewidth]{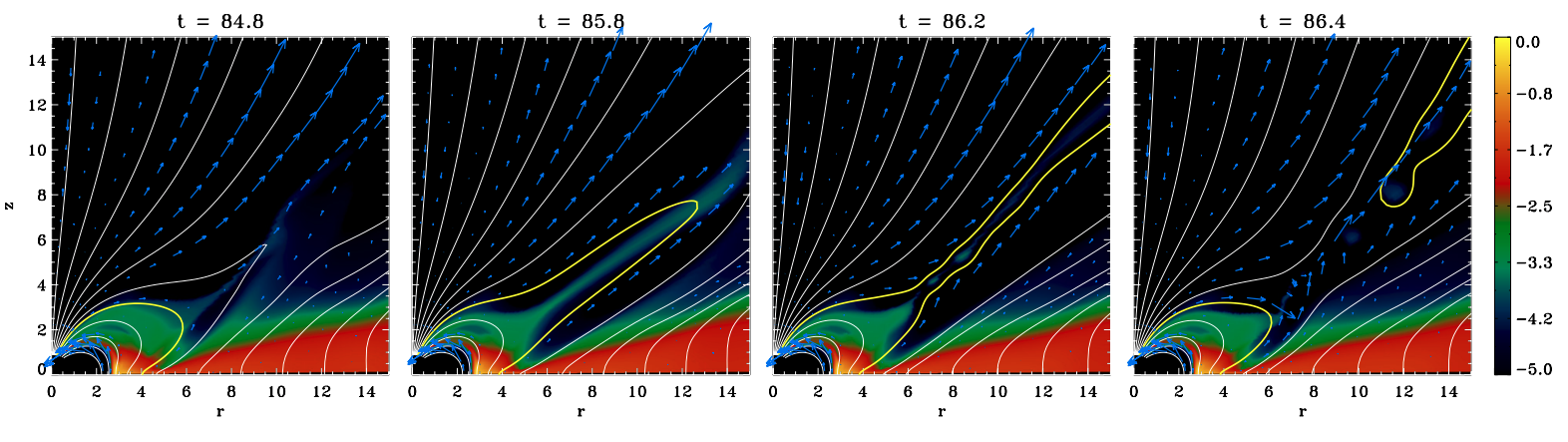}
        \caption{Temporal evolution of the periodic inﬂation/reconnection process which characterizes the dynamics of the magnetospheric ejections. The logarithmic density maps is shown with sample ﬁeld lines (white solid lines) and poloidal speed vectors (blue arrows) superimposed. The yellow solid lines follow the evolution of a single magnetic surface showing clearly the dynamics of the multi X-line reconnection regime. Time is given in units of rotation periods of the central star. The figure is reproduced from \citet{zanni2013mhd}.}
        \label{fig:MagneticReconnectionRegion}
\end{figure}

Protoplanetary discs are characterised by a complex magnetic field structure, consisting of a combination of large-scale poloidal and toroidal fields, small-scale turbulent fields, and ordered helical fields \citep{2017ApJ...845...75B}. In the inner disc the magnetic field of the central star is also interacting with the field of the disc.

Magnetic reconnection is most likely to occur at locations where the magnetic field lines are sheared, compressed, or twisted by the disc dynamics, leading to the formation of thin current sheets and enhanced local magnetic field strength \citep{2001EP&S...53..473S}. The interaction region between the star and the disc is matching these criteria. 
Figure \ref{fig:ReconnectionRegime} shows the characteristic geometry of a multi X-line reconnection zone in the magnetocentrifugal region of the simulation of \citet{zanni2013mhd} reproduced in Fig. \ref{fig:MagneticReconnectionRegion}. Their model corresponds more to the topology of the star-disc flare geometry.

\paragraph{Magnetic field intensity:}
    
The strength of the magnetic field plays a critical role in determining the efficiency of magnetic reconnection in circumstellar discs. Magnetic reconnection is more efficient in regions with strong magnetic fields, as the energy release and the reconnection rate are proportional to the magnetic field strength \citep{2017JPlPh..83e7101C}. During flares, the magnetic field strength can be enhanced by the amplification of the field by turbulence, by the compression of the field due to the disc large scale dynamics, or by the generation of the field by the disc dynamo \citep{2005PhR...417....1B, 2013A&A...550A..99Z}. The magnetic field intensity in flares is expected to be of a few kG \citep{2009ARA&A..47..333D}. We will quantify the effect of the magnetic field intensity in the next paragraphs through two plasma parameters, the magnetisation and the plasma $\beta$.

Indeed, the efficiency of magnetic reconnection in circumstellar discs is strongly influenced by the plasma properties. We discuss now the role of the local density, temperature and magnetic field intensity on magnetic reconnection in T Tauri flares. We estimate the efficiency of the reconnection by computing the power-law index of the non-thermal particles as function of these parameters.

\paragraph{Plasma $\beta$ parameter:}
As seen in Sec. \ref{sec:ModelingProtoplanetaryDiscs}, we recall that the plasma parameter, $\beta$, is a dimensionless value representing the ratio of gas pressure to magnetic pressure, defined as $\beta = p_{gas} / p_{mag}  = \frac{8 \pi n k_B T}{B^2}$. Here, $p_{gas}$ is the gas pressure, $B$ the magnetic field strength, and $n$ the plasma density. 

Magnetic reconnection tends to be more effective in plasmas where the beta parameter is low. In such cases, the magnetic pressure overwhelms the gas pressure, facilitating the dissipation of magnetic energy and increasing reconnection rates \citep{2019ApJ...884..118L,2020ApJ...894L...7Z}. The efficiency of reconnection events can be described by the power-law index $\delta$ of the non-thermal particle distribution. Smaller $\delta$ values indicate a greater population of particles at high energies, thereby denoting more efficient reconnection acceleration. As demonstrated by \citet{ball2018electron}, for $\beta<3\times 10^{-3}$, the index $\delta$ remains independent of $\beta$.

T Tauri flares suggest typical density of $n\sim 10^9$ cm$^{-3}$, temperature $T\sim 10$ MK, and a magnetic field $B\sim1$ kG. Given these parameters, and assuming equal temperatures for protons and electrons, $\beta\approx 3\times10^{-5}$. With the plasma pressure being dominated by the magnetic pressure, we can expect highly efficient reconnection and a range where the power-law index remains independent of $\beta$.

\paragraph{Magnetisation:}

Magnetisation ($\sigma_m$), another dimensionless parameter, is the ratio of magnetic energy density to the plasma rest energy density. It is defined as $\sigma_m = B^2 / (4\pi \mu n  m c^2)$, where $\mu=1/2$ stands for the mean particle weight and $c$ is the speed of light. Magnetisation can determine the efficiency of energy conversion from magnetic field to plasma kinetic energy, the acceleration of charged particles, and the creation of a non-thermal particle population \citep{2015ApJ...806..167G,2018ApJ...862...80B}.

With the same parameter set used for the plasma $\beta$ calculation, we can estimate the magnetisation of the plasma $\sigma_m\approx 5 \times10^{-2}$. So the magnetisation $\sigma$ in T Tauri flares approximately corresponds to the trans-relativistic regime $\sigma_m\sim1$. In this regime, \citet{ball2018electron} give the power law index of non-thermal electrons parameterised as $\delta_e=1.8+0.7/\sqrt{\sigma_m}$. Reconnection is more effective for high $\sigma_m$ values. Given the plasma parameters anticipated in T Tauri flares, the parameter analysis of \citet{ball2018electron} predict $\delta_e\approx4$. While this value is slightly higher than the index $\delta_e=3$ predicted by \citet{arnold2021electron}, the values of both works are consistent. This provides us with a range of indices to examine in Chapter \ref{C:PublicationI}.

\subsection{Particle emission from T Tauri flares}\label{sec:ParticleEmission}
%Flares are dynamical phenomena that convert a portion of magnetic energy into the kinetic energy of particles, specifically through magnetic reconnection. 
This section focuses on electrons and protons acceleration in T Tauri flares. All together, theory, simulations and observations agree to the fact that magnetic reconnection  produces non-thermal particles following power-law distributions. We also expect magnetic reconnection to be at the origin of T Tauri flares. Thus, the electrons and protons produced in T Tauri flares should follow a power-law distribution function. We present shortly in this section how the supra-thermal distribution can be constrained based on X-ray observations. The flux of supra-thermal particles produced by the flare is called the injection flux. The method is further explained in Chapter \ref{C:PublicationI}.

\subsubsection{Parametric model of the injection spectrum}\label{sec:parametricinjectionmodel}
We present in this section how to parameterize the non-thermal particle flux produced by the flare with quantities that can observationally be constrained. 

The energy distribution per unit volume, denoted as $F(E)$, or the flux, denoted as $j(E)$, is expressed in terms of the kinetic energy, $E$, for both electrons and protons, where 
\begin{equation}
    j(E)=F(E) \frac{v}{4\pi},
\end{equation}
and $v$ represents the particle speed. The flux and distribution, consist of a thermal and a non-thermal component, denoted by the subscript "th" and "nth", respectively. It is expressed in units of $ \rm particles ~ s^{-1} cm^{-2} sr^{-1} eV^{-1}$.

The thermal component is defined by its temperature $T$ and its normalisation $n_{\rm th}$, where $n_{\rm th}$ is the thermal particle density in cm$^{-3}$. For energetic protons and electrons ($k = e$ for electrons and $k = p$ for protons), we can assume equal temperatures. This gives us the following equation for their energy distribution,

\begin{equation}
\label{jTH}
  F_{\rm {th}}(E) =  n_{\rm th} {2 \over \sqrt{\pi}} \sqrt{{E \over k T}}\frac{1}{k T} \exp(-E/kT) \ .
\end{equation}
Here, the peak value is at $E = 3/2 k T \equiv E_{\rm th}$.

As it is usually assumed in reconnection models for solar flares  \citep{ripperda2017reconnection} and supported by observation of solar flares \citep{emslie2012global,matthews2021high} energy equipartition between non-thermal protons and electrons below 1 GeV is a good proxy of non-thermal content energetics. The energy density of non-thermal particles can be represented as,

\begin{equation}
U_{e,\rm{nt}}= U_{p,\rm{nt}}=\int_{E_{\rm c}}^\infty E F_p(E) \dd E=\int_{E_{\rm c}}^\infty E F_e(E) \dd E
\ ,    
\end{equation}

If we assume that electrons and protons have the same injection energy $E_{\rm c}$, they also share the same non-thermal energy distribution, 
\begin{equation}
    F_e(E)=F_p(E)=F_{\rm nt}(E).
    \label{eq:equipartionofenergy}
\end{equation} 

As a result of the magnetic reconnection process, we introduce the non-thermal component of the energy distribution function $F_{\rm nt}$. The non-thermal energy distribution is determined by its normalisation $N_{\rm nt}$. In the following, we show a way to derive this quantity.

We define a maximum energy for the non-thermal distribution, $E_{\rm U} > E_{\rm c}$. The distribution is a power-law between these energies. The index of the power law is called $\delta$. This model is consistent with the non-thermal electron distribution observed in solar flares. 

The injection energy, $E_{\rm c}$, is proportional to the thermal energy, $E_c=\theta E_{\rm th}$. Most T Tauri flare models exhibit a single power law. We assume an exponential cut-off beyond $E_{\rm U}$,

\begin{equation}
    F_{\rm nt}(E)=N_{\rm nt} \left(\frac{E}{E_{\rm c}}\right)^{-\delta} \exp\left(-\frac{E}{E_{\rm U}}\right),
    \label{eq:nonthermalenergydistribution}
\end{equation}

The normalisation factor $N_{nt}$ can be expressed in terms of the non-thermal particle density,
\begin{equation}
    n_{nt}=\int_{E_c}^\infty \left(\frac{E}{Ec}\right)^{-\delta}\exp\left(-\frac{E}{E_{\rm U}}\right) dE \simeq N_{nt} \frac{E_c}{\delta-1}.
\end{equation}

The density derived from X-ray observation (Eq. \ref{eq:obsdensity}) is the sum of the thermal and non-thermal components, $n= n_{\rm th}+ n_{\rm nt}$. By setting the non-thermal and the thermal energy distribution to be equal at the injection energy ($E=E_c=\theta E_{\rm th}$), we can derive $n_{\rm nt}$,

\begin{equation}
n_{\rm nt} \simeq \frac{3\theta}{\delta-1}\sqrt{\frac{3\theta}{2 \pi}} \exp\left(-\frac{3}{2}\theta\right) n_{\rm th}. 
    \label{eq:nonthermaldensity}
\end{equation}

Equation \eqref{eq:nonthermaldensity} is an expression of the non thermal particle density. It depends on the local density of the thermal plasma. We use the total plasma density $n$ from X-ray observations given by Eq. \eqref{eq:obsdensity} to deduce the thermal and non-thermal densities.
Given that $n_k=n_{k,\rm nt}+n_{k,\rm th}$, we get,

\begin{equation}
n=n_{\rm th}\left(1+\frac{3\theta}{\delta-1}\sqrt{\frac{3\theta}{2 \pi}} \exp\left(-\frac{3}{2}\theta\right)\right)\ ,
    \label{eq:totaldensity}
\end{equation}

From Eqs. \eqref{eq:totaldensity} and \eqref{eq:obsdensity}, the thermal and non-thermal particle densities can be expressed as function of the observed temperature,
\begin{equation}
    n_{\rm th}= 20.5 \times 10^{10} \frac{T_{\rm 6}^{-0.69}}{1+\frac{3 \theta}{\delta-1}\sqrt{\frac{3\theta}{2 \pi}} \exp\left(-\frac{3 \theta}{2}\right)} \quad \rm cm^{-3},
    \label{eq:thermaldensity}
\end{equation}
The non-thermal particle density $n_{nt}$ is then recovered from Eqs. \eqref{eq:thermaldensity} and \eqref{eq:nonthermaldensity},
\begin{equation}
  n_{nt}=20.5 \times 10^{10} \frac{T_{\rm 6}^{-0.69}}{1+\frac{\delta-1}{3\theta}\sqrt{\frac{2\pi}{3 \theta}} \exp\left(\frac{3\theta}{2}\right)} \quad \rm cm^{-3}
    \label{eq:nonthermalfluxnormaisation}
\end{equation}

The non-thermal particle flux of particle $k=e,p$ is thus expressed as, 
\begin{equation}
\label{eq:jNTH}
    j_{k,\rm nt}(E)=\frac{\delta-1}{E_c} n_{\rm nt} \left(\frac{E}{E_{\rm c}}\right)^{-\delta}\exp\left(-\frac{E}{E_{\rm U}}\right) \frac{\beta_k(E)c}{4\pi }
    \,,
\end{equation}
The factor $\beta_k(E)$ is defined as $v_k(E)/c =\sqrt{((1+\bar{E})^2-1)/(1+\bar{E})^2}$, with $\bar{E}=E/ m_k c^2$ and $m_k$, the mass of particle $k$.

Assuming equipartition of energy ($F_e=F_p$), the following relation emerges between the flux of protons and electrons,

\begin{equation}
 j_{p}(E)= \frac{\beta_p(E)}{\beta_e(E)} j_{e}(E) 
\ .  
\label{eq:fluxprotonfluxelectron}
\end{equation}
\paragraph{Notice:}
We highlight here the difference between Eq. \eqref{eq:totaldensity} and Eq. 15 of the article \citep{2023MNRAS.519.5673B} further presented in Chapter \ref{C:PublicationI}. In this article, Eq. (15) is non homogeneous. There is obviously a confusion between the normalisation factor of the non-thermal flux and the non-thermal particle density. We corrected the issue in this section. %This error is a typographical error and is not affecting the results presented in \citep{2023MNRAS.519.5673B}.
The wrong estimation of the normalisation factor $N_{B23}$ leads to an overestimation of the flux in \citet{2023MNRAS.519.5673B}. We compare the normalisation $N_{B23}$ with the corrected normalisation factor $N_{nt}$ of Eq. \eqref{eq:jNTH},
\begin{equation}
    \frac{N_{B23}}{N_{nt}}= \frac{E_c}{k T (\delta-1)}= \frac{3\theta}{2(\delta-1)}.
\end{equation}

Taking the fiducial values of \citet{2023MNRAS.519.5673B}, $\delta=3$ and $\theta=3$, $N_{B23}\approx 2 N_{nt}$, the fluxes and ionisation rate computed in \citet{2023MNRAS.519.5673B} (Chapter \ref{C:PublicationI}) are reduced by a factor of two. We account for this reduction in the work in preparation presented Chapter \ref{C:PublicationII}.

\subsubsection {Maximal energy of the suprathermal distribution}\label{sect:Maximalenergy}

\paragraph{Maximal energy of electrons:}
From \citet{2016A&A...590A...8P}, the maximal energy of the electron distribution is attained by equating the characteristic acceleration time with the characteristic loss time $t_{acc}^{-1}=t_{loss}^{-1}$,
where the characteristic loss time is defined in terms of the loss function $L$ as,
\begin{equation}
    \frac{dE}{dt}=n v L(E)\approx \frac{E_{max}}{t_{loss}}
    .
\end{equation}
\begin{equation}
    t_{loss}^{-1}= \frac{c \beta n L(E)}{ m_e c^2 (\gamma-1)} =5\times 10^{-3} \frac{\beta}{\gamma -1} n_9 \left(\frac{L(E)}{10^{-25} ~\rm GeV ~cm^{2}}\right)
    .
\end{equation}

At high energy the dominant losses are synchrotron losses $L_s(E)$, 
\begin{equation}
    \frac{L_s(E)}{(10^{-25}~ \rm GeV ~ cm^{2})}=2\times 10^{-1} \frac{\gamma^2}{\beta} n_9 \left(\frac{B}{ 1 \rm kG}\right)^2,
\end{equation}
the energetic losses timescale $t_{loss}^{-1}$ is thus given by,
\begin{equation}
    t_{loss}^{-1}=10^{-3} \gamma \left(\frac{B}{1 \rm kG}\right)^2.
    \label{eq:tloss}
\end{equation}

To estimate the acceleration timescale, we rely on simulations of particle acceleration by magnetic reconnection.
\citet{Li_2017} showed that the electron and proton kinetic energies are proportional to the particle gyrofrequency $\Omega$, $K(t)\sim \alpha t \Omega_{g}$, with $\alpha\approx100$, where $\Omega_g=\frac{eB}{\gamma m c}$.

So the acceleration timescale is,
\begin{equation}
    t_{acc}^{-1}=\frac{\Omega_{c,e}}{\alpha}= \frac{e B}{\alpha \gamma m_e c}=\frac{10^8}{\gamma } \left(\frac{B}{1 \rm kG}\right).
    \label{eq:tacc}
\end{equation}

Equating Eqs.\eqref{eq:tacc} and \eqref{eq:tloss}, we get,
\begin{equation}
\gamma_{e,max}=3\times10^5 \left(\frac{B}{1 ~\rm kG}\right)^{-1/2},
\end{equation}
and the maximal energy reached by the electrons is,
\begin{equation}
    E_{e,max}\approx 150 ~ \rm  GeV \left(\frac{B}{1~\rm  kG}\right)^{-1/2} .
\end{equation}

\paragraph{Maximal energy of protons:}
In the case of protons, the acceleration limiting timescale is constrained by the advection timescale of the reconnected outflow. This can be expressed as follows,
\begin{equation}
    t_{p,min}=\min\{t_{\pi},t_{s},t_{adv},t_{esc}\}=t_{adv},
\end{equation}
Here, $t_{\pi}$ and $t_{s}$ represent the timescales for pion production and synchrotron losses, respectively. Meanwhile, $t_{adv}$ denotes the time required for a proton to be conveyed out of the acceleration region, and $t_{esc}$ stands for the time it takes for a proton to reach a Larmor radius larger than the acceleration lengthscale.

The acceleration lengthscale, denoted as $l_{out}$, is the outflow lengthscale from the reconnection region. The classical MHD reconnection theory, outlined in Section \ref{sect:ClassicalMHDTheories}, links the outflow lengthscale with the inflow lengthscale $l_{in}$ and the reconnection rate $R$, as $l_{out} = l_{in}/ R $. In the context of Solar flares, $l_{in}$ corresponds to the flare loop size and the observed reconnection rate is approximately $R\sim0.01$. We will use these scaling law, valid on the Sun, to estimate the properties of T Tauri flares.

The advection timescale can be written as $ t_{adv}= \frac{l_{out}}{V_{out}}$, with $l_{out}$ and $V_{out}$ indicating the length and velocity of the outflow, respectively. The outflow velocity can be approximated by the Alfvén velocity, defined as $V_A= B/\sqrt{4\pi m_p n_{rec}}$. Therefore,
\begin{equation}
    t_{ad}^{-1}= \frac{V_{out}}{l_{out}}\approx 700 ~ n_9^{-1/2} \left(\frac{B}{1 \rm kG}\right) s^{-1}.
    \label{eq:shockadvectiontimescale}
\end{equation}
The maximal energy of the protons is attained when Eq. \eqref{eq:shockadvectiontimescale} and Eq. \eqref{eq:tacc} are equivalent. Taking $\Omega=\frac{eB}{\gamma m_p c}$, 
\begin{equation}
    \gamma_{p,max}= 140 ~\rm n_9^{1/2} l_{9,loop}.
\end{equation}
The maximal energy attained by protons is thus,
\begin{equation}
    E_{p,max}= 140 ~\rm GeV ~ n_9^{1/2} l_{9,loop}
\end{equation}
In Section \ref{sect:constrainingmaxenergy} we explained, based on our analysis of \citet{2023ApJ...944..192K}, that in solar flares, the maximum energy of electrons is likely to be comparable to that of protons. If we assume density and loop size typical of solar flares, we indeed obtain $E_{e,max}\approx E_{p,max}$.

Nevertheless, as demonstrated in Sect.\ref{sec:ConstrainingPlasmaDensity}, the density and flare loop size of T Tauri flares diverge significantly from those of solar flares. Generally, T Tauri flare loops are expected to be about a hundred times longer based on soft X-ray observations. In the previous section, we derived relationships between the flare loop size, density, and temperature. We use Eqs \eqref{eq:nedelR} and \eqref{eq:LRdeL} to estimate the product $n_{rec}^{1/2} l_{loop}$ in terms of the flare temperature $T$. We get,
\begin{equation}
   ~\rm n_{9,rec}^{1/2} l_{9,loop}(T)= 10^3 T_6^{0.1} \approx 10^3.
\end{equation}

Hence, we anticipate the maximum energy of protons generated in T Tauri flares to be independent of the flare properties, equal to,
\begin{equation}
    E_{\rm p,max} = 140 ~\rm TeV.
\end{equation}

The values of \(E_{\rm p,max}\) and \(E_{\rm e,max}\) derived here allow us to estimate the value of the exponential cut-off in the suprathermal particle flux from Eq. \eqref{eq:jNTH}.
In Chapter \ref{C:PublicationI} we discuss the effects of this cut-off energy on the ionisation rate in T Tauri disc. We will see that high energy particles propagate and ionize the disc at higher column density.

\subsubsection{Defining the fiducial injection spectra of non-thermal particles}

In this section, we established the definition of what we will refer to as the "fiducial injection spectra" which we will study in the next chapters. We still proceed to a parametric study in Chapter \ref{C:PublicationI}.

Solar flare measurements by \citet{mewaldt2005proton} and studies of non-thermal particle injection in young stellar objects by \citet{waterfall2020predicting} have suggested different spectral indices, between approximately 4 and 3, respectively. For our model, we have chosen a spectral index of 3 for the non-thermal component as often assumed for solar flares (\citealt{2008LRSP....5....1B} and \citealt{2011SSRv..159...19F}). But we will also examine the effects of softer and harder spectra in Chapter \ref{C:PublicationI}. While earlier works by \citet{arnold2021electron} and \citet{oka2018electron} have suggested a range of possible power-law indices from 2 to 9, we decided to narrow this range. We chose an upper limit that allows a substantial emergence of the non-thermal component from the thermal one. Thus, we studied the power-law indices in a range of 2 to 8. 

Injection energies are typically a few times larger than the thermal energies. We take $E_c=3 E_{\rm th}$, i.e. $\theta=$3. So $E_c$ is in the 1-10 keV range, consistent with low energy cut-offs of non-thermal particle distributions deduced from solar flare surveys. 

We have considered the same injection and maximal energy for both electrons and protons, discarding any break in energy. By default, we set the high energy limit, $E_{U}$, to 100 MeV as \citet{2019MNRAS.483..917W} and introduced an exponential cut-off energy. The fiducial non-thermal flux is given by the formula,

\begin{equation}
    j_{k,\rm nth}(E)=n_{k,\rm nt} {\beta_k(E) c \over 4\pi} \left(\frac{E}{3E_{\rm th}}\right)^{-3} \exp\left(-\frac{E}{E_{\rm max}}\right).
    \label{eq:injection spectrum}
\end{equation}
The normalisation factor $n_{\rm nt}$ is given by Eqs. \eqref{eq:nonthermalfluxnormaisation} with $\theta=3$. In the fiducial case the flare has a temperature of 1 MK. For $E<E_U$ we can write the injection flux as,
\begin{equation}
    j_{\rm inj,k}(E)=j_{\rm 0,k} \beta_k(E) \left(\frac{E}{E_{0,k}}\right)^{-3},
\end{equation}
where, for protons $j_0 =  4\times 10^4 ~ \rm part/eV/s/str/cm^{2}$ and $E_0=10^6$ eV. And for electrons, $j_0 = 4\times 10^{13}~ \rm part/eV/s/str/cm^{2}$ and $E_0=10^3$ eV. In the subsequent chapter, Sec. \ref{sec:TransportMechanisms}, we introduce the particle propagation model employed in Chapter \ref{C:PublicationI} and \ref{C:PublicationII}. It is the Continuous Slowing Down Approximation (CSDA). We will demonstrate that this approximation is valid in discs for the propagation of particles with energies less than 1 GeV. Although we find \(E_{p,max}=140\) TeV and \(E_{e,max}=150\) GeV, we set \(E_U= 1\) GeV to ensure that only particles that meet the CSDA criterion are treated.

\section{Conclusion}

This chapter aimed at establishing a framework for understanding particle production from magnetic reconnection events in T Tauri flares. We initiated the discussion with an overview of key theoretical and historical models, namely, the Sweet-Parker and the Petscheck models, and turbulent reconnection models. These studies shaped our current understanding of magnetic reconnection. We have pointed out that, despite its widespread recognition in astrophysical contexts, the development of magnetic reconnection theory and of its role in particle acceleration is still work in progress.

To offer a more concrete understanding of particle acceleration in T Tauri flares, we used parallels with the more extensively researched realm of solar flares. By adopting the well-established standard model for solar flares, we focused on translating observable X-ray parameters into constraints on the flux and energy distribution of non-thermal particles produced during flares. Specifically, we highlighted how factors such as X-ray luminosity constrains key parameters like the power-law index, injection energy, and maximum energy attainable by these particles. The standard solar flare model further afforded us insights into the geometric aspects of these flares.

Focusing the discussion onto T Tauri-specific scenarios, we introduced our own model for particle acceleration, tailored to the specifications of T Tauri flares, which energies are orders of magnitude greater than their solar counterparts. However, the adaptability of the solar model to T Tauri stars was noted to have limitations, particularly with respect to the flare geometry. Given the magnetic environment shaped by the presence of protoplanetary discs around T Tauri stars, we proposed a revised flare geometry, inspired by numerical simulations and observational data. Understanding this geometry is essential, as it determines the entry points for particle flux into the disc, thereby influencing the spatial distribution of ionisation.

Finally, this chapter serves to lay the foundation for our model of particle acceleration in T Tauri flares. This model not only predicts the energy spectrum of emitted particles but also anticipates their trajectories along magnetic field lines in both the disc and outflows. Such insights are crucial for understanding the various transport regimes and energy losses that these particles will encounter, topics that will be the focus of our next chapter.

\chapter{Energetic particles interaction and propagation in T Tauri discs}\label{C:Propagation}

%\textcolor{red}{je propose : 1 introduction : necessite de prendre en charge le transport mais les theories standards du transport du rayonnement cosmique dans le milieu interstellaire ne sont pas facilement  adaptables aux disques. On a affaire en geenral a des particules de basse energie et de petit rayons de Larmor pour lesquelles on peut en premirer approximation decrire le transport en utilisant une trajectoire ballsitique. Mais des travaux recents ont  commmence a aexplorer les effets de la diffusion : Rodgers-Lee ... 2) two main transport mechanisms : CSDA, diffusion. Pour la diffusion il faut reprendre les resultats de Rodgers Lee puis ceux de Ivlev et faire uen mini revue 3) Magnetic field topologies: pour des particules de basse énergie les effets de mirooir magnetiques (a decrire) sont importants: modele Padovani et al 2013 à decrire 4) quelles types de régimes ont doit s'attendre dans le disque ? ta section 4.4 actuelle mais avec le modele de Sun et Bai}

\section{Introduction}
In protoplanetary discs, magnetic fields play an indispensable role, orchestrating the propagation of energetic particles. These magnetic fields can have various origins: they may emanate from the central star, or be "seed fields" left over from prior stages of protostellar evolution, or even originate from the disc own dynamo mechanisms. 

Magnetic field lines essentially function as 'tracks' guiding these energetic particles. The configuration of these magnetic fields can range from simple and orderly to complex and chaotic, each resulting in unique propagation behaviours for the particles. In straightforward magnetic environments like dipolar fields, particles follow predictable paths along the field lines. However, when magnetic fields are intricate or disordered, the particles exhibit diffusion-like motion.

As these particles move through the disc, they undergo various interactions with the medium. These interactions result in energy losses due to processes like ionisation, as well as the production of secondary particles. Consequently, the overarching question that guides this chapter is how can we model the propagation of particles generated in the inner regions of T Tauri discs through magnetic reconnection events.

To tackle this question, the chapter is structured as follows. First in Sect. \ref{sec:EnergyLossesDisc}, we will elucidate the types of energetic losses that low-energy electrons and protons (with energies \(\lesssim 1\) GeV) could experience in a partially ionised plasma. Following that in Sect. \ref{sec:TransportMechanisms}, we will discuss the two principal mechanisms of particle transport, free streaming and diffusion, in the context of protoplanetary discs. An analytical model will be introduced for estimating ionisation rates based on the free streaming approximation, and we will also propose an estimation for the diffusion coefficients, grounded in various turbulence models. Finally, having examined the propagation of primary particles, we will estimate the flux of secondary particles that emerge as a result of these processes.

By the end of this chapter, the objective is to offer a comprehensive framework for understanding particle propagation within T Tauri discs, particularly when accelerated by magnetic reconnection events. This understanding will not only advance our knowledge of the propagation of particles in protoplanetary discs but also gives the reader the closing ingredients needed to grasp the publications in the next two chapters.

\section {Energy losses in circumstellar discs}\label{sec:EnergyLossesDisc}

In this section, we outline the primary energy loss processes impacting the penetration of primary cosmic rays and secondary particles in circumstellar discs, as described in previous works (e.g., \citealt{Padovani2009,Padovani18,padovani2022cosmic}). These processes often result in showers composed of secondary species (photons, electrons, and positrons) produced via pion decay, Bremsstrahlung (BS) and pair production. Both primary and secondary particles contribute to the overall ionisation rate of the medium.

\subsection{Energy loss function }
The energy loss function, $L^l_k$, for a particle of species $k$ varies based on the type of process $l$. If only a small portion of the particle kinetic energy is lost at each collision, the process can be considered as being continuous. This approximation is called the continuous slowing down approximation (CSDA). We discuss in more details its range of applicability in Sec. \ref{sec:FreeStreaming}. Under this assumption the losses are described by the loss function defined as,

\begin{equation}
L^l_k(E) = \int_0^{E_{max}} E' \frac{d\sigma^l_k(E, E')}{dE'} dE' 
\label{eq:LossFunctionDefinition}
\end{equation}
where $d\sigma^l_k/dE'$ is the differential cross section of the process and $E_{max}$ is the maximum energy lost in a collision. In contrast, if a large portion of the kinetic energy is lost in a single collision or the CR particle is annihilated post-collision, the process is classified as catastrophic and the loss function becomes

\begin{equation}
    L^l_k(E) = E\sigma^l_k(E)
\end{equation}

where $\sigma_k$ is the process cross section. We discuss the catastrophic losses range of applicability in 
Sect. \ref{sec:DiffusionPions}. We then express the total energy loss function $L_k = \sum_l L^l_k$ in terms of the loss functions for collisions with atomic hydrogen ($L_{k,H}$) or helium ($L_{k,He}$). The loss function for collisions with molecular hydrogen can be approximated as $L_{k,H_2} \approx 2L_{k,H}$, except for elastic and excitation losses. 

In circumstellar discs, one of the factors that influences the energy loss function of energetic particles, such as electrons and protons, is the composition of the disc itself. Assumptions about the disc composition can simplify the calculations and make them more tractable, while still providing a reasonably accurate estimation of the energy loss function.

The key assumption here is that the disc is composed of free electrons, atomic hydrogen, molecular hydrogen, and helium. These species reflect the observed and theoretical understanding of the composition of many circumstellar discs and interstellar medium. Hydrogen in atomic and molecular forms and helium are indeed the most abundant elements in circumstellar discs, while free electrons are expected due to various ionisation processes. Although this model does not consider every possible atomic or molecular species in the disc, it does capture the main contributions to the energy losses. It is a representative model, meaning it is designed to capture the essential physics without including every possible detail. Furthermore, this simple model applies broadly to a wide range of circumstellar discs around different types of stars and in different evolutionary stages. It is a useful "first approximation" for many different scenarios.

While this simplified model is satisfying for initial estimates and broad-brush studies, more complex models that include additional elements and molecules may be needed for more detailed studies, especially when considering specific disc environments or when high precision is required.

\subsection{Energy losses of protons}
The loss function, technically defined as the differential energy loss per unit path length, quantifies the amount of energy an energetic particle loses as it crosses a specific medium, in this case, the circumstellar disc. It essentially provides a mean to measure the change in the particle energy along a differential column density, accounting for the disc varying density along the particle path.

\begin{figure}
    \centering
    \includegraphics[width=\linewidth]{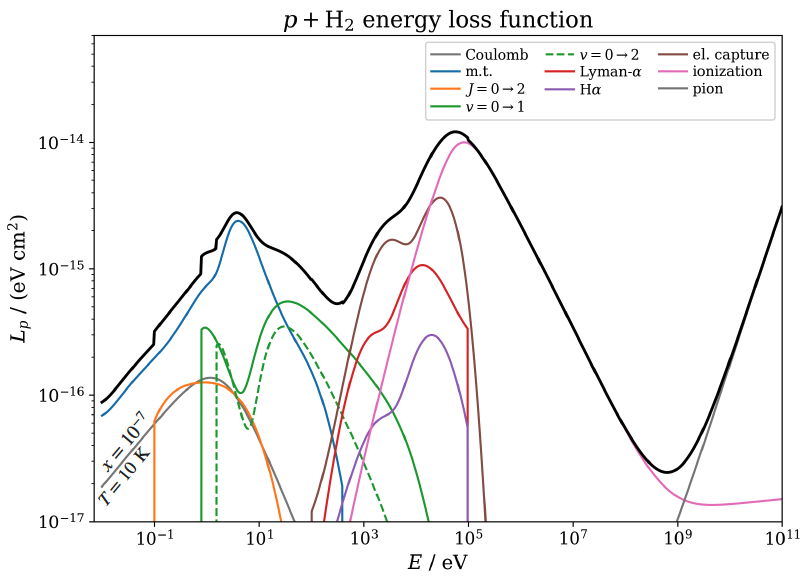}
    \caption{The energy loss function for protons colliding with H$_2$, including contributions of pion losses, is represented by a solid black line. Each coloured line refer to each individual component. The specific references tied to each line indicate the source papers from which the related cross sections were derived: momentum transfer (solid blue line ); rotational transition $J = 0 \rightarrow 2$ (solid orange line); vibrational transitions $v = 0 \rightarrow 1$ (solid green line) and $v = 0 \rightarrow 2$ (dashed green line); Lyman-$\alpha$ absorption line (solid red line); H$_\alpha$ absorption line (solid purple line); electron capture (brown line); ionisation (solid cyan line); pion production (solid grey line). The grey line at low energy illustrate the Coulomb losses at 10 K for ionisation fractions $x_E$ of $10^{-7}$ and $10^{-8}$. In the inner disc surface, temperature and ionisation fraction are expected to be much higher than those plotted on this graph. In the inner disc region, the Coulomb losses dominate at low energy, see Fig. \ref{fig:protonlossfunction1}.}
    \label{fig:ExcitationLossesProton}
\end{figure}

The proton energy loss function, denoted as $L_p(E)$, is governed by processes depending on the energy of the proton and the medium in which it is propagating. 
\begin{equation}\label{Eq:LEP}
    L_p(E)= L^C(E)+ L^{\rm ion}(E)+L^\pi(E) \ .
\end{equation}
In Eq.\ref{Eq:LEP} each term in the right hand side part refer to, the Coulomb losses ($L^C$), the ionisation losses ($L^{\rm ion}$) and the pion losses ($L^{\pi}$) respectively. All the energy losses are summarised in Fig. \ref{fig:ExcitationLossesProton} reproduced from Astrochemical Modelling by Bovino and Grassi (Chapter by Padovani \& Gaches in prep Nov. 2023).

\paragraph{Low energy processes:}
At lower energies, the loss function is dominated by Coulomb losses. Coulomb losses refer to the energy loss by a charged particle, such as a proton, due to its electromagnetic interaction with the other charged particles in the disc material. The energy loss rate of an energetic particle due to Coulomb losses is \citep{2002cra..book.....S},
\begin{equation}
    L_{s}^C=\frac{3}{2} \frac{\ln(\Lambda) \sigma_T m_e c^2 Z^2}{ \beta^2} \sum_s \frac{m_e}{m_s}Z_s^2 \frac{n_s}{n} W_s\left(\frac{\beta}{\beta_s}\right ) \quad \rm eV~ s^{-1}
\end{equation}
where $M$ is the mass of the energetic particle, $Z e$ its charge and $\beta c$ its velocity. The energetic particle propagate in a thermal plasma consisting of $s$ species, of respective mass $m_s$, charge $Z_s e$, density $n_s$ and temperature $T_s$, noting $n=\sum_s n_s$ the total plasma density. $\sigma_T$ is the Thomson cross section. We have,
\begin{equation}
    \beta_s=\left(\frac{2 k T_s}{m_s c^2 }\right)^{1/2}
\end{equation}
and 
\begin{equation}
    W_s(x)=\frac{2}{\pi^{1/2}}\left( \int_0^x e^{-y^2}dy -\left(1+\frac{m_s}{M}\right)x e^{-x^2}\right).
    \label{eq:Ws(E)}
\end{equation}
The Coulomb logarithm $\ln(\Lambda)\approx 20 $ for a wide range of plasma and temperature \citep{2002cra..book.....S}. As we are studying the propagation of protons, we then use $M=m_p$ and $Z=1$. Due to the small ratio $\frac{m_e}{m_p}= \frac{1}{1836}$ we can easily approximate the term,
\begin{equation}
    \sum_s \frac{m_e}{m_s}Z_s^2 \frac{n_s}{n} W_s\left(\frac{\beta}{\beta_s}\right )\approx n_e W_e\left(\frac{\beta}{\beta_e}\right ).
\end{equation}
This means that for all values of the metallicity of the plasma, the Coulomb collisions are dominated by scattering off thermal electrons. Hence,
\begin{equation}
    L^C(E) \simeq 30 \frac{\sigma_T m_e c^2}{\beta(E) ^2} \frac{n_e}{n} W_e\left(\frac{\beta(E)}{\beta_e}\right ).
    \label{eq:CoulombianLosses}
\end{equation}
This energy loss arises primarily from deflection and scattering processes, which become stronger when the relative velocities of the energetic particles are comparable to the thermal velocity of the plasma. 

Coulomb loss dominates at low energy over rovibrationnal losses when the ionisation fraction and the temperature of the disc are high enough. We shall show that such conditions are meet at the disc surface layer.

\paragraph{Intermediate energy processes:}
At intermediate energies, the energy loss of energetic particles in circumstellar discs is predominantly governed by ionisation losses. Ionisation losses occur when energetic particles collide with the atoms and molecules in the disc, leading to the ejection of electrons and hence ionisation of the atoms or molecules.

The key aspect to consider here is the interaction of energetic particles with the matter within the disc. As these particles cross the disc, they interact with the material (primarily hydrogen and helium) via electromagnetic forces. If the energy of the incident particle is sufficiently high, it can eject electrons from the atoms it collides with, resulting in the ionisation of those atoms.

The ionisation process inherently involves an energy transfer. The kinetic energy of the energetic particle is used to overcome the binding energy of the electron in the atom and further to provide the ejected electron with its kinetic energy. Consequently, the incident particle loses energy in the process, which is reflected in the energy loss function. We have,
\begin{equation}
    L_p^{\rm ion}(E)= \sum_i \int_0^{\frac{E-I_i}{2}} \frac{d\sigma_{p,i}^{ion}(E,E')}{dE'}(I_i+E')dE'
   \,.
\end{equation}

The fraction $\frac{d\sigma_{p,i}^{\rm ion} (E,E')}{dE'}$ is the differential ionisation cross section \citep{krause2015crime} of protons in target species $i$, where $i$ is atomic hydrogen, molecular hydrogen and helium. $E'$ is the secondary electron energy and $I_i$ the ionisation potential of species $i$.

The "Stopping and Range of Ions in Matter" (SRIM) \citep{ziegler2010srim} is a database and software for calculating the loss functions. SRIM provides ionisation cross-sections and stopping power data for a wide range of ions and materials. We use SRIM to estimate the ionisation losses of protons, providing inputs for the calculation of the energy loss function. It uses a comprehensive physical model that includes ionisation and other loss mechanisms.

However, it is important to note that while SRIM provides highly accurate estimates for ionisation losses, it may not fully account for the specific conditions in a circumstellar disc, such as variations in temperature, density, and disc composition. Therefore, while SRIM data is valuable, it may need to be supplemented or adjusted with other information specific to the circumstellar disc environment.

\paragraph{High energy processes}
Finally, at high energies, the proton energy loss function is primarily governed by pion production and subsequent decay losses. At these high energies, protons can interact with atomic nuclei in the disc, producing pions through strong interaction processes. The creation of these pions effectively represents an energy loss for the protons. The newly formed pions are unstable and quickly decay into neutrinos and gamma rays, further contributing to the energy loss of the system. The loss function for pion production is given by, \citet{2002cra..book.....S},
\begin{equation}
    L_{p,Z}^\pi(E)\simeq 2.57\times 10^{-17} \frac{A_Z^{0.79}}{\beta}\left(\frac{E}{GeV}\right)^{1.28} \left( \frac{E+E^{as}}{GeV} \right)^{-0.2} \quad \rm eV~ cm^{2},
\end{equation}
where $\beta$ represents the ratio of the proton velocity to the speed of light. Additionally, the asymptotic energy, denoted by $E^{as}$, is set to be 200 GeV. Besides, a phenomenological correction to the pion production cross section for heavier target species is introduced by the mean of the factor $A_Z^{0.79}$, where $A_Z$ is the atomic mass number of species Z.

The dominance of pion losses is noted for energies of approximately 1 GeV or greater. In these high-energy cases, pion losses become the primary driver of the energy loss function. These losses significantly influence the propagation of CRs at high column densities, effectively setting the distribution of high-energy CRs in these dense environments.

Following these considerations, the total pion production loss function can then be derived \citep{2002cra..book.....S},
\begin{equation}
  L_p^\pi(E) \simeq \left(\sum_{Z\ge1 } f_Z A_Z^{0.79}  \right) L^\pi_{p,H}(E),
\end{equation}
where $f_Z$ is the relative abundance of the species Z.

The proton energy loss function $L_p(E)$ describes how the energy of protons changes as they cross the circumstellar disc.

\subsection{Energy losses of electrons}
\begin{figure}
    \centering
    \includegraphics[width=\linewidth]{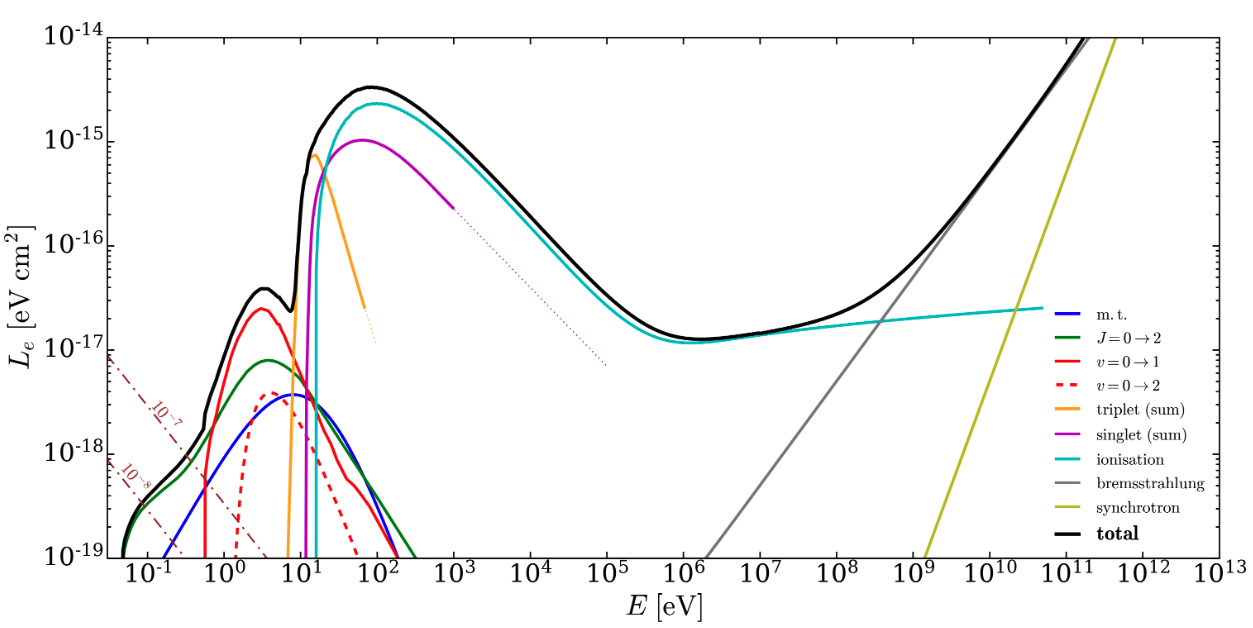}
    \caption{The energy loss function for electrons colliding with H$_2$, including synchrotron loss contributions, is represented by a solid black line. Each coloured line refers to each individual component. The specific references tied to each line indicate the source papers from which the related cross sections were derived: momentum transfer (solid blue line from \citealt{2008A&A...484...17P}); rotational transition $J = 0 \rightarrow 2$ (solid green line from \citealt{1988AuJPh..41..573E}); vibrational transitions $v = 0 \rightarrow 1$ (solid red line from \citealt{yoon2008cross}) and $v = 0 \rightarrow 2$ (dashed red line from \citealt{janev2003collision}); electronic transitions incorporating all triplet and singlet states (solid orange and magenta lines from \citealt{2021ADNDT.13701361S}); ionisation (solid cyan line from \citealt{2000PhRvA..62e2710K}); Bremsstrahlung (solid grey line from \citealt{blumenthal1970bremsstrahlung,2018A&A...614A.111P}); and synchrotron (solid yellow line from \citealt{2002cra..book.....S,2018A&A...614A.111P}). Dash-dotted brown lines illustrate the Coulomb losses at 10 K for ionisation fractions $x_E$ of $10^{-7}$ and $10^{-8}$ \citep{1971JGR....76.8425S}. In the inner disc surface, temperature and ionisation fraction are expected to be much higher than those plotted on this graph. In the inner disc region, the Coulomb losses dominate at low energy, see Fig. \ref{fig:electronlossfunction1}}
    \label{fig:ExcitationLossesElectron}
\end{figure}

The electron energy loss function, denoted as $L_e(E)$, is governed by processes depending on the energy of the electron and on the target medium in which it is propagating. 

\begin{equation}
    L_e(E)= L^C(E)+L^{\rm exc}(E)+ L^{\rm ion}(E)+L^{BS}(E)+L^{S}(E)
    \label{Eq:LEE}
\end{equation}
In Eq.\ref{Eq:LEE} each term in the right hand side part refer to, the Coulomb losses ($L^C$), the rovibrationnal losses ($L^{\rm exc}$), the ionisation losses ($L^{\rm ion}$), the Bremsstrahlung losses( $L^{BS}$) and the synchrotron losses ($L^{S}$) respectively. All the energy losses are summarised Fig. \ref{fig:ExcitationLossesElectron} reproduced from \citet{2022A&A...658A.189P}.

\paragraph{Low energy processes:}

At low energies, the function is dominated by Coulomb and rovibrationnal excitation losses. 

Coulomb losses for electrons can be described by Eqs. \eqref{eq:Ws(E)} and \eqref{eq:CoulombianLosses} where the mass of the energetic particle $M=m_e$. Rovibrational excitation losses arise from interactions of energetic particles with molecules in the disc, primarily molecular hydrogen in this context. When an energetic particle collides with a molecule, it can induce transitions in the molecule rotational and vibrational energy levels. These transitions lead to energy being transferred from the energetic particle to the molecule, effectively causing an energy loss for the particle.

In the low-energy regime, these rovibrational losses become particularly important. The energies of the particles in this regime align well with the energies required for exciting rotational and vibrational modes in molecules, making these transitions more likely. %Consequently, this process significantly contributes to the energy loss of particles in this energy range. 

\begin{equation}
    L^{\rm exc}=\sum_j \sigma_{\rm exc,j}(E) E_{\rm thr}
\end{equation}
$\sigma_{\rm exc,j}$, $j$ are the cross section of the excitation of state $j$ summarised in \citet{2022A&A...658A.189P}, $E_{\rm thr}$, $j$ is the corresponding excitation threshold energy.

\paragraph{Intermediate energy processes:}
At intermediate energies, the energy loss of electrons is predominantly governed by ionisation losses. Relativistic Binary Encounter Dipole (RBED) theory is used to describe the ionisation processes by relativistic electrons \citep{2000PhRvA..62e2710K}. 

The binary encounter dipole approximation suggests that the ionisation of an atom by an incident electron can be described as an encounter between the incident electron and a bound atomic electron, with the interaction approximated as a dipole interaction.

In the RBED model, the ionisation differential cross section, which gives the probability of ionisation for a given energy loss, is calculated by taking into account relativistic corrections to the Bethe equation. This model provides a more accurate description for high energy electrons where relativistic effects cannot be neglected.
This ionisation process results in a transfer of energy from the electrons to the ejected atomic electrons and the ionised atoms with,
\begin{equation}
    L_e^{\rm ion}(E)=\sum_i \int_0^{\frac{E-I_i}{2}} (I_i+E')\frac{d\sigma_i^{RBED}}{dE'} (E,E')dE'
    \,.
\end{equation}
The quantity $\frac{d\sigma_i^{RBED}}{dE'} (E,E')$ is the differential ionisation cross section of an electron in a target medium composed of the species $i$ \citep{2000PhRvA..62e2710K}, where $i$ can be atomic hydrogen, molecular hydrogen or helium. $E'$ is the secondary electron energy and $I_i$ the ionisation potential of species $i$.

\paragraph{High energy processes:}

\textit{Bremsstrahlung} occurs where an electron is deflected by another charged particle, such as an atomic nucleus. During this deflection, the electron velocity changes, which, according to momentum conservation, results in the emission of radiation. This emitted radiation is the Bremsstrahlung radiation, and the energy it carries away leads to an energy loss for the electron, quoted as the Bremsstrahlung loss. Bremsstrahlung losses dominate at $E \ge 100$ MeV.

In circumstellar discs, electrons moving through the disc material undergo Bremsstrahlung interactions, we consider again interactions only with atomic and molecular hydrogen as well as helium. We take into account that $L^{BS}_{e,H2}= 2L^{BS}_{e,H}$ and that the differential Bremsstrahlung cross section is proportional to $Z(Z + 1)/2$,
\begin{equation}
    L^{BS}_{e}= (f_H+2f_{H_2}+3f_{He}) L^{BS}_{e,H}, 
\end{equation}
where,
\begin{equation}
    L^{BS}_{e,H}= \int_0^E \frac{d\sigma^{BS}(E,E_\gamma)}{dE_\gamma} E_\gamma dE_\gamma
\end{equation}
is computed from \citet{Padovani18}. The quantity $\frac{d\sigma^{BS}(E,E_\gamma)}{dE_\gamma}$ is the differential Bremsstrahlung cross section \citep{blumenthal1970bremsstrahlung}, where $E_\gamma$ is the energy of the emitted photon.

\textit{Synchrotron} radiation occurs where a relativistic charged particle, such as an electron, is deflected by a magnetic field. The deflection causes the particle to undergo acceleration, which results in the emission of radiation. This emitted radiation is the synchrotron radiation, and the energy it carries away corresponds to an energy loss for the electron, quoted as synchrotron losses. They are dominating at energies above $E_{\rm syn} \approx 1$ TeV. Following \citet{2002cra..book.....S},
\begin{equation}
    \frac{L_s(E)}{\rm GeV ~ cm^{2}} \simeq 2\times 10^{-29} \frac{\gamma^2}{\beta} \left(\frac{n}{10^6 \rm cm^{-3}}\right)^{-1} \left(\frac{B}{1 \rm G}\right)^2,
\end{equation}
where gamma is the Lorentz factor of the electron, $B$ the magnetic field intensity and $n$ the medium particle density.
Assuming as in \citet{Padovani18} a relation between the magnetic field strength, $B$, and the gas number density, $n$, given by Crutcher (2012),
\begin{equation}
    B \simeq B_0 \left(\frac{n}{n_0}\right)^\kappa
\end{equation}
with $B_0\approx 10$ $\mu$G, $n_0=150$ cm$^{-3}$ and the value of $\kappa=0.5$ given by (Nakano et al. 2002; Zhao et al. 2016), we benefit from the removal of the dependence of $L^{syn}$ on $B$ and $n$, thus the loss function is independent of the position, 
\begin{equation}
    L_e^{\rm syn}(E) \simeq 5.0\times 10^{-14} \left(\frac{E}{\rm TeV}\right)^2.
\end{equation}

We did not account for inverse Compton losses.
%\textcolor{red}{pas de Compton Inverse ?}
\subsection{Averaged loss functions}

Given that the abundances of hydrogen (H), molecular hydrogen (H$_2$), and helium (He) vary throughout the depth of the circumstellar disc, we propose a method to compute the average energy loss function, denoted as ${\bar{L}(E,s)}$. This function depends on the coordinate $s$, which represents the path of the particle through the disc. 

\begin{equation}
{\bar{L}(E,s) ={1 \over s} \sum_i \int_0^s L_i \dv{N_i(s)}{N(s)} \dd s} \ ,
\end{equation}

In this equation, $i$ symbolises the particle species, in this case, H, H$_2$, and He. $ \dd N_i(s)$ is the change in column density of species $i$ corresponding to a variation $\dd s$ along the path. The total variation in column density, $\dd N(s)$, is the sum of the changes in column densities for all species, i.e., $\dd N(s)= \sum_i \dd N_i(s)$.

Importantly, the loss function for each species $L_i$ is position-independent, so $\dv{N_i}{N}$ can be represented in terms of number density as $\dv{N_i}{N}=\frac{n_i(s)}{n(s)}$, which denotes the ratio of the number density of species $i$ to the total gas density.

Consequently, the average loss function $ \bar{L}$ can be simplified and rewritten as

\begin{equation}
 \bar{L}(E,s)=\sum_i f_i(s) L_i(E),
\end{equation}

\begin{figure}[h!]
    \centering
    \includegraphics[width=0.5\linewidth]{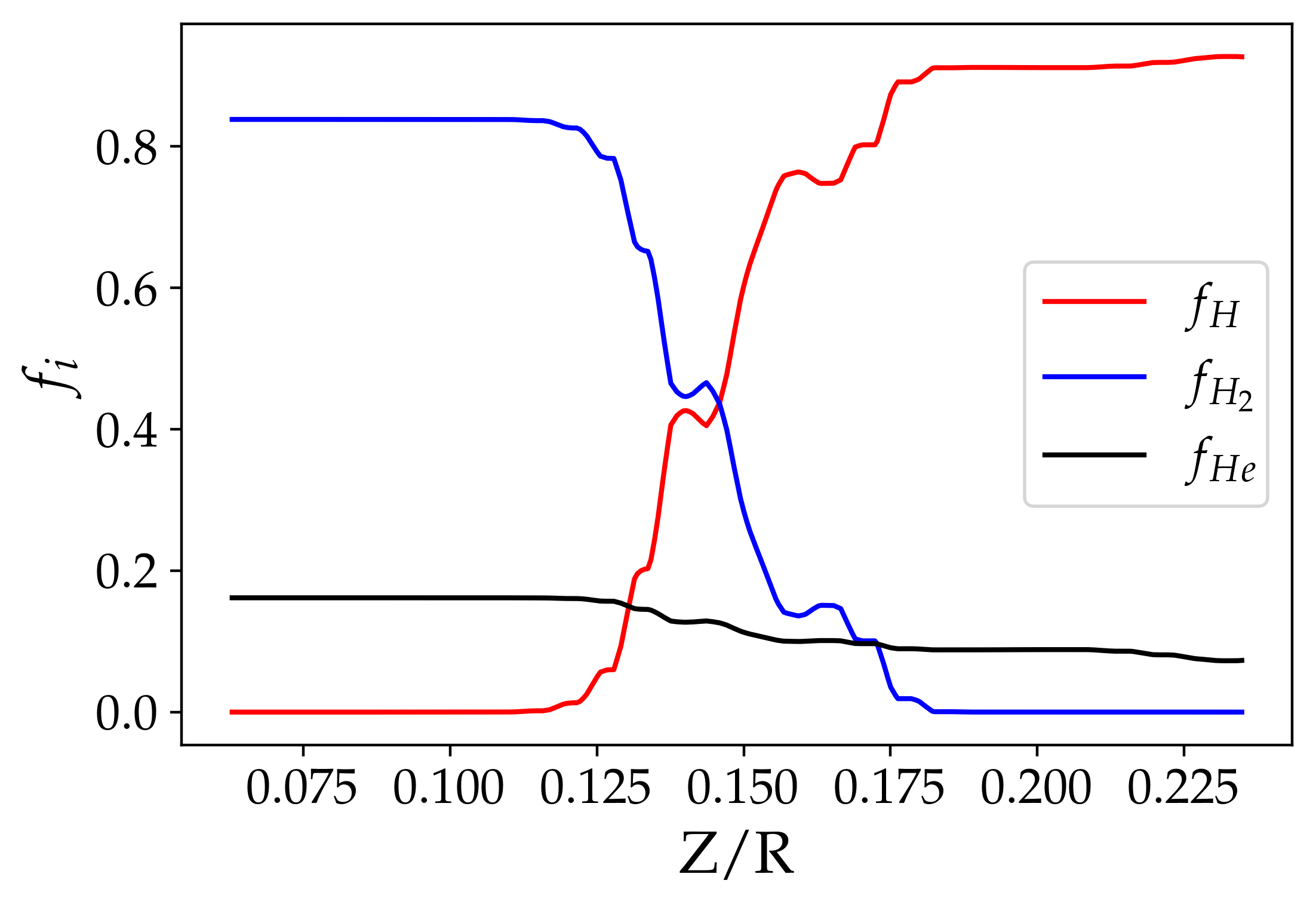}
    \caption{Plots of the fraction of chemical elements weighted along the path as function of the depth in the disc, computed from equation \eqref{Eq:Weight1}. In red we plot the fraction of atomic hydrogen, in blue the fraction of molecular hydrogen and in black the fraction of helium.}
    \label{fig:f_i1}
\end{figure}

where $f_i(s)$ is the fraction of species $i$ weighted along the path, given by
\begin{equation}\label{Eq:Weight1}
f_i=\frac{1}{s}\int_0^s \frac{n_i(s')}{n(s')}\dd s'.
\end{equation}
This method allows to take the spatial variation in the composition and the temperature of the disc into account. We hence make a new estimate of the average loss function over the particle trajectory up to point $s$ on the trajectory, or equivalently down to column density $N(s)$ crossed by the particles.

\begin{figure}[h!]
\begin{subfigure}{.5\textwidth}
  \centering
    \includegraphics[width=\linewidth]{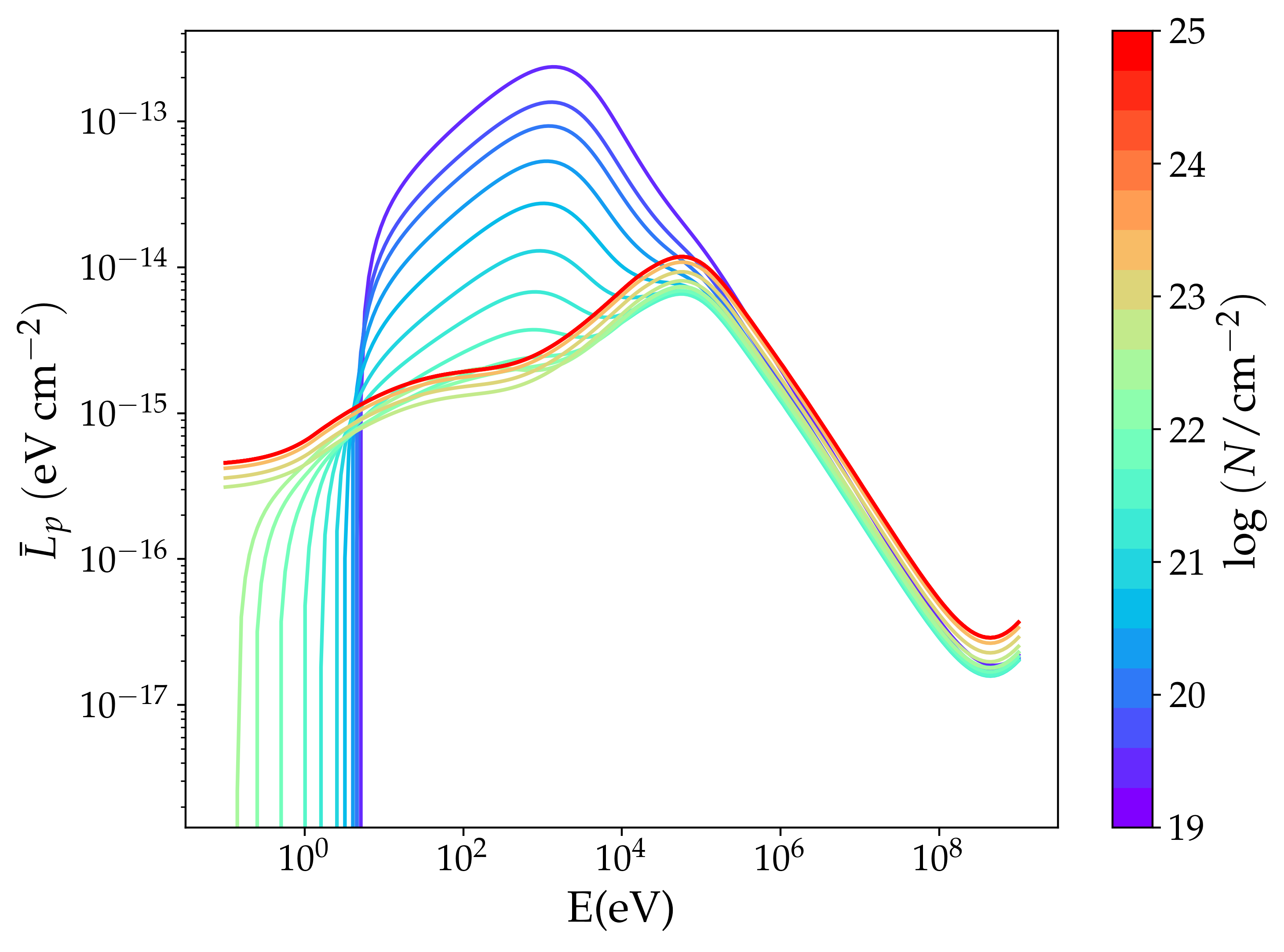}
    \caption{Mean loss functions of protons}
    \label{fig:protonlossfunction1}
\end{subfigure}
\begin{subfigure}{.5\textwidth}
  \centering
    \includegraphics[width=\linewidth]{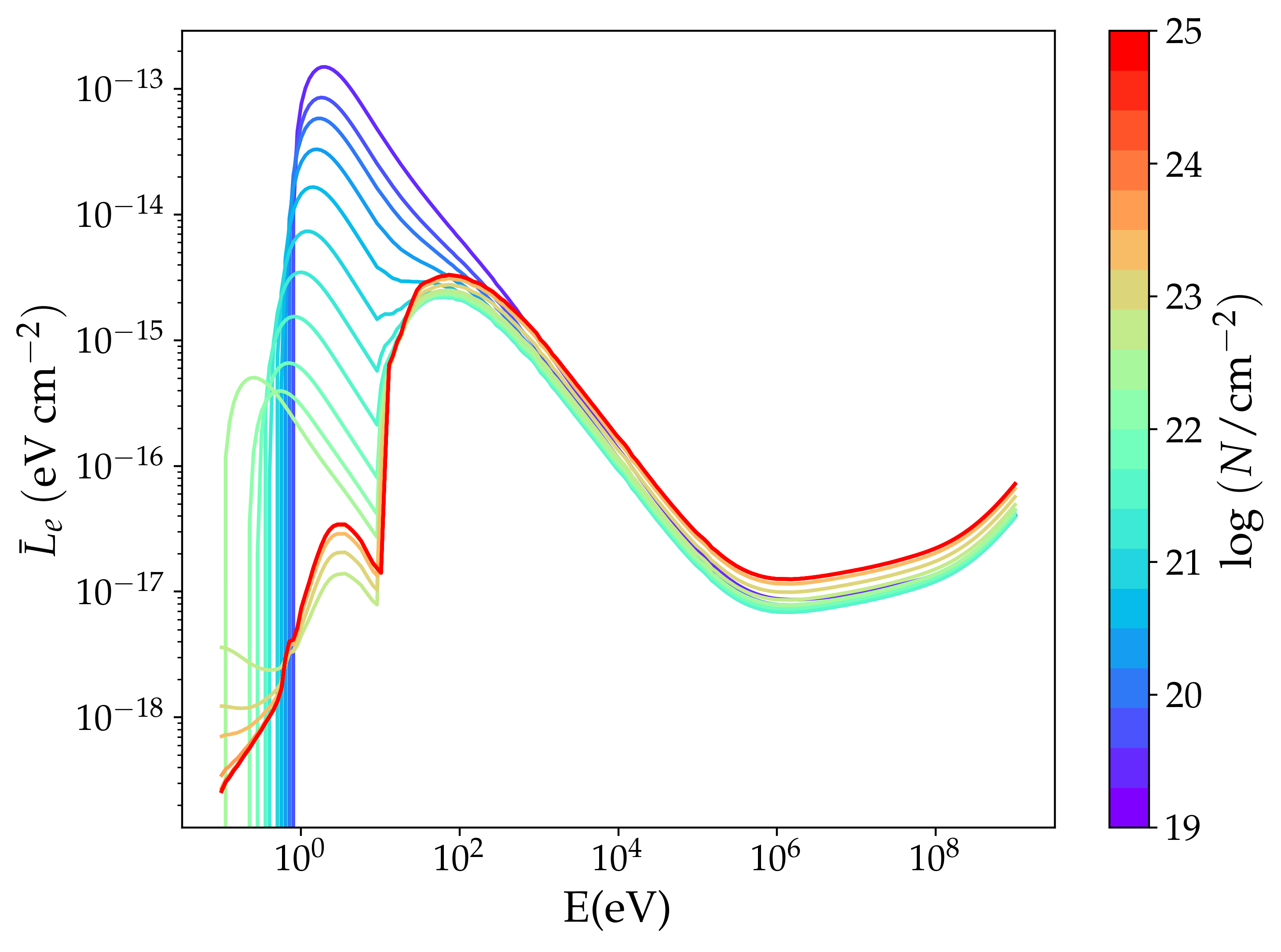}
    \caption{Mean loss functions of electrons}
    \label{fig:electronlossfunction1}
\end{subfigure} 
\caption{
Plots of the mean loss functions as function of the particle energy for protons (Fig. a) and electrons (Fig. b). Twenty loss functions at column densities N are plot, ranging from $10^{19} \rm cm^{-2}$ (blue curve) to $10^{25} \rm cm^{-2}$ (red curve).}
\label{fig:lossfunctions1}
\end{figure}

The protostellar disc model generated by {\tt ProDiMO}, presented in Sec. \ref{sec:focusonProDiMO} incorporates a comprehensive set of physical and chemical processes, including heating, cooling, radiative transfer, and a detailed treatment of the molecular chemistry. Using its results allows us to capture the variations in the abundances of species with the disc radius and its height, which are crucial for estimating the fraction of species along the path of energetic particles.
Fig. \ref{fig:f_i1} shows the species fractions and Fig. \ref{fig:lossfunctions1} the average loss functions for a set of column densities.

\section{Energetic particle transport mechanisms}\label{sec:TransportMechanisms}

The gyroradii of low-energy CRs, in the energy range of interest for ionisation and heating in discs \citep{2009A&A...501..619P,2018A&A...614A.111P}, are in principle much smaller than the spatial scales that characterise inhomogeneities in the magnetic field and gas density within these objects. 
The motion of CRs is guided by field lines, as long as the gyroradius, $r_g$, of the ion is smaller than typical magnetic coherence length. For instance, a 100 GeV CR proton and a magnetic field strength of $B = 10\, \mu G$, the gyroradius $r_g = \gamma mv_{\perp}c/(ZeB) \approx 0.002\, au$ is significantly smaller than the typical distance considered in discs $\lambda_{disc} \sim 0.1 - 1000\, au$. 
Given that the field strength is likely to be generally higher and the energy of CRs lower, this criterion holds for all particles under consideration below.

We hence use the concept of the guiding centre for CRs and consider their propagation along the local magnetic field. The coordinate $s$ is measured along the field line. 

We first introduce the effective column density of gas crossed by a CR particle for a given length $s$ 

\begin{equation}
N(s) = \int_0^s n(s') ds',
\end{equation}
where $n$ is the total number density of atoms. 

The kinetics of CRs are conveniently described in terms of their distribution function in energy space, $f_E = f (E, s, \mu)$. It is related to the distribution function in momentum space by $v f_E = p^2 f_p$, where $v(E)$ is the physical velocity of a particle and $\mu=\cos\theta$, where $\theta$ is the pitch angle, i.e. the angle between the particle velocity vector and the mean magnetic field line. The number of CRs per unit volume and energy, is then given by

\begin{equation}
F (E, s) = 2\pi \int_{-1}^{1} f_E d\mu.
\end{equation}

The transport equation for $f_E$, which describes the continuity of the distribution in the phase space, takes into account the energy losses by CRs due to their collisional interactions with gas \citep{1978A&A....70..367C}. The equation can be generally expressed as \citep{2020SSRv..216...29P},

\begin{equation}
\frac{\partial S}{\partial s} + \frac{\partial}{\partial E} \left(\dot{E}_{\text{con}}(E) f_E\right) +\nu_{\text{cat}}(E) f_E = 0,
\end{equation}
where $S(f_E)$ is the differential flux of CRs (per unit area, energy, and solid angle) and $\nu_{\text{cat}}(E)$ and $\dot{E}_{\text{con}}(E)$ denote the rates of catastrophic and continuous losses, respectively. Both $\nu_{\text{cat}}$ and $\dot{E}_{\text{con}}$ scale with $n_H$.

The distribution function generally depends on the pitch angle, which itself depends on the local strength of the magnetic field, $B(s)$, and CR scattering caused by CR interactions with MHD turbulence and gas nuclei. Depending on the degree of pitch angle scattering, there are mostly two CR transport regimes. Notice however that the transport of charged particles in electromagnetic fields, in fact, involves more regimes than free streaming and diffusion, it can be sub- or super-diffusive. But we will limit our self to the above-mentioned free streaming and then diffusive cases. We are discussing them now.

\subsection{Free Streaming approximation:}\label{sec:FreeStreaming}
The free-streaming approximation, also known as the continuous slowing-down approximation (CSDA), when it is associated with continuous losses, is often used to calculate the propagation of CRs in molecular clouds. We review here the method used in the context of molecular clouds and apply it to protostellar discs. The application of the propagation in protostellar discs is then presented in the next Chapter \ref{C:PublicationI}. In the CSDA regime, scattering processes are inefficient, and the mean squared deviation of the pitch angle along a CR trajectory is small. Thus, $\mu$ is conserved, and the differential CR flux, $F$, is given by,

\begin{equation}
S(E, s, \mu) = \mu v f_E ,
\end{equation}
where the dependence on $\mu$ is completely determined by the boundary conditions for the initial (interstellar, Galactic or flare) injection spectrum $j_{\text{inj}}(E, \mu)=v f_E$ of CRs entering a cloud or a disc. The kinetic energy $E$ of a particle after it has traversed the effective column density $N$ is related to its initial energy $E_i$ through

\begin{equation}
N = \mu \int_{E_i}^E \frac{dE'}{L(E')},
\label{eq:defColumndensity}
\end{equation}
where $L(E) = - \dot{E}_{\text{con}}/(n_H v)$ is the energy loss function. From this follows,

\begin{equation}
    \mu \frac{\partial j}{\partial N }-\frac{\partial}{\partial E}(L(E)j(E,N,\mu))=0 \ .
\end{equation}
So the conservation of the number of CR particles gives us the solution for the local (propagated) spectrum,
\begin{equation}
j(E, N, \mu) L(E) = j_{\text{inj}}(E_i,0, \mu) L(E_i),
\label{eq:CSDA}
\end{equation}
where $E_i$ is the particle energy before it started losing energy i.e. at null column density. 

To illustrate how to compute an ionisation rate in the CSDA, we present now an illustrative model, making simple assumptions on the injection spectrum and on the loss function, that can be solved analytically. A more refined model, accounting for all the energetic losses presented in Sect. \ref{sec:EnergyLossesDisc} is used in the publication of Chapter \ref{C:PublicationI}.

\paragraph{A toy model for the ionisation rate in the CSDA:}\label{sect:ToyModel}

We showed in the previous Chapter \ref{C:Reconnection}, that power-law spectrum of energetic particles are expected for magnetic reconnection events. For an initial power-law proton spectrum, considering the fiducial injection spectrum of \citet{2023MNRAS.519.5673B}, a $1~\rm MK$ flare corresponding to an X-ray luminosity $L_X=4\times10^{31}$ erg s$^{-1}$, the particle flux is

\begin{equation}
j_{k,\text{inj}}(E) = j_{k,0} \left( \frac{E}{E_{k,0}} \right)^{-a}
\,,
\label{eq:powerlawinjspectrum}
\end{equation}
where the proton flux at 1 MeV is $j_{p,0}=4 \times 10^4 ~ \rm  particles/cm^{-2}/eV/str$ and the electron flux at 1 keV is $j_{e,0}=4\times10^{13} ~ \rm  particles/cm^{-2}/eV/str$, the characteristic energies of protons and electrons are $E_{p,0}=1$ MeV, $E_{e,0}=1$ keV, respectively and the index of the power-law spectrum in the non relativistic regime is $a=2.5$. 

\begin{figure}[h!]
    \begin{subfigure}{.5\textwidth}
    \centering
    \includegraphics[width=\textwidth]{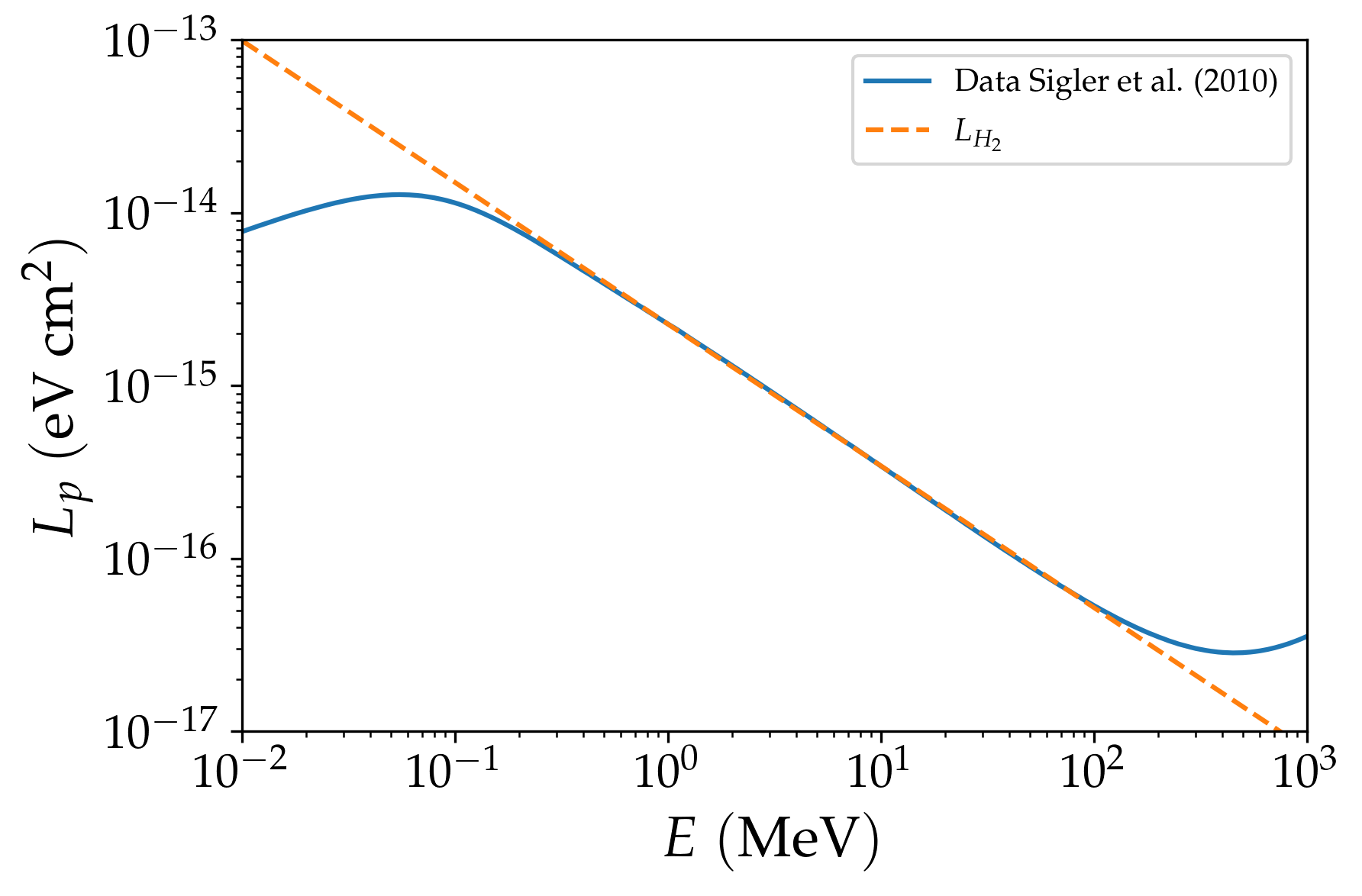}
    \label{fig:ProtonLossFunction}    
    \end{subfigure}    
    \begin{subfigure}{.5\textwidth}
    \centering
    \includegraphics[width=\textwidth]{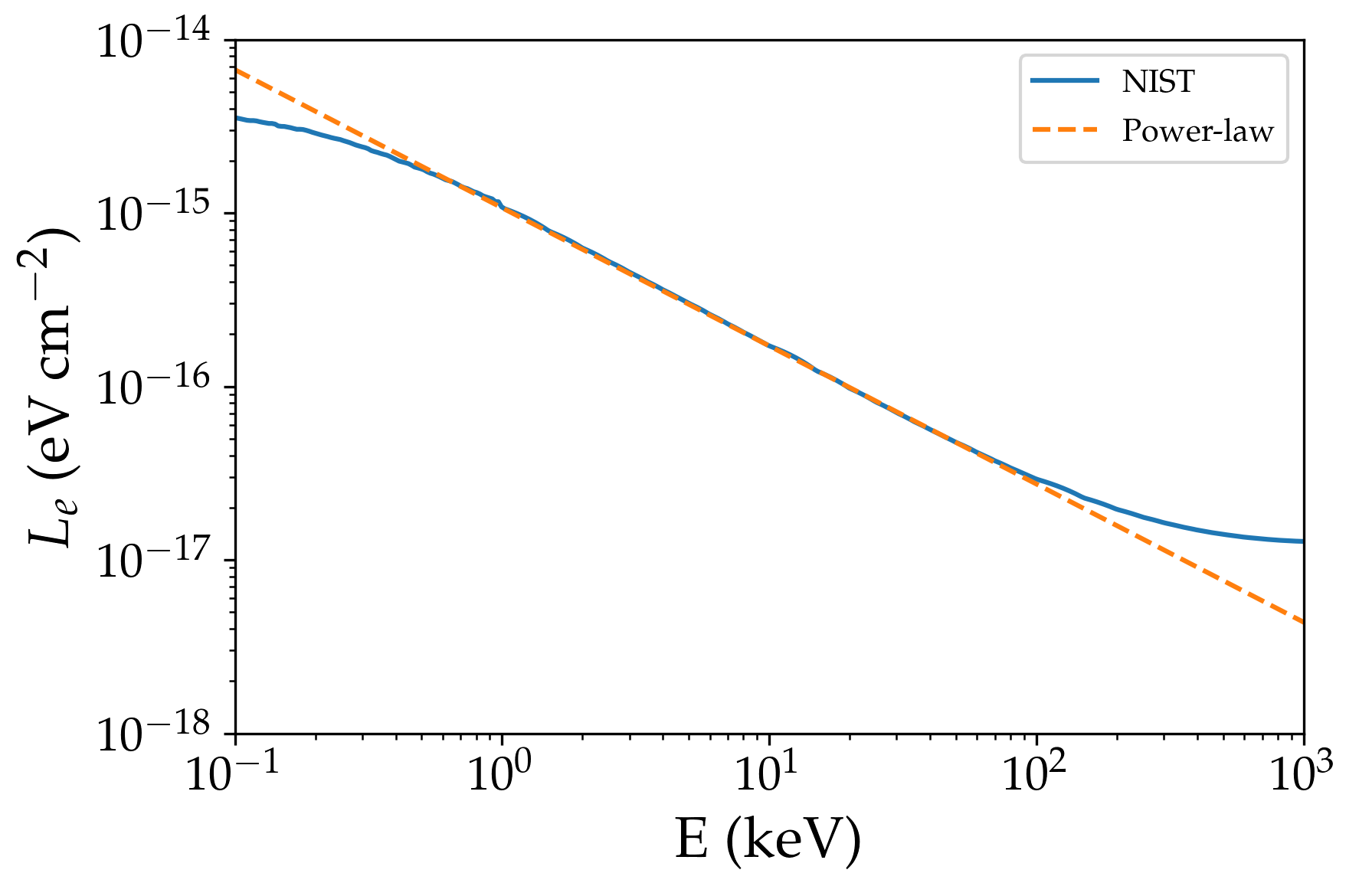}
    \label{fig:ElectronLossFunction}    
    \end{subfigure}    
    \caption{Proton (left) and electron (right) energy loss function as function of energy in the range where the dominant source of losses is ionisation. We used the data of \citet{ziegler2010srim} for protons and data from the NIST for electrons. The power-law expressions for both protons and electrons in Eq. \eqref{eq:powerlawlossfunction} are plotted in orange dashed line.}
    \label{fig:PowerlawLossFunction}
\end{figure}
We further assume for this toy model that the medium is only composed of molecular hydrogen, $n=n_{H_2}$.
The electron and proton loss functions in H$_2$ can be approximated by a power-law, as illustrated by the loss function in Fig. \ref{fig:PowerlawLossFunction}. This approximation is valid for electron in the range $\sim 0.1 \rm keV - 0.1 \rm MeV $ and for protons in the energy range $\sim 0.1 \rm MeV - 0.1 \rm GeV $. We have, 

\begin{equation}
L_k(E) = L_{0} \left( \frac{E}{E_{k,0}} \right)^{-b},
\label{eq:powerlawlossfunction}
\end{equation}
with $L_0 \approx 10^{-15}$ eV cm$^2$ and $b = 0.8$ for both electron and proton. 

From Eq \eqref{eq:defColumndensity} we define the stopping range,
\begin{equation}
R(E) = \int_0^E \frac{dE}{L(E)}= \frac{E_0}{(1+b)L_0} \left(\frac{E}{E_0}\right)^{1+b}
\end{equation}
that can be interpreted as the column density at which a particle with initial energy $E$ loses all its energy. With Eq. \eqref{eq:CSDA}, the injection energy $E_i$ of a particle attaining a column density $N$ with an energy $E$ is expressed as,
\begin{equation}
    E_i=E\left(1+\frac{N}{\mu R(E)}\right)^{\frac{1}{1+b}}.
\end{equation}
It follows,
\begin{equation}
j_k(E, N, \mu) = j_0 \left( \frac{E}{E_{k,0}} \right)^{-a} \left[ 1 + \frac{N}{\mu R_k(E)} \right]^{-\frac{a+b}{1+b}}.
\label{eq:freestreamingpropagatedspectrum}
\end{equation}

%The secondary electron flux is computed by \citet{2015ApJ...812..135I}:
%\begin{equation}
   % j_e^{sec}=\frac{E}{L_e(E)}\int_{I+E}^\infty j(E',N) \frac{d\sigma(E',E)}{dE} dE'
%\end{equation}

The local ionisation rate of H$_2$ due to protons, $\zeta_{k, H_2}$, can be calculated if the propagated proton spectrum $j(E,N)$ is known,

\begin{equation}
\zeta_{k, H2} = \int_0^1 d\mu \int_0^{\infty} j_k(E, N/\mu) \sigma_{\text{ion}}^{k,H2} (E) dE,
\label{eq:generalionisationrate}
\end{equation}
where $\sigma_{\text{ion}}^{k,H2} (E)$ is the ionisation cross section of particle $k$ with molecular hydrogen. 

Using the expression for the ionisation cross section given in \citet{1985RvMP...57..965R}, and the loss function in Equation \eqref{eq:powerlawlossfunction}, we determine this ratio to be approximately $\epsilon=\left \langle \frac{L}{\sigma_{\text{ion}}^{H2}} \right \rangle \approx 37$ eV, which is nearly constant for protons and electrons with energies between $10^5$ eV and $5 \times 10^8$ eV and $10^2$ eV and $5 \times 10^5$ eV, respectively. Note that this corresponds to an energy lost per H$_2$ ionisation event of approximately $60$ eV, which is reduced by the ratio of the Hydrogen number density to the particle number density. Thus, we can write, as in \citet{Padovani18}, $\sigma^{\rm ion}=L(E)/\epsilon$. Substituting these values into Eq.\eqref{eq:generalionisationrate}, \citet{2019ApJ...879...14S} found,

\begin{equation}
\zeta_{k} (\mu, N ) \approx \frac{1+b}{a+2b}I_f\frac{j_{k,0} E_{k,0} L_0}{\epsilon} \left( \frac{N}{\mu R_{k,0}} \right)^{-\frac{a+b-1}{1+b}} ,
\end{equation}
where $I_f$ is given by \citet{2019ApJ...879...14S}, with $a=2.5$ and $b=0.8$, $I_f=0.75$. We discuss the case of non isotropic distribution of the pitch angles in Sect. \ref{sec:mirroringfocussing} but here we assume an isotropic distribution. Integrating over $\mu$, gives

\begin{equation}
\zeta_{k} (N ) = \zeta_{k,0} \left( \frac{N}{ R_{k,0}} \right)^{-\gamma} ,
\label{eq:freestreamingionisationrate}
\end{equation}
where $\zeta_{k,0} \approx \frac{j_{k,0} E_{k,0} L_0}{\epsilon}$,  $R_{k,0} = R_k(E_0)$ and $\gamma=\frac{a+b-1}{1+b}=1.28$. 

The ionisation rate produced by electrons and protons are approximated in an analogous way. The parameters to compute $\zeta_k(N)$ are listed in Tab. (\ref{tab:ToyModelParameters}).

\begin{table}[h!]
    \centering
    \begin{tabular}{c|c|c|c|c}
         k & $E_0$ (eV) & $j_0~ \rm (part/eV/s/str/cm^{2})$  & $R_0 ~\rm (cm^{-2})$ & $\zeta_0$ (s$^{-1}$) \\
            \hline
        p & $10^6 $  & $4\times 10^4$ & $5\times 10^{20}$ & $5\times 10^{-8}$ \\
        e &  $10^3 $  & $4\times10^{13}$& $5\times 10^{17}$ & $0.1$ 
    \end{tabular}
    \caption{Parameters of the toy model to compute ionisation rates.}
    \label{tab:ToyModelParameters}
\end{table}

\begin{figure}
    \begin{subfigure}{.5\textwidth}
    \centering
    \includegraphics[width=\textwidth]{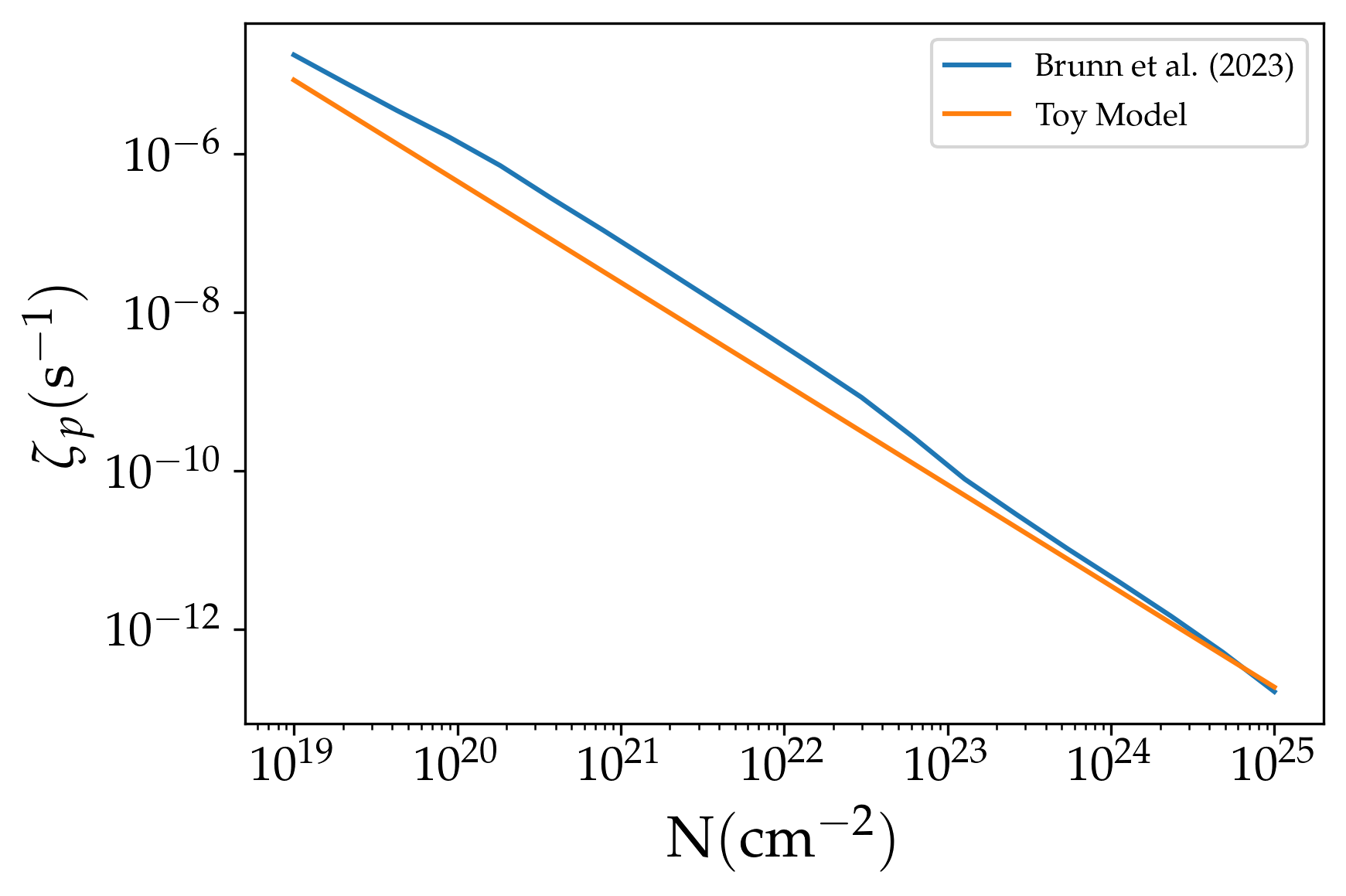}
    \label{fig:Protonionisation}    
    \end{subfigure}    
    \begin{subfigure}{.5\textwidth}
    \centering
    \includegraphics[width=\textwidth]{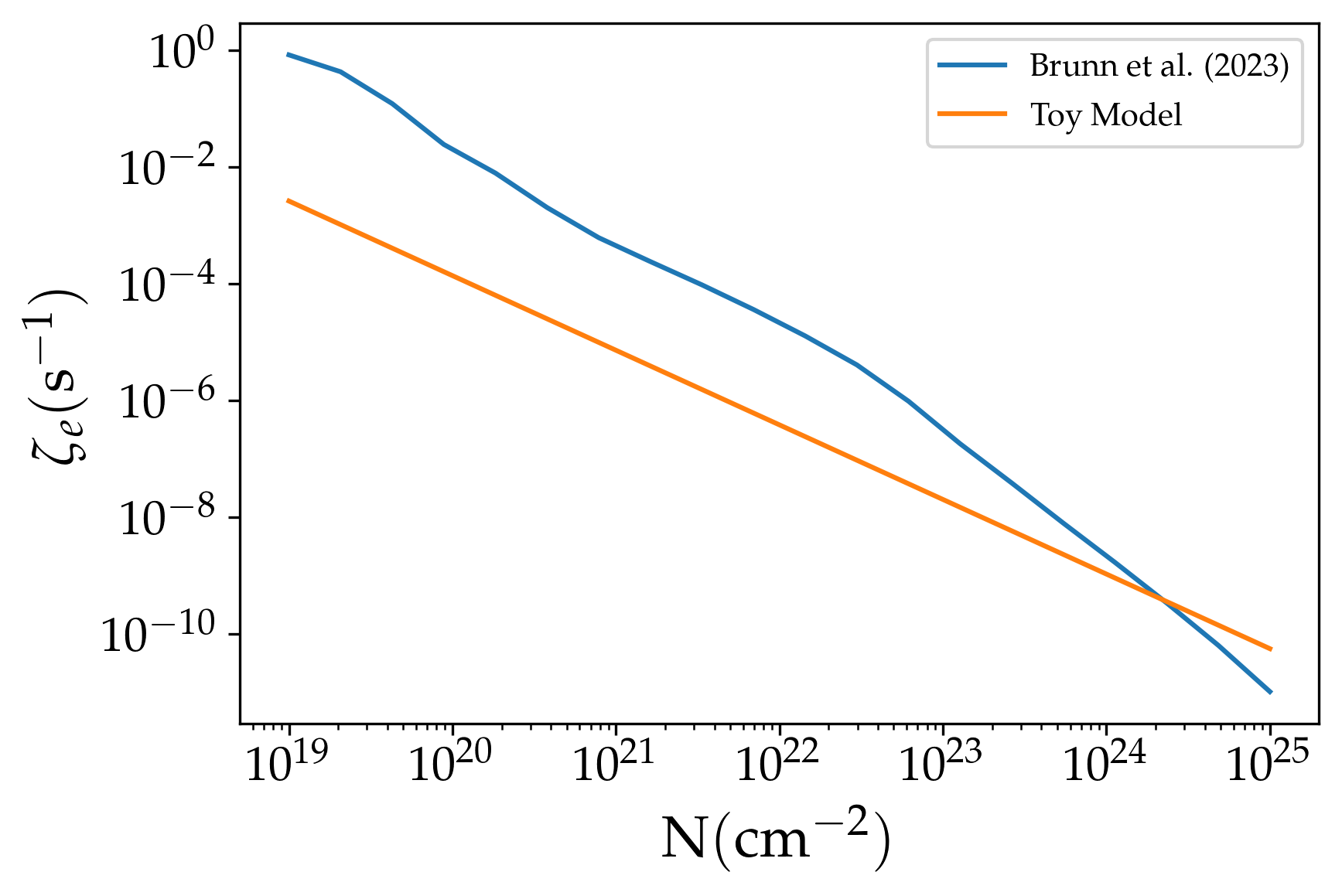}
    \label{fig:Electronionisation}   
    \end{subfigure}    
    \caption{Proton (left) and electron (right) ionisation rate as function of column density in the range where the CSDA is valid. We compare the ionisation rates of the toy model presented in this section plotted in orange with the results of \citet{2023MNRAS.519.5673B} plotted in blue. }
    \label{fig:ComparaisonToyModelProton}
\end{figure}
The results of this simplified model presented are plotted in Fig. \ref{fig:ComparaisonToyModelProton} in comparison with the more complex model of \citet{2023MNRAS.519.5673B} presented in Chapter \ref{C:PublicationI}. Indeed, this toy model leads to an underestimation of the ionisation rates due to several reasons we list here,
\begin{itemize}
    \item Neglecting Low-Energy Processes: In this model, loss and ionisation effects at low energy are not considered. Although individual low-energy CRs carry less energy, they are more numerous and more likely to interact with atoms in the disc, thus causing ionisation. By ignoring these interactions, the model misses a considerable fraction of ionising events.

    \item Excluding High-Energy Particles: The model also overlooks the contribution of high-energy particles, specifically those with energies greater than 1 MeV in the case of electrons. Such particles are particularly important because they can penetrate more deeply into dense regions of a molecular cloud or disc where low-energy particles may not reach due to energy loss. While they may ionise less frequently per unit distance than low-energy particles, their ability to penetrate further and the cumulative effect over larger path lengths can result in significant ionisation.  

    \item Ignoring Secondary Electrons: During the ionisation process by CRs, secondary electrons are often produced. These secondary electrons have sufficient energy to further ionise other atoms in the medium. Since each ionisation event can result in more than one ion-electron pair, the lack of consideration for these secondary ionisation processes significantly underestimates the electron ionisation rate.
\end{itemize}
    
%By not accounting for these factors, the model underestimates the role of cosmic rays in ionizing a disc of molecular hydrogen. To have a more accurate understanding, a more complex model is needed that accounts for the full energy spectrum of CRs, the ionisation processes at all energy levels, and secondary ionisations.

Acknowledging for these restrictions, the approximation of a power-law loss function in a restricted energy range allows for the computation of the flux of CRs and the associated ionisation rates in a relatively straightforward manner. It provides an illustrative way to understand how CR ionisation is affected by initial energy spectrum, particle types, and column density. %This can be valuable in understanding the overall trends and key processes that affect CR ionisation.

%However, in a realistic scenario, the scattering processes which the CSDA overlooks can actually be significant. These processes can affect the transport and propagation of CRs at high column density, and hence influence the spatial distribution of ionisation in the disc. 

In the forthcoming Chapter \ref{C:PublicationI},
we introduce a specific model that aims to addressing some of the limitations discussed above. While this model works within the framework of the CSDA, it attempts to incorporate additional complexities for a more accurate depiction of cosmic ray ionisation in molecular clouds and discs.

This model will take a more comprehensive approach to account for the interactions between cosmic rays presented in the previous section and a realistic mixture gas. While maintaining the ease and simplicity of the CSDA, we include processes involving low energy cosmic rays and secondary electrons in our model.

\subsection{Spatial diffusion}
The governing transport mechanisms for CR in dense cores within molecular clouds and in protostellar discs remain a contentious issue \citep{2019ApJ...879...14S}. In these regions, turbulence or collisions can cause changes in their pitch angles, leading to spatial diffusion. The characteristics of the turbulence or collision processes establish the scale of the CR diffusion coefficient and its energy dependence \citep{2019ApJ...879...14S}. 
%Furthermore, the anisotropy in the CR distribution function, which is a reaction to CR absorption in dense cores and protostellar discs, can also trigger various instabilities and contribute to MHD turbulence (Skilling and Strong, 1976; Morlino and Gabici, 2015; Ivlev et al., 2018).

We will start by outlining the basic principles of diffusive CR transport. This process is activated when the mean free path of a CR particle due to pitch-angle scattering is smaller than the spatial scale at which the CR loses all its energy. This results in an isotropic local distribution function, and the relation between the number of CRs per unit volume and energy to the distribution function becomes $F(E,s)\approx 4 \pi f_E$. The corresponding flux $S$ becomes independent of $\mu$ and generally incorporates two terms \citep{1990acr..book.....B},

\begin{equation}\label{Eq:FP}
    S(E,s)=-D\frac{\partial f_E}{\partial s}+ (u + v_s) f_E
\end{equation}
where we have adopted a simplified 1D model where $D(E)$ represents the CR diffusion coefficient along the background magnetic field. In general, it is better described by a 3D tensor to account for perpendicular transport effects. 

The second term in Eq. \ref{Eq:FP} is proportional to the advection velocity of the background plasma $u$ and the scattering centre speed $u_s$. It originates from the anisotropy of MHD turbulence. It is generally assumed that preexisting electromagnetic turbulence has no helicity (the same amount of perturbations propagate forward and backward along the mean magnetic field) on small spatial scales, which are responsible for the scattering of low-energy CRs, thereby making $v_s=0$.

We can examine the flux of particles entering the disc from the source, propagating along a magnetic field line. We can use $s$ to denote the coordinate of distance along this line. The origin of this coordinate system, where $s$ equals zero, represents the location of the source. As we move away from the source, the value of $s$ increases. The transport equation for $f_E(E, s)$, considering the elements of spatial diffusion and energy losses, writes \citep{2019ApJ...879...14S},
\begin{equation}
    \frac{\partial f_E }{\partial t}= \frac{\partial }{\partial s}\left(D \frac{\partial f_E }{\partial s}\right)-\frac{\partial  }{\partial E}\left( \frac{dE}{dt} f_E \right).
\end{equation}
Looking for steady-state distribution, we set $\frac{\partial f_E }{\partial t}=0$. We express the energy loss rate  as $\frac{dE}{dt}= L(E) n v$, where $n=n(s)$ is the plasma particle number density. We define $G(E,s)=L(E) v f_E(E,s)$ and $X(E)=\frac{L(E)v}{n D(E,s)}$. We further assume that the product $\Delta(E)=n(s)D(E,s)$ is independent of the position. We will see that this is reasonable for diffusion due to MHD turbulence or momentum transfer.

Under these assumptions, the transport equation is reduced to,
\begin{equation}
    \frac{\partial^2 G(E,N)}{\partial N^2}+\frac{\partial G(E,N)}{\partial E}=0
    \label{eq:almostlineardiffusion}
\end{equation}
By defining the pseudo time 
\begin{equation}
T(E) = -\int_0^E \frac{E'}{X(E')} dE',
\end{equation}
Equation \eqref{eq:almostlineardiffusion} reduces to a linear diffusion equation for the product $G$, with the coordinate $N$,
\begin{equation}
    \frac{\partial G}{\partial T }-\frac{\partial^2 G}{\partial N^2} = 0.
    \label{eq:energydependentdiffusionequation}
\end{equation}

The second term describes spatial diffusion process. It can be of collisionless origin due to pitch-angle scattering off magnetic perturbations, or of collisional origin due to collisions with the  particles of a dense medium.

Although Eq.\eqref{eq:energydependentdiffusionequation} is significantly simplified due to the assumptions presented above, diffusion transport processes are still considerably more complex to describe than free streaming. %For this reason, we prefer to use the free streaming approximation to describe particle transport. 
In order to determine when this approximation is applicable, we below present a criterion that will help identify the suitable transport approximation. 

%We will first discuss the collisionless scattering regime and then move to the effect of collisions.

\subsubsection{The scattering parameter: a criterion for the transport regime}
We present here a criterion to determine what is the relevant transport mechanism between diffusion and free-streaming.
The diffusion approximation is considered valid at such column densities where the particle has lost the memory of its pitch angle into the disc. The diffusion can be considered as a random walk in 3D at velocity $v=\beta c$ with step length in terms of column density, given by,
\begin{equation}
    \delta N(E) = \frac{3 n D(E)}{\beta c},
    \label{eq:diffusioncharacteristicColdens}
\end{equation}
where $n$ is the particle number density in the disc and $D(E)$  the diffusion coefficient.

In the free-streming approximation a CR of energy $E$ has lost all its energy at a column density $N=R(E)$, where $R$ is the stopping range. The stopping range was defined before, it is calculated using the equation,

\begin{equation}
    R = \int_0^E \frac{dE}{L(E)}
    \label{eq:StoppingRange}
\end{equation}

The transition from free-streaming to diffusive propagation for a particle with energy \(E\) should occur approximately at a column density \(N\) such that \(R(E) \approx \delta N(E)\).

The relative significance of diffusion versus losses can be quantified by the scattering parameter $R_{sc,k}(E)$. This criterion has been proposed by \citet{Padovani18}. It is the integral ratio of the characteristic stopping range due to energy losses to the characteristic column density that results in strong scattering. We define $\sigma$, the cross section of the diffusion process, $\sigma(E)=1/\delta N(E)$.
\begin{equation}
    R_{sc}(E)=\int_0^E \frac{\sigma(E)}{L(E)}dE
    \label{eq:ScatteringParameter}
\end{equation}
The free-streaming approximation is a suitable representation as long as $R_{sc}(E)$ is less than 1. Otherwise, scattering processes becomes important and a shift to diffusive transport is necessary.

The scattering parameter depends on the diffusion process considered. We compute now the scattering parameter for different diffusion mechanisms. They are computed assuming the physical parameters expected in circumstellar discs in the energy range of particle relevant for ionisation, $E<10$ GeV. 

\subsubsection{A parametrical study for diffusion}
Let us now turn to the effect of turbulent magnetic fields.
The diffusion coefficient, which is energy-dependent, dictates how far CRs propagate through the disc. This coefficient can be expressed in terms of the CR velocity and can be estimated from quasi-linear theory as \citep{1989ApJ...336..243S}
\begin{equation}
  D(p) = \beta c \eta(p) r_L,
  \label{eq:DiffusionRL}
\end{equation}
where $r_L$ and $p$ represent the Larmor radius and momentum of the CR, respectively and $\beta c$ is the velocity of the CR.

Here, $\eta(p)= (B/\delta B(k_{\rm res}))^2$ is associated to the turbulence level in the magnetic field $\delta B$ at a particular (resonant) wave number $k_{\rm res}(p)$ defined in the next section in Eq. \ref{Eq:kres}, with respect to the mean magnetic field strength B. Hence, the higher the turbulence level the smaller the parameter $\eta$. Strictly speaking, the quasi-linear theory on which Eq. \ref{eq:DiffusionRL} assumes the regime $\eta \gg 1$ is verified, otherwise one has to consider some renormalisation of the diffusion coefficient as it is the case in the context of Supernova remnants where strong turbulence with $\eta \sim 1$ develops \citep{2003A&A...403....1P}. 

Acknowledging from the previous limitation, $\eta$ can be parameterised as in \citet{Rodgers-Lee17},
\begin{equation}
    \eta(p) = \eta_0 \left(\frac{p}{1 \rm GeV/c}\right)^{1-\gamma}.
    \label{eq:etaDiffusionRL}
\end{equation}
The $\gamma$ parameter in Eq.\eqref{eq:etaDiffusionRL} illustrates the physical basis for turbulence in the magnetic field. If the turbulence spectrum is Kolmogorov-type, $ \gamma=5/3$. If it is a Kraichnan-type spectrum,  $\gamma=3/2$.
As $\eta$ approaches 1, we recover Bohm diffusion, where a particle with momentum $p$ scatters once per Larmor radius. The value of $\eta_0$ depends on the model of turbulence considered.

\citet{Rodgers-Lee17} examined the influence of low-energy cosmic rays emitted by a young star on its surrounding protoplanetary disc. They specifically analysed the impact of protons injected at the disc inner edge. To solve the transport equation, they modelled the CRs propagation as diffusive. They examined the impact of the CR for values of assuming the diffusion coefficient presented above. Their study showed that as the diffusion coefficients increases, CRs penetrate deeper into the disc. They varied $\eta_0$ from 3 to 300 but the parametric study is not based on any turbulence model.

First, we determine when the diffusion approximation is valid based on the values of the parameter $\eta_0$. To do that, we compute the scattering coefficient of protons and electrons varying $\eta_0$ in the range studied by \citet{Rodgers-Lee17}. Second, we estimate the value of $\eta_0$ based on the quasi linear theory. Based on this value we deduce what is the appropriate diffusion transport mechanism assuming standard physical disc properties.

The cross section of the diffusive process $\sigma_d$ needed to compute the scattering parameter is computed combining Eqs. \eqref{eq:diffusioncharacteristicColdens} and \eqref{eq:etaDiffusionRL}, giving
\begin{equation}
    \frac{1}{\sigma_d}=\eta_0 \left(\frac{p}{1 \rm GeV/c}\right)^{1-\gamma} \frac{r_L}{H} N_{\rm vert}
    \label{eq:sigmadiffRL}
\end{equation}
where the particle density is approximated by $n=N_{\rm vert}/H$ where, $N_{\rm vert}$ is the vertical column density down to the midplane and $H$ is the disc height. $N_{\rm vert}/H$ is computed assuming the {\tt ProDiMO} disc model \citep{woitke2009radiation}.

\begin{figure}[H]
    \begin{subfigure}{.5\textwidth}
    \centering
    \includegraphics[width=\textwidth]{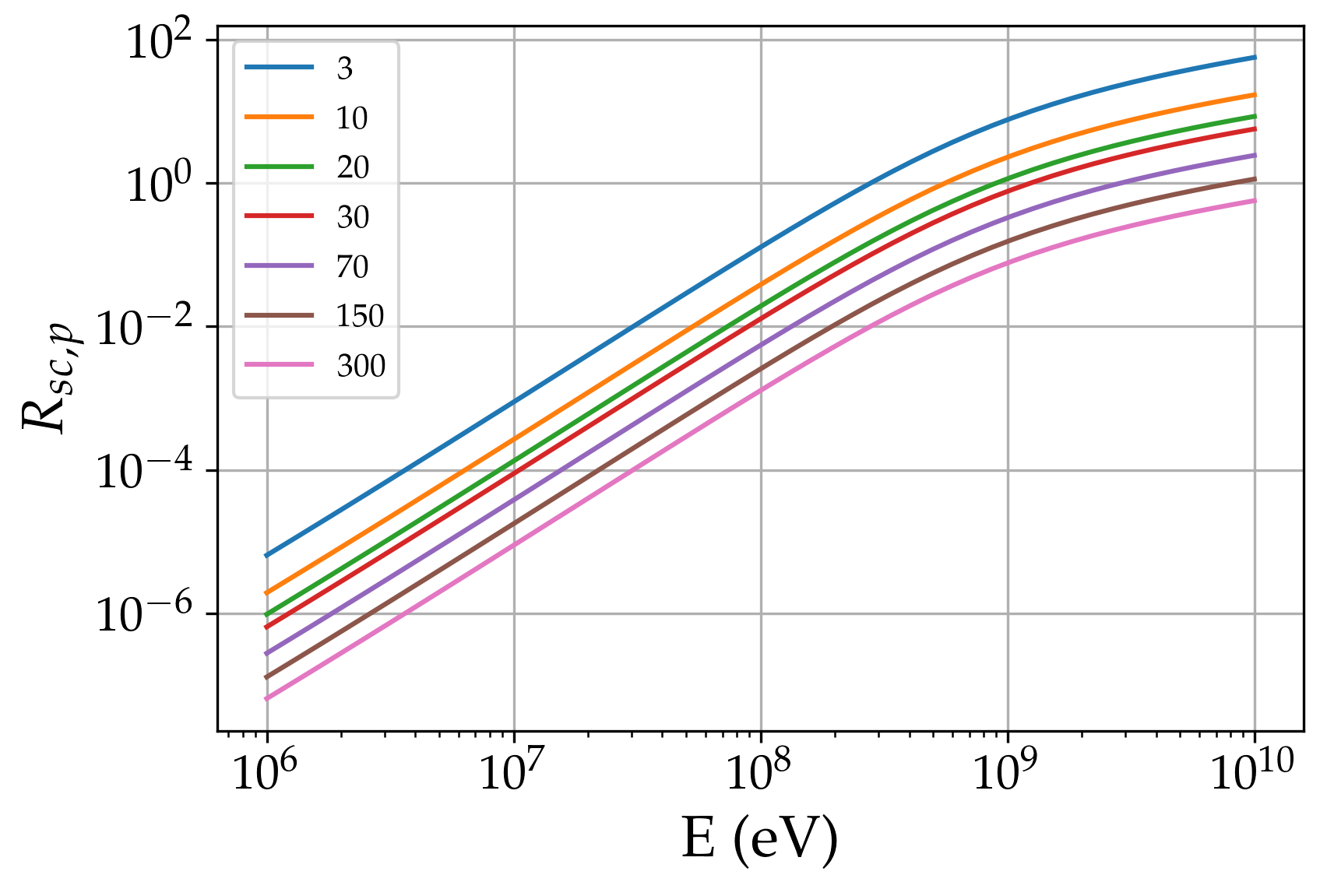}
    %\label{fig:WeakMHDTurbScatteringParameterProton}    
    \end{subfigure}    
    \begin{subfigure}{.5\textwidth}
    \centering
    \includegraphics[width=\textwidth]{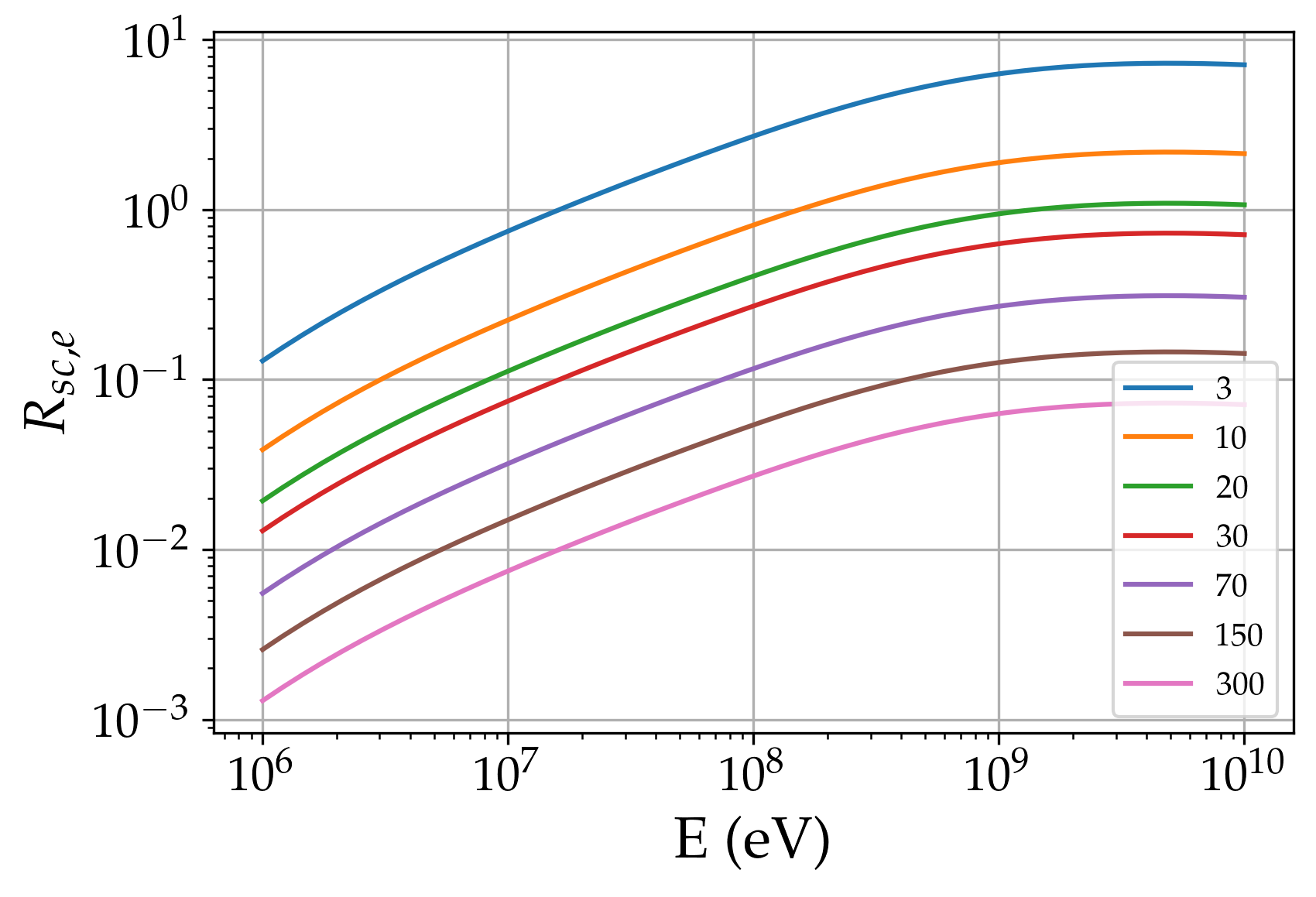}
    %\label{fig:WeakMHDTurbScatteringParameterElectron}   
    \end{subfigure} 
    \caption{Scattering parameter due to turbulence for proton (left) and electron (right) for $\eta_0$ ranging from 3 to 300 at a distance from the star $R=0.1$ au and assuming a Kolmogorov-type turbulence, $ \gamma=5/3$. The transport of particle should be treated as diffusive if the scattering parameter is greater than 1. }
\label{fig:ParametricScatteringParameter}
\end{figure}

We then insert Eq. \eqref{eq:sigmadiffRL} into the expression of the scattering parameter $R_{sc}$ of Eq. \eqref{eq:ScatteringParameter}. In Fig. \ref{fig:ParametricScatteringParameter}, we plot $R_{sc}$ for protons and electrons for $\eta_0$ ranging from 3 to 300 at a distance $R=0.1$ au from the star. We further assume a Kolmogorov-type turbulence, $ \gamma=5/3$.

From Fig. \ref{fig:ParametricScatteringParameter} we conclude that at R=0.1 au protons and electrons of energy less than 1 GeV lose all their energy before diffusing for $\eta_{0}>\eta_{0,l}$, with $\eta_{0,l}=20$. Under these conditions the transport of particle can safely be treated in the free-streaming approximation. Varying the value of $\gamma$ has no significant effect on the value of $\eta_{0,l}$.

\subsubsection{Diffusion due to MHD turbulence}

After this parametrical study we would like to give a  more physical ground to the diffusion coefficient and the parameter $\eta_0$. We propose to first look at the diffusion as a collisionless process by deriving the diffusion coefficient expected from weak MHD turbulence. We consider first a generic turbulence model, where the energy is cascading from large scales and then more specifically we will look at the diffusion coefficient produced by MRI turbulence.

To compute the CR diffusion coefficient $D(E)$,  due to the presence of weak MHD turbulence in the specific case of circumstellar discs, one can use the expression proposed by \citet{1975MNRAS.173..255S}, 
\begin{equation}
    D(E)= \beta c \frac{B^2}{6 \pi^2 \mu_* k_{\rm res}W(k_{\rm res})} r_L ,
    \label{eq:diffusioncoeff}
\end{equation}
which relates the diffusion coefficient to the spectral energy density of weak magnetic turbulence, $W(k)$, with $B$ being the strength of the magnetic field. This relationship is governed by $k_{\rm res}(E)$, the wavenumber of longitudinal MHD waves satisfying the condition of the cyclotron resonance with CR particles of energy $E$,
\begin{equation}\label{Eq:kres}
    k_{\rm res}(p)=\frac{m \Omega}{\mu_* p},
\end{equation}

where $\mu_*$ is the effective cosine of the resonant pitch angle (usually set to a constant of order unity), $p$ is the momentum of a CR particle with rest mass $m$, and $\Omega=eB/ \gamma mc$ is again the CR gyrofrequency.

From Eq. \eqref{eq:diffusioncoeff} one can identify 
\begin{equation}
    \eta(E)=\frac{B^2}{6 \pi^2 \mu_* k_{\rm res}W(k_{\rm res})}.
\end{equation}

The subsequent analysis of diffusive CR transport is crucially dependent on the source of MHD turbulence. If the turbulence is driven by CRs, its spectrum $W(k)$ is determined by the CR spectrum $f_E$ and vice versa. The case of preexisting turbulence can be analysed by equating $W(k)$ and the turbulent kinetic energy of ions,
\begin{equation}
    k W(k)\approx \frac{1}{2}\rho_i v_{\rm turb}^2(k),
\end{equation}
where $\rho_i$ is the ion mass density and $v_{\rm turb}^2$ is their mean squared turbulent velocity.
For a power-law turbulent spectrum with $v_{\text{turb}}(k) = v_* \left(\frac{k_*}{k}\right)^{\lambda}$, where $\lambda=\frac{\gamma-1}{2}$, so from the $\gamma$ values of Kolmogorov and Kraichnan-type turbulence given for Eq. \eqref{eq:etaDiffusionRL}, $\lambda = 1/3$ for a Kolmogorov spectrum and $\lambda = 1/4$ for a Kraichnan spectrum. 

\iffalse
The diffusion coefficient can be approximated as
\begin{align}
    D(p) &= \frac{ p^{2(1-\lambda)}\mu^{-2\lambda} B^2 \Omega^{2\lambda-1 }}{3 \rho_i \pi^2 v_*^2 k_*^{2\lambda} }\\
    &=D_0 \left( \frac{B}{ 1 \rm mG}\right)^2 \left( \frac{n_i}{10^9 \rm cm^{-3}}\right)^{-1} \left( \frac{p}{\rm 1 GeV/c}\right)^{2(1-\lambda)}
    \label{eq:diffusioncoeff}
\end{align}
%\begin{equation}
%\frac{v^{2(1-\lambda)}}{x_E} \cdot \frac{\mu^{-2\lambda} B^{2} \Omega^{2\lambda - 1}}{3\pi^{2}(v_{*} k_{*}^{\lambda})^{2}}
%\end{equation}

The diffusion coefficient $D_0$ is different depending on the considered CR, for protons $D_{0,p}\approx 10^{22}$ cm$^{2}$ s$^{-1}$, while for electrons $D_{0,e}\approx 10^{23}$ cm$^{2}$ s$^{-1}$. This means that protons are more confined by MHD turbulence than electrons.
\fi

To estimate $v_*$ and $k_*$, we assumed a turbulent velocity of 1 km s$^{-1}$ at a scale of 1 parsec. This requires a major extrapolation to describe the much smaller scale we are interested in. Indeed, it is possible that the turbulence is damped by ion neutral friction at intermediate scales see \citet{2013ApJ...774..128S} for a discussion.
%If we further assume that at the turbulent scale, the ionisation fraction and $B$ field are constant or more generally that $B^2/x_E$ is constant, we verify that with the diffusion coefficient defined as in Eq. \eqref{eq:diffusioncoeff}, the product $n_g(N)D(E,N)$ is independent of $N$. So under these assumptions,
%\begin{equation}
 %   \Delta(E)=\Delta_0\left(\frac{E}{E_0}\right)^{1-\lambda}.
  %  \label{eq:diffusioncoeffpowerlaw}
%\end{equation}

We can now constrain the values of the parameter $\eta$ from Eq. \eqref{eq:etaDiffusionRL}, for both electrons and protons,
\begin{equation}
    \eta_e(p)= 720 \left( \frac{B}{ 1 \rm mG}\right)^2 \left( \frac{n_i}{10^9 \rm cm^{-3}}\right)^{-1} \left( \frac{p}{\rm 1 GeV/c}\right)^{1-\gamma}.
\end{equation}
and 
\begin{equation}
    \eta_p(p)= 50 \left( \frac{B}{ 1 \rm  mG}\right)^2 \left( \frac{n_i}{10^9 \rm cm^{-3}}\right)^{-1} \left( \frac{p}{\rm 1 GeV/c}\right)^{1-\gamma}.
\end{equation}
Looking back to Fig. \ref{fig:ParametricScatteringParameter}, we conclude that for typical value of $B$ and $n_i$, the $\eta_0$ values of protons and electrons produce a scattering parameter $R_{sc}<1$. This indicates that if we assume that the dominant diffusion process is MHD turbulence then, the transport of particles can safely be treated in the free-streaming approximation in the inner disc.

\paragraph{Diffusion due to MRI turbulence:}
Turbulence in discs is often expected to originate from MRI. This section focuses on identifying the range where particle diffusion due to MRI turbulence plays a significant role in the inner disc ($R=0.1$ au). We employ shearing-box simulations from the study by \citet{2021MNRAS.506.1128S}, conducted in ideal MHD with test particles to estimate the diffusion coefficient. They identified two different regimes for the diffusion coefficient.

For particles with large gyroradii ($r_L>0.03 H$), where $H$ represents the disc scale height, the diffusion was found to be,
\begin{equation}
    D_{MRI}\approx 30 D_{\rm Bohm}= 10 v r_L.
\end{equation}

For those with smaller gyroradii ($r_L<0.03 H$), the diffusion coefficients were generally larger, scaling as,
\begin{equation}
    D_{MRI} = 0.2 D_{\rm Bohm} \frac{H}{r_L}= \frac{1}{15} H v ,
    \label{eq:MRIDiffusionCoeff}
\end{equation}
where $D_{\rm Bohm}= \frac{v r_L}{3}$ is the Bohm diffusion coefficient.
The authors observed that while varying the magnetic field intensity significantly affects turbulence, it had a minor impact on diffusion coefficients when these were considered as a function of $r_L$. Typically, a variation in the diffusion coefficient was seen within a factor of 2-3 when the magnetic field intensity was changed by two orders of magnitude. 

Assuming the disc model of {\tt ProDiMO} \citep{woitke2009radiation}, we can estimate $H$. For $E<10^{9}$ eV, $R>0.1$ au, and $B>1$ mG, the ratio always verifies $R_g/H<0.03 $. Hence, the diffusion coefficient due to MRI in a circumstellar disc can be approximated by Eq. \eqref{eq:MRIDiffusionCoeff}.

Using Eq. \eqref{eq:DiffusionRL} we can identify $\eta_{\rm MRI}$,
\begin{align}
    \eta_{\rm MRI}(p)&=\frac{1}{15} \frac{H}{r_{L}} \\
    &= 15 \left(\frac{B}{1 \rm mG}\right) \left(\frac{p}{1 \rm GeV/c}\right).
\end{align}
MRI turbulence produces $\eta_{0,MRI}=15$ for $B=1$ mG. The magnetic field intensity of $1$ mG is a lower limit in the inner disc. So if the magnetic field is greater than 1 mG, $\eta_{0,MRI}>\eta_{0,l}$, the transport of particle can safely be treated in the free streaming approximation for particles of energy less than 1 GeV.

We showed that MHD and MRI turbulence are not expected to have a significant impact on the CR propagation. On the other hand, collisions may imprint some random walk to the energetic particles. We now evaluate the effect of collisions over particle trajectories and derive the threshold energies and column density separating the CSDA and diffusive regimes.

\subsubsection{Diffusion due to collision}\label{sec:DiffusionPions}

\paragraph{Collisions induced by pion production:}

When cosmic ray protons have energies above $E_\pi=125$ MeV, they interact with their medium, producing pions. Given the significant rest mass of pions, this leads to a substantial loss of energy for CR protons at each collision. These losses are not continuous, yet they are not completely catastrophic either, as a large number of collisions are necessary for a significant reduction in energy.

In the energy range where pion production becomes the primary cause of energy losses, the elastic scattering of CR protons on target nuclei produces pions. \citet{Padovani18} assume that the cross section of pion production follows the cross section of elastic scattering of CR protons with target nuclei $\sigma_{MT}(E)$. 

%{\bf pas clair: on est dans la section des pions et tu mentionnes des energies en dessous de 1 MeV This is best explained by the concept of the momentum transfer cross-section, $\sigma_{MT}(E)$, which consists of varying contributions based on the energy range. In energy levels below approximately 1 MeV, elastic (Coulomb) scattering is predominant, while at higher energies, the interaction between the CR proton and the target nucleus becomes more significant. There's a transition region between 1 MeV and 10 MeV, where nuclear forces start to dominate the elastic scattering. At levels above around 1 GeV, the momentum transfer cross-section becomes independent of energy, indicating hard sphere-like scattering.}

The relative significance of elastic scattering due to momentum transfer (i.e pion production) for CR protons versus their attenuation can be quantified by $R_{sc}(E)$, see the criterion presented Eq. \eqref{eq:ScatteringParameter}. We note the cross section of the diffusion process $\sigma_{MT}$. It is given in \citet{Padovani18},
\begin{equation}
    R_{sc}(E)=\int_0^E \frac{\sigma_{MT}}{L(E)}dE.
    \label{eq:MTscatteringParameter}
\end{equation}

\begin{figure}
    \centering
    \includegraphics[width=0.7\linewidth]{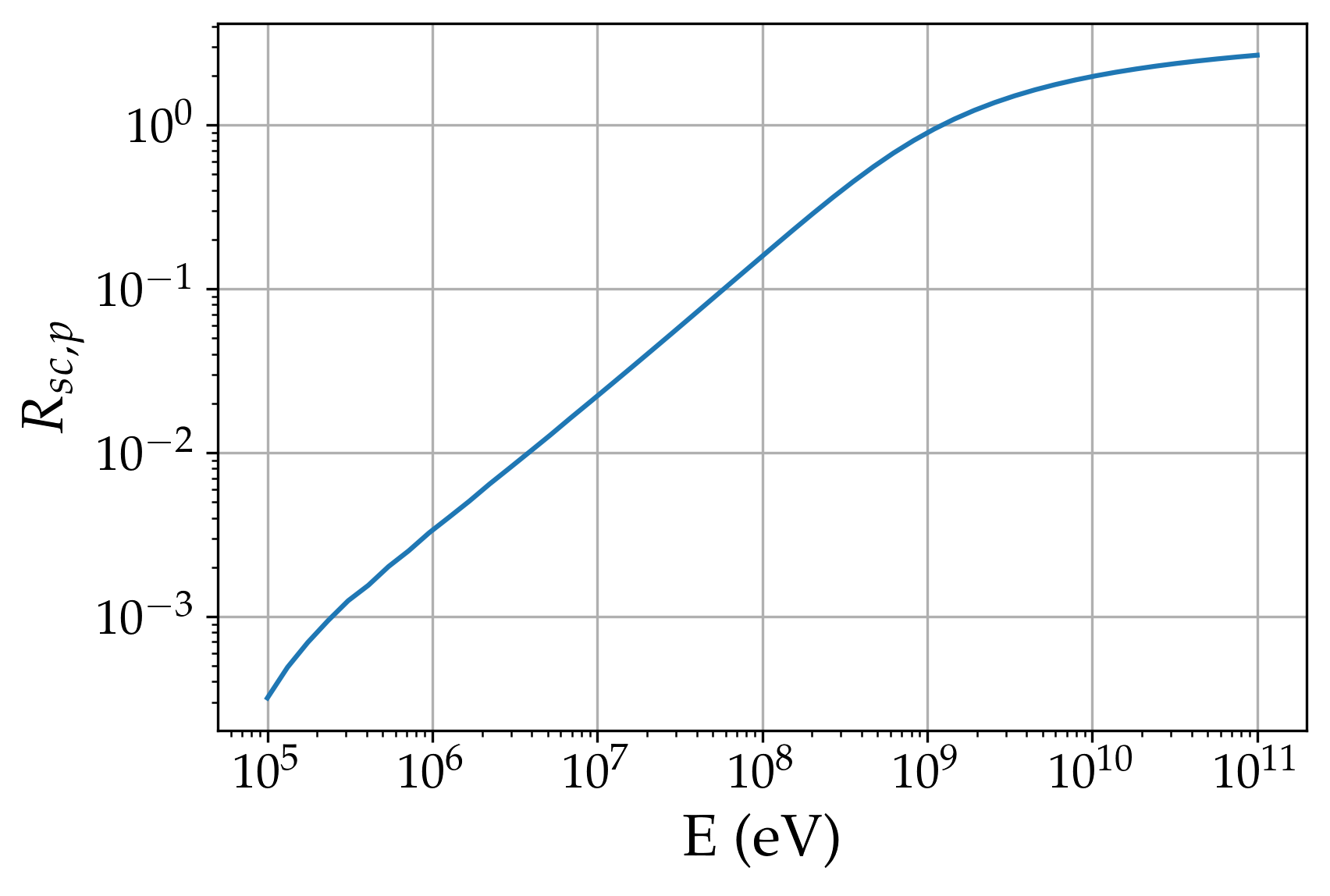}
    \caption{Scattering parameter $R_{sc}(E)$, Eq. \eqref{eq:MTscatteringParameter}. Below $10^9$ eV, where $R_{sc}(E)\ll1$ the proton scattering due to momentum transfert is unimportant.}
    \label{fig:MTScatteringParameterProton}
\end{figure}
The free-streaming or continuous slowing down approximation (CSDA) is a suitable representation as long as $R_{sc}(E)$ is less than 1. Otherwise, scattering becomes more critical, and a shift to diffusive transport occurs. As the energies increase, scattering becomes more important and a diffusive regime begins to operate at around 1 GeV (see Fig. \ref{fig:MTScatteringParameterProton}). This means that above 1 GeV, the protons experience efficient nuclei scattering before being stopped. During this scattering, the proton will produce pions, losing its energy catastrophically, breaking the CSDA assumptions. A proton with energy of 1 GeV propagate to a column densitie of $2 \times 10^{25}$ cm$^{-2}$. This means that the propagation of protons can be safely treated in the CSDA at column densities less than $10^{25}$ cm$^{-2}$.

\paragraph{Diffusion of electrons due to Bremsstrahlung scattering:}
For electrons, above an energy of about 500 MeV, their energy losses are predominantly due to Bremsstrahlung scattering. During this process, the photon energy is generally equivalent to the energy of the electron that produced it. The losses are quasi catastrophic so that above this energy, there is a deviation from the CSDA hypothesis. The effective BS cross-section corresponds to a column density of around $1/\sigma_{BS}=1.5 \times 10^{25}$ cm$^{-2}$. As a result, the CSDA tends to slightly overestimate the electron population at energies equal to or greater than this energy \citep{Padovani18}.

\subsubsection{Summary: limit between CSDA and diffusion}
To conclude about the transport mechanism of CR, our analysis estimates that the CSDA remains a valid approximation for column densities, $N$ less than $10^{25}$ cm$^{-2}$. Above this threshold, we need to consider more sophisticated transport processes, specifically diffusion processes, to provide a more accurate description of the particle transport. See \citet{Padovani18} for the treatment at higher column density.

However, understanding the role of turbulence in these diffusion processes, particularly within the context of accretion discs, remains a challenge. Turbulence in accretion discs is a complex phenomenon that is not yet fully understood, and there is no universally accepted model for it. In their work, \citet{Rodgers-Lee17} adopted some phenomenological models in which the diffusion coefficient, $D$, scales with the Larmor radius. Although this provides a way to account for turbulence, it is important to note that these models are not based on any well-established model of particle transport, they are essentially phenomenological, thus their predictive power is likely limited.

\subsubsection{Mirroring and focusing effects}\label{sec:mirroringfocussing}
However, apart diffusion some supplementary processes can induce a variation of the particle pitch angle and hence may question the CSDA adopted above, namely associated with the presence of magnetic field gradients developing over large scales along the mean magnetic field: magnetic mirroring and focusing. 

Studies like \citet{padovani2011effects,padovani2013cosmic,2018ApJ...863..188S} have examined the impact of magnetic mirroring and focusing on CR ionisation in dense molecular cores. The results suggest that mirroring is more dominant than focusing. These effects can reduce ionisation in collapsing clouds by more than an order of magnitude. This outcome could significantly impact the dynamical evolution and disc formation \citep{zhao2018decoupling}. In this section, we aim to analyse the effects of CR mirroring and focusing found in dense cores and apply them to circumstellar discs.

This section is divided into two parts. The first part deals with the fundamental principles and mechanisms causing the magnetic mirroring and focusing of particles. Here, we put aside energy losses to concentrate solely on the magnetic effects. This approach is relevant for describing cosmic ray propagation outside the disc, demonstrating that mirroring effects reduce CR fluxes even without energy losses. This process plays a role in excluding external CRs to enter in the central parts of T Tauri discs. In the second part, we incorporate the effects of energy losses. This approach is applicable to CR propagation within the disc, valid within the CSDA range, i.e., $N<10^{25}$ cm$^{-2}$. Within this range, we show that the ionisation rate is only minimally impacted by mirroring and focusing effects. At higher column densities, the primary sources of ionisation remain unaffected by magnetic fields. Our conclusion is that mirroring and focusing effects can be disregarded at any column density within the disc.

\paragraph{Basic principle:}

The shape of the environmental magnetic field can modulate the propagation of CRs through a process known as magnetic mirroring. Charged particles gyrating around magnetic field lines conserve their total kinetic energy $\propto v^2_{\parallel} + v^2_{\perp}$ and magnetic moment $\propto v^2_{\perp}/B$. As a particle enters a high magnetic field density area, it must increase its perpendicular velocity. As $v_{\perp}$ increases, $v_{\parallel}$ must decrease to keep the total kinetic energy constant. If the field is pinched to sufficiently high magnetic field strengths, the particle parallel velocity can halt and reverse, thus reflecting the CR in the opposite direction along the field line.

If $\alpha_{i}$ is the initial pitch angle between the CR velocity vector and the magnetic field, CRs with small $\alpha_{i}$ will tend not to mirror. The pitch angle $\alpha$ of a CR starting with $B_{i}$ and $\alpha_{i}$ at any given point along the field line is given by

\begin{equation}
\frac{{\sin^2 \alpha}}{{\sin^2 \alpha_{i}}} = \frac{B}{B_{i}} = \chi.
\label{eq:adiabaticinvariantconservation}
\end{equation}
This stems from the adiabatic invariance of a particle magnetic moment. Under such conditions, the steady-state kinetics of CRs with momentum, $p$, can be characterised by a distribution function $f(\mu, p, s)$. This function satisfies the following equation \citep{1978A&A....70..367C},

\begin{equation}
    \mu \frac{\partial f}{\partial s }- (1-\mu^2)\frac{d \ln \sqrt{B}}{ds}\frac{\partial f}{\partial \mu},
\end{equation}
where $\mu \equiv \cos \alpha$. The Liouville theorem implies that the distribution function, $f(\mu, s)$, in the disc, is identical to the distribution function of the interstellar CRs, $f_i(\mu_i)$,
\begin{equation}
f(\mu, s) = f_i(\mu_i)
\end{equation}
The local value of $\mu$ is determined from Eq. \eqref{eq:adiabaticinvariantconservation},
\begin{equation}
    \mu(\mu_i, s) = \pm \sqrt{1 - \chi(1 - \mu_i^2)},
\end{equation}
where $\chi(s) = \frac{B(s)}{B_i}$ denotes the magnetic "focusing factor". The convergence of the field lines induces an increase in the magnetic field intensity $B$. 

\paragraph{Mirroring and focusing along monotonically increasing magnetic field lines:}
A continual growth in the magnetic field strength results in the particle mirroring phenomenon. Interstellar CRs can only reach the position $s$ if $\sin \alpha \leq \frac{1}{\sqrt{\chi}}$.

For a disc threaded with a magnetic field intensity $B_{disc}$, there is a critical initial pitch angle such that CRs with $\alpha > \alpha_{crit}$ will be repelled before reaching the disc surface. We call $R_F=F(s)/F_{i}$ the ratio of the angle-integrated flux at position $s$ to the ISM angle-integrated flux. For particles arriving on one side of the disc, this condition corresponds to a fractional reduction of CRs by mirroring 

\begin{equation}
    R_{F, \rm mirror} = \sqrt{1 - (B_{i}/B_{\rm disc})^2},
\end{equation}

However, open magnetic field lines tend to increase density of CRs via focusing, enhancing the CR flux proportional to the increase in field line density, 
\begin{equation}
    R_{F, \rm focus} = B_{i}/B_{\rm disc}.
\end{equation}
Although mirroring generally dominates over focusing, their effects are of similar magnitude. The combined fractional removal of CRs as given by \citet{2004ApJ...602..528D}, is, 

\begin{equation}
R_{F, \rm combined} = \chi-(\chi^2-\chi)^{1/2}.
\label{eq:MirroFocusFractionalRemoval}
\end{equation}

In the limit where $B_{disc}\gg B_{i}$, \citet{2004ApJ...602..528D} showed that the combined fractional remove of CRs is $R_{F,\rm combined}=F(s)/F_{\rm ISM}=0.5$.
Mirroring is only a significant factor when the change in pitch angle due to magnetic field variations is greater than that caused by scattering. In T Tauri systems that are of interest here, large magnetic field strength gradients are expected, hence, mirroring could play a significant role. 

\paragraph{Exclusion of external CR due to magnetic mirroring:}
\citet{2013ApJ...772....5C} studied the mirroring effects for different magnetic field models. They showed that the stellar magnetic effects are capable of modulating CRs only relatively near to the star ($R < 100$ au). Observed examples of split-monopolar field strengths suggest that stellar magnetic fields can modulate CR propagation within specific distances, e.g., 96 au and 20 au.

\begin{figure}[h!]
    \centering
    \includegraphics[width=0.4\linewidth]{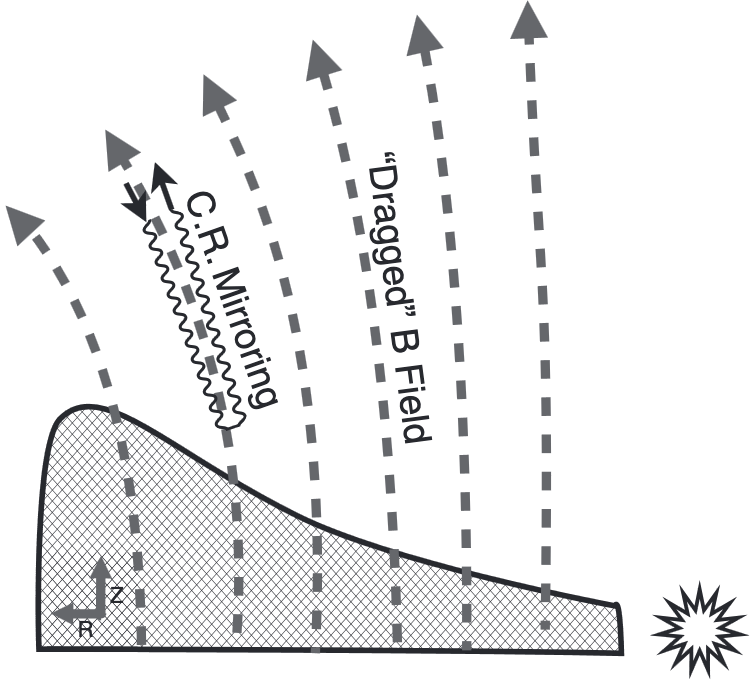}
    \caption{Illustration of the exclusion of cosmic rays by the magnetic mirror. The dotted lines indicate the hourglass-shaped background magnetic field and the hatched area indicates the disc, seen from the edge. The cosmic ray (zigzag arrow) penetrates along the magnetic field line into a region where the field lines narrow, so by conservation of magnetic flux, the field intensity increases. The particle is repelled by the magnetic mirroring before reaching the surface of the disc. }
    \label{fig:MagneticMirroringCleeves}
\end{figure}
Hourglass magnetic fields, on the other hand, can reduce incident CR fluxes on a much larger scale, of hundreds of au, see Fig. \ref{fig:MagneticMirroringCleeves}. For fields that are only moderately pinched, the entire CR rate could be reduced by half. Such a field configuration corresponds to a magnetic field strength of 0.9 mG at 100 au in the midplane.

The magnitude of magnetic modulation is less than that achieved by a stellar wind, but the mirroring effect is significant as it introduces a radial variation in the net CR flux, decreasing it from 0.5 in the inner disc, see Eq.\ref{eq:MirroFocusFractionalRemoval}. On the other hand, in Sec. \ref{sec:galacticCRExclusion}, we demonstrated the energy dependence of the radial gradient of CR flux due to wind modulation. This gradient could be mistakenly identified in observations with magnetic effects from the star or environment, especially if they are of similar magnitude. Therefore, to observationally constrain the properties of the T Tauriosphere, the exclusion of particles due to both wind modulation and magnetic mirroring needs thorough characterisation \citep{2013ApJ...772....5C}.

Large scale mirroring and focusing effects are anticipated to reduce the CR flux, particularly in the outer disc. But mirroring and focusing effect can also affect the propagation of particles at smaller scales. It would emerge from variations in magnetic intensity, possibly due to turbulence. These small scale mirroring and focusing effects could be important in the propagation of particles in the inner disc, we explore this aspect now.

\paragraph{Mirroring and Focusing in magnetic pockets:}

The impact of magnetic mirroring and focusing on local CR flux essentially balance each other out when cosmic rays move along monotonically increasing magnetic field lines, as shown above. Consequently, the CR density can be estimated by assuming a constant field strength, regardless of the complexity of the field structure.

\begin{figure}[h!]
    \centering
    \includegraphics[width=0.7\linewidth]{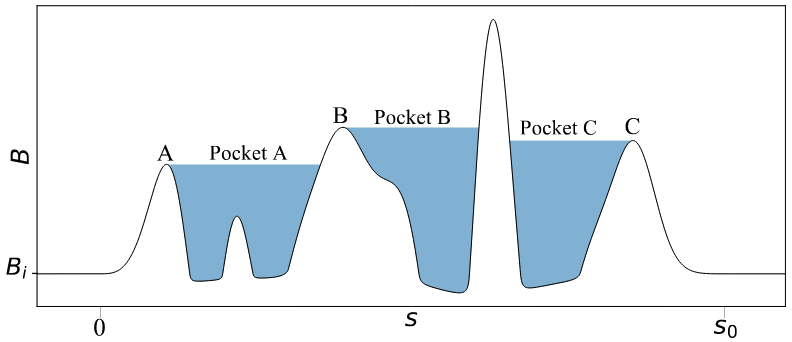}
    \caption{Sketch illustrating a situation where the magnetic field has multiple local minima along the field line. This results in multiple magnetic pockets, indicated by the shaded areas. Figure from \citet{2018ApJ...863..188S}}
    \label{fig:MagneticPocketsFocusMirror}
\end{figure}

However, this equilibrium is disrupted when the magnetic field strength has local minima along its lines. When cosmic rays enter such magnetic pockets, as depicted in Fig. \ref{fig:MagneticPocketsFocusMirror}, the equilibrium interplay between magnetic mirroring and focusing breaks down. This is because the pitch angles of particles inside the pocket only shrink compared to the values at the pocket boundary, leading to a decrease in the number of particles with pitch angles around 90 degrees, ultimately reducing the CR density.

The CR density within a magnetic pocket, in the absence of losses, is  given by,
\begin{equation}
   \frac{n(E, s)}{n_{IS}(E)} = 1 - \sqrt{1 - \frac{B(s)}{B_{lmin}}}.
\end{equation}

Here, $n_{IS}$ refers to the CR density per unit energy outside the pockets (equivalent to the interstellar density), and $B_{lmin}$ is the field strength at the minimal values of the local maxima bounding a pocket (points A, B and C in Fig. \ref{fig:MagneticPocketsFocusMirror}). This formula remains valid even if $B(s)$ exhibits minor peaks inside the pocket. 

Magnetic pockets are expected to be a common feature in the internal regions of dense molecular cloud cores and discs. These pockets are formed due to the repeated stretching and compression of magnetic field lines, a consequence of large-scale turbulence in discs. Especially where the magnetic field has a strong toroidal component induced by rotation, the occurrence of these pockets is prominent. In such dense environments, numerical models predict that the CR flux can drop by over an order of magnitude \citep{2018ApJ...863..188S}. This in turn will reduce ionisation rate leading to new implications for gas chemistry and angular momentum transport for example.
In the next Sec. \ref{sec:impactmirrorfocuseffect} we show how we take this into account in our models.

Lastly, it should be noted that this analysis specifically applies to the free-streaming regime of CR propagation. If weak, small-scale MHD turbulence is present, resonant pitch-angle scattering can effectively isotropize the distribution function, thus erasing the impact of magnetic pockets. It is worth noting, however, that very intense turbulence might further enhance mirroring, thereby reducing the overall CR density in discs \citep{2014ApJ...787...26F}.

\paragraph{Impact of mirroring and focusing effects on the ionisation rate including losses:}\label{sec:impactmirrorfocuseffect}

We start presenting the impact of magnetic mirroring and focusing using a single-peaked field profile, as illustrated in Fig. \ref{fig:MagneticMirroringCleeves}.

Ionisation is occurring in the disc, where the $B(s)$ value is significantly lower than the peak value, $B_p$. Cosmic rays from the other side of the disc are negligible due to strong damping. \citet{2018ApJ...863..188S} compute the ionisation rate increase $R$ relative to the reference value $\zeta_L$, where $\zeta_L$ is computed in Eq. \eqref {eq:freestreamingionisationrate}, not taking into account magnetic and mirroring effects. We assume constant field strength leading to an isotropic distribution of the pitch angles. In this configuration, the relative increase $R$ compared to $\zeta_L$ can be bounded \citep{2018ApJ...863..188S} in terms of the parameters of the toy model of section \ref{sect:ToyModel},

\begin{equation}
    1< R < 1+ \frac{a+b-1}{1+b}=2.27
    \label{eq:MirrorFocusIncreaseFactor}
\end{equation}
where $a = 2.5$ is the power law index of the injection spectrum and $b=0.82$ is the power law index of the ionisation loss function defined in the previous section. We conclude that for values of the spectral index of the energy distribution of particles produced in typical in flares, the total relative increase compared to $\zeta_L$ does not exceed a factor of $R_{\rm max}=2.27$.

These results remain the same in multi-peak magnetic fields  except for local field minima areas, i.e. magnetic pockets (see Fig. \ref{fig:MagneticPocketsFocusMirror}). Here, ionisation can drastically decrease. \citet{2018ApJ...863..188S} found this decrease via a reduction factor with two characteristic situations: localised pockets with small column density, thus small energy losses, compared to the distance between the pocket and disc edge, and global pockets where the column density of a magnetic pocket is much larger than that between the edge of the disc and the pocket. For localised pockets the reduction factor is,
\begin{equation}
    R_{lp}= R_{\rm max} \left(1-\sqrt{1-\frac{B}{B_l}}\right)
\end{equation}
where $B_l$ is the local maximum of magnetic field intensity, $B$ is the magnetic field in the pocket verifying $B<B_l$ and $R_{\rm max}=2.27$ is given by Eq. \eqref{eq:MirrorFocusIncreaseFactor}.
The reduction factor in global pockets is,
\begin{equation}
R_{gp}= 1-\left(1-\frac{B}{B_l}\right )^{\frac{R_{\rm max}}{2}}  . 
\end{equation}

The specific formula used is not critical for pockets with intermediate column densities since they differ at most by a factor of less than $R_{\rm max}$.

But there are considerable uncertainties in estimating the rate of ionisation in dense molecular clouds and circumstellar discs. This is primarily due to our limited understanding of the gas distribution and the layout of magnetic field lines in the discs. These limitations result in uncertainties in the column density crossed by the particles $(N)$ and, consequently, in the reference ionisation rate $\zeta_L(N)$ . \citet{2018ApJ...863..188S} concluded that due to these uncertainties, the impact on ionisation caused by CR mirroring and focusing are negligible if the field has a single-peaked profile. In magnetic pockets however, they recommend to use their analytical formulae to accurately calculate the relative decrease of the ionisation rate.

These results apply to effective column densities up to $N \sim 10^{25}$ cm$^{-2}$. Interestingly, recent research by \citet{Padovani18} showed that at $ N> 3\times 10^{25}$ cm$^{-2}$, CR ionisation is driven by secondary photons (see next Sec. \ref{sec:secondaryParticles}), making magnetic field effects irrelevant. This suggests that mirroring and focusing do not significantly affect ionisation outside magnetic pockets at any column density.

\subsection{Secondary particles}\label{sec:secondaryParticles}

The previous section explored the two key transport mechanisms involved in cosmic ray propagation and established the framework of the CSDA in discs. From the CSDA, we derived the equations for the propagated flux of particles $j(E,N)$, as per Eq. \eqref{eq:CSDA}. This equation is relevant for primary CRs. However, these primary CRs also generate secondary particles as they propagate. This section deals with the main generation channels of these secondary particles produced by CRs, with a focus on detailing the derivation of the secondary electron flux.

\begin{figure}
    \centering
    \includegraphics[width=0.7\linewidth]{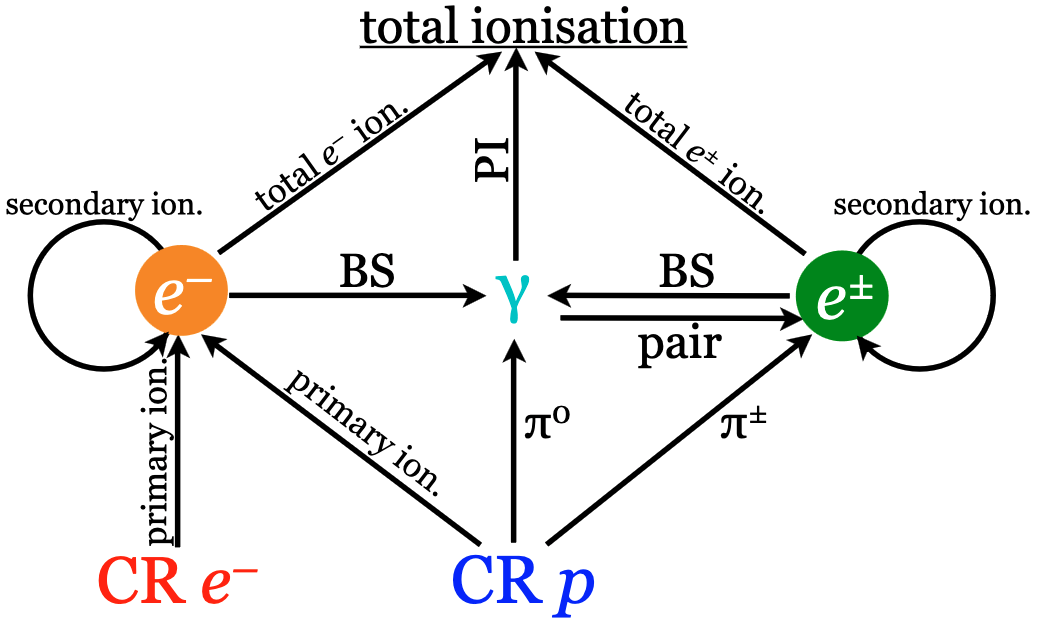}
    \caption{This is an ionisation diagram that illustrates the impact of secondary particles, which are produced either directly or indirectly by cosmic ray protons and electrons. This production occurs through several processes including ionisation, pion decay ($\pi^0$, $\pi^\pm$), Bremsstrahlung, and pair production. The secondary particles are electrons ($e^-$, from primary and secondary ionisation), positrons and electrons ($e^\pm$, arising from pair production and $\pi^\pm$ decay, including electrons created in secondary ionisation), and photons ($\gamma$, originating from BS and $\pi^0$ decay). Each of these particles contributes to their respective ionisation channel. The figure is reproduced from \citet{Padovani18}
}
    \label{fig:SecondaryionisationRoutes}
\end{figure}

\paragraph{Main ionisation channels associated with secondary particles:}
\citet{Padovani18} proposed a diagram of the main ionisation routes associated with secondary particles, reproduced in Fig. \ref{fig:SecondaryionisationRoutes}. We describe here the processes shown on this diagram.

CR protons and electrons generate secondary electrons through primary ionisation as described by the following equation,
\begin{equation}
     k_{CR} + M \rightarrow k_{CR}' + M^{+} + e^-
\end{equation}
where $k$ can be an electron or a proton and $M$ any species present in the disc.
Secondary electrons with energy higher than the ionisation potential of $M$ result in further generations of ambient electrons. CR protons colliding with protons generate charged pions which decay into electrons and positrons through muon decay, leading to secondary ionisations. Also, $p-p$ collisions generate, as we already discussed, neutral pions that decay into photon pairs. The equation for these processes is:

\begin{equation}
    \pi^{\pm} \rightarrow \mu^{\pm} \rightarrow e^{\pm} \quad \text{and} \quad \pi^{0} \rightarrow 2\gamma 
\end{equation}

In addition, Bremsstrahlung by electrons and positrons is a source of photons. Pair production by photons should also be considered,

\begin{equation}
    \gamma \rightarrow e^{+} + e^{-}
\end{equation}

\paragraph{Secondary photons:}
The behaviour of photon propagation is governed by the energy level of the photons, denoted as \(E_{\gamma}\). For energy levels below approximately \(5\) keV, photoionisation is the prevailing process, while pair production takes precedence for energies above roughly \(50\) MeV. In both cases, the processes are irreversible, meaning that the photons cease to exist after interacting with atomic nuclei. 

In the case of inverse Compton scattering, the average photon energy change in each interaction with electrons is contingent on the photon energy. When \(E_{\gamma}\) is approximately equal to \(m_e c^2\), the energy losses are irreversible (Klein-Nishina regime). However, in other energy ranges, photons only impart a minor fraction of their energy to the electrons, making the losses incremental rather than catastrophic (Thomson regime). Nevertheless, for energies below about \(1\) keV, photoionisation becomes the dominant process again, leading to irreversible losses.

\paragraph{Secondary electron-positron pairs:}
Among the  mechanisms that can generate electron-positron pairs the first involves photons possessing energy greater than \(2m_e c^2\), resulting in electron and positron energies ranging from \(0\) to \(E_{\gamma} - 2m_e c^2\). The second source of electron-positron pairs is the decay of charged pions, which are created in proton-nucleus interactions at energies exceeding \(E_{\pi} = 280\) MeV. 

In modelling the behaviour of electrons and positrons with energies above the BS threshold, approximated as \(E_{BS} \approx 500\) MeV, the CSDA can be used. While CSDA tends to slightly overestimate the flux, this overestimation is usually inconsequential for ionisation rate calculations since the contribution from electrons and positrons at such energy levels is negligible.

Therefore, when the stopping range of these particles is comparable to or greater than the local column density \(N\), the resulting energy spectrum for electrons and positrons can be described as follows \citep{Padovani18}.
\begin{equation}
    j_{e^{\pm}} (E, N) = \frac{1}{2} L_e(E) \int_{N}^{\infty} S_{e^{\pm}} (E_0, N_0) L_e(E_0) dN_0
\end{equation}

Here, $S_{e^{\pm}}$ is the total source function for electrons and positrons (including pair production and charged pion decay) and the factor $1/2$ accounts for electrons and positrons propagating in two directions. The initial energy $E_0 > E$ at $N_0$ is related to $N$ by,

\begin{equation}
    |N_0 - N| = \int_{E_0}^{E} \frac{dE'}{L_e(E')} 
\end{equation}

If the range is small, $|N_0 - N| \ll N$, the spectrum is localised,

\begin{equation}
    j_{e^{\pm}} (E, N) = \frac{E}{L_e(E)} S_{e^{\pm}} (E, N)
\end{equation}

\paragraph{Derivation of the secondary electrons flux:}

In the range of applicability of our model, i.e., the range of applicability of the CSDA where $N<10^{25}$ cm$^{-2}$, \citet{Padovani18} demonstrated that the dominant ionisation processes from secondary particles are the secondary electrons directly produced by the primary CR particles. We focus here on the secondary electrons and show the derivation of their flux as proposed by \citet{ivlev2015interstellar}.

When an electron of energy $E$ and velocity $v_e$ crosses a column density $dN = n dx$ in a medium of density $n$, it loses an energy $dE = -L_e(E) dN$, where $L_e(E)$ is the energy loss function of electrons,

\begin{equation}
    L_e(E) = -\frac{1}{n}\frac{dE}{dx} = \frac{1}{n v_e}\frac{dE}{dt}.
\end{equation}
The equation is similar for the loss function of protons. The column density $R(E)$ required to stop an electron of the initial energy $E$ is defined by Eq. \eqref{eq:StoppingRange}. It remains constant from approximately $R(E) \approx 10^{18}$ cm$^{-2}$ up to $E \approx 1$ keV, then increases weakly with energy. Secondary electrons are typically produced at energy less than 1 keV. This short range of secondary electrons, compared with typical column densities of discs ($\sim 10^{22}$ - $10^{25}$ cm$^{-2}$), justifies a local treatment of ionisation. In this approximation, the stopping time of secondary electrons is,
\begin{equation}
    \tau_{\rm stop} \approx \frac{E}{n v_e L_e(E)}
\end{equation}
and their stopping range is approximately $\tau_{\rm stop} v_e \approx  E/(n L_e(E))$. Assuming the secondary electrons are produced isotropically, the number of first-generation secondary electrons of energy $E$ (per unit energy, volume, and time) is calculated as,

\begin{equation}
    \frac{dF_{e,k}^{sec.}}{dE dV dt} = 4\pi n  \int_{E+I}^{\infty} j_k(E') \frac{d\sigma^{ion}_{k}}{dE}(E,E') dE'
\end{equation}
where $I$ is the ionisation potential of of the target species, $j_k(E')$ is the intensity of primary species $k$, and $d\sigma^{ion}_{k}/dE$ is the corresponding differential ionisation cross section. The number of secondary electrons produced per unit energy and volume is then,

\begin{align}
    &\frac{dF_{e,k}^{sec.}}{dE dV} = \frac{dF_{e,k}^{sec.}}{dE dV dt} \tau_{stop}\\
    &=\frac{4 \pi E}{v_e L_e(E)}\int_{I+E}^\infty dE' j_k(E') \frac{d\sigma^{ion}_{k}}{dE}(E,E').
\end{align}

This quantity is related to the specific intensity of secondary electrons $j^{sec.}_{e}(E)$ (number of electrons per unit energy, area, time, and solid angle) by $\frac{dF}{dE dV} = \frac{4 \pi}{v_e} j^{sec.}_{e}$, which finally yields to,

\begin{equation}
j^{\rm sec.}_{ e,k}(E) \approx  \frac{ E}{L_{e}(E)} \int_{I+E}^\infty dE' j_k(E') \frac{d \sigma^{\rm ion.}_k}{d E}(E,E').
\label{eq:secondaryelectronflux}
\end{equation}

Here, $j_k $ is the primary particle flux of $k$, $\frac{d \sigma^{\rm ion.}_k}{d E}(E,E')$ is the differential ionisation cross section. We use \citet{kim2000extension} for the differential ionisation cross section of electrons and \citet{krause2015crime} for the one of protons. Equation \eqref{eq:secondaryelectronflux} is iterated to compute intensities for the next generations of secondary electrons.

\section{Conclusion}

In this chapter we examined the complex interplay between the energetic particles  and the disc medium, focusing particularly on their trajectories and energy distribution. Two principal factors shape these characteristics, the energetic loss function and the selected transport model.

We initiated the chapter by detailing the loss functions associated with sub-GeV electrons and protons, emphasising how these are influenced by particle type, energy, and the chemical properties of the medium. This sets the stage for introducing two particle transport models, the free-streaming and diffusive models. Within the free-streaming approximation, we presented a simplified, analytical model to estimate the ionisation rate across varying column densities in the disc. This model, although rudimentary, provides valuable insights into the ionisation effects of energetic particles produced by a reconnection event on the disc medium.

We then defined the criteria for selecting the appropriate transport regime, free-streaming or diffusive, based on the column density crossed by particles within the disc. We established that for column densities less than \(10^{25} \text{cm}^{-2}\), the free-streaming model holds, while the diffusive regime is necessary beyond this threshold.

Addressing a long-standing question in the field, we proposed estimations for diffusion coefficients within the inner disc. These estimates were formulated based on different prevalent turbulence models, such as MHD resonance and MRI-induced turbulence.

We also explored magnetic mirror and focus processes. These processes are often cited for their role in modulating GCR flux in discs. However, our analysis concluded that these processes are likely insignificant in the context of the inner disc, thereby justifying our decision to not consider them in subsequent chapters.

Lastly, our examination extended beyond primary particles to encompass the impact of secondary particles, which are generated through interactions with the disc. We elaborated on how these secondary particles, particularly secondary electrons, contribute to ionisation within the disc.

In summary, this chapter builds upon the foundation laid in Chapter \ref{C:Reconnection}, where we developed a model for particle acceleration via magnetic reconnection. By detailing how these accelerated particles interact with the disc, emphasising their ionisation power, we have set the stage for the first publication of this thesis presented in next chapter.

\chapter[Stationary Model]{Stationary Model — Ionisation of inner T Tauri star discs: effects of in situ energetic particles produced by strong magnetic reconnection events \citep{2023MNRAS.519.5673B}}\label{C:PublicationI}

\section{Introduction}
\subsection{Ionisation in the inner disc}
The way T Tauri stars accrete infalling surrounding matter through the transport of angular momentum is still a widely open subject in astrophysics. Hydrodynamic or gravitational instabilities alone do not explain the dynamics of accretion-ejection of matter in T Tauri systems. To understand these phenomena, MHD must be considered. There are two main models that aim to explain the transport of angular momentum outward from the disc and the transport of matter inward, magneto-centrifugal ejection and magneto-rotational instability. Both models require a sufficient coupling between matter and the magnetic field, a condition only possible with charged matter. Therefore, this coupling occurs only if the disc matter is sufficiently ionised.

In Chapter \ref{C:ionisation}, we presented various sources of ionisation within the disc, the observational constraints on these sources, and their impacts on the dynamics and chemistry of the disc.

In this chapter, our focus is on ionisation within the inner disc, specifically where \( R < 1 \, \text{au} \). We can illustrate the importance of ionisation in the inner disc dynamics, by estimating the minimum ionisation rate to trigger MRI. With the MRI triggering condition \citep{2013ApJ...772....5C}, 

\begin{equation}
    Am = \frac{x_E}{10^{-8} }\frac{n_n}{10^{10} \rm cm^{-3}} \left( \frac{R}{1\,\text{au}} \right)^{3/2} > 0.1
\end{equation}
the minimum ionisation fraction to trigger MRI is,
\begin{equation}
   x_{\text{E,MRI}}= 10^{-9} \left( \frac{n_n}{10^{10} \rm cm^{-3}} \right)^{-1} \left( \frac{R}{1\,\text{au}} \right)^{-3/2}.
\end{equation}

Assuming the density structure calculated by {\tt ProDiMO}, at \( R = 0.1 \, \text{au} \), \( Z/R = 0.1 \), we find \( x_{\text{E,MRI}} \approx 10^{-10} \). According to Fig. \ref{fig:ThermalionisationFraction}, if $T\lesssim 1200$ K, thermal ionisation is not efficient enough to trigger MRI. At this location in the disc, accounting for X-ray and UV heating as well as the dust screening of these radiation, {\tt ProDiMO} computes \( T \sim 500\, \text{K} \), which is highly insufficient to trigger MRI.  This result is heavily influenced by the specific disc model used to compute the thermal structure and is quite approximate. Despite this, it highlights the importance of considering non-thermal ionisation sources. This consideration is essential for properly estimating the MRI active region and understanding the dynamics of the disc   inner regions.

The spatial extent of the MRI active region is a topic of much debate and difficult to constrain observationally. Based on our current understanding, the extent of the MRI zone is primarily determined by the local X-ray and UV flux in the disc. The radiative flux in this region has two major effects on ionisation. First, it has a direct effect through photoionisation. Second, it also deposits heat, which may eventually raise the temperature high enough to allow for thermal ionisation. This could be sufficient to activate MRI.

\subsection{The extent of the MRI active region in radiative simulation}
The inner disc of a star system is profoundly affected by the presence of dust grains, which play a key role in shaping the local intensity of X-ray and UV radiation. For instance, radiative MHD simulations (e.g., by \citet{2017ApJ...835..230F}) have revealed that at certain positions within the inner disc, the temperature can reach up to 1200 K or higher, thus allowing for the triggering of the MRI at those locations.

This discrepancy between different temperature calculations comes down to the precise location of the dust sublimation radius, the point where the heat from the central star causes dust grains to change from solid to gas. Inside this radius, the intense radiation sublimates the dust, leaving a dust-free region. Beyond this radius, dust particles can remain solid, shielding themselves from X-rays and UV radiation, leading to a sharp decrease in temperature and suppressing thermal ionisation.

We have seen in Eq. \eqref{eq:sublimationradius} that dust sublimation radius \( R_{\text{sub}} \) can be estimated by considering factors such as the star luminosity \( L_* \), the effective temperature \( T_{\text{eff}} \). The sublimation radius increases with increasing stellar luminosity. Observations have shown that this radius typically lies between 0.05 and 0.3 au, with a median value around 0.1 au. For instance, the {\tt ProDiMO} model, with standard values for Sun-like stars, agrees with these observations, while the simulation by \citet{flock20173d} of Herbig stars, using a much higher effective temperature, predicts a greater sublimation radius of about 0.55 au.

The position of the sublimation radius is crucial for understanding the MRI active region in the inner disc. In certain models like {\tt ProDiMO}, beyond the sublimation radius, the presence of dust completely blocks X and UV rays, causing the temperature to fall below 1000 K. As a result, in Sun-like stars, if only X-rays and UVs are considered, the inner disc is largely MRI inactive. %This highlights the determining role of dust grains in the thermal and dynamical processes of the inner regions of star systems.

\subsection{Stellar energetic particles as an additional ionisation source}

The issue of an MRI-inactive inner disc for Sun-like luminosity stars leads to the study of the effect of stellar energetic particles as an additional ionisation source \citep{2017A&A...603A..96R,Rodgers-Lee17,2018ApJ...853..112F}. There are two main reasons why the ionisation by energetic particles is interesting. First, they are not screened by dust, and second, they have a stronger penetration power into the gas. These two reasons imply that energetic particles may penetrate and ionize farther from the star and deeper into the disc than any other radiative sources, in turn extending the MRI-active region in the inner disc.

The spatial distribution of the ionisation rate produced by energetic particles depends on several factors related to the particle injection model. Essentially, these factors include the energy distribution of the particles, the total flux, the location where they enter the disc, and the losses they encounter within the disc.

In previous works studying the effect of particles on ionisation, the particles were emitted from the central star. They propagated following different transport regimes, either ballistic \citep{2017A&A...603A..96R} or diffusive \citep{Rodgers-Lee17,2018ApJ...853..112F}. These regimes of propagation lead to strong attenuation and high ionisation rates that cannot be sustained far from the central star, as the particles first interact with the very dense inner disc.

\subsection{Energetic particles produced by flares}

In this chapter, we propose a new particle emission model that has strong potential to be an important ionisation source in the inner disc. We study the ionisation produced by energetic particles accelerated by flares in the inner disc.

A study that inspired our work is the series of simulations by \citet{2011MNRAS.415.3380O, colombo2019new}. They studied the effect of heat deposition from flares occurring above the inner disc on the dynamics of accretion. Their simulations show that the pressure applied by the heated plasma on the flare side of the disc is able to trigger accretion funnels on the other side of the disc. While these studies demonstrate that flares may have a strong effect on the accretion-ejection dynamics of inner discs of T Tauri stars, their flaring model is not based on any physical assumptions. The position and intensity of the flares are chosen arbitrarily. In addition, they do not take the effects of the non-thermal particle distribution into account.

We proposed in \citet{2023MNRAS.519.5673B} (hereafter B23) to study the effect of flares based on a physical acceleration mechanism, magnetic reconnection. This assumption is widely supported by solar observations. The idea of studying the effects of particles produced by magnetic reconnection events was also inspired by the results of the simulation of magnetospheric ejection by \citet{zanni2013mhd}, reproduced in Fig. \ref{fig:MagneticReconnectionRegion}. This figure shows the magnetic field lines in the interaction zone of the magnetic field of the star and the magnetic field of the disc. We can clearly identify the typical magnetic field topology of magnetic reconnection occurring above the disc. It has been widely observed, in the Sun and in laboratory experiments, that such magnetic reconnection events are very efficient at accelerating particles to energies $\lesssim 1$ GeV. Given the geometry of the field lines above the disc, we expect the particles to propagate from the acceleration region down to the disc.

Once these particles penetrate into the disc, they encounter energy losses. At energies $\lesssim 1$ GeV, particles mainly lose their energy through ionisation losses. In other words, they are also very efficient at ionizing the disc. This supports the idea that our model could be a significant source of ionisation in the inner disc.

\subsection{The stationary model of ionisation by particle produced by magnetic reconnection}

In B23, we proposed an alternative source of ionisation. The paper examined ionisation rates resulting from the production of energetic particles in the star-disc system during magnetic reconnection flares. These particles are low-energy ($E\lesssim1$ GeV) supra-thermal particles that can ionize the discs, affecting their chemical, thermal, and dynamical evolution. The goal of this work was to investigate how energetic particles propagate in T Tauri discs and to compute the ionisation rate they produce. We collected experimental and theoretical cross-section data for the production of H$^+$, H$_2^+$, and He$^+$ by electrons and protons.

In B23, we fixed the location of penetration of the particles at R=0.1 au from the star and estimated the shape of the distribution of the injected spectra based on the expected physical properties of flares in the inner disc. In Chapter \ref{C:Reconnection}, we described how we could estimate the magnetic reconnection regime occurring in T Tauri flares from the magnetic field intensity, particle density, and Lundquist number. The inferred regime is the multi X-line collisional magnetic reconnection. We then relied on simulations of particle acceleration during reconnection in the multi X-line collisional regime to infer the energy distribution of the particles. The simulation by \citet{arnold2021electron} suggests that electrons follow a power-law energy distribution of index $\delta$, ranging from 3 to 8 depending on the magnetic configuration in the reconnection region. Thus we conducted a parametric study on $\delta$ within this range of values.

Starting from the theoretical injection spectra, we calculated the local spectra at different column densities in the disc. These local or propagated spectra are the spectra of particles attenuated by energetic losses in the disc. Considering all the energy loss processes presented in Sect. \ref{sec:EnergyLossesDisc}, we computed the propagated spectra for column densities ranging from $10^{19}$ cm$^{-2}$ to $10^{25}$ cm$^{-2}$ in order to treat the propagation of particles in the CSDA regime.

The physical properties of flares, such as their size, electron density, and temperature, were deduced from observations of the COUP (Chandra Orion Ultradeep Project) catalog. Using the radiation thermo-chemical code {\tt ProDiMO}, we calculated the structure of the disc illuminated by the X-ray field produced by a flare. This structure provides the abundance of different chemical species in the disc needed to compute the column density crossed by the particles. The column density is then used to estimate energy losses and, in turn, the propagated flux of particles. Finally, the ionisation rate is computed from the propagated flux, accounting for the ionisation of secondary particles.

The study considered various flare temperature as well as various spectral indices for the energetic particle flux and different magnetic configurations, along which the particles propagate.

B23 shows that, in a stationary configuration, energetic particles are a powerful source of local ionisation, with ionisation rates that exceed X-ray, stellar energetic particles, and radioactivity contributions in the inner disc by several orders of magnitude. For our reference case, a 1MK flare at 0.1 au propagating along a vertical magnetic field line, we found ionisation rates $\zeta=10^{-9} ~\rm  s^{-1}$ at column densities of $10^{25} ~\rm cm^{-2}$, while the ionisation rate at this depth is of the order of $10^{-17} ~\rm  s^{-1}$ due to X-rays produced in a $1~\rm MK$ stellar flare and GCRs and $10^{-18} ~\rm  s^{-1}$ by radionuclides.

Although we are aware that our assumptions may lead to overestimate the ionisation rate (for the reasons listed in the next section \ref{sec:limitationstationnarymodel}), we show that this process can be a dominant one among the ionisation processes in the inner disc of T Tauri stars. As there are several parameters in our model that are difficult to constrain either experimentally or observationally, we have conducted a comparative analysis of these parameters. The aim of this analysis is to define a range of flare parameters so that the ionisation rates produced are dominant over other ionisation sources. We anticipate that this will be the case for: 
\begin{itemize}
    \item a reconnection process that accelerates particles following an injection flux with a power law $j \propto E^{-\delta}$ for $\delta < 6$,
    \item  flares with temperatures above 1 MK,
    \item particles propagating along the field line with a ratio of the toroidal component to the poloidal component $b_{\rm g}=B_{\phi}/B_{\rm pol} < 1$.
\end{itemize}

\subsection{Limitations of the stationary model}\label{sec:limitationstationnarymodel}

The results obtained in B23 are inspiring; the idea that the effects of particles produced by flares may have a significant impact on the ionisation state of the inner disc is promising. However, we are aware that the results of this model are likely overestimated, and this overestimation arises from two main reasons.

First, the model is stationary, meaning that no temporal effects are taken into account. A more realistic model must consider the recombination rates. We need to estimate whether, under the thermal conditions of the inner disc, the free electrons produced by the flare immediately recombine, or if they remain free long enough to impact the overall inner disc dynamics. Another temporal effect that must be estimated is the rate of occurrence of these flares. We need to determine whether the occurrence rate of strong flares is high enough to sustain a high level of disc ionisation.

Second, the model is local, i.e., the ionisation rates are computed just below the flare. A more realistic model would estimate the area of the disc illuminated by the particles produced by the flare. To do this, a model of flare geometry is needed that provides the area of the illuminated disc as well as the radial position of penetration of the particles.

Based on observations of the occurrence rate, the luminosity distribution, and models of flare geometry, we tackle these issues in a spatial and time-dependent model, presented in Chapter \ref{C:PublicationII}.

\section{Publication}
V. Brunn, A. Marcowith, C. Sauty, C. Rab, M. Padovani \& C. Meskini, 2023, MNRAS, \underline{519}, 5673 : Ionization of inner T Tauri star discs: effects of in situ energetic particles produced by strong magnetic reconnection events.\\
\includepdf[pages=-]{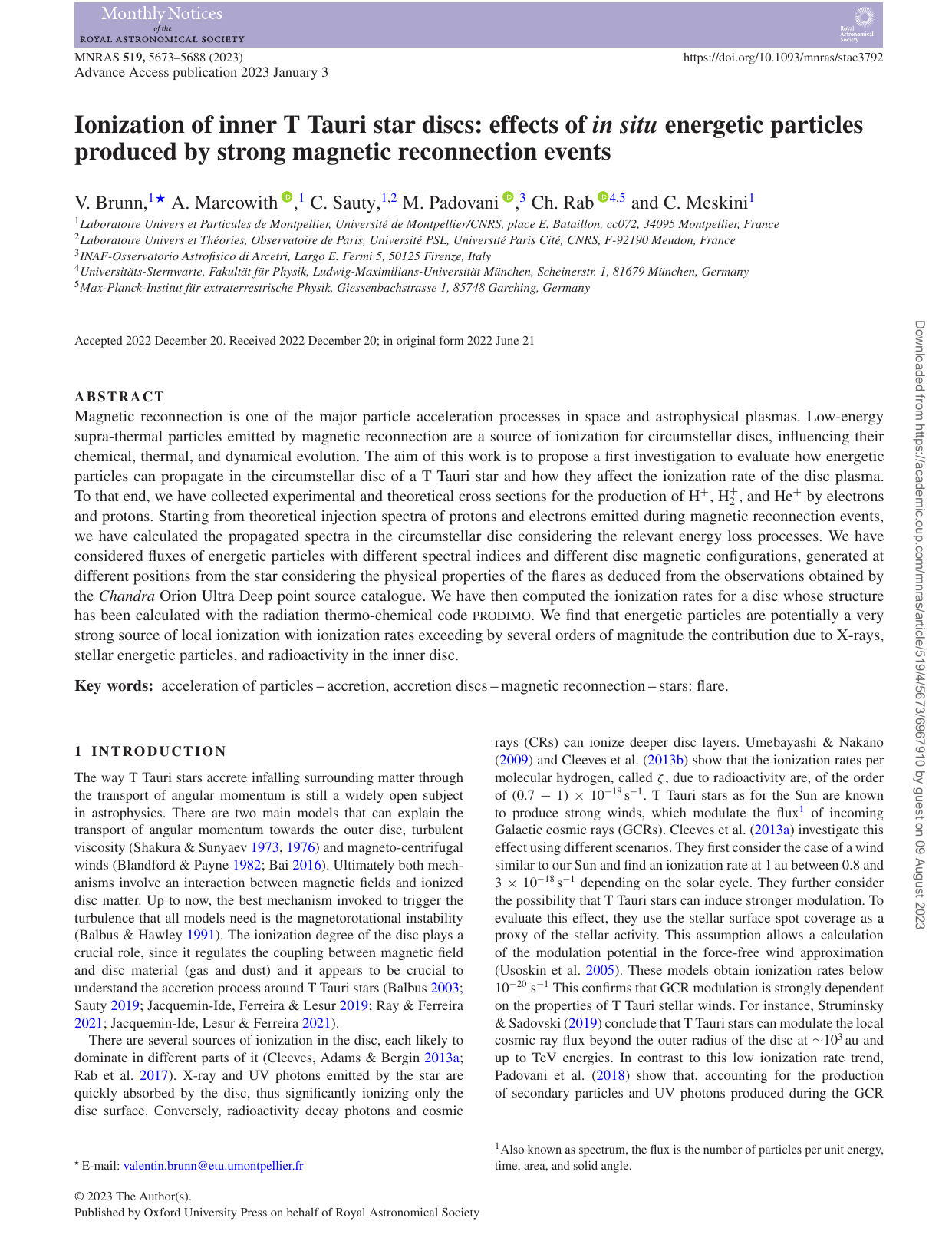}

\chapter[Time-Dependent Model]{Time-Dependent Model – Impacts of Energetic Particles from T Tauri Flares on Inner Protoplanetary Discs}\label{C:PublicationII}
\section{Summary of the publication}
\subsection{Introduction}
The ionisation rate in the accretion discs of T Tauri stars plays a crucial role in the development of the MHD instabilities essential for explaining the accretion of surrounding matter onto the central star. Furthermore, it influences the disc chemistry, which must be thoroughly understood to interpret observations and to understand the formation of the building blocks of life. We have seen in Chapter \ref{C:ionisation} that until now, the ionisation sources considered were either radiative sources, such as X-ray or UV from the nearby stellar environment or emitted from the central star itself, or energetic particles of galactic or stellar origin. While galactic energetic particles are not anticipated to significantly affect the inner disc due to the exclusion of these particles by the T-Tauriosphere, those emitted from the central star have proven effective at ionising the inner disc. Previous research on the impact of stellar particles has generally focused on a continuous emission of particles accelerated at the stellar surface without considering the microphysics of the acceleration processes \citep{Rab17,Rodgers-Lee17} or has studied particles accelerated by shocks in the accretion columns \citep{2019ApJ...883..121O}. In Chapter \ref{C:PublicationI}, we examined the effects of an additional ionisation source based on microphysical and observational considerations. We explored the ionisation of particles accelerated by magnetic reconnection events resulting from the interaction between stellar and disc magnetic fields. The efficiency of this ionisation source lies in the particles being accelerated just above the disc, ensuring the particle flux entering the disc is not diminished by energetic losses before reaching the disc, unlike particles accelerated in accretion columns. This efficiency is further enhanced as these particles encounter minimal perpendicular magnetic diffusion since they can propagate ballistically along nearly vertical magnetic fields to penetrate deeper regions of the disc. In our previous paper, we estimated the ionisation rates produced by protons, electrons, and secondary electrons accelerated by these flares, determining the ionisation rate dependence on column densities ranging from \(10^{19} \rm cm^{-2}\) to \(10^{25} \rm cm^{-2}\). The results of the work presented in Chapter \ref{C:PublicationI} assume a stationary state of the particle emission leading to a likely overestimation of the ionisation rate. 

The paper presented in this chapter was submitted, underwent minor revisions, and is currently in the process of being reviewed.  It addresses the stationary issues by incorporating time-dependent considerations. We first estimate the waiting time distribution between two flares based on the frequency of occurrence of Solar flares. Then, using soft X-ray observations of T Tauri stars, we estimate the temporal luminosity profile of a single flare and the time profile of the ionisation rate generated by a flare. This allows us to deduce the luminosity distribution of the flares from X-ray observations. 

We then introduce the two flare models presented in Chapter \ref{C:Reconnection}, one based on the geometry of solar flares with both footpoints anchored on the stellar surface, which we call Star-Star flares (SSF), and another, more specific to T Tauri stars, based on simulation results with one footpoint on the stellar surface and the other on the inner disc, which we call Star-Disc flares (SDF). For each of these two models, we propose a relation that links the flare luminosity to the distance from the central star where particles produced by that flare are penetrating into the disc.

\subsection{Results from the time-dependent model}
\paragraph{Impact of a single flare:}
We begin by examining the average ionisation rate generated by a flare to compare it to the ionisation rate from the stellar X-rays. We conclude that both sources produce similar average ionisation rates at low column densities (\(N < 10^{22} \text{cm}^{-2}\)). However, particles generated by the flares produce an ionisation rate between 1 and 3 orders of magnitude higher than the ionisation rate from X-rays at high column densities (\(N > 10^{22} \text{cm}^{-2}\)). We also estimated the pressure of non-thermal particles produced by a flare based on its temperature. We find that this is dominant over other pressure components, such as the disc gas thermal pressure or the magnetic pressure. Thus, we anticipate that non-thermal particles play a role in the MHD dynamics of the inner disc. This issue will need to be addressed with simulations.

\paragraph{Impact of multiple flares:}
Using the distributions of luminosity, waiting time and luminosity profile estimated from observations, we conduct a Monte Carlo analysis to determine the spatial distribution of a temporally averaged ionisation rate. We average over 1,000 years, which roughly corresponds to the characteristic timescale of chemical reactions in discs. This Monte Carlo study allows us to study the effects of multiple flares. To facilitate the reuse of our numerical results, we propose a parametric expression for the spatial distribution of the ionisation rate produced by both SSF and SDF.

The spatial distribution of the ionisation rate produced by the particles is added to the ionisation sources of the {\tt ProDiMo} code. As such, {\tt ProDiMo} recalculates the disc structure with this additional ionisation source. This new disc structure enables us to study the impact of energetic particles produced by flares on the chemistry, viscosity, accretion rate, and heating rate within the disc.

\paragraph{Impact on Chemistry:} The ionisation fraction increases by at least one order of magnitude up to \(N\sim 10^{25} \text{cm}^{-2}\) in the region which particles penetrate. The distribution of cations is also significantly altered, especially the density of H\(^+\) increases in the photodissociation region. Additionally, a layer emerges where HCNH\(^+\) is the most abundant cation at \(Z/R\approx0.12\). Upon recombination, this cation forms hydrogen cyanide, HCN. The {\tt ProDiMo} models, accounting for the additional ionisation rate from flares, calculate a drastic multi-order magnitude increase in the abundance of HCN within the discs. This aspect is certainly of interest for understanding prebiotic chemistry since HCN is a foundational molecule for the synthesis of amino acids.

\paragraph{Impact on Viscosity and Accretion:} We introduce a model to calculate the viscous parameter \( \alpha \) based on steady-state disc models such as {\tt ProDiMo}, as detailed by \citet{2019A&A...632A..44T}. This enables us to compare the viscosity in the disc models with and without the additional ionisation rate due to flares. We observe that the increase in the ionisation fraction in the disc due to flares makes the disc region unstable to the magnetorotational instability (MRI). As a result, the extent of the MRI-stable region, often referred to as the "dead zone", decreases. Based on the distribution of viscosity, we extract an approximate estimation of the ionisation rate. We then compare the accretion rates inferred from models with and without flares. We estimate that the mass accretion rate increases approximately by one order of magnitude in regions where the energetic particles produced by flares penetrate.

\paragraph{Impact on the Heating Rate:} By ionising chemical species, particles release a given amount of heat into the disc. For instance, an average of 4 eV is released following the ionisation of a hydrogen atom, and 18 eV is released after ionising a dihydrogen molecule. We avaluate the heating rate deposited by the particles produced by flares. This heating rate will alter the disc thermal structure. The rate of heating may also play a significant role in the launching processes  of winds and jets.
%especially affecting their collimation and velocity. CS Not sure about the collimation
Numerical simulations have explored the effects of heating on outflows, but the microphysical origin of this heating rate remains elusive. In this study, we determine the heating rate resulting from the ionisation of disc matter by energetic particles produced by flares. To facilitate the use of our numerical results, we provide a parametric expression for the heating rate at the disc surface.

\subsection{Model Limitations}\label{sec:modellimitationpubII}
\paragraph{Particle model:}
Our energetic particle model has some limitations. 

First, there are aspects of the particle propagation model we have not addressed, especially when it comes to propagation at column densities greater than \(10^{25} \text{cm}^{-2}\). Beyond this column density, the dominant loss for protons is through pion production. During pion production, protons lose a significant portion of their energy and alter their angle of incidence. Due to these factors, particle transport can no longer be approached using the CSDA framework. Diffusive processes need to be considered. For a more detailed discussion, see Chapter \ref{C:Propagation}. Although we have not dealt with particle transport in the diffusive regime, we were still able to estimate the ionisation rate at \(N>10^{25} \text{cm}^{-2}\) using the parametric expression suitable for high column densities, as per \citet{Padovani18}.

Second, we need to clarify, in the current state of the model, what is the proportion of particles, accelerated in the flare reconnection zone that penetrate the disc. This aspect heavily relies on the flare geometry and the magnetic field structure within the star-disc interaction zone. For now, we will proceed with a parametric study that scales down the incoming particle flux to estimate the minimum proportion of particles that must enter the disc to significantly impact its chemistry and dynamics. For a quantitative estimate, we should integrate a test particle distribution in a reconnection zone of an MHD simulation of the inner disc. This would then allow us to ascertain the proportions of particles that penetrate the disc, fall onto the star surface, and propagate into the outflows.

Besides, we are aware that the particle acceleration model we use, as described in Sect. \ref{sec:parametricinjectionmodel}, is somewhat over simplified. However, as we have demonstrated throughout Chapter \ref{C:Reconnection}, the models and simulations of particle acceleration by magnetic reconnection remain an active research field. The magnetic structure of the star-disc interaction region does not only influence the location and proportion of particles entering the disc, but also, as detailed in B23, the particle acceleration process itself. As shown in B23, the presence of a guide field, the magnetic field component that is not involved in the reconnection process, decreases the particle acceleration efficiency. We did not consider a guide field here, placing us in an optimal configuration for particle acceleration. In a model with a more realistic magnetic structure, featuring a toroidal component that could act as a guide field, the acceleration process would be less effective, thus resulting in lower ionisation rates.

To summarise, the particle model used here provides an upper limit to their impact on discs. A model that describes large-scale particle acceleration by magnetic reconnection in the specific magnetic environment of T Tauri stars is currently not available. Nevertheless, such model would be crucial for a comprehensive understanding of the effects of particles produced by magnetic reconnection on discs. For a more realistic picture, a first step could be to add particles in an MHD code in the test-particle limit.

\paragraph{Disc model:}
A crucial point for interpreting our results is that all our conclusions about the impact of particles are specific to the chemical structure computed by {\tt ProDiMo}. For a more complete picture, it would be interesting to examine the impact of the particles by varying the disc structure in {\tt ProDiMo}. For instance, we anticipate that the particle impact would differ by modifying the radial surface density profile. Specifically, assuming a lower surface density would lead to higher ionisation rates at the same height above the equatorial plane. Conversely, a more massive disc with higher surface densities would have lower ionisation rates at the same height. In short, our results depend on the radial and vertical structure of the disc mass, and this dependence should be evaluated. 

To be more specific, we would like to caution about the increase in the density of HCNH$^+$ and HCN due to the additional ionisation rate from particles. The {\tt ProDiMo} code, although it considers hundreds of chemical species and thousands of reactions, does not take into account molecules with more than four atoms. Therefore, the rise in the abundance of HCNH$^+$ might not be physical but rather a dead-end in the reactions calculated by {\tt ProDiMo}. The recombination of HCNH$^+$ would then lead to an increase in HCN which might not be realistic. To assess the realism of increasing the abundance of HCN, we would need to conduct a more in-depth analysis of the chemical reactions leading to this increase and subsequently assess the influence of particles from flares in a chemical model that considers more complex molecules with additional atoms. Nevertheless, we can highlight that our conclusion on the complexity increase in molecules remains valid in all scenarios. If the rise in HCN abundance is due to a chemical reaction dead-end,
the chemistry of more complex species will surely be boosted if included in the model.

Another limitation concerns the calculation of the viscous parameter \(\alpha\) inside the disc. As discussed in Sect. \ref{sec:ModelingProtoplanetaryDiscs}, although \(\alpha\) was initially treated as a parameter to study accretion processes in discs, to this parameter can be given a physical origin from the stress tensor of the Navier-Stokes equations. The physical origin of this parameter is thus inherently MHD. In the paper presented in this chapter, we compute the value of this parameter based on the properties of a static disc. Hence, the derived \(\alpha\) value is inevitably approximate. The intent of this computation is more a proof of concept, illustrating that the size of a disc dead zone is reduced when considering ionisation by flares. From the \(\alpha\) calculation, we also have estimated the mass accretion rate, taking flare ionisation into account. This estimation too should be viewed as a proof of concept, given that it relies on the thin disc model of \citet{Shakura73}, applicable far from the disc inner edge, a criterion not necessarily met in our study.

In this paper, we have demonstrate that particles produced by flares appear to have multiple significant impacts on both, the chemistry and the dynamics of accretion-ejection processes in T Tauri stars. We also highlighted several limitations inherent to our model. Therefore, it is essential to compare our model with actual observations. In our second article, we will propose a way to refine our model presented in B23 using synthetic mid-infrared spectra from JWST. In this spectral range, CO and OH lines are apparent. We find that accounting for the additional ionisation rate due to flare particles enhances the intensity of these lines. Thus, our model could be tested by comparing real spectra observed by the JWST with the synthetic spectra generated by {\tt PRODIMO}. However, before making this comparison, we need both, real data and further refinement of the synthetic spectrum generation model of {\tt ProDiMo}.

\section{Publication}
Brunn, V., Rab, C., Marcowith, A., Sauty, C., Padovani, M., \& Meskini, C. (2024). Impacts of energetic particles from T Tauri flares on inner protoplanetary discs. \textit{Monthly Notices of the Royal Astronomical Society}, 530(4), 3669-3687. 
\includepdf[pages=-]{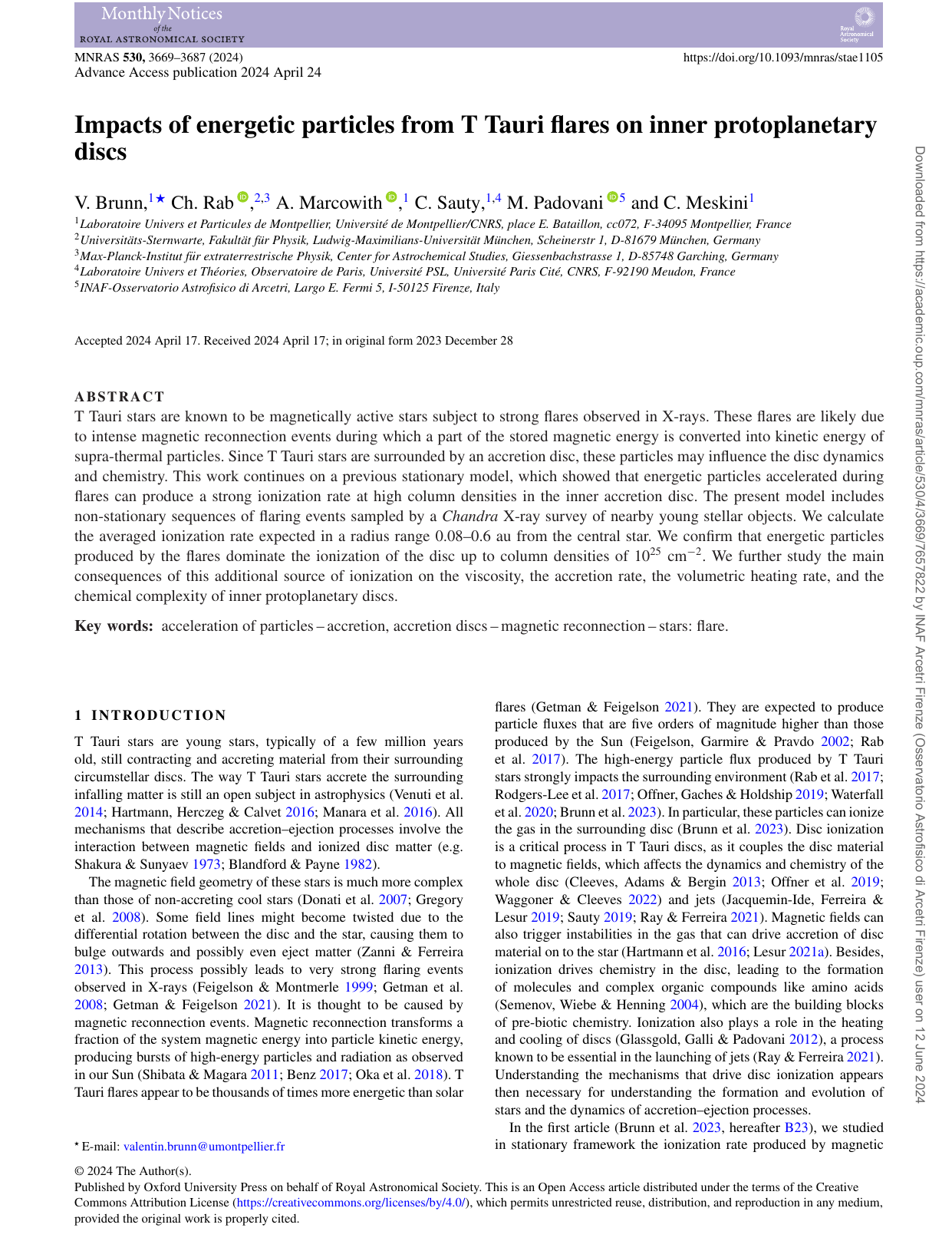} 

\chapter{Conclusion and Perspectives}\label{C:Conclusion}

\section{Conclusion}
The complex characteristics of protoplanetary discs around T Tauri stars necessitate an interdisciplinary strategy to improve our comprehension of these objects. This thesis has contributed to construct a framework that combines observational methodologies, protoplanetary disc chemical and dynamical models, and the mechanics of energetic particle acceleration and transport. This synergy is aimed at showing their collective impact on the dynamics and chemistry of both the disc and possibly to its associated jets.

Chapter \ref{C:TTAURI} establishes the foundational parameters to model T Tauri stars and protoplanetary discs. Observational constraints like mass distribution and thermal structure were also brought into focus to constrain the models.

Chapter \ref{C:ionisation} explores the critical nature of ionisation in disc dynamics. The ionisation state of the disc plays a key role in mass accretion rates, ejection processes, and chemical compositions. Ionisation appears to result from various energy inputs, from stellar radiation to Galactic Cosmic Rays (GCRs). However, we have shown that these sources of ionisation are not the only ones and that potential non-thermal ionisation mechanisms from reconnection events may also be essential.

Chapter \ref{C:Reconnection} extends this notion further by examining magnetic reconnection events in T Tauri flares as an alternative ionisation source. In contrast to well-established solar models, T Tauri systems were found to demand a unique set of theoretical considerations due to their higher luminosities and unique magnetic environments. These aspects demand the development of a specific model for particle acceleration within these flares.

Chapter \ref{C:Propagation} explores the behaviour of energetic particles in the accretion disc, focusing on their trajectories and energy losses. Two transport models, free-streaming and diffusive, are introduced and applied based on particle column density within the disc. The chapter also provides diffusion coefficient estimates for the inner disc. It concludes by highlighting the role of secondary particles in disc ionisation, setting the groundwork for the publications.

Chapter \ref{C:PublicationI} consolidates these findings by presenting the first publication. We use the {\tt ProDiMO} code to quantify the ionisation rate in discs due to magnetic reconnection events. Notably, our results indicate that these particles could be a dominant source of ionisation in the inner regions of the disc, dominating contributions from other commonly used sources. However, we acknowledge that the stationary and local nature of the model presented in this first publication may not capture the complete dynamics, particularly temporal factors like recombination rates and flare occurrence rate and duration.

Chapter \ref{C:PublicationII} tackles, in a second publication, these temporal factors relying on X-ray observations of T Tauri flares. By using spatial and temporal averages over flare distributions, we successfully capture the long-term influence of energetic particles generated by magnetic reconnection events on the inner disc. Our results demonstrate that when these temporal factors are taken into account, the increase of the ionisation rate leads to an accretion rate increase in the inner disc by one order of magnitude. Moreover, we propose a physical mechanism for the heating rate required to initiate stellar outflows. Surprisingly, we also observe an increased abundance of some of the chemical species in the inner disc, which opens up new perspectives for the study of prebiotic chemistry. \\

In conclusion, this thesis suggests a possible shift in our understanding of ionisation sources in inner protoplanetary discs around young T Tauri stars. Magnetic reconnection events not only serve as efficient particle accelerators but their resultant energetic particles could be an essential source of ionisation in the inner part of the accretion disc. These results open significant implications for disc dynamics, chemistry, and ultimately, the processes that lead to planetary formation. 

\section{Perspectives}
The results presented in this thesis and its publications provide an insightful glimpse into the potential effects of energetic particles produced by magnetic reconnection on T Tauri discs. It is clear that the ways for future research on this topic are abundant and promising.

\paragraph{Diffusive transport and Secondary particles:}
To begin with, after establishing diffusion coefficients based on physical arguments, our next step is to develop a diffusive propagation model. This model will enable us to describe particle propagation at column densities greater than \(10^{25} \, \text{cm}^{-2}\). In doing so, we will also need to account for the processes responsible for the generation of secondary particles. With this refined model, we will be more accurately able to estimate the impact of particles generated by magnetic reconnection in the deeper regions of the disc, close to the equatorial plane. This is particularly interesting since, in these regions, aside from short-lived radionuclides (SLRs), energetic particles are the sole source of ionisation. Introducing a new source of ionisation could have strong implications for the chemistry and dynamics of these embedded regions. In this thesis, we have laid out the theoretical foundations required to tackle this issue. To practically develop this model, we will draw upon the work of \citet{Padovani18}. We have continuously referred to this research in building our model, but this study still contains elements we intend to incorporate in a future particle propagation model. Specifically, we should take into account the effects of secondary photons produced either by Bremsstrahlung or by the decay of $\pi^0$, as well as the influence of electron-positron pairs generated by these secondary photons or the decay of $\pi^\pm$. Recent works calculate the UV photon flux emitted by the de-excitation of H$_2$, which was initially excited by cosmic rays (or, more accurately, by secondary electrons originating from primary cosmic rays). This phenomenon is referred to as the Prasad-Tarafdar effect \citep{1983ApJ...267..603P}. Consequently, this UV photon flux can now be evaluated based on various factors: the model of Galactic cosmic ray flux, the H$_2$ column density, the composition of H$_2$ isomers, and the characteristics of dust, changing the maximum energy of injected particle energies, especially we anticipate that flares could produce protons with maximum energies exceeding 100 TeV (see sect. \ref{sect:Maximalenergy}). One immediate application related to the work presented in this thesis involves computing this UV photon flux using the local cosmic ray fluxes from magnetic reconnection events. We could compare it with the stellar UV. We expect that as cosmic rays travel deeper in the disc than stellar UV, the secondary UV photons from Prasad-Tarafdar effect could propagate locally, in deeper regions of the disc. As we know that UV photons are very efficient ionisation sources this could significantly affect the ionisation in these regions.

\paragraph{Turbulent reconnection:}
We have shown in Chapter \ref{C:PublicationII} that particles produced by reconnection events have strong impacts on the chemistry and the accretion ejection processes of the inner disc. To extend the scope of our work, we are developing a particle acceleration model due to magnetic reconnection that could occur further out (up to 1-10 au), at the surface of the discs. We have estimated in Chapter \ref{C:PublicationII} the values of the viscous Shakura-Sunyaev parameter $\alpha$. This estimation, despite its limitations, see Sect. \ref{sec:modellimitationpubII}, indicates a high level of turbulence in the disc surface layers. Turbulence in these areas could give rise to smaller-scale magnetic reconnection events than those explored in our present study. 
\citet{2004ApJ...603..180L} first examine stochastic reconnection in a partially ionised, magnetised environment. But more recently, \citet{2023ApJ...942...21X} focused on particle acceleration during fast 3D turbulent reconnection. Particles bounce back and forth between the reconnection-driven inflows, a result of the mirror reflection and convergence of magnetic fields. Through consecutive direct collisions, the kinetic energy of the inflows is transformed into accelerated particles. Turbulence moderates the inflow speed and presents varied inflow angles relative to the local turbulent magnetic fields. Given that both the energy boost and the likelihood of particle escape depend on the inflow speed, the particle energy distribution index appears to be variable, falling between roughly 2.5 and 4 according to the authors, with the sharpest spectrum anticipated under strong guide field. Power-law index in this range have been shown in \citet{2023MNRAS.519.5673B} (Chap. \ref{C:PublicationI}) to produce high ionisation rate in disc. In order to study the impact of these distributions of particles, these turbulent reconnection models should be adapted to the specific turbulence and magnetic conditions of T Tauri discs. This will constrain the proportion of particles accelerated. 

While these events might be too weak in luminosity or a too low in temperature to be detected with current X-ray sensitive instruments, the particles they produce could have a substantial impact on the chemical and dynamical properties of the discs, farther than the ones studied so far. 
It is worth noting that, sources of ionisation even though dominant in the inner disc, such as thermal ionisation, are completely suppressed farther in the disc ($R\gtrsim 0.2$ au) due to low temperatures, while X rays are geometrically reduced. Therefore, extending the study to examine the impact of energetic particles produced by turbulence-induced reconnection events appears naturally of significant interest for the ionisation of the mid disc.

\paragraph{Observational constrains}
After the development of these models, it is crucial to compare our theoretical framework with observations. We will once more turn to {\tt ProDiMO} to generate synthetic spectra, and compare them with observed lines. In our second publication (see chapter \ref{C:PublicationII}), we explore the impacts of various heating mechanisms and particle from flares on the disc spectra. Given that particles from flares affect both the chemical and the thermal structure of the disc in regions that JWST/MIRI is capable of tracing, it is reasonable to expect that these flares would impact JWST spectra.

\begin{figure}[h!]
    \centering
    \includegraphics{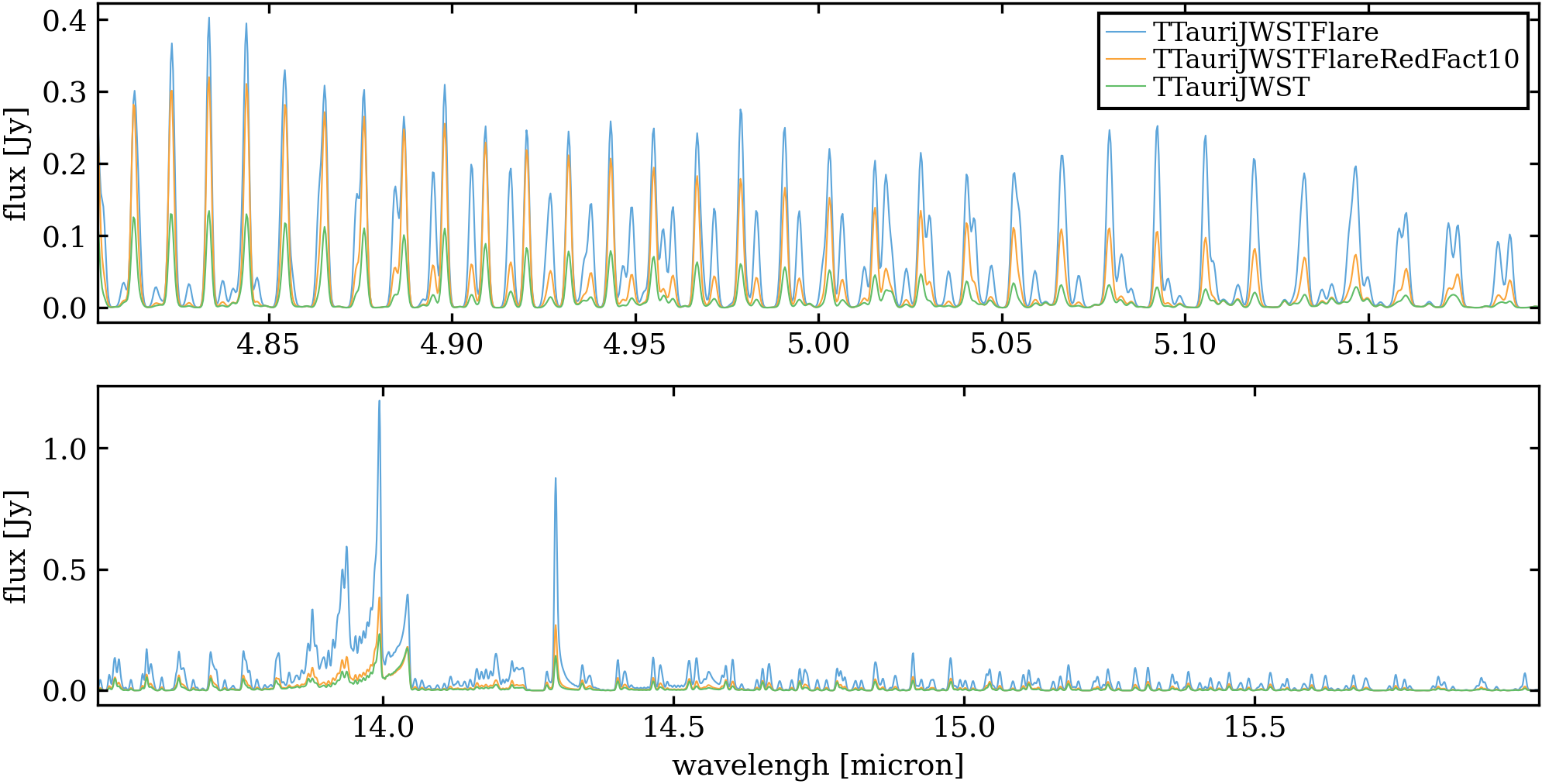}
    \caption{Synthetic mid-infrared spectra are produced by three different models. The solid green line, TTauri JWST, represents the ProDiMo model without flare. The solid blue, (resp. orange) line, TTauri JWST Red Fact 10 (resp. TTauri JWST Flare), represents the spectral lines accounting for the additional ionisation rate due to the SSF model (resp. reduced by a factor of ten). On the top panel, we show the CO $5 \mu$m emission lines, while on the bottom panel, the two distinct blocks represent HCN lines.}
    \label{fig:JWSTSpectraComparisonBasicModels}
\end{figure}

In Fig. \ref{fig:JWSTSpectraComparisonBasicModels}, we show the spectral lines accounting for the additional ionisation rate due to the SSF model in blue, the spectral lines accounting for the additional ionisation rate due to the SSF model reduced by a factor of ten in orange and the spectral lines produced by the reference model, without flares in green. The shape of spectral lines are notably impacted, particularly those around 5 microns corresponding to CO, are noticeably stronger. This is most likely due to the additional heating of the gas, leading to the increase of the size of the emitting region of CO, leading to increased line fluxes. This impact is significant enough to be observable with the JWST. 
We plan to refine our flare model using the ProDiMo framework. However, this may take some time as multiple aspects related to synthetic JWST spectra production are currently under development within ProDiMo.

In {\tt ProDiMO} present state, we cannot estimate with confidence the flux of the synthetic spectrum lines. Once this overcame, the task will be to pinpoint which lines are most influenced by our models. Our preliminary results already suggest that both CO and OH are impacted, though this assertion needs further validation. Other species, such as HCN, are notably affected by the presence of energetic particles, and could serve as effective markers. 

For the following step, we will of course need real spectra observed by JWST in inner discs. With these spectra, we can calibrate our model parameters to fit observations. One of the crucial parameters that can be constrained in this way is the proportion of energetic particles that penetrate the disc. Constraining this parameter is important both in the flare model presented in this thesis and in the turbulent reconnection model under development, since the ionisation rate and its subsequent incidences correlate with this quantity. In the work presented here, we have postulated that all particles penetrate the disc, providing an upper limit to their impact. Observational constraints will help to moderate this influence. So it will be crucial to take advantage of the new generation instruments such as the JWST or forthcoming instruments like the Extremely Large Telescope (ELT) to resolve the regions impacted by these particles.

\paragraph{Effects of particles from flares on disc, winds and jets dynamics}
Above the disc, magnetised jets and winds may experience heating through non-ideal MHD processes like Alfv\'{e}n wave damping heating, Ohmic heating and ambipolar diffusion heating, especially at altitudes where photo-heating is less effective, as showed by various studies.
Building on the results detailed in this thesis, focusing on discs, it would be interesting to explore the influence of particles produced by flares, which move in the opposite direction to the disc, within winds and jets. With the theoretical groundwork we laid out, on particle propagation in an MHD environment and the loss processes they undergo (Chap. \ref{C:Propagation}), we have the expertise to study particle propagation in outflows. Upcoming work by Meskini et al.  already indicates, via MHD simulations, that an additional heat source in outflows can significantly impact their collimation and ejection speed. A detailed estimation of the heating rate distribution is outflow appears interesting for the understanding of the dynamics of winds and jets. About the pressure of non-thermal particles, one intriguing phenomenon we would like to investigate is the triggering of Alfv\'{e}n waves in the magnetic fields of the collimated jet. The damping of these waves might present an additional heat source further up in the jet. Furthermore, there remains the enigma of the origin of synchrotron radiation in protostellar jets. We could model the synchrotron emission in jets generated by particles produced by flares to determine if they align with observations.

As highlighted in Chapter \ref{C:PublicationII}, particles produced by flares are a strong heat source and also exert pressure in the same order of magnitude as the magnetic pressure. It could be of great interest to restart from the set up proposed in \citet{orlando2011mass} but then include energetic particles more explicitly. We can use various approaches either using bi-fluid models where energetic particles are treated as a fluid (eg see \citep{2019A&A...631A.121D, 2020LRCA....6....1M} for details about this model) or use a combined kinetic and MHD approach like the so-called particle-in-cell-Magnetohydrodynamics (PIC-MHD) model (see \citet{2018MNRAS.473.3394V, 2015ApJ...809...55B}. The latter approach has the benefit to be able to handle energy-dependent particle populations and the back reaction of these particles over the MHD solutions through a modified implementation of the Ohm's law.

\paragraph{Enrichment of radioactive isotopes}

Among various short lived isotopes, aluminium-26 ($^{26}$Al), with a half-life of 0.717 Myr, stands out for its importance. It has been suggested that within planetesimals, $^{26}$Al can radiate sufficient heat through its radioactive decay to give rise to bodies characterised by metal cores and silicate mantles \citep{2011AREPS..39..351D}. The origin of the $^{26}$Al abundance measurement can be traced back to the calcium-aluminium-rich inclusions (CAI) in meteorite by \citet{1976GeoRL...3..109L}. Ever since its identification in meteorites, considerable efforts have been invested to pinpoint a convincing enrichment mechanism. The ideal mechanism aims to produce the observed $^{26}$Al/$^{27}$Al ratio and also to explain why this ratio is uniform across the solar system, as noted by \citet{2011AREPS..39..351D}.

It is worth noting that all external injection methods come with the necessity that the enriched gas mixes efficiently in short times, scenario that appears improbable \citep{2018ApJ...855...81S}. In contrast, a local enrichment source might address the blending issue by directly introducing $^{26}$Al into the disc right before the condensation of dust particles. As a potential mechanism, energetic particles generated during flares in T-Tauri stars have already been suggested by \citet{1998ApJ...506..898L}. However, the propagation of these particles within the disc was not treated at that time. We could follow the method of \citet{2020ApJ...898...79G} that studied the enrichment via proton irradiation of the surface of protostellar discs by CRs accelerated in the accretion shocks. Indeed, the production rate of $^{26}$Al, is $\zeta \sim \int j_p(E) \sigma_i(E) dE$. Where $j_p(E)$ is the proton flux and $\sigma_i$ is the interaction cross section for process, $i$. For example the dominant process $i$ is $\rm{CR}$$+^{27}$Al$ \rightarrow (^{26}$Al$+$ $n) + \rm{CR}$, and the cross section of this reaction is known from laboratory experiments. The quantity hard to estimate is the proton flux $j_p(E)$. However, in Chapter \ref{C:PublicationI} we estimated this quantity, so the only step left to estimate the production rate of $^{26}$Al is to account for the relevant $^{26}$Al production processes, these are given in \citet{2020ApJ...898...79G}. To compute the $^{26}$Al enrichment, we would need in addition a model that estimates the distribution of $^{27}$Al and the right isotopic evolution equation, also provided in \citet{2020ApJ...898...79G}. \\

In summary, until now, the focus of our work has been solely on the influence of energetic particles on the ionisation rates, which subsequently affect the disc dynamics and chemistry. However, energetic particles can also interact with stable isotopes, creating radioactive ones. The increased presence of these radioactive species acts as a heat source within the disc and planetesimals. And previous research has shown that planet formation is significantly influenced by this heat source. Although the formation of planets was not the subject of this thesis, our model could perfectly be applied to this broader subject.

\paragraph{Compact objects:}
The structural similarities between T Tauri stars and Active Galactic Nuclei (AGN), X-ray binaries (XRB) or microquasars present another intriguing avenue for future research. With their central objects emitting extreme X-ray and gamma-ray radiation, highly magnetised accretion discs, winds and jet structures, AGNs/ XRBs resemble T Tauri systems, though on a far larger scale in terms of mass, luminosity, and magnetic intensity. \citet{2010A&A...518A...5D} propose that the relativistic ejections observed in microquasars could be produced by violent magnetic reconnection episodes at the inner disc coronal region. They discussed the role of magnetic reconnection and associated heating and particle acceleration in different jet/disc accretion systems, namely young stellar objects (YSOs), microquasars, and active galactic nuclei (AGNs). In microquasars and AGNs, \citet{2010A&A...518A...5D} suggest that violent reconnection episodes between the magnetic field lines of the inner disc region and those anchored in the black hole can heat the coronal/disc gas and accelerate the plasma to relativistic velocities. This acceleration is due to a diffusive first-order Fermi-like process within the reconnection site, producing intermittent relativistic ejections. \citet{2010A&A...518A...5D} find that the resulting power-law electron distribution aligns with the synchrotron radio spectrum observed during the outbursts of these sources. They indicate that the magnetic reconnection power is sufficient to account for the observed radio luminosities of outbursts from microquasars to low luminous AGNs. Moreover, magnetic reconnection events lead to the heating of the coronal gas, potentially conducting heat back to the disc, thereby amplifying its thermal soft X-ray emission observed during microquasar outbursts. 

After the work done by \citet{2010A&A...518A...5D}, models of particle acceleration by magnetic reconnection in relativistic plasmas have made significant progress. On the basis of these advances and relying on our particle propagation model, it would be interesting to study the effect of these particles on the accretion/ejection processes of compact objects. The upscale parameters of these objects lead us to speculate that violent magnetic reconnection events could have strong impact on the disc and jets of these objects, thereby offering a new class of objects to which our model could be applied.\\

In summary, the journey toward understanding the role of energetic particles in shaping T Tauri stars environments is far from over, and the future offers a plethora of exciting opportunities for further discovery and improvement.


%\bibliographystyle{a} % Use unsrt style for numerical order
\setlength{\bibsep}{0pt plus 0.2ex} % Adjust spacing between references
\renewcommand{\bibfont}{\footnotesize}
%\printbibliography[title={References}]
%\bibliographystyle{abbrvnat}
\bibliography{biblio}
%\bibliography{biblio1}
%\bibliography{biblio2}
%\bibliography{biblio3}
%\bibliography{biblio4}
%\bibliography{biblio5}
%\bibliography{biblio6} % if your bibtex file is called example.bib

% End of mnras_template.tex

\appendix

\chapter{Technical Appendix: Internship Report Mathieu Venet}\label{app:Internshipreport}
\includepdf[page=-]{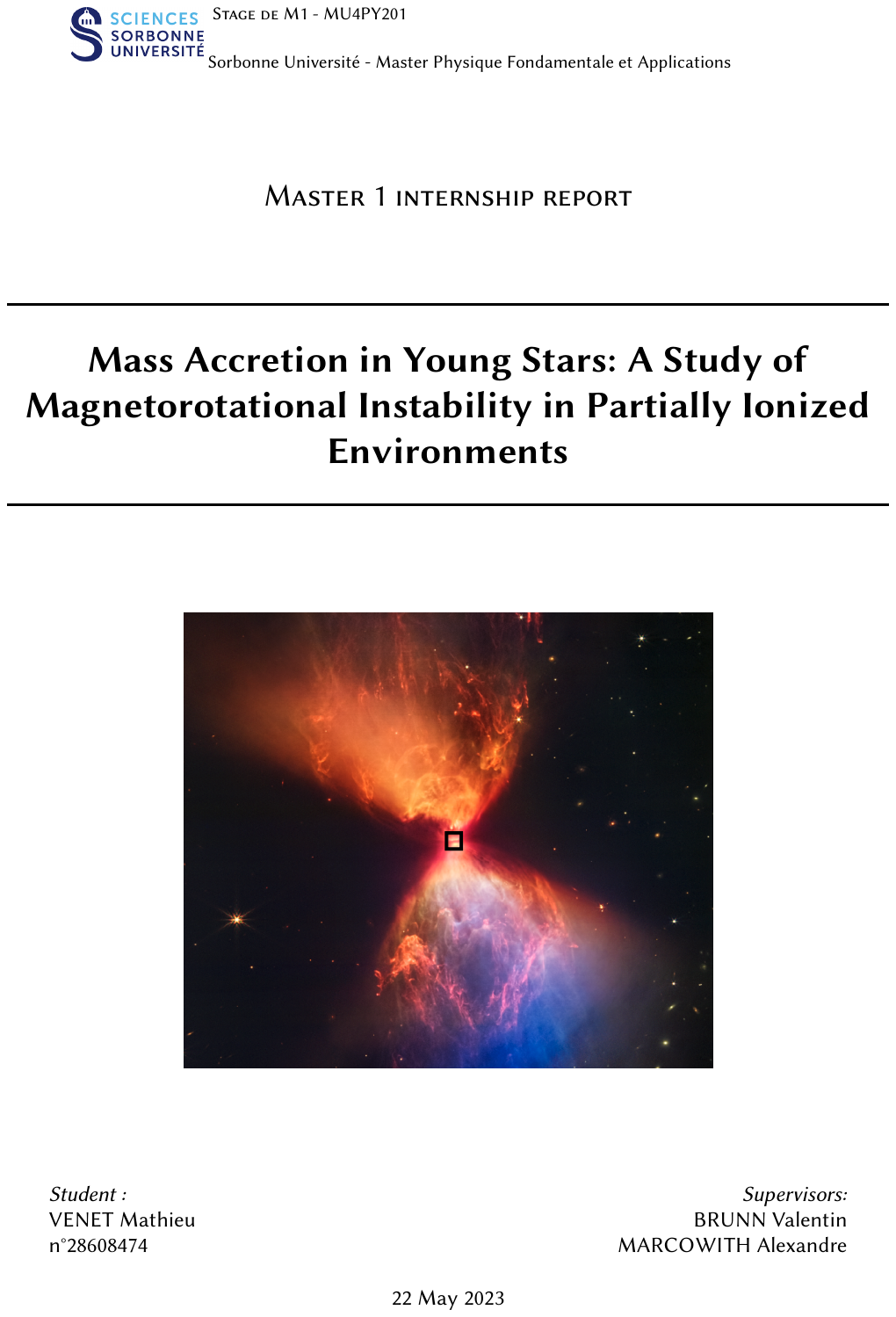} 
\chapter{Résumé en français - Summary in French}
Les étoiles T Tauri, considérées comme les homologues jeunes des étoiles de type solaire, sont sujettes à des questions persistantes en astrophysique. Ces corps célestes jeunes sont des objets stellaires entourés de disques protoplanétaires de poussière et de gaz. Les étoiles T Tauri et leurs disques environnants détiennent les indices des processus qui régissent à la fois l'évolution stellaire et la formation des planètes dans les systèmes stellaires de faible masse. Comme ces étoiles sont extrêmement actives, c'est l'interaction entre l'accrétion, l'éjection, le magnétisme et les interactions chimiques qui est au centre de notre étude. Ce processus d'accrétion-éjection n'est pas un simple écoulement continu de matière, mais il est très variable, souvent modulé par des champs magnétiques et une cinématique complexe. L'accrétion est indissociablement liée aux phénomènes d'éjection, tels que les jets et les vents, qui semblent être ses conséquences naturelles, voire même une condition préalable à l'accrétion. Ces processus d'éjection servent non seulement de mécanisme de libération de pression, mais emportent également le moment angulaire, contribuant ainsi d’avantage à l'accrétion.

Les étoiles T Tauri servent également de fenêtre unique sur les premières étapes du développement des systèmes planétaires. Ces jeunes objets stellaires offrent l'occasion d'étudier les processus complexes qui conduisent à la formation des planètes. Les mécanismes d'accrétion et d'éjection dans ces systèmes influencent non seulement la croissance de l'étoile centrale, mais affectent également de manière significative la répartition de la matière dans le disque environnant. Cela prépare le terrain pour la formation des corps planétaires. Les processus d'accrétion dans les systèmes T Tauri sont fondamentaux pour réguler la disponibilité de la matière qui pourrait éventuellement coalescer en planètes. Comprendre ces processus offre un aperçu de la quantité de masse qui est alimentée dans les régions internes du disque protoplanétaire, ce qui est crucial pour la formation et la migration des noyaux planétaires potentiels. Ces mécanismes ne sont pas simplement de nature gravitationnelle, mais sont souvent contrôlés par des interactions magnétiques complexes qui canalisent la matière vers l'étoile. Les phénomènes d'éjection sont tout aussi importants pour la formation des planètes. Ces écoulements, en éliminant le moment angulaire du disque interne, facilitent davantage l'accrétion, mais peuvent également redistribuer la matière et créer des conditions pour l'espacement et la stabilité des planètes. La formation des planètes, même si elle suscite un grand intérêt dans la communauté des spécialistes des étoiles T Tauri, ne sera cependant pas traitée ici. Nous discuterons cependant à la toute fin de cette thèse de certaines applications prospectives de nos modèles dans le domaine de la formation des planètes.

En comparaison avec les nuages moléculaires parentaux, le gaz du disque protoplanétaire est beaucoup plus dense, ce qui écrante le rayonnement ionisant et entraîne un degré d'ionisation plus faible, c'est-à-dire une abondance relative d'électrons par rapport aux molécules neutres. Cependant, même avec ce degré d'ionisation réduit, on s'attend toujours à ce qu'il soit suffisamment élevé pour permettre un couplage partiel avec les champs magnétiques dans les disques. Ce couplage pourrait déclencher des instabilités MHD, contribuer au lancement de vents dans le disque et permettre le transfert nécessaire du moment angulaire pour l'accrétion de masse. Le degré d'ionisation est le paramètre clé qui détermine l'état d'ionisation d'un disque, influençant à la fois la dynamique physique et chimique des disques protoplanétaires, ce qui à son tour façonne la formation des systèmes planétaires. Les principaux porteurs de charge positive varient selon les différentes régions du disque. Par conséquent, nous pouvons utiliser les ions atomiques et moléculaires comme traceurs observationnels de la structure du disque protoplanétaire. Une meilleure compréhension de ces structures nécessite des modèles numériques détaillés de la chimie du disque. Les modèles astrochimiques, nécessaires pour interpréter les abondances moléculaires observées, et les codes MHD non idéaux, qui simulent la dynamique du disque, utilisent tous deux les taux d'ionisation comme paramètre central. Ce taux mesure le nombre d'ionisations par unité de temps et est déterminé par l'interaction entre les sources d'ionisation (UV, rayons X, CR) et les molécules à l'intérieur du disque. Comprendre la distribution spatiale des taux d'ionisation dans les disques T Tauri est crucial, car cela a un impact direct sur la chimie, l'équilibre thermique et la dynamique du disque. Les taux d'ionisation varient considérablement en fonction de la distance radiale par rapport à l'étoile centrale, ainsi que de la hauteur verticale par rapport au plan médian du disque, en raison des contributions variables des différentes sources d'ionisation. Comme nous le verrons, l'ionisation du disque externe est principalement dominée par des sources externes, telles que les rayons cosmiques galactiques et le rayonnement interstellaire, tandis que dans les régions internes du disque (à quelques unités astronomiques), les taux d'ionisation sont principalement déterminés par les sources internes, telles que le rayonnement continu X ou UV stellaire.
Les étoiles T Tauri sont connues pour présenter une variabilité significative en photométrie et en spectroscopie. Elles sont particulièrement réputées pour émettre d'intenses éruptions de rayons X. Une éruption de rayons X fait référence à une augmentation soudaine et forte des émissions de rayons X d'une étoile. Chez les étoiles T Tauri, ces éruptions peuvent être plus de trois ordres de grandeur plus lumineuses que celles observées chez les étoiles de la séquence principale. Le mécanisme à l'origine de ces éruptions est supposé être analogue aux éruptions solaires, mais beaucoup plus intense. À la fois dans le Soleil et chez les étoiles T Tauri, on pense que des événements de reconnexion magnétique déclenchent ces éruptions. Pendant les événements de reconnexion magnétique, une partie de l'énergie magnétique de la couronne stellaire est libérée sous forme d'énergie cinétique de particules supra-thermiques. Certaines de ces particules chauffent ensuite le plasma de la chromosphère sous-jacente à des dizaines de millions de degrés, émettant les rayons X observés. L'autre partie des particules est censée s'échapper de la chromosphère et interagir avec l'environnement circumstellaire environnant.

La question qui se pose est d'estimer l'impact des particules produites par les événements de reconnexion magnétique sur les disques des jeunes étoiles.\\

Pour répondre à cette question, nous discutons d'abord des modèles et des contraintes observationnelles actuelles que nous avons sur les systèmes T Tauri. Dans le chapitre \ref{C:TTAURI}, bien que nous abordions les contraintes concernant l'étoile centrale, nous nous concentrons principalement sur les modèles et les contraintes du disque environnant. Nous verrons dans ce chapitre que l'ionisation est une propriété cruciale guidant la dynamique et la chimie du disque. Le chapitre \ref{C:ionisation} se penche sur les sources et les effets de l'ionisation sur la dynamique et la chimie du disque, dans ce chapitre, nous proposons {\it une source d'ionisation du disque jusqu'ici non prise en compte, à savoir les éruptions provoquées par les événements de reconnexion magnétique}. Le chapitre \ref{C:Reconnection} explore l'accélération des particules par la reconnexion magnétique dans les systèmes T Tauri dans le but d'étudier les effets que ces particules peuvent avoir sur le disque. Le chapitre \ref{C:Propagation} présente les processus de transport et d'interaction des particules au sein des disques protostellaires/planétaires. Le chapitre \ref{C:PublicationI} offre une étude paramétrique stationnaire des taux d'ionisation produits par les éruptions dans la région interne du disque. Le chapitre \ref{C:PublicationII} étend ce modèle stationnaire en tenant compte des effets temporels, nous permettant d'estimer les taux d'ionisation moyens produits par les éruptions et, en même temps, en utilisant le code thermo-chimique {\tt ProDiMO}, de mieux comprendre l'effet de ces taux sur la chimie et la dynamique du disque. Les résultats du chapitre \ref{C:PublicationI} ont été publiés en 2023, et ceux du chapitre \ref{C:PublicationII} le seront prochainement. Nous concluons avec le chapitre \ref{C:Conclusion} en discutant des perspectives futures offertes par cette étude sur l'impact des événements de reconnexion magnétique sur les disques et les jets des jeunes étoiles.

\section*{Étoiles T Tauri}
Une étoile T Tauri est un type d'étoile jeune en pré-séquence principale qui est en train de se contracter sous l'effet de la gravité avant d'atteindre la phase de séquence principale de son évolution. Ces étoiles se caractérisent par la présence d'un disque protoplanétaire environnant. Les étoiles T Tauri présentent une forte activité magnétique et une structure magnétique complexe qui canalise la matière du disque vers l'étoile. Comprendre ces processus offre un aperçu de la quantité de masse alimentée dans les régions internes du disque protoplanétaire, ce qui est crucial pour la formation et la migration des noyaux planétaires potentiels. Les phénomènes d'éjection sont tout aussi importants pour la formation des planètes, car ils éliminent le moment angulaire du disque interne, facilitent davantage l'accrétion et peuvent également redistribuer la matière du disque. Ainsi, les processus d'accrétion et d'éjection chez les étoiles T Tauri sont directement liés aux perspectives et aux conditions de la formation des planètes. Ces mécanismes jouent un rôle crucial dans la formation de l'environnement physique et chimique du disque protoplanétaire. Cet environnement, riche en molécules complexes, est le berceau même où se forment de nouvelles planètes et les éléments constitutifs de la vie.

L'objectif de ce chapitre est de poser une base solide pour la modélisation des étoiles T Tauri et de leurs disques protoplanétaires en se basant sur des contraintes observationnelles.

Dans ce chapitre, notre exploration est double. Tout d'abord, dans la section \ref{sec:ConstrainingCentralStar}, nous nous concentrons sur la caractérisation des propriétés physiques de l'étoile T Tauri centrale elle-même. Nous cherchons à établir le contexte de notre compréhension actuelle de ces objets. L'aperçu historique introductif se transformera en un examen détaillé des caractéristiques physiques des étoiles T Tauri, notamment leur masse, leur rayon, leur luminosité, leurs caractéristiques spectrales et leurs champs magnétiques. Comprendre la nature de ces étoiles est crucial pour toute discussion ultérieure sur leur environnement environnant. Deuxièmement, dans la section \ref{sec:ProtoplanetaryDiscs}, nous plongeons dans le monde complexe des disques protoplanétaires. Nous discutons de la manière dont les observations astrophysiques nous ont permis de contraindre leur structure chimique et thermique. De telles contraintes sont fondamentales pour façonner nos théories et nos modèles. De plus, nous explorons les modèles computationnels et analytiques existants qui simulent la structure et l'évolution de ces disques.
Dans ce chapitre, nous préparons le terrain pour notre exploration des disques protoplanétaires dans les systèmes T Tauri en identifiant les composants essentiels et les techniques d'observation qui permettent notre compréhension de ces objets. Nous avons commencé par nous concentrer sur l'étoile centrale et son disque protoplanétaire environnant. Nous avons souligné l'importance des paramètres observables clés de l'étoile centrale, tels que la luminosité et la température effective, qui servent d'entrées critiques pour estimer la masse stellaire et l'âge grâce aux modèles d'évolution des jeunes étoiles. Comprendre ces propriétés stellaires est crucial pour notre enquête ultérieure sur les disques protoplanétaires car elles contrôlent le potentiel gravitationnel, le champ de rayonnement et les processus d'accrétion à l'intérieur du disque.

Nous avons ensuite déplacé notre attention de manière plus spécifique vers les objets d'intérêt de notre thèse, les disques protoplanétaires. Plus précisément, nous avons analysé les contraintes observationnelles qui guident notre compréhension de ces structures complexes, notamment la distribution de masse de gaz et de poussière, ainsi que la structure thermodynamique radiale et verticale du disque. Nous avons fourni un aperçu de la structure hydrostatique du disque du code {\tt ProDiMO}, qui servira de cadre de référence pour nos analyses ultérieures.

Reconnaissant les limites des modèles purement hydrostatiques, nous avons également souligné la nécessité d'incorporer les théories de la magnétohydrodynamique (MHD) dans notre analyse. Nous avons souligné que les mécanismes de la MHD sont actuellement les explications les plus plausibles pour les processus d'accrétion-éjection dans les étoiles T Tauri. Cependant, nous avons également souligné que l'efficacité de ces théories de la MHD est étroitement liée au niveau d'ionisation à l'intérieur du disque, ce qui ouvre la voie à l'étude de l'ionisation du disque dans le prochain chapitre.

En somme, ce chapitre sert de pierre angulaire fondamentale pour notre thèse, mettant en évidence à la fois la complexité et la nature interdisciplinaire du sujet. Il offre un survol des connaissances actuelles et des techniques d'observation, préparant le terrain pour notre examen approfondi ultérieur des sources d'ionisation et de leur rôle dans les disques protoplanétaires entourant les étoiles T Tauri.

\section*{Ionisation dans les disques}
Malgré le degré d'ionisation réduit attendu dans les disque protoplanétaire, on s'attend toujours à ce qu'il soit suffisamment élevé pour permettre un couplage partiel avec les champs magnétiques dans les disques. Ce couplage pourrait déclencher des instabilités de la MHD, contribuer aux vents du disque et permettre le transfert nécessaire du moment angulaire pour l'accrétion de masse \citep{2013ApJ...767...30B,suzuki2014magnetohydrodynamic}. Le taux d'ionisation, noté $\zeta$ en s$^{-1}$, représente la production d'électrons par atomes de H$_2$ par seconde, et le degré d'ionisation, noté $x_E=n_e/n_{n}$, est l'abondance relative d'électrons par rapport aux molécules neutres. Ce sont les paramètres clés qui déterminent l'état d'ionisation d'un disque. Ils influencent à la fois la dynamique physique et chimique des disques protoplanétaires, ce qui façonnent à leurs tour la formation de systèmes planétaires \citep{1988PThPS..96..151U}.
Le degré d'ionisation varie spatialement à l'intérieur du disque, influençant sa structure physique \citep{2017A&A...600A..75B}. Cette variabilité spatiale a également un impact sur les réactions ion-molécule, car les taux de réaction ion-molécule sont directement liés aux niveaux d'ions. De plus, l'ionisation catalyse des interactions chimiques complexes dans les disques T Tauri, conduisant à une multitude d'espèces moléculaires qui ont un impact sur la composition chimique globale du disque. La connaissance des mécanismes d'ionisation est également essentielle pour prédire la synthèse et la distribution de molécules organiques, potentiellement cruciales pour la chimie prébiotique dans la formation des systèmes planétaires. De plus, le chauffage induit par l'ionisation module l'équilibre thermique dans les disques T Tauri, affectant leur structure verticale et influençant le comportement à la fois de la poussière et du gaz. Cette chaleur pourrait également être l'une des sources d'énergie pour le lancement de vents et de jets.
Les études sur l'ionisation des disques visent généralement à deux choses : déterminer le degré d'ionisation (voir la section \ref{sec:ionisationConstrains}) et identifier les principales sources d'ionisation (voir la section \ref{Sec:ionisationSources}). Ce degré d'ionisation est influencé par un équilibre entre les taux d'ionisation et de recombinaison, qui dépendent tous deux de facteurs tels que la composition chimique du disque et la propagation du rayonnement ionisant \citep{2004A&A...417...93S,woitke2009radiation,2019ApJ...872..107X}. En raison de la stratification chimique de ces disques, les principaux porteurs de charge positive diffèrent entre différentes régions du disque (comme le montre la couche colorée du côté droit de la figure \ref{fig:ionisationTracersDistribution}). Par conséquent, l'utilisation d'ions atomiques et moléculaires comme indicateurs des niveaux d'ionisation nécessite des modèles numériques détaillés et des observations de la chimie du disque. Sur le front de la modélisation, les simulations utilisant à la fois le transfert radiatif et des modèles chimiques comme {\tt ProDiMO}, ou des simulations MHD radiatives comme celles proposées par \citet{flock20173d}, génèrent des observables nécessaires pour interpréter les données observationnelles, permettant d'estimer l'état d'ionisation des disques T Tauri. Ce chapitre présente les bases nécessaires à notre compréhension des processus d'ionisation dans les disques protoplanétaires entourant les étoiles T Tauri. Nous avons initié la discussion en élucidant les techniques d'observation utilisées pour contraindre l'ionisation dans ces disques. En examinant divers traceurs, pouvant être des ions moléculaires et atomiques ou des espèces neutres, nous avons établi que le choix du meilleur traceur dépend de la région spécifique du disque examinée. Pour quantifier l'étendue complète de l'ionisation, nous avons démontré la nécessité d'utiliser des modèles chimiques de disque, car ils permettent d'estimer l'abondance d'espèces non observables. Notre étude a ensuite porté sur un examen exhaustif des sources d'ionisation qui ont été modélisées à ce jour. Dans le disque interne, l'ionisation est principalement gouvernée par le rayonnement UV et X de l'étoile, qui ionise non seulement directement mais chauffe également le disque, maintenant ainsi un haut niveau d'ionisation thermique. En revanche, le disque externe a souvent été considéré comme étant principalement ionisé par les rayons cosmiques galactiques, bien que cette hypothèse nécessite une étude plus approfondie en raison de l'exclusion potentielle des rayons cosmiques galactiques par les vents de la Tauriosphère. Nous avons également introduit des études récentes explorant l'impact des particules énergétiques produites localement dans le système T Tauri, préparant ainsi le terrain pour notre recherche pionnière, telle que présentée dans les chapitres \ref{C:PublicationI} et \ref{C:PublicationII}, qui se concentrera sur les effets de l'ionisation induite par les éruptions dans le disque interne.
Nous avons démontré le rôle essentiel de l'ionisation dans la détermination de la dynamique et de la chimie du disque ainsi que des jets associés. Il est important de souligner que l'ionisation est un facteur clé dans le déclenchement des processus d'accrétion, et nous avons fourni une formule pour calculer les taux d'accrétion de masse en fonction des fractions d'ionisation. Cela est essentiel pour comprendre des mécanismes tels que l'instabilité magnéto-rotationnelle, qui ne peut se produire que dans des régions suffisamment ionisées. De plus, nous avons discuté des effets potentiels de chauffage des sources d'ionisation non thermiques, une caractéristique qui pourrait être importante pour expliquer les mécanismes à l'origine des vitesses observées des jets et des vents. Ces sources non thermiques ont également des répercussions sur le profil chimique du disque, influençant les traceurs observationnels et déclenchant potentiellement la formation de molécules complexes telles que les acides aminés, qui présentent un intérêt du point de vue de la chimie prébiotique.

En résumé, contraindre la distribution spatiale de l'ionisation est crucial pour comprendre ses innombrables effets sur la dynamique du disque, sa chimie et les processus associés.

À mesure que nous nous tournons vers le prochain chapitre, nous introduirons une source d'ionisation alternative grâce à un modèle d'accélération de particules généré par la reconnexion magnétique lors des éruptions solaires. Ce modèle à venir cible la région interne du disque, où l'activité des éruptions solaires est élevée et peut être contrainte par l'observation.

\section*{Reconnexion magnétique et accélération des particules}
Dans ce chapitre, nous étudions en détail le sujet critique de l'accélération des particules, un processus omniprésent dans le plasma spatial. Divers mécanismes, tels que l'accélération stochastique dans des champs électriques turbulents ou l'accélération par choc diffusif, facilitent ce gain d'énergie dans des conditions et des paramètres différents.

Un mécanisme essentiel d'accélération des particules est la reconnexion magnétique, un processus omniprésent dans les plasmas au cours duquel les lignes de champ magnétique subissent une reconfiguration brutale. Cela entraîne une conversion locale de l'énergie magnétique en énergie cinétique et thermique. Se produisant dans des environnements allant de la magnétosphère terrestre aux éruptions solaires, en passant par les disques d'accrétion et les jets, la reconnexion magnétique joue un rôle crucial dans la dynamique du plasma, entraînant des phénomènes tels que les éjections de masse coronale et l'accélération des jets de plasma.

La pertinence astrophysique de ces processus est particulièrement frappante chez les étoiles T Tauri, qui présentent une activité accrue dans le spectre des rayons X. On conjecture que ces émissions de rayons X proviennent d'éruptions de type solaire amplifiées en raison de l'activité magnétique intense des T Tauri. Notamment, ces éruptions stellaires peuvent avoir des boucles magnétiques qui s'étendent au-delà du rayon de troncation du disque, les ancrant dans le disque circumstellaire T Tauri.

La question qui se pose est donc de savoir comment modéliser l'accélération des particules par la reconnexion magnétique afin d'étudier leur impact sur les propriétés chimiques et dynamiques du disque interne entourant les étoiles T Tauri.

Nous commençons par présenter le modèle théorique de reconnexion existant dans le cadre de la MHD (magnétohydrodynamique) dans la section \ref{sect:ClassicalMHDTheories}. Ensuite, nous explorons divers mécanismes d'accélération des particules particulièrement efficaces dans les régions de reconnexion turbulente dans la section \ref{sec:kineticapproach}. Nous examinons ensuite comment les données d'observation des éruptions solaires peuvent être utilisées pour contraindre ces mécanismes dans la section \ref{sec:solarflares}. Par la suite, nous étendons ces conclusions pour construire un modèle d'éruptions solaires T Tauri dans la section \ref{sec:ttauriflares}, pour finalement modéliser la distribution des particules supra-thermiques basée sur les observations de rayons X de ces éruptions solaires T Tauri dans la section \ref{sec:ParticleEmission}.

Ce chapitre visait à établir un cadre pour comprendre la production de particules à partir d'événements de reconnexion magnétique dans les éruptions solaires T Tauri. Nous avons initié la discussion par un aperçu des modèles théoriques et historiques clés, à savoir les modèles Sweet-Parker et Petscheck, et les modèles de reconnexion turbulente. Ces études ont façonné notre compréhension actuelle de la reconnexion magnétique. Nous avons souligné que, malgré sa reconnaissance généralisée dans les contextes astrophysiques, le développement de la théorie de la reconnexion magnétique et de son rôle dans l'accélération des particules est encore en cours.

Pour offrir une compréhension plus concrète de l'accélération des particules dans les éruptions solaires T Tauri, nous avons établi des parallèles avec le domaine plus largement étudié des éruptions solaires. En adoptant le modèle standard bien établi des éruptions solaires, nous nous sommes concentrés sur la traduction des paramètres observables des rayons X en contraintes sur le flux et la distribution en énergie des particules non thermiques produites pendant les éruptions. Plus précisément, nous avons mis en évidence comment des facteurs tels que la luminosité des rayons X contraignent des paramètres clés tels que l'indice de loi de puissance, l'énergie d'injection et l'énergie maximale atteignable par ces particules. Le modèle standard des éruptions solaires nous a également permis d'obtenir des informations sur les aspects géométriques de ces éruptions.

En recentrant la discussion sur des scénarios spécifiques aux étoiles T Tauri, nous avons introduit notre propre modèle d'accélération des particules, adapté aux propriétés des éruptions stellaires T Tauri, dont les énergies sont des ordres de grandeur supérieur à celles de leurs homologues solaires. Cependant, nous avons noté que l'adaptabilité du modèle solaire aux étoiles T Tauri présentait des limites, notamment en ce qui concerne la géométrie des éruptions. Compte tenu de l'environnement magnétique façonné par la présence de disques protoplanétaires autour des étoiles T Tauri, nous avons proposé une géométrie d'éruption révisée, inspirée des simulations numériques et des données d'observation. Comprendre cette géométrie est essentiel, car elle détermine les points d'entrée du flux de particules dans le disque, influençant ainsi la distribution spatiale de l'ionisation.

Enfin, ce chapitre sert de base pour notre modèle d'accélération des particules dans les éruptions solaires T Tauri. Ce modèle prédit non seulement le spectre d'énergie des particules émises, mais anticipe également leurs trajectoires le long des lignes de champ magnétique dans le disque et les vents. De telles informations sont cruciales pour comprendre les différents régimes de transport et les pertes d'énergie que ces particules rencontreront, des sujets qui seront au centre de notre prochain chapitre.

\section*{Interaction et propagation des particules énergétiques dans les disques T Tauri}

Dans les disques protoplanétaires, les champs magnétiques jouent un rôle indispensable en orchestrant la propagation des particules énergétiques. Ces champs magnétiques peuvent avoir diverses origines : ils peuvent émaner de l'étoile centrale, être des "champs initiaux" laissés par des stades antérieurs des étoiles en formation, ou même provenir des mécanismes dynamo propres au disque.

Les lignes de champ magnétique fonctionnent essentiellement comme des "voies" guidant ces particules énergétiques. La configuration de ces champs magnétiques peut varier d'une structure simple et ordonnée à une structure complexe et chaotique, chaque configuration donnant lieu à des comportements de propagation uniques pour les particules. Dans des environnements magnétiques simples comme les champs dipolaires, les particules suivent des trajectoires prévisibles le long des lignes de champ. Cependant, lorsque les champs magnétiques sont complexes ou désordonnés, les particules adoptent un mouvement diffusif.

À mesure que ces particules se déplacent à travers le disque, elles subissent diverses interactions avec le milieu environnant. Ces interactions entraînent des pertes d'énergie en raison de processus tels que l'ionisation, ainsi que la production de particules secondaires. 
L'objectif est de fournir un cadre complet pour comprendre la propagation des particules au sein des disques T Tauri, en particulier lorsqu'elles sont accélérées par des événements de reconnexion magnétique. Cette compréhension permet non seulement d'avancer dans notre connaissance de la propagation des particules dans les disques protoplanétaires, mais donne également aux lecteurs les éléments nécessaires pour appréhender les publications des deux chapitres suivants.

Par conséquent, la question centrale qui guide ce chapitre est la suivante : comment pouvons-nous modéliser la propagation des particules générées dans les régions internes des disques T Tauri à travers les événements de reconnexion magnétique ?

Dans ce chapitre, nous avons examiné l'interaction complexe entre les particules énergétiques et le milieu du disque, en mettant particulièrement l'accent sur leurs trajectoires et leur distribution d'énergie. Deux facteurs principaux façonnent ces caractéristiques, la fonction de perte d'énergie et le modèle de transport sélectionné.

Nous avons entamé ce chapitre en détaillant les fonctions de perte associées aux électrons et aux protons de sous-GeV, en mettant l'accent sur la façon dont elles sont influencées par le type de particule, l'énergie et les propriétés chimiques du milieu. Cela a permis d'introduire deux modèles de transport des particules, les modèles de libre diffusion et de diffusion. Dans le cadre de l'approximation du \textit{free-streaming}, nous avons présenté un modèle analytique simplifié pour estimer le taux d'ionisation à travers différentes densités de colonnes dans le disque. Bien que ce modèle soit rudimentaire, il offre des informations précieuses sur les effets d'ionisation des particules énergétiques produites par un événement de reconnexion sur la matière dû disque.

Nous avons ensuite défini les critères de sélection du régime de transport approprié, \textit{free-streaming} ou diffusion, en fonction de la densité de colonnes traversée par les particules dans le disque. Nous avons établi que pour des densités de colonnes inférieures à \(10^{25} \text{cm}^{-2}\), le modèle de libre diffusion s'applique, tandis que le régime de diffusion est nécessaire au-delà de ce seuil.

Abordant une question de longue date dans le domaine, nous avons proposé des estimations des coefficients de diffusion à l'intérieur du disque interne. Ces estimations ont été formulées en fonction de différents modèles de turbulence prévalents, tels que la turbulence de résonance MHD et la turbulence induite par l’instabilité magnétorotationnelle.

Nous avons également exploré les processus de miroir magnétique et de focalisation. Ces processus sont souvent cités pour leur rôle dans la modulation du flux de rayons cosmiques galactiques dans les disques. Cependant, notre analyse a conclu que ces processus ne sont probablement pas significatifs dans le contexte du disque interne, justifiant ainsi notre décision de ne pas les prendre en compte dans les chapitres suivants.

Enfin, notre examen s'est étendu au-delà des particules primaires pour englober l'impact des particules secondaires, générées par des interactions avec le disque. Nous avons expliqué comment ces particules secondaires, en particulier les électrons secondaires, contribuent à l'ionisation dans le disque.

En résumé, ce chapitre s'appuie sur les bases posées au chapitre \ref{C:Reconnection}, où nous avons développé un modèle d'accélération des particules par reconnexion magnétique. En détaillant comment ces particules accélérées interagissent avec le disque, en mettant l'accent sur leur pouvoir d'ionisation, nous avons préparé le terrain pour la première publication de cette thèse présentée dans le chapitre suivant.

\section*{Modèle stationnaire - Ionisation des disques internes des étoiles Tauri : effets des particules énergétiques produites localement par de forts événelents de reconnexions magnétiques \citep{2023MNRAS.519.5673B}}
Dans l’article \citet{2023MNRAS.519.5673B} noté B23, nous avons proposé une source alternative d'ionisation. L'article a examiné les taux d'ionisation résultant de la production de particules énergétiques dans le système étoile-disque lors des éruptions de reconnexion magnétique. Ces particules sont des particules supra-thermiques de basse énergie ($E\lesssim1$ GeV) capables d'ioniser les disques, ce qui affecte leur évolution chimique, thermique et dynamique. L'objectif de ce travail était d'étudier comment les particules énergétiques se propagent dans les disques de T Tauri et de calculer le taux d'ionisation qu'elles produisent. Nous avons recueilli des données expérimentales et théoriques sur les sections efficaces de production de H$^+$, H$_2^+$ et He$^+$ par des électrons et des protons.

Dans B23, nous avons fixé l'emplacement de pénétration des particules à R=0.1 ua de l'étoile et estimé la forme de la distribution des spectres injectés en fonction des propriétés physiques attendues des éruptions dans le disque interne. Au chapitre \ref{C:Reconnection}, nous avons décrit comment nous pouvions estimer le régime de reconnexion magnétique se produisant dans les éruptions de T Tauri à partir de l'intensité du champ magnétique, de la densité de particules et du nombre de Lundquist. Le régime déduit est la reconnexion magnétique collisionnelle multi-X-line. Nous nous sommes ensuite appuyés sur des simulations de l'accélération des particules lors de la reconnexion dans le régime collisionnel multi-X-line pour déduire la distribution d'énergie des particules. La simulation réalisée par \citet{arnold2021electron} suggère que les électrons suivent une distribution d'énergie de loi de puissance d'indice $\delta$, variant de 3 à 8 en fonction de la configuration magnétique dans la région de reconnexion. Nous avons donc mené une étude paramétrique sur $\delta$ dans cette plage de valeurs.

À partir des spectres théoriques d'injection, nous avons calculé les spectres locaux à différentes densités de colonnes dans le disque. Ces spectres locaux ou propagés sont les spectres des particules atténuées par les pertes énergétiques dans le disque. En tenant compte de tous les processus de perte d'énergie présentés dans la section \ref{sec:EnergyLossesDisc}, nous avons calculé les spectres propagés pour des densités de colonnes allant de $10^{19}$ cm$^{-2}$ à $10^{25}$ cm$^{-2}$ afin de traiter la propagation des particules dans le régime CSDA.

Les propriétés physiques des éruptions solaires, telles que leur taille, la densité électronique et la température, ont été déduites à partir des observations du catalogue COUP (Chandra Orion Ultradeep Project). En utilisant le code thermo-chimique de rayonnement {\tt ProDiMO}, nous avons calculé la structure du disque éclairé par le champ de rayons X produit par une éruption solaire. Cette structure fournit l'abondance de différentes espèces chimiques dans le disque nécessaires pour calculer la densité de colonne traversée par les particules. La densité de colonne est ensuite utilisée pour estimer les pertes d'énergie et, par conséquent, le flux propagé de particules. Enfin, le taux d'ionisation est calculé à partir du flux propagé, en tenant compte de l'ionisation des particules secondaires.

L'étude a pris en compte diverses températures d'éruption ainsi que divers indices spectraux pour le flux de particules énergétiques et différentes configurations magnétiques le long desquelles les particules se propagent.

Dans le document B23, il est démontré que, dans une configuration stationnaire, les particules énergétiques sont une source puissante d'ionisation locale, avec des taux d'ionisation qui dépassent de plusieurs ordres de grandeur les contributions des rayons X, des particules énergétiques stellaires et de la radioactivité dans le disque interne. Pour notre cas de référence, une éruption à 1MK à 0,1 ua se propageant le long d'une ligne de champ magnétique verticale, nous avons trouvé des taux d'ionisation $\zeta=10^{-9} ~\rm  s^{-1}$ à des densités de colonne de $10^{25} ~\rm cm^{-2}$, tandis que le taux d'ionisation à cette profondeur est de l'ordre de $10^{-17} ~\rm  s^{-1}$ en raison des rayons X produits lors d'une éruption stellaire à $1~\rm MK$ et des GCR, et de $10^{-18} ~\rm  s^{-1}$ en raison des radionucléides.

Bien que nous soyons conscients que nos hypothèses peuvent entraîner une surestimation du taux d'ionisation, nous montrons que ce processus peut être dominant parmi les processus d'ionisation dans le disque interne des étoiles T Tauri. Comme il existe plusieurs paramètres dans notre modèle difficiles à contraindre, nous avons mené une analyse comparative de ces paramètres. L'objectif de cette analyse est de définir une plage de paramètres d'éruption pour que les taux d'ionisation produits soient dominants par rapport aux autres sources d'ionisation. Nous anticipons que cela sera le cas pour :
\begin{itemize}
    \item un processus de reconnexion qui accélère les particules selon un flux d'injection avec une loi de puissance $j \propto E^{-\delta}$ pour $\delta < 6$,
    \item des éruptions avec des températures supérieures à 1 MK,
    \item des particules se propageant le long de la ligne de champ avec un rapport de la composante toroïdale à la composante poloidale $b_{\rm g}=B_{\phi}/B_{\rm pol} < 1$.
\end{itemize}

\section*{Modèle dépendant du temps - Impact des particules énergétiques produites par les flares de T Tauri sur leur disque interne}

Le taux d'ionisation dans les disques d'accrétion des étoiles T Tauri joue un rôle crucial dans le développement des instabilités MHD essentielles pour expliquer l'accrétion de la matière environnante sur l'étoile centrale. De plus, il influence la chimie du disque, qui doit être parfaitement comprise pour interpréter les observations et comprendre la formation des éléments constitutifs de la vie. Nous avons vu dans le chapitre \ref{C:ionisation} que jusqu'à présent, les sources d'ionisation prises en compte étaient soit des sources radiatives, telles que les rayons X ou les UV provenant de l'environnement stellaire proche ou émis par l'étoile centrale elle-même, soit des particules énergétiques d'origine galactique ou stellaire. Bien que l'on ne s'attende pas à ce que les particules énergétiques galactiques affectent de manière significative le disque interne en raison de leur exclusion par la T-Tauriosphère, celles émises par l'étoile centrale se sont avérées efficaces pour ioniser le disque interne. Les recherches antérieures sur l'impact des particules stellaires se sont généralement concentrées sur une émission continue de particules accélérées à la surface stellaire sans tenir compte de la microphysique des processus d'accélération \citep{Rab17, Rodgers-Lee17} ou ont étudié les particules accélérées par des chocs dans les colonnes d'accrétion \citep{2019ApJ...883..121O}. Dans le chapitre \ref{C:PublicationI}, nous avons examiné les effets d'une source d'ionisation supplémentaire basée sur des considérations microphysiques et observationnelles. Nous avons exploré l'ionisation de particules accélérées par des événements de reconnexion magnétique résultant de l'interaction entre les champs magnétiques stellaires et du disque. L'efficacité de cette source d'ionisation réside dans le fait que les particules sont accélérées juste au-dessus du disque, ce qui garantit que le flux de particules entrant dans le disque n'est pas diminué par des pertes énergétiques avant d'atteindre le disque, contrairement aux particules accélérées dans les colonnes d'accrétion. Cette efficacité est encore renforcée car ces particules rencontrent une diffusion magnétique perpendiculaire minimale, car elles peuvent se propager de manière balistique le long de champs magnétiques presque verticaux pour pénétrer plus profondément dans le disque. Dans notre précédent article, nous avons estimé les taux d'ionisation produits par les protons, les électrons et les électrons secondaires accélérés par ces éruptions, déterminant la dépendance du taux d'ionisation en fonction des densités de colonne allant de \(10^{19} \rm cm^{-2}\) à \(10^{25} \rm cm^{-2}\). Les résultats du travail présenté dans le chapitre \ref{C:PublicationI} supposent un état stationnaire de l'émission de particules, ce qui conduit probablement à une surestimation du taux d'ionisation.

Dans l'article présenté dans ce chapitre qui sera bientôt publié, nous abordons ce problème en incorporant des composantes dépendantes du temps dans notre modèle. Nous estimons d'abord la distribution des temps d'attente entre deux éruptions en fonction de la fréquence des éruptions solaires. Ensuite, en utilisant les observations des rayons X mous des étoiles T Tauri, nous estimons le profil de luminosité temporelle d'une seule éruption ainsi que le profil temporel du taux d'ionisation généré par une éruption. Cela nous permet de déduire la distribution de luminosité des éruptions à partir des observations des rayons X.

Nous introduisons ensuite les deux modèles d'éruptions présentés dans le chapitre \ref{C:Reconnection}, l'un basé sur la géométrie des éruptions solaires avec les deux points d'ancrage sur la surface stellaire, que nous appelons "éruptions coronales", et l'autre plus spécifique aux étoiles T Tauri, basé sur les résultats de simulations avec un point d'ancrage sur la surface stellaire et l'autre sur le disque interne, que nous appelons "éruptions étoile-disque". Pour chacun de ces deux modèles, nous proposons une relation qui lie la luminosité de l'éruption à la distance par rapport à l'étoile centrale où les particules produites par cette éruption pénètrent dans le disque.

Nous commençons par examiner le taux d'ionisation moyen généré par une éruption pour le comparer au taux d'ionisation provenant des rayons X stellaires. Nous concluons que les deux sources produisent des taux d'ionisation moyens similaires à de faibles densités de colonne (\(N < 10^{22} \text{cm}^{-2}\)). Cependant, les particules générées par les éruptions produisent un taux d'ionisation entre 1 et 3 ordres de grandeur plus élevé que le taux d'ionisation des rayons X à des densités de colonne élevées (\(N > 10^{22} \text{cm}^{-2}\)). Nous avons également estimé la pression des particules non thermiques produites par une éruption en fonction de sa température. Nous constatons que cette pression est dominante par rapport à d'autres composantes de pression, telles que la pression thermique du gaz du disque ou la pression magnétique. Ainsi, nous prévoyons que les particules non thermiques jouent un rôle dans la dynamique MHD du disque interne. Cette question devra être abordée avec des simulations.

En utilisant les distributions de luminosité, de temps d'attente et de profil de luminosité estimées à partir des observations, nous effectuons une analyse de Monte Carlo pour déterminer la distribution spatiale d'un taux d'ionisation moyenné dans le temps. Nous effectuons une moyenne sur 1 000 ans, ce qui correspond approximativement à l'échelle de temps caractéristique des réactions chimiques dans les disques. Cette étude en Monte Carlo nous permet d'étudier les effets de plusieurs éruptions. Pour faciliter la réutilisation de nos résultats numériques, nous proposons une expression paramétrique pour la distribution spatiale du taux d'ionisation produit à la fois par les éruptions "coronales" et "étoile-disque".

La distribution spatiale du taux d'ionisation produit par les particules est ajoutée aux sources d'ionisation du code {\tt ProDiMo}. Ainsi, {\tt ProDiMo} recalcule la structure du disque avec cette source d'ionisation supplémentaire. Cette nouvelle structure de disque nous permet d'étudier l'impact des particules énergétiques produites par les éruptions sur la chimie, la viscosité, le taux d'accrétion et le taux de chauffage dans le disque.

La fraction d'ionisation augmente d'au moins un ordre de grandeur jusqu'à \(N\sim 10^{25} \text{cm}^{-2}\) dans la région où les particules pénètrent. La distribution des cations est également significativement modifiée, en particulier la densité de H\(^+\) augmente dans la région de photodissociation. De plus, une couche émerge où HCNH\(^+\) est le cation le plus abondant à \(Z/R\approx0.12\). Lors de la recombinaison, ce cation forme le cyanure d'hydrogène, HCN. Les modèles {\tt ProDiMo}, prenant en compte le taux d'ionisation supplémentaire dû aux éruptions, calculent une augmentation drastique de plusieurs ordres de grandeur de l'abondance de HCN dans les disques. Cet aspect est certainement intéressant pour comprendre la chimie prébiotique car HCN est une molécule fondamentale pour la synthèse des acides aminés.

Nous introduisons un modèle pour calculer le paramètre de viscosité \( \alpha \) basé sur des modèles de disque à l'état stationnaire tels que {\tt ProDiMo}, comme détaillé par \citet{2019A&A...632A..44T}. Cela nous permet de comparer la viscosité dans les modèles de disque avec et sans le taux d'ionisation supplémentaire dû aux éruptions. Nous observons que l'augmentation de la fraction d'ionisation dans le disque due aux éruptions rend la région du disque instable à l'instabilité magnétorotationnelle (MRI). En conséquence, l'étendue de la région stable à la MRI, souvent appelée "zone morte", diminue. Sur la base de la distribution de la viscosité, nous extrayons une estimation approximative du taux d'ionisation. Nous comparons ensuite les taux d'accrétion déduits des modèles avec et sans éruptions. Nous estimons que le taux d'accrétion massique augmente d'environ un ordre de grandeur dans les régions où les particules énergétiques produites par les éruptions pénètrent.

En ionisant des espèces chimiques, les particules libèrent une certaine quantité de chaleur dans le disque. Par exemple, en moyenne, 4 eV sont libérés après l'ionisation d'un atome d'hydrogène, et 18 eV sont libérés après l'ionisation d'une molécule de dihydrogène. Nous évaluons le taux de chauffage déposé par les particules produites par les éruptions. Ce taux de chauffage modifiera la structure thermique du disque. Le taux de chauffage peut également jouer un rôle significatif dans les processus de lancement des vents et des jets.
%en particulier en ce qui concerne leur collimation et leur vitesse. CS Pas sûr de la collimation
Des simulations numériques ont exploré les effets du chauffage sur les écoulements, mais l'origine microphysique de ce taux de chauffage reste énigmatique. Dans cette étude, nous déterminons le taux de chauffage résultant de l'ionisation de la matière du disque par des particules énergétiques produites par les éruptions. Pour faciliter l'utilisation de nos résultats numériques, nous proposons une expression paramétrique du taux de chauffage à la surface du disque.

\end{document}